\documentclass[draft,numberedheadings]{aipproc}

\usepackage{graphicx}

\newcommand{\dn}{\downarrow}
\newcommand{\up}{\uparrow}
\newcommand{\half}{\hbox{$\frac{1}{2}$}}
\newcommand{\code}{\null\vskip-2mm\noindent}
\newcommand{\br}{\hfill\break}
\newcommand{\cia}{\null\hskip5mm}
\newcommand{\cib}{\null\hskip10mm}
\newcommand{\cic}{\null\hskip15mm}
\newcommand{\cid}{\null\hskip20mm}
\newcommand{\cie}{\null\hskip25mm}

\layoutstyle{6x9}

\begin{document}

\title{Computational Studies \break of Quantum Spin Systems}

\classification{75.10.Jm, 75.40.Mg, 75.40.Cx, 02.70.Ss}

\keywords{Quantum spin system, antiferromagnet, valence-bond solid, quantum phase transition, finite-size scaling,
quantum Monte Carlo, exact diagonalization, Lanczos method} 

\author{Anders W. Sandvik}
{address={Department of Physics, Boston University, \\
590 Commonwealth Avenue, Boston, Massachusetts 02215, USA}}

\begin{abstract}
These lecture notes introduce quantum spin systems and several computational methods for studying their ground-state and finite-temperature 
properties. Symmetry-breaking and critical phenomena are first discussed in the simpler setting of Monte Carlo studies of classical spin systems, 
to illustrate finite-size scaling at continuous and first-order phase transitions. Exact diagonalization and quantum Monte Carlo (stochastic
series expansion) algorithms and their computer implementations are then discussed in detail. Applications of the methods are illustrated by 
results for some of the most essential models in quantum magnetism, such as the $S=1/2$ Heisenberg antiferromagnet in one and two dimensions, 
as well as extended models useful for studying quantum phase transitions between antiferromagnetic and magnetically disordered states.
\vskip5mm

\noindent
{\bf Publication:} {\it Lectures on the Physics of Strongly Correlated 
Systems XIV}, proceedings of the 14th Training Course in Physics of 
Strongly Correlated Systems, edited by A. Avella and F. Mancini.
AIP Conference Proceedings {\bf 1297}, 135-338 (2010).
\vskip2mm

\noindent
{\bf Cite as:}  A. W. Sandvik, AIP Conf. Proc. {\bf 1297},  135 (2010).
\vskip2mm

\end{abstract}

\maketitle
\newpage

\tableofcontents

\newpage

\section{Introduction}
\label{sec_intro}

One of the primary goals of theoretical physics is to provide the simplest models capturing various complex physical
phenomena. In condensed matter physics, as well as in statistical mechanics more broadly, spin systems often 
serve in this role. As the Ising model and other idealized classical spin models have been invaluable in forming our understanding 
of thermal phase transitions and critical phenomena, so are various quantum spin models now instrumental in developing a theoretical 
framework for exotic quantum many-body states and quantum phase transitions, i.e., phase transitions driven by quantum fluctuations 
(controlled by some tunable interaction parameter) at temperature $T=0$ \cite{sachdevbook,sachdev1}. With their many possible ordered and 
disordered ground states and different types of excitations arising from them, quantum spin systems also provide rich opportunities to 
study other manifestations of collective quantum behavior \cite{sachdev2,auerbachbook}.

Although often used as simplified prototypical model systems for various phenomena, and not always intended for describing 
fully all the details of specific real materials, quantum spin systems have also been very successful in explaining quantitatively 
the antiferromagnetic properties of a variety of Mott insulators with localized electronic spins. A prime example of this is provided 
by the undoped parent compounds of the high-temperature superconductors and other related quasi-two-dimensional and quasi-one-dimensional 
copper oxides. Their experimentally measured magnetic response functions can be remarkably well reproduced by the $S=1/2$ Heisenberg 
model on two-dimensional (2D) planes \cite{chn,manousakis}, isolated chains \cite{eggert94,eggert96}, and ``ladders'' \cite{dagotto1} 
consisting of two or more coupled chains. In agreement with model calculations, 2D layered systems exhibit an exponentially divergent correlation 
length as $T$ is lowered (until ordering sets in below some critical temperature due to 3D couplings or anisotropies), while chain and ladder
compounds exhibit only short-range (power-law or exponentially decaying) correlations. In addition to cuprates, many other inorganic and organic 
antiferromagnets also show similarly good agreement between theory and experiments \cite{miller}. 

A prominent research theme in contemporary condensed matter physics is to model and explain {\it magnetically disordered ground states} of 2D or 
quasi-2D materials with non-uniform or frustrated (competing) antiferromagnetic interactions \cite{schollwock1,diep,gardner,balents10}. Quantum phase 
transitions in 2D spin systems challenge the classical Ginzburg-Landau framework \cite{cardy} for understanding and classifying phase transitions 
based on order parameters, as exemplified by the recent theory of ``deconfined'' quantum critical points \cite{sachdev1,senthil1}, which separate 
antiferromagnetic (N\'eel) and non-magnetic valence-bond solid (VBS) ground states \cite{sandvik1}. In a field-theory proposed to describe this quantum 
phase transition (the non-compact CP$^1$ model), deconfined spinons (collective $S=1/2$ degrees of freedom) are the ``elementary particles'', out of 
which the two order parameters can be formed due to condensation (in the N\'eel state) or confinement (in a VBS state) \cite{senthil1}. Apart from 
the interest in such unusual phase transitions in condensed matter physics, there are also intriguing connections to deconfinement in gauge theories 
in particle physics \cite{sachdev3}. Interacting quantum spins have also recently become interesting in the context of ultra-cold atoms in optical 
lattices \cite{bloch,kim10}, as well as in quantum information theory \cite{verstraete1}. Fundamental many-body concepts such as entanglement 
entropy \cite{levin} are currently explored in various ground states of quantum spin systems \cite{hastings1}.

Exact solutions of quantum spin systems are very rare beyond one dimension, where there are several important cases (enough to fill
a whole encyclopedia \cite{lieb}, in fact). In two dimensions there are also some examples \cite{shastry,aklt}, but normally analytical 
calculations rely on approximations or assumptions that cannot be rigorously justified. Purely computational studies of model hamiltonians
are therefore also essential. Unbiased numerical results are important for testing theories and analytical calculations (in particular, continuum 
field theories for the low-energy physics). Moreover, numerical ``simulations'' can also in their own right serve as laboratories for exploration
and discovery, and may thus stimulate further theoretical and experimental developments. 

In classical statistical physics, almost any model can be studied in detail using Monte Carlo or molecular dynamics simulations (although
there are also challenging classical systems, e.g., ones with very slow, ``glassy'' dynamics \cite{fernandez}). The situation is different in quantum mechanics. 
There are still enormous hurdles limiting computational studies of generic quantum spin hamiltonians, especially ones with frustrated 
interactions and, going beyond pure spin models, strongly correlated fermion systems. Devising efficient practically useful numerical algorithms 
for these types of systems is one of the greatest challenges in theoretical physics. Thanks to a series of significant developments over the 
past couple of decades, large-scale computational studies have already become possible for some important classes of quantum lattice models. 
Very large 1D systems can be studied using the density matrix renormalization group (DMRG) method \cite{white1,schollwock2} or 
related methods formulated using matrix-product states \cite{verstraete1,mcculloch}. Quantum Monte Carlo (QMC) methods with loop-cluster updates 
\cite{evertz1,prokofev96,syljuasen02} can be used to study a wide range of spin and boson models in any number of dimensions, typically on lattices with up 
$10^4$ sites or more in the ground state, and much larger still at elevated temperatures. In addition to breakthroughs in efficient algorithms, the 
impressive improvements in computer performance have of course played an important role in recent progress, too. The most dramatic gains have, however, 
been achieved as a result of better algorithms, and there is reason to believe that this will continue to be the case in the future as well.

\paragraph{Topics}

Two classes of computational methods will be discussed in these lecture notes: Numerical (exact) diagonalization and quantum Monte Carlo simulation. Exact 
diagonalization methods will be developed primarily for 1D systems, followed by some discussion of extensions to 2D square-lattice systems. 
The use of symmetries for block-diagonalization will be developed and used in both complete diagonalization ($T>0$ 
calculations) and with the Lanczos method (for obtaining the ground state and low-energy excitations in a given symmetry sector). QMC simulations based on the 
series expansion of the partition function (stochastic series expansion; SSE) will be developed for $T>0$ calculations (and applied also in the limit $T\to 0$). 
Computer programs written in close correspondence with the pseudocodes are available on-line \cite{sandvikwebsite}. 

Beyond describing the technical aspects of the numerical methods, an integral goal of these lecture notes is to introduce the most essential quantum spin 
models (hamiltonians) and to present some of their physical properties from a computational perspective. While the discussion is largely self-contained as far 
as the algorithms and implementations are concerned, the physics of the systems is for the most part discussed in a rather ``light'' fashion, in the form 
of elementary calculations (e.g., spin-wave theory) and qualitative descriptions illustrated by numerical results. Connections to complementary analytical 
approaches (e.g., predictions based on field theories) are also pointed out, with key references for further study. The topics range in maturity from 
well established basics to very recent and ongoing research on exotic quantum phase transitions. 

\paragraph{Outline}

The quantum spin models to be discussed in the subsequent sections are first introduced in Sec.~\ref{sec_heisenberg}, along with a brief summary of various 
types of ground states and quantum phase transitions. Classical phase transitions, Monte Carlo simulations, and finite-size scaling techniques are 
reviewed Sec.~\ref{transitions} in order to set the stage for quantum-mechanical finite-lattice calculations and data analysis. Exact diagonalization 
techniques and their applications to 1D spin systems are discussed in Sec.~\ref{sec_diag}. Basic properties of the Heisenberg chain and its 
extension with frustrated interactions are illustrated with numerical results (including the frustration-driven quantum phase transition into a dimerized 
VBS state). Extensions of the methods to 2D systems are also summarized, and used to study the low-energy states (quantum-rotor states) of small 
antiferromagnetic systems. Sec.~\ref{sec_sse} begins with a general discussion of path integrals, followed by the alternative series-expansion formulation 
of quantum statistical mechanics, on which the SSE QMC method is based. The SSE method is then developed in detail for the $S=1/2$ Heisenberg model. Illustrative 
results for chains, ladders, and 2D planes are presented, including a study of quantum-criticality in dimerized 2D systems. Applications of the SSE method to 
``J-Q'' models with four- and six-spin interactions are also discussed, and the N\'eel--VBS transitions occurring in these systems as a function of the
strength of the multi-spin iteractions are studied. Sec.~\ref{sec_survey} concludes with a brief survey of other recent works related to the topics of the 
lecture notes. 

\section{Quantum spin models, their ground states and quantum phase transitions}
\label{sec_heisenberg}

In solid-state physics, quantum spin hamiltonians describe the effective magnetic interactions between localized electronic spins. As such, they can be 
derived starting from the full problem of interacting electrons \cite{anderson,mattis,delannoy}. Here we will not discuss their relationships with real materials 
in detail, but just introduce some well established models with Heisenberg couplings and other, similar interactions. To motivate and prepare for the quantitative 
numerical calculations in the later sections, in this introductory part we will first survey some of the possible types of ground states and quantum phase 
transitions. We will discuss the nature of the quantum fluctuations in spin systems from different perspectives, including spin wave theory and singlet
pairing (valence bonds). 

The Heisenberg exchange is the most important spin-spin interaction and forms the starting point for understanding many materials and phenomena 
in quantum magnetism. Two spins are coupled according to the hamiltonian 
\begin{equation}
H_{ij}=J_{ij}{\bf S}_i \cdot {\bf S}_{j} = J_{ij}(S^x_iS^x_{j} + S^y_iS^y_{j} + S^z_iS^z_{j}).
\label{hamij1}
\end{equation}
Often this pair interaction is summed over only nearest-neighbor sites $(i,j)$, but longer-range interactions can also be included. The type of 
ground state, the nature of the excitations, and the finite-temperature properties of a system with Heisenberg interactions depend strongly on 
the underlying lattice. The dimensionality plays a crucial role. According to the Mermin-Wagner-Hohenberg theorem \cite{mermin,hohenberg}, a continuous 
symmetry of a quantum system with short-range interactions, here the global SU($2$) spin rotation symmetry, can be broken neither at $T \ge 0$ in 
one dimension nor at $T>0$ in two dimensions. This normally rules out magnetic order in 1D Heisenberg models (unless there are long-range interactions, 
in which case the theorem does not apply---we will discuss an example of this in Sec.~\ref{sec_longrange}), but in two dimensions the ground state 
can be magnetic, i.e., $\langle {\bf S}_i\rangle \not=0$ (with all these vector expectation values parallel in a ferromagnet and alternating between 
two opposite directions in a bipartite antiferromagnet). Beyond dimensionality, the microscopic details of the lattice and the coupling strengths $J_{ij}$ (e.g., 
uniform, modulated in some periodic pattern, or in some random, disordered way) are also decisive, and many different types of ground states and quantum
phase transitions can be realized. Some of these states and transitions are still not very well understood and subjects of ongoing research. 

Apart from a brief review of spin wave theory for general $S$, we will in these lecture notes focus on the simplest
case of $S=1/2$ spins, corresponding to individual uncompensated electronic spins. This is often the most interesting case, as $S \to \infty$ is the 
classical limit, and $S=1/2$, thus, maximizes the effects of quantum fluctuations (although in some cases, $S=1$ or higher can actually lead to even 
larger quantum effects, e.g., in the case of the ``Haldane state'' of the $S=1$ chain \cite{aklt}). We will only consider antiferromagnetic interactions, 
$J_{ij}>0$ in (\ref{hamij1}), which from a theoretical perspective are more interesting than ferromagnetic couplings. Antiferromagnetic interactions
in strongly-correlated systems are also more prevalent in nature. A much broader area of quantum many-body physics can be entered by also allowing 
anisotropies in spin space, i.e., different $x$, $y$, and $z$ coupling strengths in (\ref{hamij1}). In addition to the relevance of such anisotropies 
in many real magnetic materials, the mapping between $S=1/2$ spins and ``hard-core'' bosons makes such models interesting also for other reasons. 
The spin-isotropic Heisenberg interaction can be considered the essence of quantum magnetism, however, and we will focus almost exclusively on this 
case here.

\paragraph{Section Outline}

After discussing the antiferromagnetic (N\'eel) state and the nature of its quantum fluctuations based on spin wave theory in Sec.~\ref{neel}, two important 
classes of non-magnetic states---spin liquids and valence-bond solids---will be discussed in Sec.~\ref{vbsrvb}. In Sec.~\ref{chains} the special properties 
of 1D systems are briefly reviewed. Sec.~\ref{qtrans} introduces the extended quantum Heisenberg models and the types of quantum phase transitions 
that we will study in much more detail in the later sections, in applications of the various computational methods.

\subsection{The N\'eel state and its quantum fluctuations}
\label{neel}

On the 2D square lattice (and other bipartite 2D and 3D lattices with uniform interactions) the ground state of the Heisenberg model with only 
nearest-neighbor interactions is antiferromagnetically (N\'eel) ordered, with neighboring spins being oriented, on average, in an antiparallel
(staggered) fashion. Note that a maximally ordered antiferromagnetic state, e.g., $|\up\dn\up\dn,\ldots\rangle$ on a chain (or a checkerboard pattern 
of $\up$ and $\dn$ spins on the 2D square lattice), is not an eigenstate of the Heisenberg hamiltonian, whereas a fully polarized ferromagnetic state, 
e.g., $|\up\up\up\up\ldots\rangle$, is an eigenstate (and, in the case of ferromagnetic interactions, has the minimum energy). This can be easily 
seen with the pair interaction (\ref{hamij1}) written as
\begin{equation}
H_{ij}=J_{ij}(S^x_iS^x_{j} + S^y_iS^y_{j} + S^z_iS^z_{j})=J_{ij}[S^z_iS^z_{j} + \half(S^+_iS^-_{j}+S^-_iS^+_{j})].
\label{hamij2}
\end{equation}
When acting on the perfect N\'eel state, the raising and lowering operators flip pairs of spins, causing local defects, whereas they destroy the 
perfect ferromagnetic state. The antiferromagnetic order must therefore always be reduced by quantum fluctuations, whereas the fully polarized 
ferromagnetic state is the ground state also in the presence of the off-diagonal interactions. The amount of magnetic order (if any) remaining 
in the true ground state of a system with antiferromagnetic interactions depends sensitively on details of the lattice and the interactions included 
in the hamiltonian. 

Note again that the magnetic order parameter of a system with Heisenberg interactions (e.g, the magnetization of a ferromagnet or the sublattice 
magnetization of an antiferromagnet) is a vector in spin space. The sublattice magnetization operator is
\begin{equation}
{\bf m}_s=\frac{1}{N} \sum_{i=1}^N \phi_i {\bf S}_i,
\label{msublattdef}
\end{equation}
where $\phi_i = \pm 1$ is the staggered phase factor, e.g., on the 2D square lattice $\phi_i = (-1)^{x_i+y_i}$, where $x_i$ and $y_i$ are the lattice 
(integer) coordinates of site $i$. In a N\'eel state the expectation value $\langle {\bf m}_s\rangle = m_s^x{\hat {\bf x}} + m_s^y{\hat {\bf y}} 
+ m_s^z{\hat {\bf z}}$ is non-zero in the thermodynamic limit. The order can form---the spin-rotational symmetry can be broken---in any direction in spin 
space. For convenience one normally associates the $z$ spin components with this direction, so that the staggered magnetization 
$\langle m_s\rangle=|\langle S^z_i\rangle|$. In a
finite system a non-zero magnitude $m_s$ of the N\'eel order can form, even though the direction of the vector remains fluctuating over all angles and, thus,
$\langle {\bf m}_s\rangle=0$. In a calculation for a finite system one should therefore detect the presence of order using a quantity which is independent 
of the direction, e.g., $\langle m_s^2\rangle $ or $\langle| m_s|\rangle$. We will see many examples of finite-lattice calculations later. First, let us 
discuss some of the basic properties of the symmetry-broken N\'eel state in more detail.

\subsubsection{spin wave theory}
\label{spinwave}

The N\'eel state and its excitations in 2D and 3D systems can be understood within a simple linear spin wave theory (discussed in more detail in many 
standard works, e.g., the review article by Manousakis \cite{manousakis} and the book by Mattis \cite{mattis}). Such a calculation starts from a maximally 
ordered state of staggered spins, which is the exact ground state in the classical limit where the spin magnitude $S \to \infty$. This is regarded as a vacuum 
state, on top of which quantum effects are included systematically by adding bosons, representing the deviations of the spins from $|S^z_i|=S$, in such a 
way as to obtain a good approximation to the ground state of the system for finite $S$.

\begin{figure}
\includegraphics[width=12cm, clip]{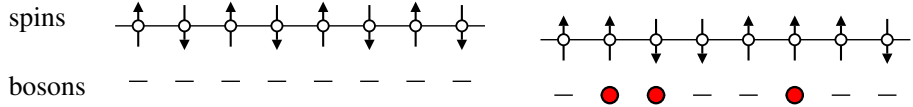}
\caption{Correspondence between $S=1/2$ spins and bosons, here illustrated for a 1D chain. The fully staggered reference state (left) is the vacuum for bosons
(with -- indicating zero boson occupation). For each spin flipped with respect to the reference state there is a boson ($\bullet$) at the corresponding 
site (right). For $S=1/2$ the boson occupation numbers are $0$ or $1$ on each site, while for arbitrary spins they are $0,\ldots,2S$. Linear spin wave 
theory corresponds to non-interacting bosons, and the occupation number constraints are not enforced. Higher-order calculations in $1/S$ gradually 
restore the constraints.} 
\label{spinbosons}
\end{figure}

The relationship between spins and bosons for $S=1/2$ is illustrated in Fig.~\ref{spinbosons}. The reason for using this 
mapping is that it is easier to carry out calculations with the bosons, due to their simpler commutation relations. Linear spin wave theory corresponds 
to non-interacting bosons and is, by construction, exact for $S\to \infty$. Results for finite $S$ can be systematically improved by including 
interactions in the form of an $1/S$ expansion. Here we just outline the lowest-order calculation.

If we neglect the constraint that the boson number $n_i$ for each site should be within the range $0,\ldots,2S$ (the physical subspace), we can use 
the following simple (lowest order in $1/S$) mapping between the spin operators in the original hamiltonian (\ref{hamij2}) and boson creation and destruction 
operators $a^+_i$ and $a_i$ (and the number operators $n_i=a^+_ia_i$);\footnote{Often two species of bosons, $a_i$ and $b_j$, are used, corresponding to the 
two sublattices. In momentum space, the use of a single boson species implies that we are here working in the full Brillouin zone of the original lattice 
with $N$ momenta, instead of one with $N/2$ momenta corresponding to a two-site unit cell.} 
\begin{equation}
\begin{array}{llll}
i~ \in \hskip0.8mm\up{\rm sublattice:} & S^z_i=S-n_i,~~& S^+_i = \sqrt{2S}a_i,~~  & S^-_i = \sqrt{2S}a^+_i \\
j~ \in \hskip0.8mm\dn{\rm sublattice:} & S^z_j=n_j-S,~~& S^+_j = \sqrt{2S}a^+_j,~~& S^-_j = \sqrt{2S}a_j.
\end{array}
\label{sbmapping}
\end{equation}
It is useful to look at Fig.~\ref{spinbosons} to understand this mapping---apart from the obvious way in which the off-diagonal operators 
can affect the states, one just has to be careful with the different factors associated with boson creation/destruction and spin raising/lowering (discussed
in standard quantum-mechanics texts). The factors in (\ref{sbmapping}) are correct for $n_i \ll S$, but note that they are also exact for $S=1/2$ in the physical 
subspace. In principle, one can write down more complicated expressions that are formally correct for any $S$ and $n_i$ in the physical subspace (and also 
automatically decouple the physical and unphysical subspaces), but these are only relevant when going beyond linear spin wave theory.

We now transform the terms in the hamiltonian (\ref{hamij2}) using the mapping (\ref{sbmapping}). Because we are considering a bipartite system, where the 
two sites $i$ and $j$ are always on different sublattices, we obtain the off-diagonal term
\begin{equation}
\half(S^+_iS^-_{j}+S^-_iS^+_{j}) \to S(a_ia_j + a^+_ia^+_j),
\label{offboson}
\end{equation}
and the diagonal interaction is
\begin{equation}
S^z_iS^z_j \to -S^2+S(n_i+n_j)-n_in_j.
\label{diaboson}
\end{equation}
spin wave theory should formally be regarded as a large-$S$ calculation. In the lowest-order calculation we should therefore discard the interaction 
term $n_in_j$ in (\ref{diaboson}), because it is a factor $1/S$ smaller than the non-interacting contributions from (\ref{offboson}) and (\ref{diaboson}). 
Let us for definiteness consider the two-dimensional square lattice with $N=L^2$ sites. We then have the effective hamiltonian
\begin{equation}
H = -2NS^2J+4SJ\sum_{i=1}^N n_i + SJ\sum_{\langle ij\rangle} (a_ia_j + a^+_ia^+_j).
\label{swham1}
\end{equation}
Here it should be noted that we consider a finite system but assume that the symmetry is broken (the direction of the staggered magnetization 
has been locked) which can be strictly true only in the limit $N\to \infty$ (unless some symmetry-breaking field is added to $H$). 
This is fine, however, because we will anyway take the limit $N\to \infty$ at the end.

The boson hamiltonian (\ref{swham1}) can be easily diagonalized (i.e., written in terms of number operators). To construct this 
solution, we first Fourier transform,
\begin{equation}
a_{\bf k} = N^{-1/2} \sum_{\bf r} {\rm e}^{i {\bf k}\cdot {\bf r}}a_{\bf r},~~~~~~
a_{\bf r} = N^{-1/2} \sum_{\bf k} {\rm e}^{-i {\bf k}\cdot {\bf r}}a_{\bf k},
\end{equation}
where the real-space operators have been labeled by their lattice coordinate vectors ${\bf r}$ instead of just the site index $i$, and the momentum
${\bf k}$ is in the first Brillouin zone of the square lattice (i.e., the reciprocal square lattice of $N$ sites). The hamiltonian is then
\begin{equation}
H = -2NS^2J+4SJ\sum_{{\bf k}} n_{\bf k} + 2SJ\sum_{\langle ij\rangle} \gamma_{\bf k}(a_{\bf k}a_{\bf -k} + a^+_{\bf k}a^+_{\bf -k}),
\label{swham2}
\end{equation}
where, with the lattice constant set to $1$ (i.e., $k_x,k_y=n2\pi/L$, $n \in \{ 0,\ldots,L-1\}$),
\begin{equation}
\gamma_{\bf k} = \half [\cos(k_x)+\cos(k_y)].
\label{gammak}
\end{equation}
The next step is to carry out a {\it Bogolubov transformation} to boson operators mixing the original $+{\bf k}$ and $-{\bf k}$ operators;
\begin{equation}
\alpha_{\bf k} = \cosh(\Theta_{\bf k})a_{\bf k} + \sinh(\Theta_{\bf k})a^+_{-\bf k},
\end{equation}
which has the inverse
\begin{equation}
a_{\bf k} = \cosh(\Theta_{\bf k})\alpha_{\bf k} - \sinh(\Theta_{\bf k})\alpha^+_{-\bf k}.
\end{equation}
It is easy to verify that the operators $\alpha_{\bf k}$ obey the standard bosonic commutation relations for any $\Theta_{\bf k}$. The trick is to choose 
these ``angles'' for each ${\bf k}$ such that all operators of the form $\alpha_{\bf k}\alpha_{-\bf k}$ and $\alpha^+_{\bf k}\alpha^+_{-\bf k}$ 
cancel out in the hamiltonian (\ref{swham2}). This is the case if
\begin{equation}
\frac{2\cosh(\Theta_{\bf k})\sinh(\Theta_{\bf k})}{\cosh^2(\Theta_{\bf k})+\sinh^2(\Theta_{\bf k})}=\gamma_{\bf k}.
\end{equation}
The Bogolubov-transformed hamiltonian (the spin wave hamiltonian) is then diagonal in the occupation numbers;
\begin{equation}
H = E_0 + \sum_{{\bf k}} \omega_{\bf k} \alpha_{\bf k}^+\alpha_{\bf k},
\label{hsw}
\end{equation}
where, after some algebra making use of Eq.~(\ref{gammak}), the constant can be written as
\begin{equation}
E_0 = -2SJ \sum_{\bf k}\frac{\gamma^2_{\bf k}}{1+\sqrt{1-\gamma^2_{\bf k}}}-2NS^2J,
\label{swe0}
\end{equation}
and the dispersion relation in (\ref{hsw}) for the spin wave states $\alpha_{\bf k}^+|0\rangle$ is given by
\begin{equation}
\omega_{\bf k} = 4SJ\sqrt{1-\gamma^2_{\bf k}}.
\end{equation}
For momenta close to $(0,0)$ and $(\pi,\pi)$, this dispersion is linear, $\omega_{\bf k} = ck$ and $\omega_{\bf k} = c|(\pi,\pi)-{\bf k}|$, respectively,
with velocity (the spin wave velocity) $c=2\sqrt{2}SJ$. 

The ground state $|0\rangle$ of (\ref{hsw}) is the vacuum for spin waves, where the energy is just $E_0$ given by (\ref{swe0}). The sum can be evaluated 
numerically, most easily by a straight-forward summation over the momenta on large lattices, and extrapolating $E_0/N$ to $N\to \infty$ (or converting 
the sum divided by $N$ to an integral, the numerical evaluation of which gives the $N=\infty$ value directly). The result is $E_0/JN = -0.65795$ 
for $S=1/2$. 

Note that while the ground state does not contain any Bogolubov $\alpha$-bosons (spin waves), it does contain some amount of the original 
$a$-bosons. The sublattice magnetization is directly related to the density of these bosons, which is uniform and can be computed at any site
or averaged over the sites;
\begin{equation} 
\langle m_s\rangle = S - \langle 0|a^+_ia_i|0 \rangle = S - \frac{1}{N} \sum_{i=1}^N \langle 0|a^+_ia_i|0 \rangle.
\end{equation}
Using the Bogolubov transformation, this becomes
\begin{equation}
\langle m_s\rangle = S - \frac{1}{N} \sum_{\bf k} \sinh^2(\Theta_{\bf k}).
\end{equation}
In the most interesting case of $S=1/2$, this evaluates to $\langle m_s\rangle = 0.3034$, or $\approx 61\%$ of the ``classical'' value $1/2$. 
Thus, the quantum effects (zero-point fluctuations, represented by the presence of of bosons) reduce, but do not destroy, the long-range order. 

In principle it is not clear whether spin wave theory should be reliable for small $S$. There has been much discussion of this issue, but the method 
does in fact give a good description of the 2D Heisenberg model on the square lattice. As we will see later, unbiased QMC calculations 
give results for $E_0$ and $\langle m_s\rangle$ differing from those quoted above by only $1-2$\%. This can be traced, {\it a posteriori}, to the 
true value of $\langle m_s\rangle$ being quite large (i.e., spin wave theory can be expected to be accurate when the density of bosons is low). In cases 
where the true sublattice magnetization is small or vanishes, spin wave theory normally does not work that well, even when going to higher orders in 
$1/S$ (which increases the complexity of the calculation very significantly \cite{canali,hamer,igarashi}).

\subsubsection{Destruction of the N\'eel order}

When going beyond bipartite lattices with uniform interactions, or by supplementing the Heisenberg model with additional interactions (e.g., including 
more than two spins) the quantum fluctuations can become so significant that the ground state loses its long-range N\'eel order (retaining only short-range 
antiferromagnetic correlations). There are several other possible types of ordered and disordered ground states, some of which have no classical counterparts. 
Much of the current interest in quantum spin models is related to the existence of non-magnetic states and quantum phase transitions between them and the N\'eel state 
\cite{sachdev1}. This is also the main theme of the numerical calculations to be discussed in these lecture notes. One long-standing motivation for studying 
such transitions stems from the cuprate high-T$_{\rm c}$ superconductors, the undoped parent compounds of which are antiferromagnets corresponding closely 
to weakly coupled Heisenberg layers (many properties of which can be understood based on a single layer) \cite{manousakis}. In these systems the magnetic 
order is destroyed upon doping by mobile charge carriers. This is a very challenging electronic many-body problem, where computational studies are also  
playing an important role \cite{dagotto2,aimi,white2}. While the full solution of the high-T$_{\rm c}$ problem will of course require more complicated models 
(perhaps some variety of the t-J or Hubbard model), some generic aspects of the physics close to a quantum phase transition out of the N\'eel state can, however, 
be understood based on spin-only models \cite{sachdev4}. 

Apart from the cuprates and related antiferromagnetic systems, there are also many materials with non-uniform or frustrated 
spin interactions \cite{schollwock1,diep,gardner}, which can lead to non-magnetic low-temperature states. Many of these states, and the possible quantum phase 
transitions between them, are still not well understood. It is therefore useful to search for and study prototypical quantum spin models that realize various 
types of ground states and quantum phase transitions. Studies of quantum phase transitions also have a broader context of understanding ``exotic'' manifestations of 
quantum mechanics at the collective many-body level \cite{sachdev1}. There are even interesting connections with particle physics---close analogies exist 
between phase transitions in supersymmetric gauge theories in 2+1 dimensions and ``deconfined'' quantum-critical points that may separate the 2D N\'eel and 
VBS ground states \cite{sachdev3}.

\subsection{Spin liquids and valence-bond solids}
\label{vbsrvb}

Within linear spin wave theory, a non-magnetic state corresponds to boson density $\langle n_i\rangle = S$. This kind of state is, however, not
a good representation of actual non-magnetic ground states of Heisenberg and related quantum spin models, because it does not contain any correlation 
effects. To discuss more interesting non-magnetic states, it is useful to first look at the quantum fluctuations from a different 
perspective. 

For two isolated spins $i,j$ (a dimer, $N=2$), the ground state of the $S=1/2$ antiferromagnetic Heisenberg hamiltonian (\ref{hamij1})
is the singlet; 
\begin{equation}
|\phi^s_{ij}\rangle = \frac{|{\up}_i{\dn}_j\rangle - |{\dn}_i{\up}_j\rangle}{\sqrt{2}}.
\label{singeltij}
\end{equation}
Although the two spins in such a singlet are always perfectly anti-correlated (entangled), the individual spins are strongly (maximally) fluctuating, 
and there is no static spin order; $\langle {\bf S}_i\rangle=\langle {\bf S}_j\rangle=0$. In contrast, the perfect N\'eel states for $N=2$, 
$|{\up}_i{\dn}_j\rangle$ and $|{\dn}_i{\up}_j\rangle$, are product states (i.e., of the form $|\phi_i\rangle \otimes |\phi_j\rangle$) with no fluctuations 
(and no entanglement---loosely speaking, the degree of entanglement corresponds to the deviations from a product state). Note that for $N=2$ the (eigen) 
energy of the singlet is $-3J_{ij}/4$, whereas the expectation value of the energy in the N\'eel states is much higher, $-J_{ij}/4$ (and the states are not 
eigenstates). The tendency of interacting spins to entangle by forming pair-wise singlets to minimize the energy remains in multi-spin systems, but when 
$N>2$ a spin cannot simultaneously form pure singlet pairs with all its neighbors. The system can instead be thought of as a superposition of 
different singlet pairings. No pair is then in a pure singlet, and the energy contribution from each interaction $H_{ij}$ is therefore always higher 
than the singlet energy $-3J_{ij}/4$. This can be regarded as a reduction of quantum fluctuations (leading to less than maximal two-spin entanglement) for 
$N>2$, bringing the state (or, more correctly, the density matrix) of each interacting pair closer to a product state. A state with antiferromagnetic long-range 
order, breaking the rotational invariance of the hamiltonian, can form in the thermodynamic limit if there are enough fluctuations in the singlet pairings 
(and note that, from the perspective of singlets, the N\'eel state has larger fluctuations than, e.g., a state of fixed singlet pairings---to be precise, 
fluctuations should always be specified with respect to some reference state). If the system is one-dimensional, or if the interactions are strongly 
{\it frustrated} (competing), or if the lattice geometry and couplings favor the formation of singlets in a specific pattern (e.g., in a system of weakly 
coupled dimers), antiferromagnetic order may not be present in the ground state when $N\to \infty$.

\paragraph{Valence-bond states}

The above intuitive picture of a state of fluctuating singlets can in fact be made rigorous. Any singlet state can be expanded in basis states that are 
products of singlet pairs, or {\it valence bonds}. Denoting by $(i,j)$ a singlet of spins $i$ and $j$, as in Eq.~(\ref{singeltij}), a normalized 
valence-bond basis state  for $N$ (even) spins is of the form 
\begin{equation}
|\phi\rangle = N^{-1/4}|(i_1,j_1)(i_2,j_2)\cdots (i_{N/2},j_{N/2})\rangle,
\label{psivbprod}
\end{equation}
where each site index appears exactly once (i.e., each spin belongs to one singlet). This basis is {\it over-complete} in the singlet subspace, and, 
thus, any singlet state can be expanded in these states, but the expansion coefficients are not unique. The valence-bond basis and computational methods
using it are discussed in Refs.~\cite{vbmethod1,vbbasis,vbmethod2,awshg}. Here we just note that some types of states are more naturally expressed in the
valence-bond basis than in the standard basis of individual $\up$ and $\dn$ spins.

\begin{figure}
\includegraphics[width=10.5cm, clip]{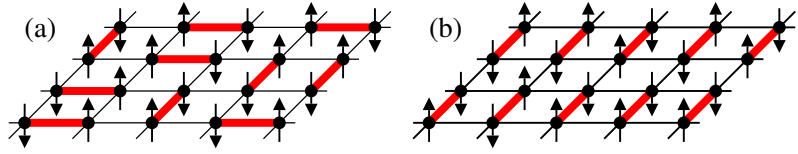}
\caption{Valence-bond states in two dimensions. The thick lines represent singlets. The arrows show a spin configuration compatible 
with the valence bonds (i.e., the spins on each bond are antiparallel---the valence-bond state is a superposition of all such spin
configurations). A valence-bond state with only short bonds (or any superposition of such states), as shown here in the extreme case 
of all bonds of length one, has no magnetic order in two dimensions. There can, however, be order in the valence bonds. (a) and (b) 
represent typical configurations of a disordered RVB spin liquid and a columnar VBS, respectively.} 
\label{rvbfig}
\end{figure}

For the valence-bond basis to be (over-)complete, arbitrary lengths of the bonds must be allowed. Restricting the lengths renders the basis incomplete 
(although one can actually restrict the bonds to only connect sites on two different sublattices). Some types of states are still completely 
dominated by short bonds (i.e., the probability of a bond of length $r$ decreases rapidly with $r$). Such short-bond 
states in two dimensions have no magnetic long-range order (while in three dimensions they can be N\'eel ordered) and are often called spin liquids or 
{\it resonating valence-bond} (RVB) states. Various types of crystalline order can also form in the bond configurations, leading to periodic modulations 
in observables such as $\langle {\bf S}_i \cdot {\bf S}_j\rangle$ (where $i,j$ are nearest-neighbor sites). Such ordered states (which break lattice 
symmetries) are called {\it valence-bond solids} (VBSs), or valence-bond crystals. Representative configurations of valence bonds in RVB and VBS states 
are illustrated in Fig.~\ref{rvbfig}.

\subsection{One-dimensional systems}
\label{chains}

1D systems are rather special and deserve their own introduction. Studies of quantum spin chains date back to Bethe's exact solution 
of the $S=1/2$ Heisenberg model, which was published in 1931 \cite{bethe} and worked out in greater detail some time thereafter \cite{hulthen}. The solution 
is very complicated (often requiring complex numerical calculations \cite{caux,maillet}), however, and many properties of the heisenberg chain were 
only obtained much later and with complementary methods (in particular, renormalization-group treatments of effective low-energy field-theories 
\cite{haldane1,affleck1,affleck2,singh1,giamarchi}). 

As we have already noted, there can be no magnetic order in a 1D Heisenberg system (but VBS order is allowed, since it breaks a discrete symmetry). 
The spin correlation function $C(r)=\langle {\bf S}_{i}\cdot {\bf S}_{i+r}\rangle$ of the Heisenberg chain decays 
with the distance $r$ as $(-1)^r/r$ (with a multiplicative logarithmic correction, which we will discuss later). Thus, the ground state is critical (or 
quasi long-range ordered, on the verge of ordering). Including a frustrated second-nearest-neighbor interaction causes, when sufficiently strong, a 
quantum phase transition into a VBS state (some times also called the spin-Pierls state), where the spin correlations decay exponentially with 
distance \cite{majumdarghosh}.

While quasi-1D antiferromagnets were actively studied experimentally already in the 1960s and 70s, these efforts were further stimulated by 
theoretical developments in the 1980s. Haldane conjectured \cite{haldane1}, based on a field-theory approach, that the  Heisenberg chain has completely 
different physical properties for integer spin ($S=1,2,\ldots$) and ``half-odd integer'' spin ($S=1/2,3/2,\ldots$). It was known from Bethe's 
solution that the $S=1/2$ chain has a gapless excitation spectrum (related to the power-law decaying spin correlations). Haldane suggested the possibility 
of the $S=1$ chain instead having a ground state with exponentially decaying correlations and a gap to all excitations; a kind of spin liquid state 
\cite{aklt}. This was counter to the expectation (based on, e.g., spin wave theory) that increasing $S$ should increase the tendency to ordering. 
Haldane's conjecture stimulated intense research activities, theoretical as well as experimental, 
on the $S=1$ Heisenberg chain and 1D systems more broadly. There is now completely conclusive evidence from numerical studies that Haldane 
was right \cite{nomura89,liang90,schollwock96}. Experimentally, there are also a number of quasi-one-dimensional $S=1/2$ \cite{tennant93} and 
$S=1$ \cite{regnault94} (and also larger $S$ \cite{granroth96}) compounds which show the predicted differences in the excitation spectrum. 
A rather complete and compelling theory of spin-$S$ Heisenberg chains has emerged (and includes also the VBS transitions for half-odd integer $S$), 
but even to this date various aspects of their unusual properties are still being worked out \cite{pereira}. There are also many other variants of 
spin chains, which are also attracting a lot of theoretical and experimental attention (e.g., systems including various anisotropies, external fields 
\cite{caux05}, higher-order interactions \cite{lauchli06}, couplings to phonons \cite{sandvik99b,michel07}, long-range interactions 
\cite{laflorencie,awslr}, etc.). In Sec.~\ref{sec_diag} we will use exact diagonalization methods to study the $S=1/2$ Heisenberg chain, as well as 
the extended variant with frustrated interactions (and also including long-range interactions).  In Sec.~\ref{sec_sse} we will investigate longer 
chains using the SSE QMC method. We will also study ladder-systems consisting of several coupled chains \cite{dagotto1}, which, for an even number 
of chains, have properties similar to the Haldane state (i.e., exponentially decaying spin correlations and gapped excitations).

\subsection{Models with quantum phase transitions in two dimensions}
\label{qtrans}

\begin{figure}
\includegraphics[width=9cm, clip]{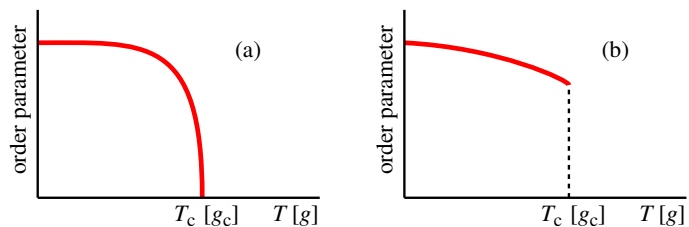}
\caption{Temperature ($T$) or coupling ($g$) dependence of the order parameter (e.g., the magnetization of a ferromagnet) at a continuous (a) 
and a first-order (b) phase transition. A classical, thermal transition occurs at some temperature $T=T_{\rm c}$, whereas a quantum
phase transition occurs at some $g=g_c$ at $T=0$.}
\label{ptcases}
\end{figure}

The existence of different types of ground states implies that phase transitions can occur in a system at $T=0$ as some parameter in the hamiltonian is varied 
(which experimentally can be achieved in quantum magnets, e.g., as a function of pressure or external magnetic field \cite{ruegg08}---it should be 
noted, however, that the possibilities 
are more limited than in models). As in classical, temperature driven phase transitions, such {\it quantum phase 
transitions} can be continuous or first-order, as illustrated in Fig.~\ref{ptcases}. Normally a phase transition is associated with an {\it order parameter}, 
which is zero in the disordered phase and non-zero in the ordered phase. e.g., the magnetization $m$ of a ferromagnet or the sublattice magnetization $m_s$ of 
an antiferromagnet. In Secs.~\ref{sec_diag} and \ref{sec_sse} we will also discuss examples of order-order (N\'eel-VBS) transitions. Continuous phase transitions 
are associated with scaling and universality, which we will discuss in more detail in Sec.~\ref{transitions}. In these lecture notes we will not discuss 
{\it topological phase transitions}, which are not associated with any local order parameter. The different states across such transitions are distinguishable 
only through some global topological quantity \cite{wenbook}. 

In this section we introduce some spin models exhibiting quantum phase transitions in two dimensions. As we have already seen, the 2D $S=1/2$ Heisenberg model 
with nearest-neighbor interactions has N\'eel order at $T=0$. The question is then how to destroy this order and create some different kind of ground state. 
Specific studies based on QMC calculations will be presented in Sec.~\ref{sec_sse}. The discussion here serves as an introduction to the kind of models and 
physical quantities that we will have in mind when discussing computational techniques. Some QMC results will be shown for illustration purposes already in 
this section, but at this point we do not need to worry about how they were obtained---it suffices to say that the data are numerically exact (to within 
statistical errors that are in most cases too small to be discerned in the figures). 

There are also 1D analogues of most of the transitions discussed here and some of them will be discussed in Sec.~\ref{sec_results1d}. Systems in three dimensions 
can exhibit transitions similar to those in two dimensions, but most computational research is currently focused on 1D and 2D systems, partially because 3D systems
are much more challenging. The quantum fuctations are normally also more prominent in 1D and 2D systems, and some of the most interesting open questions
related to real materials are associated with the quasi-2D nature of the systems (although there are also interesting 3D systems \cite{ruegg08}).

\subsubsection{Dimerized systems}
\label{dimsystems}

\begin{figure}
\includegraphics[width=11cm, clip]{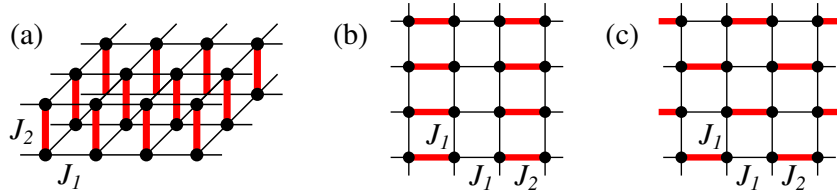}
\caption{Dimerized systems with two different coupling strengths between nearest neighbors; a bilayer (a) with the dimers across the layers and
single layers with columnar (b) and staggered (c) dimers.}
\label{dlattices}
\end{figure}

The perhaps simplest way to obtain a non-magnetic ground state of a 2D Heisenberg model is to {\it dimerize} the system \cite{singh88}, 
i.e., to introduce weak and 
strong antiferromagnetic couplings (bonds), $J_1>0$ and $J_2>J_1>0$, respectively, in a pattern such that each spin belongs exactly to one strong bond. There are 
several ways to do this, three of which are illustrated in Fig.~\ref{dlattices}. When $J_1=0$, these systems consist of $N/2$ independent pairs of spins (dimers) 
with intra-dimer coupling $J_2$, and, as we have already discussed, the ground state of each such a pair is a singlet; thus the ground state of the whole system is a 
singlet-product (valence-bond) state, which clearly has no magnetic order. On the other hand, for $J_2=J_1$ the ground state has N\'eel order. The question is 
then how the ground state evolves as a function of the coupling ratio $g=J_2/J_1$. One might perhaps think that some amount of N\'eel order should appear once the 
dimers are coupled, i.e., for any $g<\infty$. It turns out, however, that there is actually a phase transition at some critical value $g_c$ (which depends on
the dimer arrangement). While this transition can be analyzed using several analytical and numerical approaches, finite-size scaling of QMC data 
is currently the only way to obtain unbiased quantitative results (see, e.g., \cite{morr} for a discussion of how spin wave theory breaks 
down close to the phase transition). 

\begin{figure}
\includegraphics[width=12.5cm, clip]{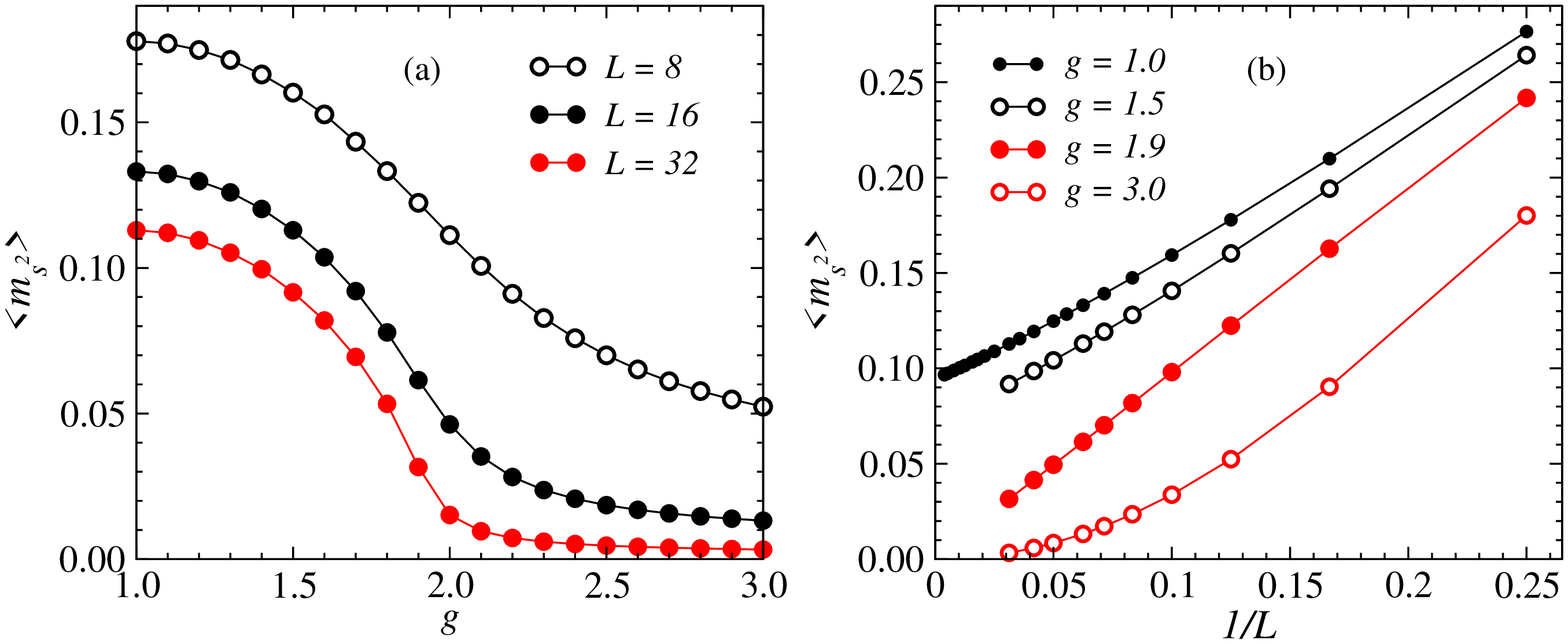}
\caption{QMC results for the squared sublattice magnetization in the two-dimensional Heisenberg model with columnar dimerization. 
(a) shows results versus the coupling ratio $g$ for different lattice sizes and (b) shows the size dependence for several values of $g$.
A quantum phase transition where the N\'eel order vanishes occurs at $g\approx 1.9$.}
\label{jjmag2}
\end{figure}

Fig.~\ref{jjmag2} shows some QMC results for the columnar dimer model of Fig.~\ref{dlattices}(b). The order parameter is the staggered magnetization, 
with the corresponding operator defined in Eq.~(\ref{msublattdef}). This is a vector operator, and, for a finite lattice, its expectation value vanishes due 
to the spin-rotational symmetry of the hamiltonian. Its square, $\langle m_s^2\rangle$, was computed in the simulations. Fig.~\ref{jjmag2}(a) shows results 
for $L\times L$ lattices versus the coupling ratio $g$, and Fig.~\ref{jjmag2}(b) shows data for several values of $g$ graphed versus $1/L$ (which is often 
the most convenient way of graphing data when examining the convergence to a non-zero value for $L \to \infty$). Here it is clear that the behavior changes 
at $g\approx 1.9$; below this coupling the sublattice magnetization extrapolates to a non-zero value in the thermodynamic limit, whereas for larger $g$ 
it decays to zero. 

In the N\'eel state, the leading finite-size corrections 
to $\langle m^2_s\rangle$ are $\propto 1/L$ (a result which can be obtained using spin wave theory \cite{huse,reger88}). 
This can be seen quite clearly in Fig.~\ref{jjmag2}(b) for $g=1$ and $1.5$. The extrapolation for $g=1$ (see Ref.~\cite{awshg}, from which the results for
this coupling are taken) gives $\langle m_s\rangle= 0.3074$. This is only about $1\%$ higher than the linear spin wave result, $\langle m_s\rangle = 0.3034$. 
As discussed above in Sec.~\ref{spinwave}, the success of spin wave theory for this model can be explained {\it a posteriori} by the fact that the quantum 
fluctuations are rather weak, reducing the sublattice magnetization only by about $40\%$ from the classical value. For larger $g$, the agreement between
spin wave theory and QMC results quickly becomes much worse. In this kind of dimerized model linear spin wave theory typically over-estimates the critical 
$g$ \cite{monienbilayer,awsbilayer,morr}, while in other cases, such as a system with uniform couplings with different strengths in the $x$ and $y$ lattice 
directions (coupled chains) linear spin wave theory predicts a critical point at non-zero coupling when, in fact, there is none \cite{mchains}. 
Various improvements can be made of the spin wave theory (going to higher order in $1/S$, incorporating self-consistency, 
etc.~\cite{manousakis,canali,hamer,monienbilayer,morr,shevchenko}), but a system at or close to a quantum phase transition cannot be captured correctly 
by these approximative methods.

An important question regarding the quantum phase transition in dimerized systems is its {\it universality class} (a concept to be discussed further in
Sec.~\ref{transitions}). A quantum many-body system in $d$ dimensions can formally be mapped, using path integrals, onto an effective classical system in 2+1 
dimensions (as we will discuss in Sec.~\ref{sec_sse}), and effective continuum field-theories can be constructed for the low-energy behavior, including
the quantum phase transition. Many properties have been predicted in this way based on renormalization-group treatments of one such field theory---the
nonlinear $\sigma$-model in 2+1 dimensions \cite{chn,chubukov}. Based on symmetry arguments alone, one would then expect the transition to be in the universality 
class of the 3D classical Heisenberg model. There are, however, subtle issues in the quantum-classical mapping, and QMC simulations are therefore needed 
to test various predictions. We will see examples of such comparisons between results of simulations and field theories in Sec.~\ref{sec_sse}. While results 
for the transition in the bilayer (a) \cite{wang} and columnar dimer (b) \cite{matsumoto} systems in Fig.~\ref{dlattices} (and several other cases 
\cite{troyer,wenzel1}) are in good agreement with the expectations, recent studies of the staggered dimers (c) show unexpected deviations \cite{wenzel2} 
that are still not understood.

\subsubsection{Frustrated systems}

\begin{figure}
\includegraphics[width=5cm, clip]{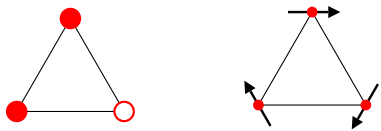}
\caption{Ising (left) and planar (right) spin configurations with minimum energy on a triangle with antiferromagnetic nearest-neighbor interactions.}
\label{triangles}
\end{figure}

The prototypical example of frustration is a system with antiferromagnetic interactions on a triangular lattice. Looking at this 
problem first within the Ising model, the spins on a single triangle cannot simultaneously be anti-parallel to both their neighbors---there are six configurations 
with minimum energy, and these all have one ``frustrated'' bond (two parallel neighbors), as shown in Fig.~\ref{triangles}. Being a consequence of the lattice, 
this is often referred to as {\it geometric frustration}. Upon increasing the system size, the ground-state degeneracy grows with the system size, and in the ensemble 
including all these configurations there is no order of any kind \cite{stephenson,blote}. In the case of the classical XY (planar vector) or Heisenberg (vectors in 
three dimensions) model, there is, however, order at $T=0$ (but not at $T>0$, according to the Mermin-Wagner theorem). The energy is minimized by arranging the spins 
in a plane at 120$^\circ$ angle with respect to their neighbors on the same triangle, as shown for a single triangle in Fig.~\ref{triangles}. This is referred 
to as a three-sublattice N\'eel state. There have been many studies of the $S=1/2$ variant of this model. This was, in fact, the system for which the RVB spin-liquid 
state was initially proposed \cite{fezekas}. There is now, however, compelling numerical evidence for the three-sublattice N\'eel order actually surviving the 
quantum fluctuations \cite{bernu,chernyshev}.

\begin{figure}
\includegraphics[width=11.5cm, clip]{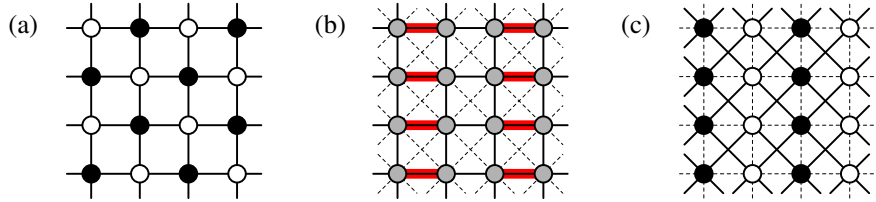}
\caption{The square-lattice $S=1/2$ Heisenberg model with only nearest-neighbor interactions $J_1>0$ has an antiferromagnetic ground state, in (a) illustrated 
by open and solid circles for $\langle S^z_i\rangle>0$ and $\langle S^z_j\rangle<0$. In (b), next-nearest-neighbor 
interactions $J_2>0$ are shown as dashed lines. When $0.4 < J_2/J_1 < -0.6$ (approximately) the ground state may be a columnar VBS, with $\langle S^z_i\rangle=0$ 
on all sites but modulations in the bond correlations $\langle {\bf S}_i\cdot {\bf S}_j\rangle $ (where $i,j$ are nearest-neighbors) as shown with thicker lines 
for the more strongly correlated bonds forming a columnar pattern. For $J_2/J_1>0.6$, the ground state has collinear (striped) magnetic order, as shown in (c).}
\label{j1j2states}
\end{figure}

On the square lattice, frustration effects can be investigated by adding interactions beyond nearest neighbors. A well studied case is the J$_{\rm 1}$-J$_{\rm 2}$ 
model, where $J_{\rm 1}$ and $J_{\rm 2}$ refer, respectively, to the strengths of the nearest- and next-nearest-neighbor interactions. The system is frustrated if 
both $J_1>0,J_2>0$ or if $J_1<0,J_2>0$. Even with Ising spins, this is a highly non-trivial system, with unresolved questions still attracting interest 
\cite{yin,kalz,jin}. We will discuss the frustrated Ising system further in Sec.~\ref{firstordertrans}.

In the case of quantum spins, the $S=1/2$ J$_{\rm 1}$-J$_{\rm 2}$ Heisenberg model with all antiferromagnetic couplings is one of the prototypical 
models with a quantum phase transition out of the standard N\'eel state, in this case as a function of the coupling ratio $g=J_2/J_1$. While many different calculations 
show rather consistently that the N\'eel order vanishes at a critical coupling ratio $g_c \approx 0.4$ \cite{dagotto3,schulz1,singh,sushkov,kruger,beach1,isaev}, the 
order of the phase transition and the nature of the non-magnetic ground state are still controversial issues. Most studies indicate some type of VBS state, 
with a columnar one being the prime candidate, but an RVB spin liquid state has also been proposed \cite{capriotti}. 

For larger $g$, above $g \approx 0.6$, there is again magnetic order. This can be understood in the limit $g \to \infty$, where the system decouples into two separate
square-lattice Heisenberg antiferromagnets. At $g=\infty$ ($J_1=0$) the relative direction in spin space of the antiferromagnetic order within these subsystems is 
arbitrary, but for any $J_1>0$ the subsystems lock to each other and form {\it  collinear} spin order, with vertical or horizontal stripes of parallel spins (a state
breaking the $90^\circ$ lattice rotation symmetry, in addition to the global spin rotation symmetry). The transition between the non-magnetic and collinear states 
is most likely first-order.

The three different ground states of the  $J_{\rm 1}$-$J_{\rm 2}$ Heisenberg model with $S=1/2$ spins are illustrated in Fig.~\ref{j1j2states}. Note that a 
classical version of this model (Ising, XY, or Heisenberg) has a direct (first-order) N\'eel-collinear $T=0$ transition exactly at $g=1/2$ (which can be easily 
verified just by computing the energies of those states for $g<1/2$ and $g>1/2$). The magnetically disordered state is thus induced by quantum fluctuations, and 
it has no direct classical analogue. It is not clear whether this kind of state persists for $S=1$ or higher spins---for some large $S$, perhaps already $S=1$, 
it must give way to a first-order N\'eel-Collinear transition, as in the classical system.

The reason why it has been so difficult to reach a firm conclusion on the nature of the non-magnetic state and the quantum phase transition between it and
the N\'eel state is that large-scale unbiased computational studies of the $S=1/2$ $J_{\rm 1}$-$J_{\rm 2}$ model  are currently not possible, because of 
``sign problems'' affecting QMC calculations of frustrated systems. We will discuss this issue further in Sec.~\ref{sec_sse}. There are no other unbiased 
method that can reach sufficiently large lattices, e.g., exact diagonalization can reach only $N \approx 40$ spins.\footnote{Recently developed variational 
methods based on tensor-product states in principle become unbiased in the limit of large tensors \cite{verstraete1}. They have been applied to various 
frustrated spin models \cite{murg1}. The computational complexity of such calculations is, however, currently too demanding to reach tensors sufficiently 
large to produce unbiased results in practice.}

\subsubsection{The J-Q class of models}
\label{jqmodelintro}

VBS states of quantum spin systems in two dimensions were predicted theoretically more than two decades ago \cite{read}. The VBS formation is associated with a 
spontaneously broken translational lattice symmetry. The VBS state and quantum phase transitions into it are therefore quite different from the non-magnetic 
states and transitions in the ``manually'' dimerized systems discussed above in Sec.~\ref{dimsystems}. Although there is a pattern of ``strong'' and ``weak'' 
spin correlations in both cases, the quantum fluctuations in systems with manually and spontaneously broken translational symmetry are different (being much 
more interesting in VBS states). Until recently, large-scale computational studies of VBS states and N\'eel-VBS quantum phase transitions had not been carried 
out starting from microscopic hamiltonians, because of the QMC sign problems affecting frustrated Heisenberg models (which at first sight are the most natural 
systems in which to explore the physics of non-magnetic states). 

VBS states and the N\'eel-VBS transition have come into renewed focus with a proposal by Senthil {\it et al.} \cite{senthil1} that this transition is 
generically continuous and, thus, violates the ``Landau rule'', according to which an order-order transition (between states breaking unrelated symmetries) 
should be first-order (except at fine-tuned multi-critical points). In the ``deconfined'' quantum-criticality scenario, the VBS and N\'eel order parameters 
are manifestations of spinon confinement and condensation, respectively. Spinons can be thought of as $S=1/2$ degrees of freedom, but not just corresponding 
to the bare individual spins on the lattice sites, but more complex collective objects ``dressed'' by interactions. A non-compact CP$^1$ field theory was 
proposed to describe such spinons coupled to an emergent gauge field (which corresponds to the properties of the valence-bond background in which the spinons 
exist) \cite{senthil1}. A central question is then whether this low-energy physics of a continuum field theory really can arise starting from a reasonable 
microscopic hamiltonian. Answering this question requires large-scale computational studies of models exhibiting N\'eel-VBS transitions.

Since the QMC sign problem prohibits large-scale studies of the $J_{\rm 1}$-$J_{\rm 2}$ Heisenberg model and other similar frustrated systems, we have to try 
something else. In the ``J-Q'' class of models \cite{sandvik1,lou1,arnab}, the N\'eel order is destroyed by an interaction (Q) which is not frustrated, in the 
standard sense, but still competes with the Heisenberg (J) interaction. To understand these J-Q models, note first that the Heisenberg interaction is, up to a 
constant, equal to a singlet projector operator: $H_{ij}=-S_{ij}+ \frac{1}{4}$, where
\begin{equation}
S_{ij} = \hbox{$\frac{1}{4}$} - {\bf S}_i \cdot {\bf S}_j.
\label{singproj}
\end{equation}
The pair-singlet, Eq.~(\ref{singeltij}), is an eigenstate of this operator with eigenvalue $1$, whereas a triplet state is destroyed by it;
\begin{equation}
S_{ij}|\phi^s_{ij}\rangle=|\phi^s_{ij}\rangle,~~~~~S_{ij}|\phi^{t,m}_{ij}\rangle=0,~~~ (m=0,\pm 1).
\end{equation}
Thus, when $S_{ij}$ acts on a singlet-triplet superposition, only the singlet component survives (is ``projected out''---note that the property 
$S_{ij}^2=S_{ij}$ required of a projection operator is satisfied). The standard Heisenberg interaction thus favors the formation of singlets on pairs of 
nearest-neighbor sites, but, as we discussed in Sec.~\ref{vbsrvb}, the fluctuations of these singlets among many different pairings of the spins leads to 
N\'eel order in the ground state. The idea behind the J-Q models is to project singlets on two or more bonds in a correlated fashion, using products of 
several $S_{ij}$ operators on a suitable set of different bonds. This favors a higher density of short valence bonds, thereby reducing or completely 
destroying the antiferromagnetic order.

\begin{figure}
~~~~~~~\includegraphics[width=9.5cm, clip]{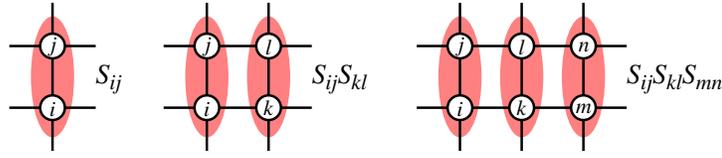}
\caption{Graphical representation of possible arrangements of products of singlet-projector operators $S_{ij}$ in the J-Q model and its generalizations. 
(a) is the Heisenberg exchange, (b) the four-spin interaction of the original J-Q model, and (c) a six-spin interaction which leads to more robust VBS
order. These operators, and their $90^\circ$-rotated analogues, are summed over all positions on the square lattice.}
\label{jqops}
\end{figure}

The original J-Q hamiltonian \cite{sandvik1} on the square lattice can be written as
\begin{equation}
H = -J\sum_{\langle ij\rangle}S_{ij} -Q\sum_{\langle ijkl\rangle}S_{ij}S_{kl},
\label{jqham}
\end{equation}
where both the $J$ and $Q$ terms are illustrated in Fig.~\ref{jqops}. The $Q$ interaction involves four spins on a $2\times 2$ plaquette. An interaction 
with three singlet projectors in a columnar arrangement is also shown, and operators with even more projectors, or with the projectors arranged on the lattice 
in different (non-columnar) patterns, can also be considered \cite{arnab}. With $J>0$ and $Q>0$ [and the minus signs in front of the interactions in 
Eq.~(\ref{jqham})], correlated singlets are favored on the lattice units formed by the product of singlet projectors. It is still not clear just from the 
hamiltonian whether a VBS state is realized for large $Q/J$---since the hamiltonian does not break any symmetries (the interactions in Fig.~\ref{jqops} are 
summed over all distinct lattice translations and rotations), a VBS state only forms if the hamiltonian also contains in it, implicitly, some effective 
interactions that favor singlets in some ordered pattern. The J-Q model (\ref{jqham}) {\it does} exhibit a N\'eel-VBS transition, at $Q_{\rm c}/J\approx 22$ 
\cite{sandvik1,lou1,sandviklogs}, and at lower $Q_c$ if more than two singlet projectors are used \cite{lou1,arnab}. We will discuss this in more detail in 
Sec.~\ref{sec_sse}. Here we just look briefly at examples of correlation functions useful for characterizing the N\'eel and VBS states and the quantum 
phase transition between them. 

\paragraph{Order parameters}

\begin{figure}
\includegraphics[width=12.5cm, clip]{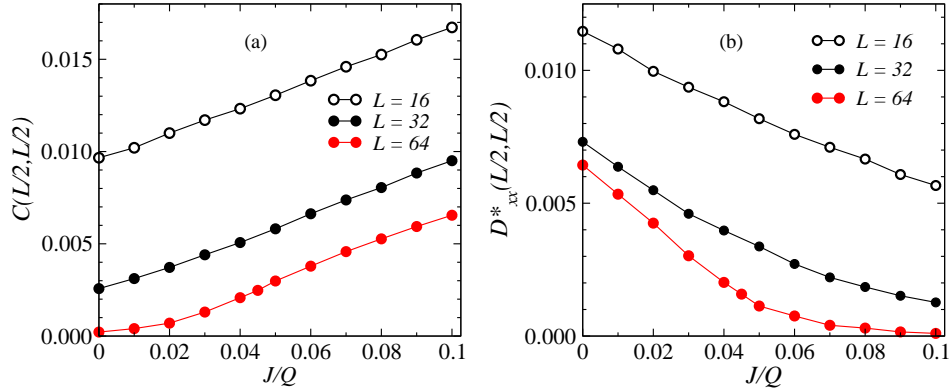}
\caption{QMC results for the long-distance spin (a) and staggered dimer (b) correlation functions (corresponding the N\'eel and VBS order,
respectively) in the ground state of the J-Q model (\ref{jqham}) versus the coupling ratio $J/Q$. A careful analysis of these correlation functions, as well 
as other quantities, indicates a single critical point $(J/Q)_c\approx 0.045$ where both the N\'eel 
order and VBS order vanish continuously.} 
\label{jqmagdim}
\end{figure}

As an alternative to the square of the full spatially averaged sublattice magnetization (\ref{msublattdef}), 
we can detect the presence or absence of of N\'eel order using the spin correlation function,
\begin{equation}
C({\bf r}_{ij}) = \langle {\bf S}_i \cdot {\bf S}_j\rangle,
\end{equation}
at long-distances. Fig.~\ref{jqmagdim}(a) shows J-Q model results for the largest separation of the spins, ${\bf r}_{\rm max}=(L/2,L/2)$, on periodic $L\times L$ 
lattices. The results have to be analyzed carefully to determine the transition point, but already this raw data suggest that the N\'eel order vanishes, i.e., 
$C({\bf r}_{\rm max}) \to 0$ when $L \to \infty$, for $J/Q<0.04$.

VBS order can be detected in the dimer (or bond) correlation function, defined as
\begin{equation}
D_{xx}({\bf r}_{ij}) = \langle B_x({\bf r}_i)B_x({\bf r}_j)\rangle,
\label{dxxdimorder}
\end{equation}
where the bond operator is given by
\begin{equation}
B_x({\bf r}_i) = {\bf S}({\bf r}_i) \cdot {\bf S}({\bf r}_i+\hat {\bf x}).
\label{bxxdef}
\end{equation}
Here, instead of using a subscript $i$ on the spin operator for some arbitrary site labeling, it is more convenient to use the corresponding lattice position
vector ${\bf r}_i$. Then ${\bf r}_i+\hat {\bf x}$ corresponds to the site immediately next to ${\bf r}_i$ in the positive $x$ direction. The subscripts $xx$ 
in (\ref{dxxdimorder}) indicate that the two bond operators $B_x$ are both oriented in the $x$ direction, and this correlation function is, by symmetry, 
equal to $D_{yy}$ on an $L\times L$ lattice. One can also consider bond correlations $D_{xy}$, where the two bond operators are oriented differently. In a VBS 
state such as the one illustrated in Fig.~\ref{j1j2states}, one expects $D_{xx}({\bf r})$ to exhibit a columnar pattern of smaller and larger values. A VBS order 
parameter can then be defined as a suitable difference between these modulated correlations, e.g.,
\begin{equation}
D^*_{xx}({\bf r}) = D_{xx}({\bf r}) - \half [D_{xx}({\bf r}-\hat {\bf x})  + D_{xx}({\bf r}+\hat {\bf x})]. 
\end{equation}
This correlation function is shown in Fig.~\ref{jqmagdim}(b) at the largest lattice separation for several different system sizes. 
Here it is clear that VBS order exists for $J=0$, up to $J/Q \approx 0.04$, roughly where the N\'eel order sets in. As we will discuss in 
Sec.~\ref{sec_sse}, all calculations so far point to a single critical point, without any intervening third phase or region of coexistence of 
both N\'eel and VBS order.

\section{Classical phase transitions, Monte Carlo simulations, and finite-size scaling}
\label{transitions}

Many aspects of quantum phase transitions, and how to analyze them based on numerical finite-lattice data, are very similar to classical phase 
transitions. Here we discuss this common formalism in the simpler context of classical phase transitions, before turning to calculations and data analysis 
for quantum systems.

In classical statistical mechanics the prototypical example of a system with a continuous phase transition is the 2D Ising ferromagnet, the 
exact solution \cite{onsager,baxter} of which shows rigorously that such a phase transition exists. This model is defined by 
\begin{equation}
E_\sigma = -J\sum_{\langle ij\rangle} \sigma_i\sigma_j - h\sum_{i=1}^N \sigma_i,
\label{isingh}
\end{equation}
which is just the (potential) energy as a function of the spins $\sigma_{i}=\pm 1$. We use $\sigma$ to collectively denote an entire spin configuration; 
$\sigma=(\sigma_1,\ldots,\sigma_N)$. With the interaction bonds $\langle ij\rangle$ restricted to nearest-neighbors on the simple 2D square lattice, 
the exact critical temperature for an infinite system is $T_c/J=2/\ln(1+\sqrt{2}) \approx 2.269$. The order parameter is the magnetization 
$\langle m\rangle=\langle \sigma_i\rangle$. It has the asymptotic $T\to T_c$ ($T<T_c$) form $m \propto |t|^\beta$, where $t$ is the reduced temperature, 
$t=(T-T_c)/T_c$, and the exponent $\beta=1/8$. Such {\it critical exponents} and other aspects of scaling behavior at continuous phase transitions 
will be discussed in this section.

The exact solution of the Ising model is very special, and normally one studies phase transitions in other ways. Critical exponents appear already in 
simple mean-field theories, but their values are typically not correct. Mean-field theory is nevertheless essential as a starting point, which we here
outline for the Ising model. The most important theoretical framework for phase transitions is the {\it renormalization group}, which explains how 
universal (depending only on symmetries and dimensionality, not details of the interactions) non-trivial exponents (i.e., different from the generic 
mean-field values) can arise (see, e.g., the book by Cardy \cite{cardy}). To compute critical exponents and other properties in an unbiased way, one 
normally uses {\it Monte Carlo simulations}, where spin configurations are stochastically sampled according to the Boltzmann distribution.\footnote{Other 
important methods include high-temperature series expansions \cite{series} and $\epsilon$-expansions (where $\epsilon=d_u-d$) around systems at their upper 
critical dimension $d_u$ \cite{epsilon}, where the exponents take their mean-field values. Monte Carlo simulation is normally the most reliable method, 
however.} In this section the basics of Monte Carlo simulations are outlined (for more details, see, e.g., the books by Landau and Binder \cite{binderlandau} 
and Newman and Barkema \cite{newman}). To motivate and illustrate finite-lattice calculations in general, we will also discuss how properties in 
the thermodynamic limit can be extracted from Monte Carlo data. 

\paragraph{Section outline}

The standard mean-field treatment of the Ising model is discussed in Sec.~\ref{meanfield}. In Sec.~\ref{montecarlo} the Monte Carlo method is
introduced and used to illustrate how phase transitions and symmetry-breaking can occur in practice for large systems (formally in the limit of infinite system size). 
In Sec.~\ref{scaling} we review the key 
aspects of critical behavior (defining exponents, etc.) in the thermodynamic limit, then discuss the {\it finite-size scaling} hypothesis and demonstrate 
its usage by analyzing Monte Carlo results for the 2D Ising model. First-order transitions are discussed in Sec.~\ref{firstordertrans}, using a frustrated 
2D Ising model as an example of finite-size scaling methods for detecting discontinuities. In Sec.~\ref{stiffkt} the important concept of {\it spin 
stiffness} (which corresponds to an elastic modulus of a solid) is introduced in the context of XY (planar vector) spin models. The scaling properties 
of the spin stiffness are illustrated with Monte Carlo results for the 3D and 2D XY models, the latter of which does not exhibit a normal phase transition 
into an ordered state, but exhibits a different {\it Kosterlitz-Thouless} transition into a critical low-temperature phase. In Sec.~\ref{qcpintro} 
we briefly discuss how the classical criticality concepts are generalized to quantum phase transitions.

\subsection{Mean-field theory of the Ising model} 
\label{meanfield}

In mean-field theories the environment of a subsystem of an infinite lattice is replaced by an external field representing the average interactions between
the subsystem and the environment. The subsystem can be a single spin or a cluster of several spins. Here we will just consider the simplest case 
of a single-spin calculation for the Ising model, i.e., the infinite environment of a single spin is replaced by an effective field.

It is convenient to write the hamiltonian (\ref{isingh}) in an extended form with general interactions $J_{ij}$ between all the spins (on an arbitrary
lattice), and also include an external magnetic field;
\begin{equation}
E_\sigma = -\frac{1}{2}\sum_{i=1}^N\sum_{j=1}^N J_{ij} \sigma_i\sigma_j - h\sum_{i=1}^N \sigma_i.
\label{isingh2}
\end{equation}
Note the factor $1/2$, which compensates for each interacting pair being included twice in the sum (and $J_{ii}=0$). We do not impose any restrictions of
the signs and magnitudes of $J_{ij}$, but for simplicity we assume that if there is a phase transition, the ordered state is ferromagnetic. To construct and 
motivate the single-spin approximation, we first group together all the interactions in (\ref{isingh2}) involving an arbitrary spin $i$;
\begin{equation}
E_i = -\sigma_i \left (\sum_j J_{ij} \sigma_{j} + h \right ).
\label{isingemf3}
\end{equation}
Note that there is no factor $\half$ in front of the sum here, and we have used $J_{ij}=J_{ji}$. 

We now assume that $h>0$, so that $m=\langle \sigma_i\rangle > 0$. We will investigate spontaneous ordering in the absence of the field, by eventually 
letting $h \to 0$. Adding and subtracting a constant, we can write the terms within the parentheses in (\ref{isingemf3}) as
\begin{equation}
\sum_j J_{ij} \sigma_j + h ~=~ m\sum_j J_{ij} + h + \sum_j J_{ij}(\sigma_j - m).
\label{isingemf2}
\end{equation}
In its most basic formulation, mean-field theory amounts to neglecting the second sum in (\ref{isingemf2})---the fluctuation term---after which we are 
left with the easily solvable problem of a single spin in an effective magnetic field of strength $J_sm+h$;
\begin{equation}
E_\sigma = -(J_sm + h)\sigma,
\label{emfallspins}
\end{equation}
where $J_s$ is the sum of the original couplings
\begin{equation}
J_s = \sum_j J_{ij},
\label{jsumdef}
\end{equation}
and we assume a translationally invariant system, so that this sum is independent of $i$. The magnetization $m$ in (\ref{emfallspins}) is at this 
stage unknown and will be determined through a self-consistency condition; $\langle \sigma\rangle=m$.

Mean-field theory in the present formulation can be justified if the neglected fluctuations are much smaller, on average, than the other terms in 
(\ref{isingemf2}), i.e., if
\begin{equation}
\delta_m =\frac{\left \langle \left | \sum_j J_{ij}(\sigma_j - m) \right | \right \rangle }{J_sm + h} \ll 1.
\end{equation}
Here $\langle \rangle$ denotes the expectation value under the actual probability distribution (the Boltzmann distribution) of the spins, which we cannot 
compute exactly. In principle the fluctuations can also be calculated within mean-field theory, as an internal consistency check. Even without doing any 
calculations, we can  roughly deduce the conditions under which $\delta_m$ will be small. Clearly, $\delta_m$ is small if there is substantial order, i.e., when 
$(1-m)=(1-\langle \sigma_j\rangle) \ll 1$, because then most $\sigma_j=1 \approx m$. This is the case if $h$ is large. It is also true for $h=0$ if the symmetry 
is spontaneously broken and $m$ is close to $1$, i.e., for $T \ll T_c$ if there is a phase transition. Moreover, $\delta_m$ can be small even if $m$ is not very 
large, if the sum over $j$ involves many non-zero (and relatively large) coupling constants $J_{ij}$---because of cancellations of fluctuations, the sum 
$\sum_j J_{ij} \sigma_j$ will then typically (in most of the statistically important spin configurations) be close to $m\sum_j J_{ij}$. In the extreme case of 
infinite-range uniform interactions, $J_{ij}=J/N$ for all $i,j$, and $N \to \infty$ (where we regard $J$ as a finite constant, e.g., $J=1$, in order to have a 
finite energy) all the fluctuations cancel out exactly and $\delta_m=0$ (and the mean-field theory is then exact). This holds true exactly also for short-range 
interactions on an infinite-dimensional lattice. Thus, in general, even if $m$ is not large, the fluctuation measure $\delta_m$ can be expected to be small for 
systems in high dimensions and/or for long-range interactions. These are then the conditions under which mean-field theory can be expected to be quantitatively 
accurate. Even in cases where it is not quantitatively accurate, mean-field theory can still provide valuable insights qualitatively.

\begin{figure}
\includegraphics[width=11cm, clip]{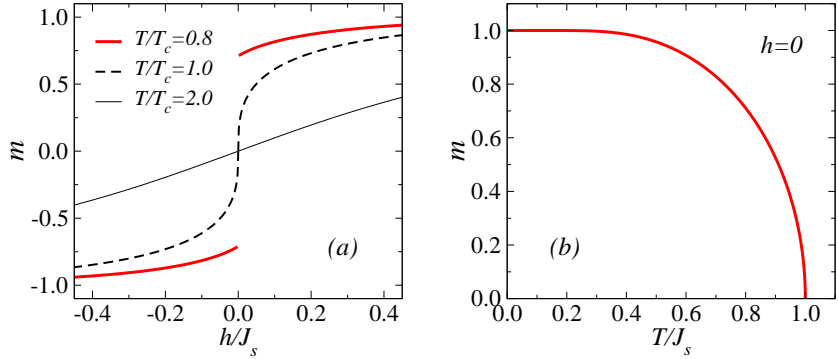}
\caption{Mean-field solution of the Ising model. (a) Magnetization versus external field for temperatures above, at, and below $T_c$.
The discontinuity for $h\to 0^+$ and $h \to 0^-$ corresponds to the spontaneous magnetization at $h=0$. (b) Temperature dependence of the 
spontaneous magnetization.}
\label{mfising}
\end{figure}

Let us now actually solve the mean-field problem (\ref{emfallspins}), i.e., the single-spin problem (\ref{emfallspins}) under the self-consistency 
condition $\langle \sigma\rangle=m$. The magnetization is
\begin{equation}
\langle \sigma\rangle = \frac{\sum_{\sigma} \sigma {\rm e}^{\sigma(J_sm +h)/T}}{\sum_{\sigma}{\rm e}^{\sigma(J_sm +h)/T}} = \tanh[(J_sm + h)/T],
\end{equation}
and thus the self-consistency condition reads
\begin{equation}
m=\tanh[(J_s m + h)/T].
\label{mftanh}
\end{equation}
This equation in general has to be solved numerically, which can be done easily using successive bracketing of the solution. For small $m$ and $h$, we can 
proceed analytically by expanding to leading order in $m,h$. First, when the external field $h=0$, $m=0$ is a solution for all $T$. Looking for other possible 
solutions, expanding (\ref{mftanh}) to third order in $x=J_sm+h$, $\tanh(x)=x-x^3/3$, we have
\begin{equation}
m^2=3\frac{T^2}{J_s^2} \frac{J_s - T}{J_s}, 
\end{equation}
from which we can identify the critical temperature $T_c=J_s$ below which the magnetization can be non-zero. The asymptotic $T \to T_c$ behavior is 
\begin{equation}
m = \left (3\frac{T_c-T}{T_c}\right )^{1/2},~~~~(h=0,~T<T_c). 
\label{mtmfform}
\end{equation}
We can also obtain the $h \to 0$ field dependence of the magnetization at $T_c$ from (\ref{mftanh}). Keeping the leading-order terms in $h$ of
the third-order expansion at $T=J_s$ we get
\begin{equation}
m = \left (3\frac{h}{J_s}\right )^{1/3},~~~~(T=T_c). 
\label{mhmfform}
\end{equation}
It is instructive to also look at the full numerical solution for $m$. Fig.~\ref{mfising}(a) shows some examples of the field dependence of $m$ at different
temperatures. For $T>T_c$ the behavior is analytic across $h=0$, with a singularity described by (\ref{mhmfform}) developing as $T\to T_c$. Below $T_c$ the 
behavior is discontinuous, which corresponds to a first-order transition versus $h$ between the $m<0$ and $m>0$ states. The discontinuity corresponds to a
spontaneous magnetization at zero field, the full numerical solution of which is graphed in Fig.~\ref{mfising}(b) and the asymptotic $T \to T_c^-$ behavior of 
which is given by (\ref{mtmfform}). Here the role of $h \to 0^+$ or $h \to 0^-$ in the calculation is to break the degeneracy between the $\pm m$ solutions. 
The degeneracy for $h=0$ also corresponds to co-existence of the two ordered states exactly at the first-order transition, as we will discuss more
generally in Sec.~\ref{firstordertrans}.

Translating the result $T_c=J_s$ with $J_s$ given in Eq.~(\ref{jsumdef}) gives $T_c=4J$ for the 2D Ising model with nearest-neighbor 
interactions of strength $J$. This is well above the correct value, $T_c/J \approx 2.269$. An over-estimation of $T_c$ can be expected on account of the
neglected fluctuations (which naturally lower $T_c$).

The exponents $\beta=1/2$ and $\delta=1/3$ in (\ref{mtmfform}) and (\ref{mhmfform}) are generically the values of these critical exponents within mean-field 
theories. Other exponents can also be computed \cite{cardy}. While power laws are indeed correct generic features of continuous phase transitions, the mean-field 
values are not correct in general. As already mentioned, for the 2D Ising model, $\beta=1/8$ from Onsager's exact solution. In three dimensions its 
value in the Ising universality class is $\beta \approx 0.33$, as determined using Monte Carlo simulations and series expansion methods (and also, less precisely, 
using field-theoretical methods and other analytical approaches). The mean-field critical exponents are exact in four and higher dimensions (four being the 
{\it upper critical dimension} for the Ising model---the dimensionality above which the mean-field critical exponents become exact). We will discuss critical 
behavior in greater detail further below, after developing the technical aspects of Monte Carlo simulations to study criticality in practice.

\subsection{Monte Carlo simulations of the Ising model} 
\label{montecarlo}

In a Monte Carlo simulation, the goal is to generate a sequence of spin configurations, $\sigma(1),\sigma(2),\ldots,\sigma(K)$, representing 
a statistically unbiased sample from the Boltzmann distribution, i.e., the probability $P(\sigma)$ of an arbitrary configuration $\sigma$ to be among 
the sampled ones should be proportional to the Boltzmann {\it weight} at temperature $T$;
\begin{equation}
W_\sigma = {\rm e}^{-E_\sigma/T},
\label{wsigma}
\end{equation}
where we work in units such that $k_B=1$. The actual 
(properly normalized) Boltzmann probability is $P_\sigma=W_\sigma/Z$, where $Z$ is the partition function
\begin{equation}
Z = \sum_{\sigma} {\rm e}^{-E_\sigma/T}.
\end{equation}
In Monte Carlo simulations we do not need the full partition function, only the un-normalized weights (\ref{wsigma}). It is also important to note that 
the sequence $\sigma(1),\ldots,\sigma(K)$ can be correlated (as we will discuss further below), but we only need the probability $P(\sigma) \propto W_\sigma$ 
and for now do not need to worry abot joint probabilities such as $P(\sigma_1,\sigma_2)$. 

\paragraph{The Metropolis algorithm}

The simplest way to generate a valid  sequence of configurations is with the Metropolis algorithm \cite{metropolis}. This kind of simulation starts 
from an arbitrary spin configuration $\sigma(1)$ (e.g., randomly generated). Thereafter, each successive $\sigma(k+1)$ is obtained stochastically from its 
predecessor $\sigma(k)$ according to a few simple steps based on flipping randomly selected spins with a probability related to the desired distribution.
We only need to store the current configuration and from now on suppress the ``time'' index $k$. We denote by $\sigma^{-i}$ the configuration obtained 
when the $i$th spin of $\sigma$ has been flipped; $\sigma^{-i}=(\sigma_1,\ldots,-\sigma_i,\ldots,\sigma_N)$. Normally one defines a Monte Carlo {\it sweep} 
as $N$ such flip attempts, so that $\propto N$ spins are flipped, on average, during this size-normalized unit of simulation time. A Monte Carlo sweep 
can be carried out according to the following simple algorithm:\footnote{This pesudocode segment resembles a computer language such as Fortran, but with 
mathematical notation making it easier to read. Such pseudocodes will be used throughout these notes to illustrate implementations of algorithms (showing 
the essential steps, but without trivial details that may in practice make real working code somewhat longer). Numbers in curly brackets, here $\{1\}$, 
will be used to label codes. In most cases the syntax will be rather self-explanatory and does not require any detailed discussions of definitions. In 
the segment above, and in many other ones to follow, {\bf random} denotes a random-number generator, with the square brackets $[1,\ldots,N]$ and $[0-1]$ 
indicating the range of uniformly distributed integers and floating-point numbers, respectively.}

{\code
\cia {\bf do} j=1,N \br
\cib    $i=\mathbf{random}[1,\ldots,N]$ \hfill \{1\}\break
\cib    {\bf if} ($\mathbf{random}[0-1]) < W_{\sigma^{-i}}/W_\sigma$) $\sigma_i=-\sigma_i$ \br
\cia {\bf enddo}
\code}

\noindent
This {\it Metropolis algorithm} is based on the {\it detailed balance principle}, which is a general theorem for a stochastic process (a Markov chain in some 
arbitrary configuration space) that should generate a probability distribution $W$. By this we mean, roughly speaking, 
that the set of sampled configurations should approach the distribution $W$ as the number of configurations is increased, independently of the initial
condition. Denoting by $P(A\to B)$ the transition probability of ``moving'' to configuration $B$ if the current one is $A$, the detailed balance principle 
states that the desired distribution is generated if all the transition probabilities satisfy the condition
\begin{equation}
\frac{P(A \to B)}{P(B \to A)} = \frac{W(B)}{W(A)},
\label{dbalance}
\end{equation}
for all pairs of configurations $A,B$ for which $P(A \to B) >0$. In addition, the sampling should be {\it ergodic}, i.e., any configuration $C$ with non-zero 
weight $W(C)$ must be reachable, in principle, with non-zero probability through a series of moves starting from an arbitrary configuration.

In the Metropolis algorithm, as implemented in code $\{1\}$, the transition probability consists of two factors;
\begin{equation}
P(\sigma \to \sigma^{-i})=P_{\rm select}(i) P_{\rm accept}(\sigma^{-i}), 
\label{pselpacc}
\end{equation}
where $P_{\rm select}(i)$ is the probability of randomly selecting spin $i$, which here always equals $1/N$. The full transition probability in (\ref{dbalance})
can therefore be replaced by the probability $P_{\rm accept}$ of actually carrying out (accepting) the spin flip. This probability is not unique.
In the Metropolis algorithm it is taken as
\begin{equation}
P_{\rm accept}(\sigma^{-i})={\rm min}\left [\frac{W_{\sigma^{-i}}}{W_\sigma},1\right ].
\end{equation}
One can easily confirm that this satisfies the detailed balance condition (\ref{dbalance}). A probability formally has to be $\le 1$, and 
this is taken care of above with the ``minimum of'' function. In code $\{1\}$, comparing the weight ratio with a random number in the range 
$[0,1)$ automatically achieves the same result. Note that the ratio of the Boltzmann factors depends only on the spins interacting with the flip-candidate
$\sigma_i$ (in the simplest case just its nearest neighbors) and can be rapidly evaluated. In more physical terms, a spin flip leading to a 
lower energy is always accepted, whereas a configuration with higher energy is accepted only with probability 
$P_{\rm accept}(\sigma^{-i})={\rm exp}[(E_{\sigma}-E_{\sigma^{-i}})/T]$.

The Metropolis algorithm leads to the correct distribution after some transient time, which depends on the initial configuration, the temperature, 
and the system size. In a  simulation one should therefore discard some number of configurations before ``measuring'' physical observables. After this 
{\it equilibration}, measurements are normally carried out after every or every few sweeps. 

In the case of the ferromagnetic Ising model, the most important quantity to calculate is the magnetization. It should be computed using all the spins to 
take advantage of {\it self-averaging} to improve the statistics. Thus, we will normally use
\begin{equation}
m = \frac{1}{N} \sum_{i=1}^N \sigma_i.
\label{mag1ising}
\end{equation}
We will here consider only simulations with the external field $h=0$. We have already discussed that fact that relevant symmetries of the 
hamiltonian (here the discrete spin-inversion symmetry) are not broken in simulations of finite systems. For the Ising model, one should therefore 
compute spin-inversion invariant expectation values such as $\langle m^2\rangle$ or $\langle |m|\rangle$ in order to detect the phase transition into the 
ferromagnetic state. 

\begin{figure}
\includegraphics[width=12.5cm, clip]{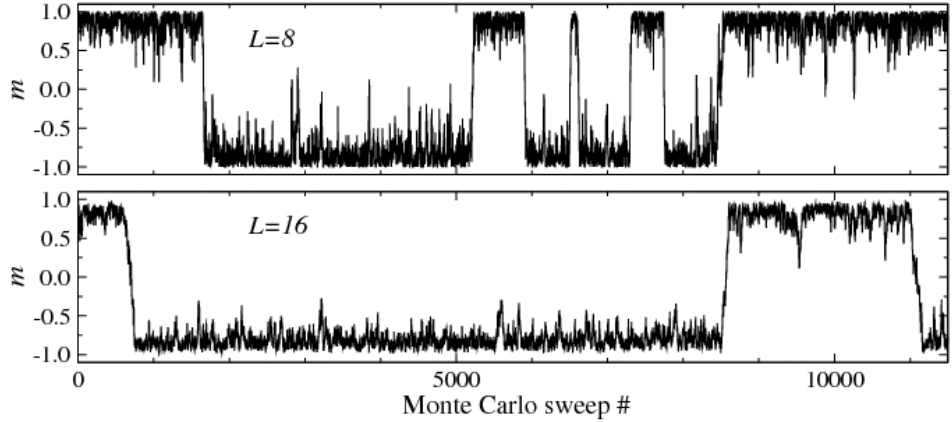}
\caption{Time series for the magnetization generated in Monte Carlo simulations of the Ising model on $L\times L$ lattices with $L=8$ (top) 
and $L=16$ (bottom) at temperature $T/J=2.2$ ($< T_c \approx 2.269$). In both cases the starting configuration was fully polarized, $m=1$, and 
the subsequent points are separated by $N=L^2$ Metropolis spin flip attempts (constituting one Monte Carlo sweep).}
\label{mtime}
\end{figure}

\paragraph{Symmetry-breaking and finite systems}

\begin{figure}
\includegraphics[width=10cm, clip]{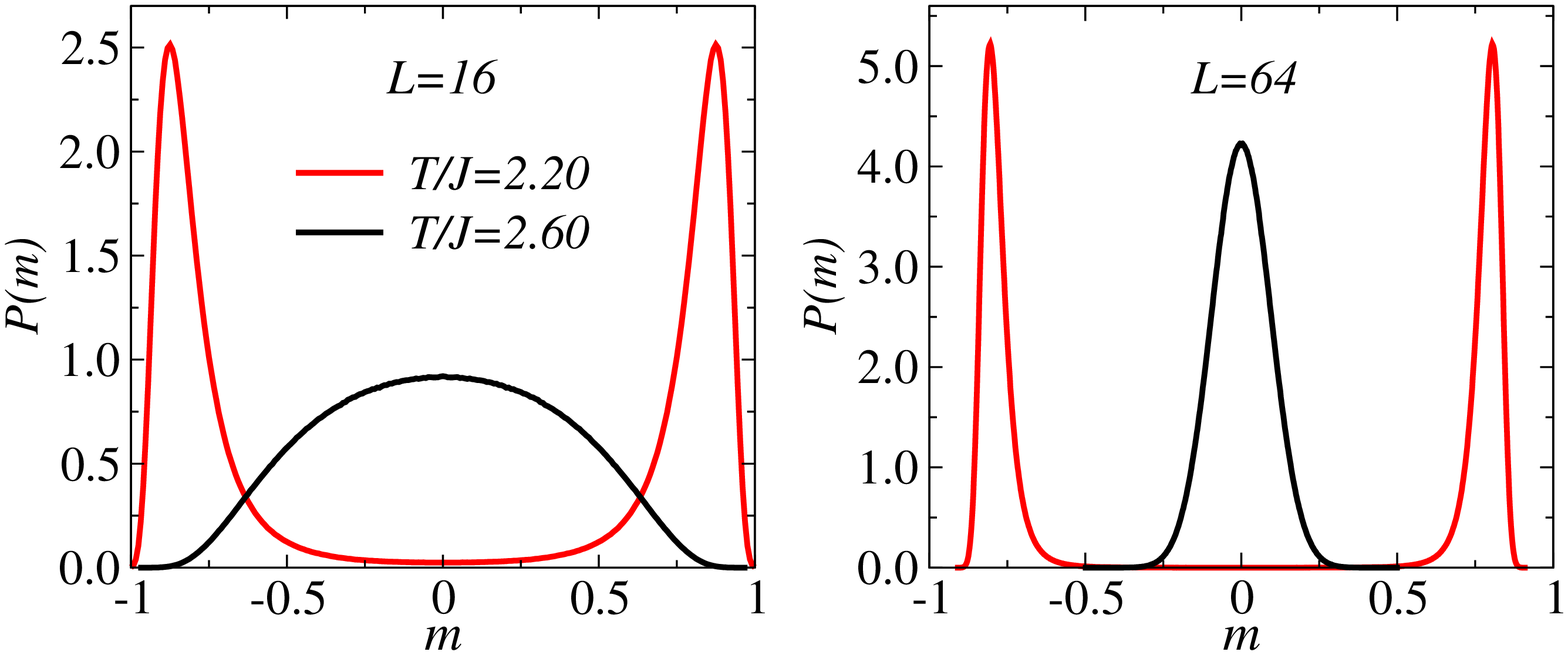}
\caption{Magnetization distributions for $L\times L$ Ising models with $L=16$ (left) and $64$ (right) at two temperatures below 
($T/J=2.2$) and above ($T/J=2.6$) the critical temperature $T_c/J \approx 2.27$.}
\label{mdist2d}
\end{figure}

To obtain a qualitative understanding of how symmetry-breaking in the thermodynamic limit is manifested in practice for large lattices, it is
useful to first look at an actual Monte Carlo time series for $m$. Examples for two small systems at a temperature below $T_c$ are shown 
in Fig.~\ref{mtime}. Here one can see that the magnetization fluctuates between positive and negative values, and that the typical time taken to 
reverse the sign of $m$ is longer for the larger system. Plotting the series over a longer time makes this clearer, but note that the time the 
system spends close to $m=0$ is visibly much smaller for $L=16$ than $L=8$, and this is of course directly related to a typically longer reversal 
time. It is clear from results such as these that the distribution of $m$ values is peaked at non-zero positive and negative values for $T<T_c$.
For $T>T_c$ the distribution is instead a single peak centered at $m=0$. Examples of such distributions are shown in Fig.~\ref{mdist2d}.
In thermodynamics language, these two qualitatively different distributions can be understood as a consequence of the free energy $F(m)=E(m)-TS(m)$ 
at low $T$ being dominated by the internal energy $E$ (which is low in large-$|m|$ configurations), and at high $T$ by the entropy $S$ 
(which is large for small $|m|$).

A double-peaked magnetization distribution at low temperature suggests that even a finite system can be considered as ordered, although the 
spin-inversion symmetry is not broken in the simulations. It appears very plausible that the typical reversal time should diverge with 
$L$ (and it is not difficult to check this with simulation data for several $L$ values and a suitably definition of a reversal), and then no 
reversals would take place for large $L$, even during very long simulations. The reason for this divergent time scale is that, 
in order for the magnetization to reverse, a series of local spin flips must necessarily take the system through many configurations 
with $m\approx 0$, which have increasingly high energy for increasing system size (and, thus, lower Boltzmann probability). For a large system 
the distribution for $T<T_c$ is therefore only sampled among the subset of configurations with fixed sign of $m$---the stochastic process in 
practice becomes {\it non-ergodic}. Broken symmetry and non-ergodic sampling are manifested strictly only for $N=\infty$, but in practice also for large 
but finite systems on time scales less than the typical magnetization reversal time. This time scale of course depends on the details of how the spin 
configurations are thermally sampled---in Monte Carlo simulations and in a real magnets---but diverges as $N\to \infty$ for any local sampling scheme. 

Normally, in Monte Carlo simulations one does not investigate the time series and the full distribution of physical quantities (although some
times this is useful). By computing $\langle m^2\rangle$ or $\langle |m|\rangle$, we do not have to worry about the time scale of reversals. We 
normally want to extrapolate finite-$N$ results to the thermodynamic limit, where $|\langle m\rangle| = \langle |m|\rangle = \langle m^2\rangle^{1/2}$ 
(but note that $\langle m^2\rangle^{1/2}\not=\langle |m|\rangle$ for finite $N$, because of the finite width of the peaks in the $m$-distribution).

\paragraph{Autocorrelations and statistical errors}

Before calculating expectation values, we have to discuss how to analyze Monte Carlo data statistically. Consecutive configurations generated with the 
Metropolis algorithm are not statistically independent---only configurations separated by a number of sweeps much larger than the {\it autocorrelation time} 
are statistically independent. The {\it autocorrelation function} for a quantity $Q$ is defined as
\begin{equation}
A_Q(t)=\frac{\langle Q(i+t)Q(i)\rangle - \langle Q\rangle^2}{\langle Q^2\rangle - \langle Q\rangle^2},
\end{equation}
where $t$ and $i$ denote simulation time, normally in units of the Monte Carlo sweeps defined above, and averages are over the reference time $i$. The
normalization is such that $A_Q(0)=1$ and $A_Q(t \to \infty)=0$. The asymptotic decay is exponential, $A_Q(t) \sim {\rm e}^{-t/\tau_Q}$, which can be used 
to the define the autocorrelation time $\tau_Q$. Normally one instead uses the {\it integrated autocorrelation time}, which also contains contributions from 
the often dominant non-asymptotic behavior;
\begin{equation}
\tau^{\rm int}_Q =\frac{1}{2} + \sum_{t=1}^\infty A_Q(t).
\end{equation}
Here we will not discuss autocorrelations at length, but only summarize their underlying role in determining the statistical precision 
(``error bars'') of computed quantities.

The autocorrelation time for $m$ of the Ising model roughly corresponds to the typical time between magnetization reversals (as in Fig.~\ref{mtime}). 
Other quantities, such as $m^2$ and $|m|$, that are not sensitive to the sign of $m$, have shorter autocorrelation times (however, often much longer 
than a single Monte Carlo sweep). A long autocorrelation time does not bias a computed average (i.e., it is not wrong to measure $Q$ after every Monte 
Carlo sweep even if $\tau_Q \gg 1$), provided that the total simulation time is much longer than $\tau_Q$. The autocorrelation time does, however, 
come into play (explicitly or implicitly) when computing statistical errors.

To calculate the statistical errors, one can subdivide a simulation into a number $B$ of {\it bins}, each containing some number 
$M$ of Monte Carlo sweeps. For some quantity $Q$, averages  $\bar Q_b$, $b=1,\ldots,B$ are computed over each bin, and the final average $\bar Q$ and 
error bar $\sigma_Q$ (one standard deviation of the average of the bin averages) are calculated according to
\begin{equation}
\bar Q = \frac{1}{B}\sum_{b=1}^B \bar Q_b,~~~~~~~\sigma^2_Q = \frac{1}{B(B-1)}\sum_{b=1}^B (\bar Q_b - \bar Q)^2.
\end{equation}
The final estimate of the true expectation value $\langle Q\rangle$ should then be quoted as $\bar Q \pm \sigma_Q$. 

The reason for binning the data is that, according to the central limit theorem, the distribution of bin averages is Gaussian for large $M$ (unlike 
the distribution of individual measurements, as seen clearly for the magnetization in Fig.~\ref{mdist2d}), and the computed error bar then
has a well defined unique meaning (e.g., we know that the probability of $\langle Q\rangle$ being within one error bar of $\bar Q$ is about 
$2/3$). This is only true if the bin length $M$ is also much longer than the autocorrelation time, so that the bin averages can be regarded as 
statistically independent. If that is the case, the error bar should depend only on the total number of sweeps; $\sigma_Q \propto 1/\sqrt{MB}$, 
where the factor of proportionality is $\propto \sqrt{\tau_Q}$ (and of course also depends on the detailed form of the distribution of the 
individual measured $Q$ values). It is not necessary to calculate $\tau_Q$ explicitly. If there is any doubt about the bins being sufficiently long,
one can check this by using a rather large number of bins (e.g., in the range 100-1000) and saving all the bin averages on disk. The data can then 
be  {\it re-binned} into longer bins post-simulation, and the convergence of $\sigma_Q$  as a function of the bin length can be tested. Saving the 
bin averages on disk is always advisable, not only for the purpose of analyzing the error bars, but also in order to make it easy to add more data 
at some later time (to improve the results, if needed). 

An important point that has to be mentioned is that the autocorrelation time of any local updating scheme, e.g., the Metropolis algorithm, diverges for 
$T\to T_c$ and $N\to \infty$. We previously saw an example of this in the magnetization reversal time of the Ising model, but divergent autocorrelation 
times also affect any quantity (in particular $m^2$ and $|m|$) that is sensitive to the fluctuations of the magnitude of $m$ (which in practice is the case 
for most quantities of interest). It is therefore difficult to obtain good results for large systems close to a critical point. In many cases, including 
the Ising model, this problem can be solved (or almost solved) by using {\it cluster algorithms} \cite{swendsenwang,wolff}, where clusters of spins 
(constructed so as to satisfy detailed balance) are flipped collectively (instead of flipping individual spins one-by-one). While we will consider an analogous 
{\it loop-cluster} algorithm for QMC simulations in Sec.~\ref{sec_sse}, we do not have to discuss classical cluster algorithms here; they have been 
described extensively in the literature (e.g., in \cite{binderlandau} and \cite{newman}). It should be noted, however, that the Ising 
results to be discussed below, for systems with up to $1024^2$ spins, were in fact obtained using a cluster algorithm. It would be very difficult to 
generate (within reasonable time) data of the same quality using the Metropolis scheme for such large systems in the critical region.

\subsection{Finite-size scaling and critical exponents}
\label{scaling}

\begin{figure}
\includegraphics[width=12cm, clip]{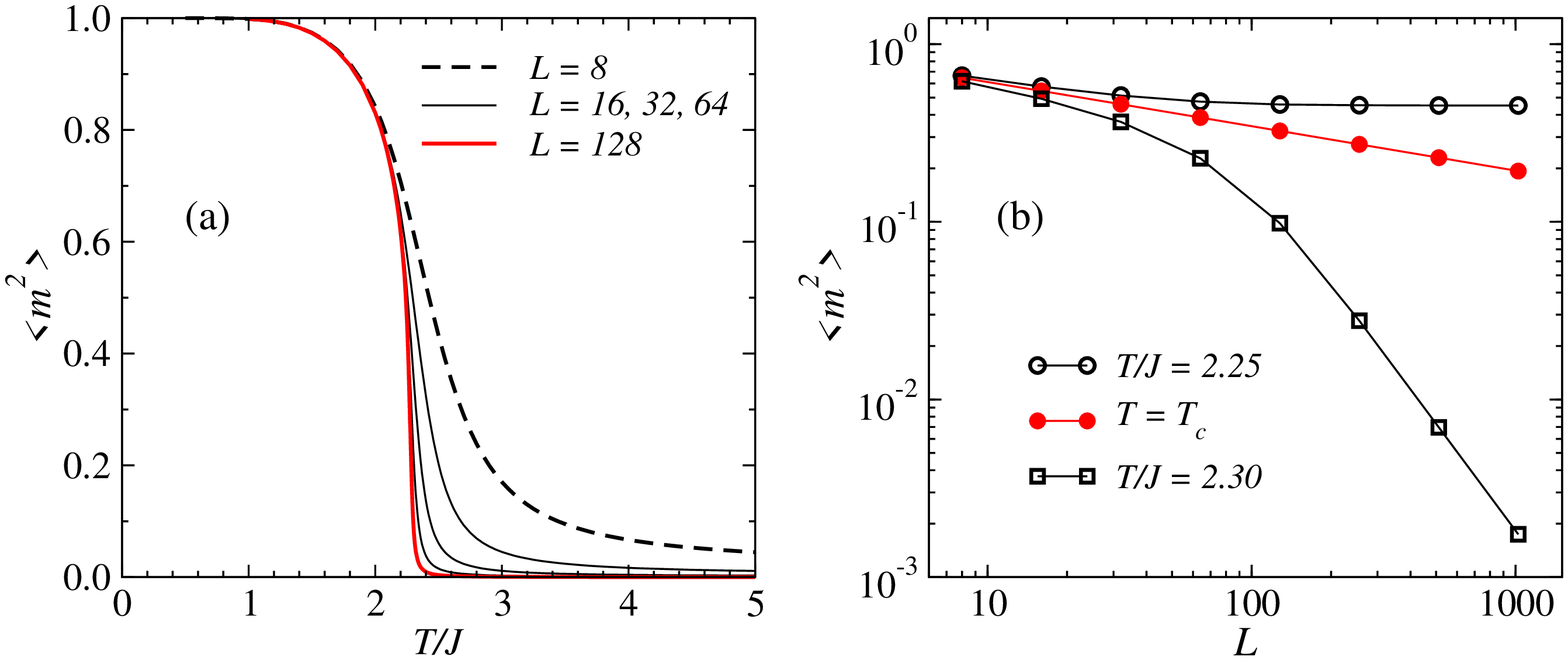}
\caption{(a) The squared magnetization as a function of temperature for several $L\times L$ Ising systems. (b) The finite-size dependence at, below, 
and above $T_c$. At $T_c$ the behavior is a power law, $m^2 \sim L^{-1/4}$, which on this log-log scale corresponds to a line with slope $-1/4$. For
$T>T_c$ the decay to zero is of the form $L^{-2}$, and this is also the rate of convergence for $T<T_c$.}
\label{ismag2}
\end{figure}

Fig.~\ref{ismag2}(a) shows the temperature dependence of the squared magnetization $\langle m^2\rangle$ of the 2D Ising model for several lattice sizes.
An increasingly sharp feature develops with increasing $L$ in the neighborhood of the known $T_c$, and it appears very plausible that $\langle m^2\rangle$ 
vanishes for $N\to \infty$ for $T>T_c$ and remains non-zero for $T<T_c$. In this plot, error bars are not visible because they are much smaller than the 
line widths. The data were generated on a very fine $T$-grid to produce continuous-looking curves.

The size dependence of $\langle m^2\rangle$ at three different temperatures is shown in Fig.~\ref{ismag2}(b). For $T < T_c$ it converges with 
$L$ to a non-zero value, while for $T>T_c$ it decays to $0$ as $L^{-2}$. Exactly at $T_c$ the decay follows a non-trivial power-law; 
$\langle m^2\rangle \propto L^{-2\beta/\nu}$, where $\beta=1/8$ is the same exponent as in the $N=\infty$ magnetization for $T<T_c$, and $\nu$ (the 
correlation-length exponent, to be discussed below) equals $1$ for this model. This behavior is an example of finite-size scaling. A perfect power-law 
behavior at criticality holds strictly only when $L\to \infty$, while for small systems there can be significant 
{\it corrections to scaling} (which are 
unusually small in the case of the Ising model discussed here). Normally we do not know the exact value of $T_c$ (which may of course be one reason why a 
simulation is carried out), and procedures for locating the critical point then have to be devised. There are many ways to do this, all building in some 
way or another on the fact that the order parameter and related quantities should behave as non-trivial power laws (with known or unknown exponents) 
for large lattices at a critical point.

Note that the squared magnetization is related to the spin correlation function,
\begin{equation}
C({\bf r}_{ij})=\langle \sigma_i\sigma_j\rangle,
\end{equation}
where ${\bf r}_{ij}$ is the distance between the spins. If there is long-range order, then $C(r) \to \langle m^2\rangle$ when $r\to \infty$ in an
infinite lattice. For finite systems, the same is true for $C(r_{\rm max})$ in the limit $L\to \infty$, where $r_{\rm max}$ is the longest distance on a 
periodic lattice, e.g., $r_{\rm max} = \sqrt{2}L$ for an $L\times L$ lattice. If there is no order, then $C(r) \to 0$, according to a power-law at $T_c$ 
and exponentially for $T>T_c$. Below $T_c$, in the infinite syztem the ``connected'' correlation function,
\begin{equation}
C^*(r) = C(r)-\langle m\rangle^2,
\label{connectedcorr}
\end{equation}
decays to zero exponentially 
as $r \to \infty$. The exponential forms of both $C(r)$ and $C^*(r)$ are characterized by a correlation length $\xi$, which diverges as $T\to T_c$ 
from either side. Note also that $\langle m^2\rangle$ can be written exactly as a sum over the spin correlations;
\begin{equation}
\langle m^2\rangle =\frac{1}{N}\sum_{\bf r} C({\bf r}).
\end{equation}
The $L^{-2}$ convergence of $\langle m^2\rangle$ (both below and above $T_c$) is simply related to the finite correlation length.

Here it should be pointed out that the nature of the spin correlations below $T_c$ depends on the symmetry of the order parameter we are dealing with. 
For a vector order parameter, e.g., in the Heisenberg model (with classical or quantum spins), the correlations of the spin components parallel 
(longitudinal) to the order parameter of a symmetry-broken system decay exponentially, as in the Ising model. The transverse correlations decay as 
a power law, however. This is related to the continuous symmetry of the order parameter. In the Ising model the symmetry is discrete---the ordered state 
breaks the spin-inversion symmetry and the free-energy cost of a local magnetization fluctuation is large (proportional to the boundary of a flipped 
domain). This leads to the exponentially decaying (connected) correlation function in the ordered phase. In the Heisenberg model, on the other hand, there
are gapless spin-wave excitations (which are excitations of the spins in the directions transverse to the ordering direction and are more generally called 
Goldstone modes). Fluctuations due to these lead to a power-law form of the transverse spin correlations. The transverse correlation length is then formally 
infinite, and in a finite-lattice calculation in which the rotational symmetry is not explicitly broken (so that the computed $\langle m^2_s\rangle$ contains 
contributions from both longitudinal and transverse correlations), the size corrections to $\langle m^2_s\rangle$ are $\propto 1/L$. We already discussed 
this behavior in the ground state of the 2D $S=1/2$ Heisenberg antiferromagnet in Sec.~\ref{qtrans}, and Fig.~\ref{jjmag2}(b) shows the $1/L$ corrections 
very clearly.

\subsubsection{Scaling and critical exponents}

To discuss finite-size scaling close to a critical point in more detail, we first have to review some basic aspects of critical phenomena in the thermodynamic 
limit. We will only list some of the key results and definitions here; see any standard text on critical phenomena (e.g., the book by Cardy \cite{cardy}) for 
more details.

The correlation length is one of the most important concepts underlying the theory of phase transitions and critical phenomena. In an infinite system, 
as the critical temperature is approached, the correlation length diverges according to a power law;
\begin{equation}
\xi \sim |t|^{-\nu}.
\label{xidiv}
\end{equation}
The exponent $\nu$ is the same upon approaching $T_c$ from above or below, but the prefactor in (\ref{xidiv}) is in general different for $t \to 0^+$ and 
$t \to 0^-$ (however, ratios of these prefactors---called amplitude ratios---are often universal). Within mean-field theory, $\nu=1/2$ . Exactly at $T_c$, 
although the correlation length is formally infinite, the system is not yet ordered. Instead, for an infinite system the correlation function exactly at $T_c$ 
has the power-law form, 
\begin{equation}
C({\bf r}) \sim r^{-(d-2+\eta)},
\label{cofreta}
\end{equation}
where $d$ is the dimensionality of the system and $\eta$ is another critical exponent (also called the anomalous dimension, because it can be related
to the fractal dimensionality of ordered domains at the critical point), the mean-field value of which is $\eta=0$. Long-range order sets in only infinitesimally 
below $T_c$, where the asymptotic long-distance correlation approaches a constant; $C(r \to \infty) = \langle m^2\rangle$.

We have already discussed the behavior of the order parameter in the symmetry-broken state as $t \to 0^-$;
\begin{equation}
\langle m\rangle \sim |t|^\beta.
\end{equation}
We will also be interested in the corresponding susceptibility, defined as
\begin{equation}
\chi = \frac{d\langle m\rangle}{dh}\Big |_{h\to 0} = \frac{N}{T}\bigl ( \langle m^2\rangle - \langle |m|\rangle^2 \bigr ),
\label{chidef}
\end{equation}
where $h$ is the strength of a field coupling to the order parameter, e.g., in the case of the Ising ferromagnet a term $-h\sum \sigma_i$ in the hamiltonian. 
The last expression in (\ref{chidef}) shows explicitly how $\chi$ is directly related to the fluctuations of the order parameter. The susceptibility 
diverges at $T_c$;
\begin{equation}
\chi \sim |t|^{-\gamma}.
\label{chidiv}
\end{equation}
Thus, upon approaching a critical point, the system becomes infinitely sensitive to a field $h$ coupling to the order parameter, and exactly at $T_c$ 
the linear response form $\langle m\rangle = \chi h$ ceases to be valid. Instead, at $T_c$, the order parameter depends on the (weak) field as 
$h^{1/\delta}$. According to the result in Eq.~(\ref{mhmfform}), the mean-field value of the exponent here is $\delta=3$.

The specific heat is also singular at $T_c$,
\begin{equation}
C \sim |t|^{-\alpha},
\end{equation}
where $\alpha$ can be positive or negative. In mean-field theory $\alpha=0$. When $\alpha < 0$, there is no divergence, only a cusp singularity at $T_c$. 
In some cases, e.g., for the 2D Ising model, $\alpha=0$ but there is still a weak, logarithmic divergence of the specific heat.  

The critical exponents $\nu,\beta,\eta$, etc., that we have encountered above are not all independent of each other. Relationships between the exponents are 
explained by the renormalization group theory, which, in the type of order-disorder transitions we are discussed here, shows that there are two more fundamental 
exponents, in terms of which the physically observable exponents can be written \cite{cardy}. Exponent relations had been found using other arguments even before 
the advent of this ultimate theory of phase transitions (in the most complete form by Widom in the mid 1960s), e.g.,
\begin{eqnarray}
&& \gamma = \nu (2-\eta), \nonumber \\
&& \nu d = 2-\alpha, \\
&& \alpha + 2\beta + \gamma = 2. \nonumber
\label{exprelations}
\end{eqnarray}
Such relations are very useful for checking the consistency of numerical calculations of the exponents. Exponent relations involving the dimensionality $d$ are 
called {\it hyper-scaling relations} and are less generic than the other relations. They are not applicable, e.g., within mean-field theory, and, therefore, for any 
system at or above the upper critical dimension. 

\subsubsection{The finite-size scaling hypothesis}
\label{finitsizescaling}

The basic assumption underlying finite-size scaling theory \cite{fisher72} is that deviations from the infinite-size critical behavior should occur 
when the correlation length 
$\xi$ (of the infinite system) becomes comparable with the finite system length $L$. If $L \gg \xi$, the fact that the system is finite should be irrelevant and 
the infinite-size behavior applies. On the other hand, if $L \ll \xi$, then $L$, not $\xi$, should be the most relevant length-scale. In order to see how the two 
length scales come into play, it is useful to express quantities of interest in terms of the correlation length. Consider a quantity $Q$ which exhibits a 
power-law divergent behavior at $T_c$ (reduced temperature $t \to 0$),
\begin{equation}
Q \sim |t|^{-\kappa},
\label{adepont}
\end{equation}
e.g., the susceptibility (in which case $\kappa=\gamma$). We can use Eq.~(\ref{xidiv}) to express $|t|$ as a function of the 
correlation length; 
\begin{equation}
|t| \sim {\xi}^{-1/\nu},
\label{tdeponxi}
\end{equation}
and using this we can write $Q$ as
\begin{equation}
Q \sim {\xi}^{\kappa/\nu}.
\label{adeponxi}
\end{equation}
This form should apply for $\xi \ll L$, but when $\xi \sim L$ the divergence can no longer continue on the finite lattice. The maximum value $Q_{\rm max}$ 
attainable by $Q$ on the finite lattice should then be obtained by replacing $\xi \to L$ in Eq.~(\ref{adeponxi}), giving
\begin{equation}
Q_{\rm max}(L) \sim {L}^{\kappa/\nu}.
\label{amaxl}
\end{equation}
In the same way, from Eq.~(\ref{tdeponxi}) we can also deduce the scaling of the reduced temperature at which $\xi$ reaches $L$, which should also be the 
temperature at which the maximum value of the quantity $Q$ is reached;
\begin{equation}
|t_{\rm max}(L)| \sim L^{-1/\nu}.
\label{tmaxl}
\end{equation}
It should be noted that this is just a proportionality, and the number in front of $L^{-1/\nu}$ depends on the quantity. 
The shift also applies to non-divergent quantities---any feature which develops singular behavior as $T \to T_c$ should shift at the rate $L^{-1/\nu}$. If there 
is a well-defined maximum or other distinguishable feature in some quantity at $T=T^*(L)$, then this temperature can be used as a size-dependent critical 
temperature (and, again, this temperature is not unique but depends on the quantity considered).

\begin{figure}
\includegraphics[width=13cm, clip]{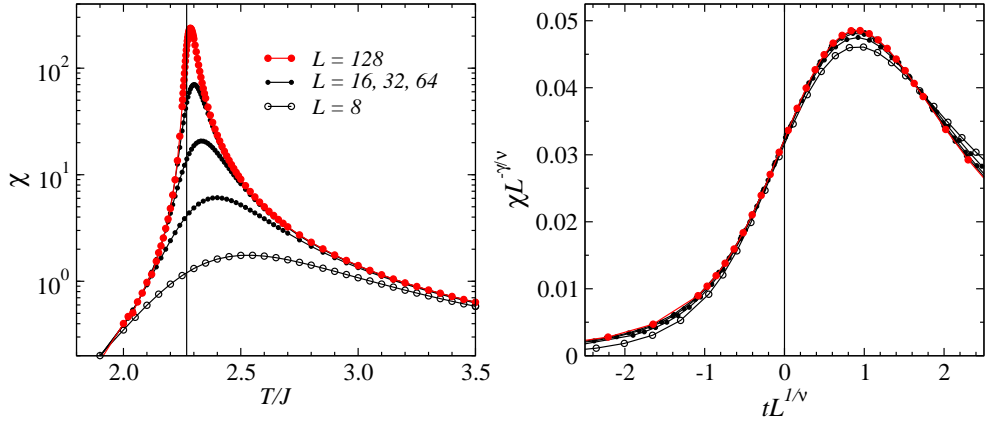}
\caption{Monte Carlo results for the susceptibility (\ref{chidef}) of the Ising model on several different $L\times L$ lattices. (a) shows the temperature 
dependence, with the vertical line indicating $T_c$. Note the vertical log scale. In (b) the data has been scaled using the exact values of the Ising exponents,
$\gamma=7/4$ and $\nu=1$, and the exact value of $T_c$ in $t=(T-T_c)/T_c$.}
\label{isingsusc}
\end{figure}

The finite-size scaling laws (\ref{amaxl}) and (\ref{tmaxl}) follow from a more general finite-size scaling hypothesis \cite{fisher72}, which, like the 
scaling theory for infinite systems, was initially proposed based on phenomenological considerations, but was later derived using the renormalization group 
theory. The hypothesis is that an observable which is singular at $T_c$ in the thermodynamic limit scales with the system size close to $T_c$ as a power of $L$ 
multiplied by a non-singular function of the ratio $\xi/L$. Any singular quantity (not necessarily divergent) should thus be of the form
\begin{equation}
Q(t,L) = L^\sigma f(\xi/L),
\label{fsxilhypo}
\end{equation}
which, using $\xi \sim |t|^{-1/\nu}$, we can also write as
\begin{equation}
Q(t,L) = L^\sigma g(tL^{1/\nu}).
\label{afscaling}
\end{equation}
This scaling law should hold both above ($t>0$) and below ($t<0$) the critical point. Exactly at $T_c$, we recover the size-scaling $Q(0,L) \sim L^\sigma$. To 
relate $\sigma$ to the standard critical exponents, we can use the fact that, for fixed $t$ close to $0$, as the system grows the behavior for any $t\not=0$ 
eventually has to be given by Eq.~(\ref{adepont}); $Q(t,L\to \infty) \sim |t|^{-\kappa}$ (where $\kappa$ is negative for a singular non-divergent quantity, 
e.g., the for the order parameter we have $\kappa=-\beta$). To obtain this form, the {\it scaling function} $g(x)$ in (\ref{afscaling}) must asymptotically 
behave as $g(x) \sim x^{-\kappa}$ for $x \to \infty$. In order for the size-dependence in (\ref{afscaling}) to cancel out, we therefore conclude that 
$\sigma  = \kappa/\nu$, i.e.,
\begin{equation}
Q(t,L) = L^{\kappa/\nu} g(tL^{1/\nu}).
\label{atlscaling}
\end{equation}
To extract the scaling function $g(x)$ using numerical data, one can define
\begin{equation}
y_L=Q(t,L)L^{-\kappa/\nu},~~~~~x_L=tL^{1/\nu},
\label{xlyldef}
\end{equation}
and plot $y_L$ versus $x_L$ for different system sizes. If the scaling hypothesis is correct, data for different (large) system sizes should fall onto the same 
curve, which then is the scaling function (this is referred to as curves {\it collapsing} onto each other); $g(x)=y_{L\to \infty}(x)$. Fig.~\ref{isingsusc} 
illustrates this using Monte Carlo data for the magnetic susceptibility of the 2D Ising model. The peak location in panel (a) clearly moves toward the known 
$T_c$ with increasing $L$. After scaling the data according to the above procedures, as shown in panel (b), the curves indeed  collapse almost onto each other 
close to $t=0$, but further away from the critical point deviations are seen for the smaller systems. These are due to corrections to scaling, which in 
principle can be described with subleading exponents.

We can apply the scaling form (\ref{atlscaling}) to the correlation length itself, for which $\kappa=\nu$ and the $L$-scaling is independent of 
model-specific exponents. In cases where the universality class is not known {\it a priori}, this is useful for extracting the exponent $\nu$ by curve-collapsing
$\xi/L$ data without having to simultaneously adjust another exponent $\kappa$ in (\ref{atlscaling}). 

\paragraph{Practical definitions of the correlation length}

The correlation length can be defined in various ways, not necessarily just based on the asymptotic decay of the correlation function (which is often 
difficult to extract reliably). One practical and commonly used correlation length definition is based on the Fourier transform of the correlation 
function, often called the (static) structure factor,
\begin{equation}
S({\bf q})=\langle \sigma_{\bf -q}\sigma_{\bf q}\rangle = \sum_{{\bf r}}{\rm e}^{-i{\bf q}\cdot {\bf r}}C({\bf r})
=\sum_{{\bf r}}\cos({\bf q}\cdot {\bf r})C({\bf r}),
\label{sqdefising}
\end{equation}
where $\sigma_{\bf q}$ is the Fourier transform of an individual spin configuration,
\begin{equation}
\sigma_{\bf q}=\frac{1}{\sqrt{N}}\sum_j \sigma_j {\rm e}^{-i{\bf q}\cdot {\bf r}_j}.
\label{sigmaqdefising}
\end{equation}
We denote by ${\bf Q}$ the wave-vector of the dominant correlations---for a ferromagnet $Q=0$, for a 2D antiferromagnet 
${\bf Q}=(\pi,\pi)$, etc. To simplify the notation, ${\bf q}$ will be used for the {\it deviation} from ${\bf Q}$. Then $q_1=2\pi/L$ corresponds 
to one of the wave-vectors closest to ${\bf Q}$, e.g., ${\bf Q} + (2\pi/L)\hat {\bf x}$, where $\hat {\bf x}$ is the reciprocal-space unit vector in 
the $x$-direction. 

A correlation length $\xi_a$ can be defined using the structure factors at $q=0$ and $q_1$;
\begin{equation}
\xi_a = \frac{1}{q_1} \sqrt{\frac{S(0)}{S(q_1)}-1}.
\label{xiadef}
\end{equation}
One can show that, for a $d$-dimensional lattice,
if the correlation function is given by the {\it Ornstein-Zernike form} (obtained in mean-field treatments \cite{cardy}),
\begin{equation}
C_{OZ}({\bf r}) \sim r^{-\frac{1}{2}(d-2)}{\rm exp}(-r/\xi),
\label{ozform}
\end{equation}
then $\xi_a$ is related to the original correlation length $\xi$ appearing in the exponential decay of this correlation function according to
\begin{equation}
\xi_{\rm a} = \xi \sqrt{\frac{(1+d)(3+d)}{8d}}.
\label{xiaxi}
\end{equation}
Thus, for $d=1$ and $3$, $\xi_a=\xi$, whereas the 2D case is special, with $\xi_{\rm a} = \xi (15/16)^{1/2}$ (or, one may say that 
$d=1,3$ are special cases, since the factor differs from one also for all $d>3$). The Ornstein-Zernike form is normally valid in disordered phases 
for $r \gg \xi$ \cite{cardy}. Deviations from this form at short distances imply that (\ref{xiaxi}) is not true exactly, but regardless of the 
short-distance behavior this relation holds exactly when $\xi \to \infty$.

In the case of a long-range ordered classical system, $\xi_a$ normally diverges as $T\to 0$, for any $L$, because there are no fluctuations
in the ground state (and, thus, the structure factor vanishes for $q\not =0$. In order to remove the contributions from the non-decaying part 
of the correlation function in an ordered system, we can use the connected correlation function (\ref{connectedcorr}). Although $\langle m\rangle^2$ 
is not uniquely defined for a finite system, one can, e.g., subtract $C({\bf r}_{\rm max})$. As an alternative, we can use a different correlation 
length definition, based on the structure factor at $q_1$ and $q_2=2q_1=4\pi/L$;
\begin{equation}
\xi_{\rm b} = \frac{1}{q_1}\sqrt{\frac{S(q_1)/S(q_2)-1}{4-S(q_1)/S(q_2)}}.
\label{xibdef}
\end{equation}
One can show that (\ref{xiaxi}) also holds for this definition if $C({\bf r})$ is given by the Ornstein-Zernike correlation function, also when a constant 
corresponding to the long-range order is added for $T<T_c$ [since this affects only $S({0})$].

\begin{figure}
\includegraphics[width=12.5cm, clip]{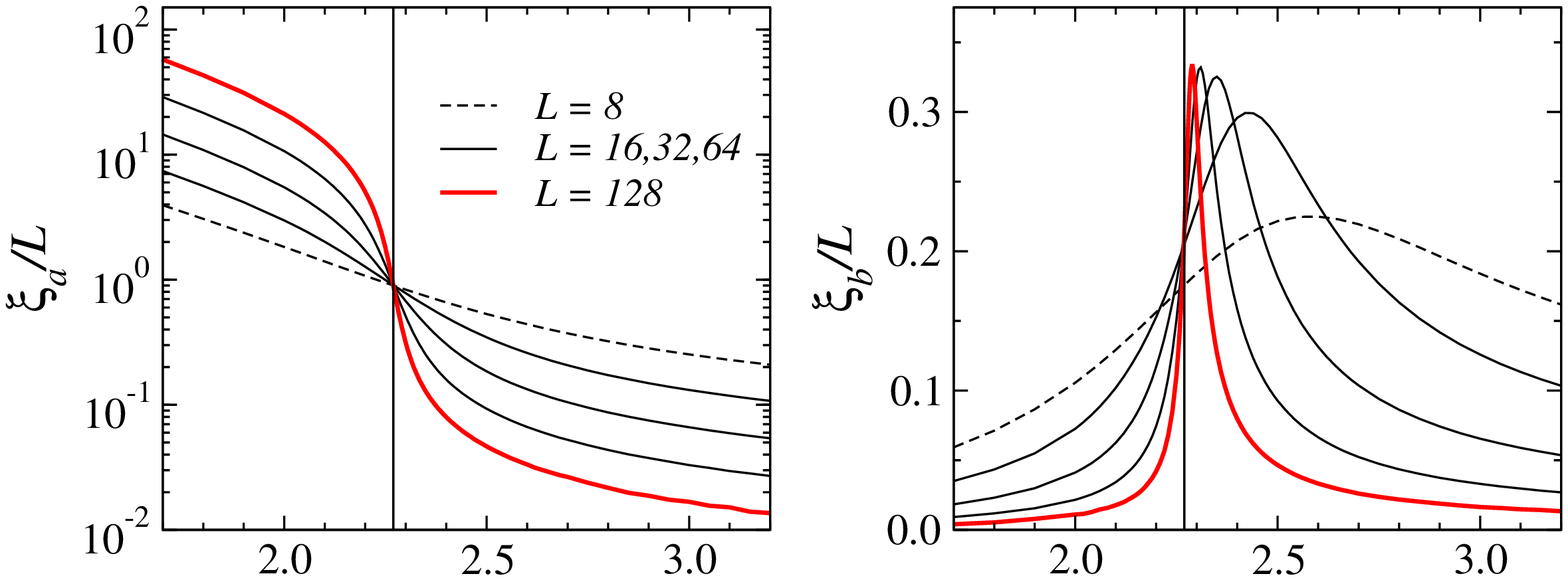}
\caption{Temperature dependence of the two correlation length definitions, Eqs.~(\ref{xiadef}) and (\ref{xibdef}), normalized by the size $L$ 
for $L\times L$ Ising models. The vertical lines indicate $T_c$.}
\label{isingxi2d}
\end{figure}

Fig.~\ref{isingxi2d} shows results for $\xi_a/L$ and $\xi_b/L$ for the 2D Ising model. Note that a logarithmic scale is used in the case of the 
$T\to 0$ divergent $\xi_a/L$, while $\xi_b/L$ is convergent and graphed on a linear scale. Both quantities exhibit size independence (curves crossing 
each other) at $T_c$, but the values there are clearly different. This is because the Ornstein-Zernike form of the correlation function applies 
asymptotically only for $T>T_c$, and there is no reason why the two definitions $\xi_a$ and $\xi_b$  should agree exactly at $T_c$ (although their
values should be related). Their values are very similar for large systems close to $T_c$ in the disordered phase. The crossing points (or peak location 
of $\xi_b$) can be used to extract $T_c$ in systems where it is not known. The temperature axis can also be scaled in the same way as in Fig.~\ref{isingsusc} 
to extract the correlation-length exponent.

Note that for the small number of ${\bf q}$-points needed,  $S({\bf q})$ can be efficiently evaluated using the Fourier transform 
(\ref{sigmaqdefising}) of the spin configurations generated in Monte Carlo simulations. Since the structure factor is real-valued we have
\begin{equation}
S({\bf q}) = \langle {\rm Re}\{\sigma_{\bf q}\}^2\rangle + \langle {\rm Im}\{\sigma_{\bf q}\}^2\rangle.
\label{sqimre}
\end{equation}
Computing the full correlation function $C({\bf r})$ and Fourier-transforming it post-simulation is much more time-consuming.

\paragraph{Binder ratio and cumulant}

Besides $\xi/L$, there are also other dimensionless quantities that are size-independent at the critical point and useful for extracting $T_c$ independently 
of the values of the critical exponent. The perhaps most frequently used one is the {\it Binder ratio} \cite{binder81,binder84}:
\begin{equation}
R_2 = \frac{\langle m^4\rangle}{\langle m^2\rangle^2}.
\label{r2def}
\end{equation}
At $T_c$, the power laws cancel out, and the ratio is $L$-independent (and also universal), up to subleading finite-size corrections. Normally, 
graphing $R$ versus $T$ for different system sizes produces curves that intersect each other close to $T_c$. Locating the points where $R_2$ for 
pairs of system sizes (e.g., $L$ and $2L$) cross each other, one obtains a size dependent critical point which typically converges faster than 
the $L^{-1/\nu}$ shift in (\ref{tmaxl}). One can think of this as being a results of the leading corrections canceling in a quantity involving two 
system sizes, and one is then left with something which approaches $T_c$ according to a faster, higher-order scaling correction. One can also define ratios 
similar to (\ref{r2def}) based on other powers of $m$, e.g., $R_1 = \langle m^2\rangle/\langle |m|\rangle$. The curve-crossing method for
locating $T_c$ can also be applied in the same way with $\xi/L$.

The Binder ratio also has other interesting properties. In the case of a scalar order parameter (e.g., for the Ising model), the Binder {\it cumulant} 
is defined as 
\begin{equation}
U_2 = \frac{3}{2} \left (1- \frac{1}{3}R_2 \right ).
\label{u2def}
\end{equation}
In an ordered state, $U_2 \to 1$ for $N\to \infty$, because the magnetization distribution $P(m)$ then approaches two delta-functions at 
$\pm \langle |m|\rangle$, and, therefore, $R_2\to 1$. In contrast, in a disordered phase, the fluctuations of $m$ are Gaussian around $m=0$ 
(following from the central limit theorem, since fluctuations in regions separated by distance $\gg \xi$ in a large system are uncorrelated). 
Based on Gaussian integrals, $R_2\to 3$ and $U_2\to 0$. 

\begin{figure}
\includegraphics[width=12.5cm, clip]{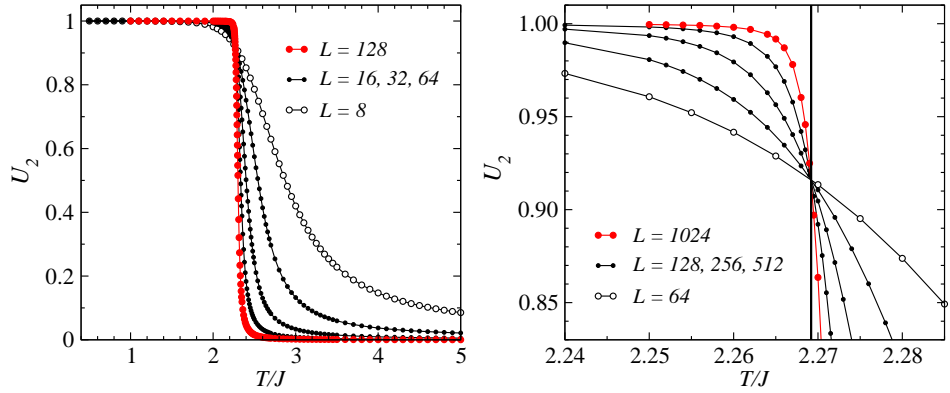}
\caption{The Binder cumulant (\ref{u2def}) for $L\times L$ Ising models. In (a) on can see the approach to the limiting values $U_2\to 0$ 
(for $T>T_c$) and $U_2 \to 1$ (for $T<T_c$) for increasing $L$. In (b) the data close to $T_c$ (vertical line) are graphed on a more detailed 
scale and for larger $L$ to show the crossings of the curves.}
\label{isingbinder}
\end{figure}

Generalizing the Binder cumulant to an $n$-component order parameter (where $n=1$ for the Ising magnetization, $n=2$ for the XY model, etc.), one should keep in 
mind that $m^2= {\bf m} \cdot {\bf m}$ and $m^4$ in (\ref{r2def}) are insensitive to angular fluctuations of the order parameter. Integrating a Gaussian 
distribution of $|m|$ over the $n$-dimensional space to compute the averages in (\ref{r2def}) does, however, introduce $n$-dependent factors. To reproduce 
the above properties of $U_2$ one therefore has to define it for a general order parameter as
\begin{equation}
U_2 = \frac{n+2}{2} \left (1- \frac{n}{n+2}R_2 \right ).
\label{u2defn}
\end{equation}

Fig.~\ref{isingbinder} shows Monte Carlo results for $U_2$ as a function of $T$ for several 2D lattices. The evolution into a step-function 
at $T_c$ with increasing system size can be seen clearly. In this case all the curves cross each other very close to the known $T_c$, reflecting very small
subleading corrections in this model. It is common in other systems that the crossings exhibit some drift, and a careful extrapolation of
crossing points have to be carried out (e.g., based on data sets for sizes $L$ and $2L$, or some other aspect ratio).

\paragraph{Finite-size scaling in practice}

\begin{figure}
\includegraphics[width=12.5cm, clip]{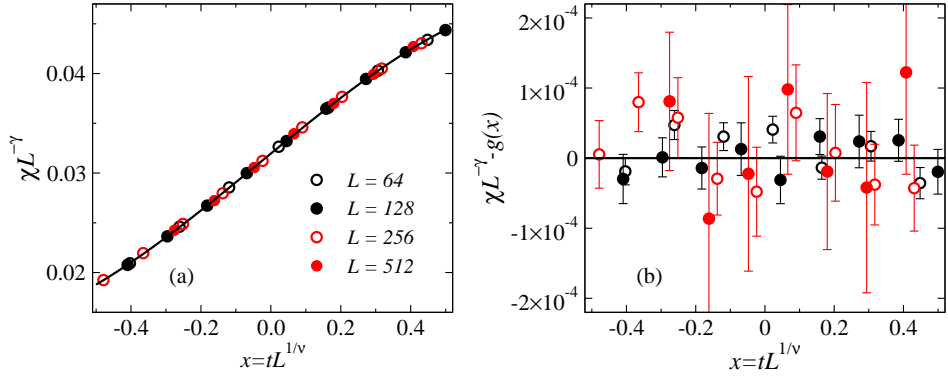}
\caption{(a) Best-fit scaling collapse of the susceptibility of the 2D Ising model; the same data as in Fig.~\ref{isingsusc}, but including larger lattices 
and adjusting $T_c$ in $t=(T-T_c)/T_c$ as well as the exponents $\nu,\gamma$ to minimize $\chi^2$ with respect to a scaling function $g(x)$ in the form of a
polynomial (here of fourth order, shown as the solid curve). The optimal values for this data set are: $T_c/J=2.26921 \pm 0.00002$, $\nu=0.9985\pm 0.0011$, 
and $\gamma=1.750\pm 0.002$, where the error bars (one standard deviation) were computed by repeating the fit several times with Gaussian noise (of magnitude
equal to the Monte Carlo error bars) added to the data. (b) shows the data with the fitted scaling function $g(x)$ subtracted, so that the 
error bars become visible. The fit is statistically sound, with $\chi^2 \approx 0.9$ (per degree of freedom).}
\label{xscalefit}
\end{figure}

We briefly discuss how to carry out finite-size  data-collapse in practice. The number of parameters involved (i.e., $T_c$ as well
as one or two exponents, and possibly also exponents of subleading corrections to be discussed below) is rather small, and normally one has some rough knowledge of 
their values just from looking at raw data and doing some initial experimentation, e.g., by just locating a non-trivial power-law behavior as in Fig.~\ref{ismag2} 
[and this may often be enough to determine the exponent ratio $\kappa/\nu$ in Eqs.~(\ref{amaxl}) and (\ref{atlscaling})]. An analysis of the Binder cumulant or 
$\xi/L$ may already have given $T_c$ to adequate precision, but it may still be useful to check the sensitivity of other fits to its value. Thanks to the power 
of modern computers, as an alternative to using some complicated multi-dimensional optimization procedure, one can write a simple brute-force computer program to 
search for the best set of parameters on a suitable finite mesh. The goodness of the data collapse produced by a set of parameters can be
quantified as the $\chi^2$-value obtained by fitting a single high-order polynomial through all the scaled data points $(x_L,y_L)$, defined in (\ref{xlyldef}), 
simultaneously for all $L$ for which data are available. Normally there is a large number of data points, for different couplings and system sizes, and, since 
the scaling function should be well behaved, a polynomial of reasonable order (3th-8th, as a rough guideline) should work well within the window where the data can 
be collapsed. The size of this window also has to be adjusted until the data collapse well. As an example, for the data in Fig.~\ref{isingsusc}, $x$ in the range 
$(-0.5,0.5)$ should be appropriate, although the window also depends on the system sizes included in the analysis and the the error bars (which determine the 
sensitivity to neglected subleading scaling corrections). 

It is not always easy to determine reliable error bars for parameters obtained in this kind of fitting. Beyond the purely statistical errors (which can 
be determined, e.g., by repeating the $\chi^2$ minimization several times with Gaussian noise, of magnitude equal to the error bars, added to the data), 
there are also systematical errors due to scaling corrections, which can be difficult to estimate. If $\chi^2$ is statistically reasonable (i.e., close 
to $1$ per degree of freedom), one would normally assume that the neglected corrections have not influenced the parameters beyond the statistical uncertainties. 

Fig.~\ref{xscalefit} shows an example of data collapse. The critical exponents and $T_c$ of the 2D Ising model were determined using data for 
$L \in \{64-512\}$ in the window $|t|L^{1/\nu}<0.5$ (where $t$ contains the variable $T_c$ adjusted in the procedure). The results,
listed in the figure caption, are in excellent agreement with the known $T_c$ and the 2D Ising exponents, and the fit is also statistically very good. Using the 
same fitting window, a statistically acceptable fit cannot be obtained if much smaller lattices are included, e.g., including $L=32$ gives $\chi^2\approx 2$ 
(per degree of freedom), which is marginally too large for the number of degrees of freedom of the fit. In this case $T_c=2.26924 \pm  0.00002$, about 3 error 
bars above the true value, while the exponent are correct to within two error bars.

\paragraph{Corrections to scaling }

As a check, it is often a good idea to include some corrections to the leading scaling forms, and some times it is even necessary to do so in order to obtain 
good fits. The most commonly used method is to do the data collapse with 
\begin{equation}
y_L=Q(t,L)L^{-\kappa/\nu}(1+aL^{-\omega})^{-1},~~~~~x_L=tL^{1/\nu},
\end{equation}
where $\omega$ is a subleading exponent and $a$ is a constant. Here there is no correction to the $x_L$ scaling (the finite-size shift of the critical point), but 
such a correction can in principle be included as well \cite{wang}. Normally, good fits with corrections can be obtained also when lattice sizes are included that 
have to be left out if no corrections are used. A larger range of system sizes can partially compensate for the fact that the statistical uncertainties of all 
parameters increase when more parameters are included. If consistent leading critical exponents are obtained in fits both with and without corrections, 
then one can be reasonably certain that the results are correct. 

\subsection{First-order transitions}
\label{firstordertrans}

The scaling properties we discussed in the preceding section apply at continuous phase transitions, where the correlation length diverges.
At first-order (discontinuous) transitions, the correlation length remains finite at the transition point and the order parameter, as well as
other quantities, exhibit discontinuous jumps. The discontinuities  develop in the limit of infinite system size, normally according to power-laws 
which can also be studied using finite-size scaling techniques. The exponents associated with these powers-laws are typically trivially related to the 
dimensionality of the system \cite{fisher82,vollmayr91,lee91}. For instance, the specific heat diverges with the system size as $L^d$ at a first-order 
transition, instead of $L^{\alpha/\nu}$, with a typically very small (or even negative) $\alpha$, at a continuous transition. The shift of the critical 
point with the system size scales as $L^{-d}$, instead of the $L^{-1/\nu}$ shift of a continuous critical point.

Although finite-size scaling with exponents equal to $d$ in principle makes it easy to recognize a first-order transition, studies of {\it weak} first-order 
transitions are difficult, because they exhibit large corrections to the leading scaling forms. It may then be difficult, with system sizes accessible in
practice, to clearly distinguish slowly developing discontinuities from weaker singular behavior at a continuous transition. 

Strongly first-order transitions are also difficult to study, for a completely different reason. A Monte Carlo simulation may get stuck in a 
meta-stable state, in which case computed quantities do not correspond to correct thermal averages (which, on the other hand, is completely analogous
to real systems, for which metastability and hysteresis effects are hall-marks of first-order transitions). It may then even be difficult to accurately
locate the transition point. To alleviate such problems, various {\it multi-canonical} Monte Carlo methods have been developed in which the configurations 
are sampled in an extended ensemble where the temperature of the system is also fluctuating (tempering or parallel tempering \cite{marinari,hukushima}
methods), or with a distribution different from the Boltzmann probability \cite{berg,wanglandau,trebst} (to which the measurements are re-weighted), 
which makes it easier for the system to explore the configuration space. 

\subsubsection{Phase coexistence}

\begin{figure}
\includegraphics[width=14cm, clip]{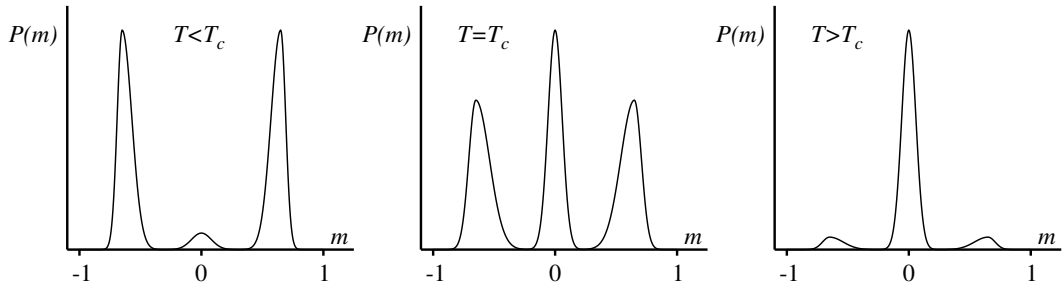}
\caption{Evolution (schematic) of the magnetization distribution of a finite Ising ferromagnet close to a first-order transition. There is 
a rapid transfer of weight between the central peak (corresponding to the disordered phase) and the two peaks at non-zero magnetization (the 
two ferromagnetic states) as the temperature is tuned through the transition region. The peaks become delta-functions in the limit of infinite 
system size. The transition temperature $T_c$ can be defined as a point with some specific (but essentially arbitrary) feature of the three peaks, 
e.g., equal weight or equal height of the peaks.}
\label{mdistfirst}
\end{figure}

One of the most important characteristics of a first-order transition is phase coexistence. In Monte Carlo simulations at the transition point
(in practice in a small window which shrinks to a single temperature with increasing system size) this is manifested in the generation of two types 
of configurations, corresponding to the two distinct phases existing just above and below the transition temperature. This is provided that the full 
configuration space can be ergodically sampled, which, as discussed above, is not always possible in practice. In small systems, there will also be 
configurations that cannot be clearly associated with one of the phases---they correspond to the fluctuations (domain walls) required for the system to 
transition between the two phases. Fig.~\ref{mdistfirst} illustrates this schematically for an Ising order parameter. The transition window in which a 
three-peak distribution can be observed narrows rapidly with increasing system size, and the peaks develop into delta-functions. When the weight between 
the peaks becomes very small, it may in practice not be possible to ergodically sample the configurations, as the system gets trapped within the sub-space
corresponding to just one of the peaks (in a way exactly analogous to the symmetry-breaking discussed in Sec.~\ref{montecarlo}).

Note that the type of order-parameter distribution in Fig.~\ref{mdistfirst} does not apply to the field-driven first-order transition of the Ising model, 
with magnetization curves illustrated in Fig.~\ref{mfising}. In that case there would be just two peaks, with weight transferring between them as the field 
is tuned through $h=0$. The phase coexistence at $h=0$ (and $T<T_c$) corresponds to the two symmetric peaks in the distributions in Fig.~\ref{mdist2d}. The 
similarity between this case and a paramagnetic-magnetic transition (with a general $n$-component order parameter ${\bf m}$) can be made clearer by considering 
the one-dimensional distribution of $|m|$. This distribution has two peaks, one at $|m|=0$ and one at $|m|>0$, when an ordered and disordered phase coexist. 
Note again that in continuous transitions (Fig.~\ref{mdist2d}) the central peak continuously splits into two peaks below $T_c$, in contrast to the two 
ordering peaks emerging at a non-zero value of $|m|$ in the first-order case.

The Binder ratio, Eq.~(\ref{u2defn}) in the case of a generic $n$-dimensional order parameter depends only on the distribution of $|m|$. It has a very
interesting property at a first-order transition. One can easily check, using, e.g., an idealized co-existence distribution of the type 
$P(|m|)=(1-p)G(|m|)+p\delta(|m|-m^*)$, where $G$ is a normalized ``half-Gaussian'' (defined for $|m|\ge 0$) and $0< m^* \le 1$, 
that $U_2$ exhibits a negative divergence when the ordered-phase weight $p$ is tuned from $0$ to $1$, and if the width of the Gaussian vanishes 
(corresponding to infinite system size). While the full $m$ distribution of course contains more information, locating a window of negative Binder cumulant 
and checking for a divergence of the peak value can be a practical way to analyze a first-order transition. 

\subsubsection{Frustrated Ising model}
\label{j1j2ising}

Here we look at a particular example of a first-order transition, in the frustrated 2D square-lattice Ising model with hamiltonian
\begin{equation}
E_\sigma = -J_1\sum_{{\langle ij\rangle}_1} \sigma_i\sigma_j  + J_2\sum_{{\langle ij\rangle}_2} \sigma_i\sigma_j, 
\label{isinghj1j2}
\end{equation}
where both couplings $J_1,J_2>0$ (but note the different signs in front of the parameters). The first term (where ${\langle ij\rangle}_1$ denotes 
nearest-neighbor spins) is then the standard Ising ferromagnet, whereas the second term (where ${\langle ij\rangle}_2$ refers next-nearest-neighbor spins, 
i.e., across the diagonals on $2\times 2$ plaquettes, as in Fig.~\ref{j1j2states}) is antiferromagnetic and causes frustration. For coupling ratios 
$g=J_2/J_1<1/2$ the ground state of the system is ferromagnetic (fully polarized). In the limit $g \to \infty$ the system reduces to two decoupled 
antiferromagnets, with four striped (or collinear) ground states, such as the one illustrated in Fig.~\ref{j1j2states}(c). These remain the ground 
states for all $g>1/2$. At the point $g=1/2$ the ground state is massively degenerate. Note that the case of $J_1<0$ is equivalent to $J_1>0$, through 
the invariance of $E_\sigma$ under the transformation $\sigma_i \to -\sigma_i$ on one of the checker-board sublattices; thus in this case the 
ferromagnetic phase is replaced by an antiferromagnetic one, whereas the striped phase is not changed.

The frustrated Ising model has been studied for a long time, but some features of its phase diagram are still debates or were resolved only 
recently \cite{yin,kalz,jin}. For $g<1/2$, the transition is of the standard Ising type, but is difficult to study close to $g=1/2$, due to 
large scaling corrections and long Monte Carlo autocorrelation times. Here we will consider only $g>1/2$, for which 
there is a first-order transition between a high-temperature paramagnet and a low-temperature striped phase, up to a coupling $g^* \approx 0.8$ \cite{jin} 
after which the transition becomes continuous. We analyze results obtained with the standard single-spin Metropolis algorithm (as cluster Monte Carlo 
methods do not work in the presence of frustration). Temperatures will be quoted in units of $J_1$.

\paragraph{Order-parameter distribution}

\begin{figure}
\includegraphics[width=14cm, clip]{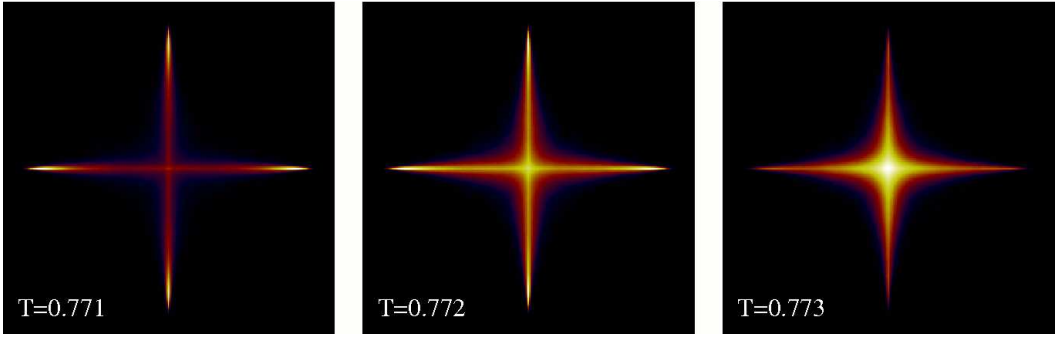}
\caption{Order parameter distributions in the $(m_x,m_y)$ plane (in the full space $|m_x|\le 1,|m_y|\le 1$) of the 2D frustrated Ising model of size $L=128$ 
at coupling ratio $g=0.55$. Brighter features correspond to higher probability density. The temperatures (indicated in the panels in units of $J_1$) 
are in the first-order transition region, with the phase coexistence (a central peak as well as four peaks corresponding to $x$- and $y$-oriented 
stripes) seen most clearly at $g=0.772$.}
\label{j1j2hist}
\end{figure}

The ordered phase can have its stripes oriented either along the $x$ or $y$-axis, with the corresponding order parameters
\begin{equation}
m_x=\frac{1}{N}\sum_{i=1}^N\sigma_i (-1)^{x_i},~~~~~~m_y=\frac{1}{N}\sum_{i=1}^N\sigma_i (-1)^{y_i},
\end{equation}
where $x_i$ and $y_i$ are the (integer) lattice coordinates of site $i$. Let us first look at the order-parameter distribution 
$P(m_x,m_y)$. If the four possible ordered states are sampled equally below $T_c$, the distribution should be four-fold symmetric, with peaks located 
on the negative and positive $x$- and $y$-axis. In the paramagnetic phase there should be a single central peak. Coexistence at a first-order transition 
should hence be reflected in the presence of five peaks in this case (instead of the 3-peak distribution for the scalar order parameter in 
Fig.~\ref{mdistfirst}). Fig.~\ref{j1j2hist} shows results for an $L=128$ system at $g=0.55$, for three temperatures in the transition window. 
A very distinct four-fold symmetry can be seen in all cases. Four symmetric peaks clearly indicate an ordered phase at $T=0.771$, but there are
strong fluctuations, reflected in weight extending to the center of the distribution. At $T=0.773$, the distribution is peaked at the center, but 
weight also extends far into the ordered regions. Between these temperatures, at $T=0.772$, one can discern both the four ordering peaks and a central 
peak. All these plots show the hall-marks of coexistence; even when there are not five peaks present, there is still significant probability
for both ordered and paramagnetic configurations. Away from the narrow transition window, the distribution rapidly turns into one with either
distinctly ordered or disordered features.

\paragraph{Binder cumulant}

\begin{figure}
\includegraphics[width=12cm, clip]{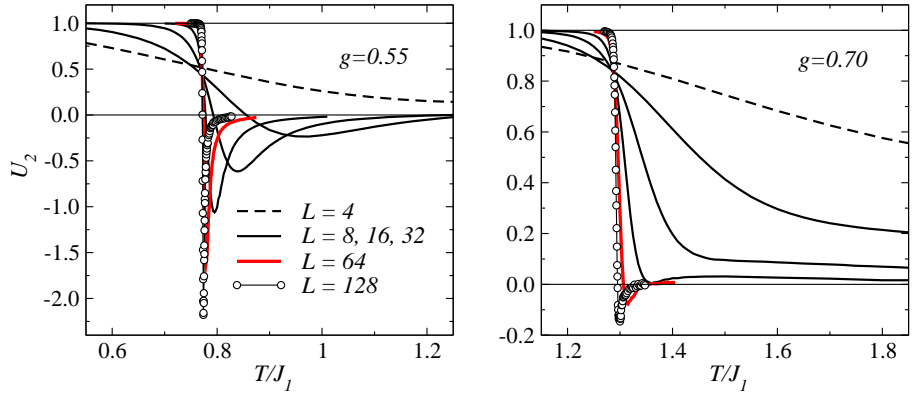}
\caption{Binder cumulant across the transition to the striped phase of the frustrated Ising model for coupling ratios $g=0.55$ (left) 
and $g=0.70$ (right). The divergent negative peak developing with increasing $L$ is an unambiguous signal of a first-order transition.}
\label{j1j2binder}
\end{figure}

Let us analyze the Binder cumulant, using $m^2=m_x^2+m_y^2$ in Eq.~(\ref{u2defn}). The question now is what order-parameter dimensionality $n$ to use in
this definition. At high temperatures, fluctuations of $m_x$ and $m_y$ in different parts of a large system are uncorrelated. This implies a 
rotationally-symmetric (circular) distribution $P(m_x,m_y)$. The $n$-dependent factors are intended to make $U_2 \to 0$ for $T\to \infty$, and in this 
case we should therefore use $n=2$. This is correct also at low temperatures (to guarantee $U_2 \to 1$), because in an ordered phase the Binder ratio, 
Eq.~(\ref{u2def}), $R_2 \to 1$ when the system size diverges, regardless of the order parameter structure. Fig.~\ref{j1j2binder} shows results as a function
of temperature at two coupling ratios; $g=0.55$ and $0.70$. For $g=0.55$, a negative Binder cumulant can be seen for $L\ge 8$, with the negative peaks 
becoming very narrow and apparently diverging as the system size is increased. At $g=0.70$, the transition is still first-order, but with weaker 
discontinuities that start to manifest themselves as co-existence and a negative Binder cumulant only around size $L=32$. In both cases, the Monte Carlo
simulations were still ergodic, but a large number of updating sweeps ($\approx 10^8$ for each case) had to be carried out
to obtain smooth curves.

\paragraph{Discontinuities and finite-size scaling}

\begin{figure}
\includegraphics[width=12.5cm, clip]{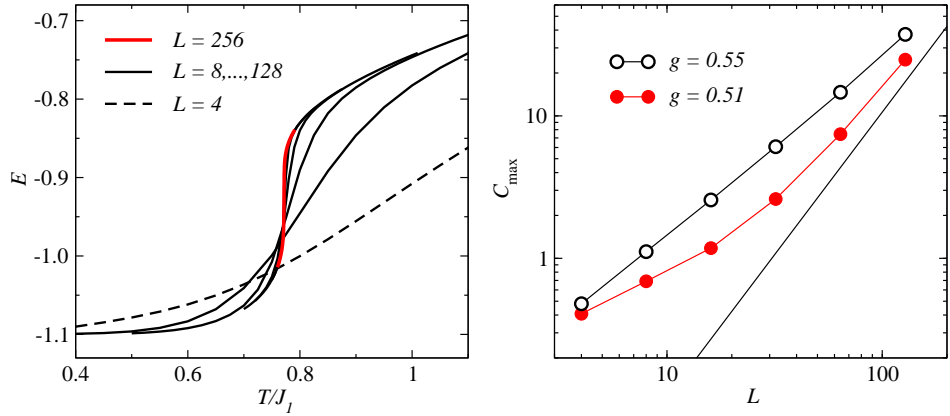}
\caption{Left: Internal energy versus temperature for the frustrated Ising model at $g=0.55$. The discontinuity developing
with increasing lattice size corresponds to the latent heat. Right: Finite-size scaling of the peak value of the specific heat
at $g=0.51$ and $0.55$. The line shows the expected asymptotic $L^2$ scaling at a first-order transition.}
\label{j1j2latent}
\end{figure}

In first-order transitions that are not very strong, as in the systems above, it may not be easy to accurately extract
the size of discontinuities in physical observables. An example is shown in Fig.~\ref{j1j2latent}, where the temperature dependence of the internal 
energy is graphed for several lattice sizes. While a discontinuity (latent heat) clearly develops with increasing $L$, it is not easy to determine 
exactly between which two energies the jump will eventually take place. This requires a careful analysis using larger lattices.

Fig.~\ref{j1j2latent} also shows the maximum value of the specific heat versus the system size for two different coupling ratios. At $g=0.51$, the
transition is rather strongly first-order, and a scaling consistent with the expected $C_{\rm max} \sim L^2$ can be observed for the largest lattices.
In contrast, at $g=0.55$ the behavior appears to follow a different power law (with an exponent close to $1.2$) up to $L \approx 64$. For larger lattices 
the results start to show a somewhat more rapid divergence, however, and eventually, for very large lattices, one would expect the exponent to be 
equal to $d=2$ also in this case. 

\subsection{Spin stiffness and the Kosterliz-Thouless transition}
\label{stiffkt}

An important aspect of a system with long-range magnetic order is that it exhibits a non-zero {\it spin stiffness}. For a system with continuous
vector spins (XY or Heisenberg models), the spin stiffness is the analogue of an elastic modulus of a solid. It is also often refereed to as the 
{\it helicity modulus} \cite{fisher73}. To study this concept, we here consider the XY model, with the hamiltonian written as
\begin{equation}
H=-J\sum_{\langle i,j\rangle}\cos(\Theta_i-\Theta_j),
\end{equation}
where $\Theta_i \in [0,2\pi]$ is the angle characterizing spin $i$. For simplicity we assume only nearest-neighbor interactions, but generalizations 
to arbitrary interactions are straight-forward. When deriving an expression for the spin stiffness, we will consider the 2D square lattice (and some 
times use a 1D chain for simplicity), but later will study also a 3D cubic system. In 2D XY systems, the spin stiffness is an important quantity
characterizing the unconventional (topological) Kosterlitz-Thouless transition exhibited by this system, which we also discuss briefly in this section.

\subsubsection{Definition of the spin stiffness}

Loosely speaking, the spin stiffness $\rho_s$ characterizes the tendency of ordered spins to adapt in response to perturbations imposing modulations 
of the direction of the order parameter (in contrast to the susceptibility, which measures the tendency of the order parameter to change in response 
to a field applied in a fixed direction). It is analogous to the shear modulus in continuum mechanics, which characterizes the tendency to shape 
deformation of an elastic object (while the compressibility corresonds to the tendency to volume change with maintained shape). The definition of the 
spin stiffness is easiest to understand at $T=0$, which we consider first, before generalizing to $T>0$.

\paragraph{The stiffness at $T=0$}

\begin{figure}
\includegraphics[width=6.75cm, clip]{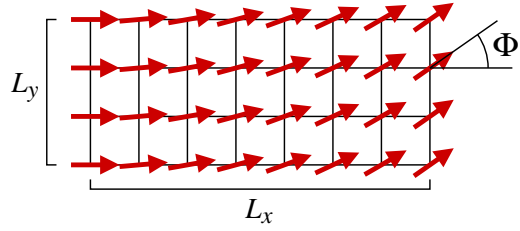}
\caption{A 2D classical XY model with a phase twist in the $x$-direction imposed by fixing the spins in the boundary column at a relative angle 
$\Phi$. To minimize the energy the total twist $\Phi$ is distributed evenly, so that the spins in neighboring columns are twisted by 
$\phi=\Phi/L_x$ relative to each other.}
\label{xystiff}
\end{figure}

We will first consider a system with open boundaries in the $x$ direction, while periodic boundaries in the $y$ direction can be assumed (and later we will 
generalize to periodic boundaries also in the $x$ direction). Fig.~\ref{xystiff} illustrates how ferromagnetic XY spins at $T=0$ adapt in order to minimize 
the energy when an over-all change in the spin angle $\Phi$ between the left and right boundaries is imposed (e.g., due to strong magnetic fields applied at the
boundary columns). To minimize the energy, each column is twisted with respect to the following column by an angle $\phi=\Phi/(L_x-1)$, where $L_x$ is the 
length of the system in the $x$ direction (the number of columns). At $T=0$, we are interested in the energy in the presence of this twist, which for 
the 2D XY model is simply given by
\begin{equation}
E(\phi)=-J(L_x-1)L_y\cos(\phi) = E(0)+J(L_x-1)L_y[1-\cos(\phi)]. 
\end{equation}
For small $\phi$ we get $E(\phi)-E(0) = (J/2)(L_x-1)L_y\phi^2$ to leading-order. Motivated by this result, the $T=0$ spin 
stiffness is defined as
\begin{equation}
\rho_s = \frac{1}{N} \frac{\partial^2 E(\phi)}{\partial \phi^2},
\label{stiffe0def}
\end{equation}
and we have $\rho_s=J$ for large $N$ [or for any $N$ if we normalize by the number $L_y(L_x-1)$ of interacting $x$ bonds, over which the energy cost 
due to the twist is distributed].

Normally it is more convenient to consider a periodic system. To derive an expression for the spin stiffness in this case, a phase
twist is first imposed at the boundary. For simplicity we work this out for a 1D chain, with spin angles $\Theta_x$, $x=0,\ldots,N-1$, but the calculation can 
be trivially generalized to higher dimensionality. The interaction energy for each bond is $E_x=-J\cos(\Theta_{x+1}-\Theta_x)$, except at the boundary, 
where a twist $\Phi$ corresponds to $E_N=-J\cos(\Theta_{0}-\Theta_{N-1}+\Phi)$. The configurations minimizing the energy have $\Theta_{x+1}-\Theta_x=\delta$, 
where $\delta$ is independent of $x$, which gives the total energy 
\begin{equation}
E(\Phi)=-(N-1)\cos(\delta)-\cos(\Phi-[N-1]\delta),
\end{equation}
which is minimized by $\delta=\phi\equiv \Phi/N$, with $E(\phi)=E(0)+JN[1-\cos(\phi)]$, and the spin stiffness defined according to (\ref{stiffe0def}) 
is again $\rho_s=J$.

It is useful to consider also another way of twisting the spins in a periodic system, by introducing a {\it twist field} $\Phi_x=x\phi$ in the hamiltonian. 
For a 1D system the energy for each bond in the presence of this field is $-J\cos(\Theta_{x+1}-\Theta_x+\Phi_{x+1}-\Phi_{x})$. To treat the boundary correctly, 
$\Phi_x$ should be considered as a function of a continuous variable $x$, which jumps discontinuously from $N\phi$ to $0$ at $x=N$. The phase difference 
appearing in the XY interaction should then be interpreted as
\begin{equation}
\Phi_{x+1}-\Phi_x=\int_{x}^{x+1} \Phi_x dx = \int_{x}^{x+1} \phi dx=\phi,
\end{equation}
which holds also at the boundary ($x+1=N$). Apart from factors, $\phi$ is analogous to a flux threading the ring. The energy in the presence of this twist 
is minimized for $\Theta_x=\Theta$ independently of $x$, giving $E(\phi)=E(0)+JN[1-\cos(\phi)]$ as before in the case of the twisted boundary condition. One 
can consider the twist field as a way of transforming away the twisted boundary condition (rotating to the reference frame of the spins in the presence of 
the boundary twist), which makes many calculations easier in practice. We will work with the twist field from now on.

\paragraph{The stiffness at $T>0$}

Now we consider non-zero temperatures, in which case the spin stiffness is defined as
\begin{equation}
\rho_s = \frac{1}{N} \frac{\partial^2 F(\phi)}{\partial \phi^2},
\label{stifffdef}
\end{equation}
where $F(\phi)$ is the free energy in the presence of a twist field (or, equivalently, a twisted boundary condition), which in turn 
is related to the partition function according to
\begin{equation}
F(\phi)=-\frac{1}{\beta}\ln[Z(\phi)].
\label{freeneergyphi}
\end{equation}
For $T \to 0$, Eq.~(\ref{stifffdef}) clearly reduces to the ground-state energy definition (\ref{stiffe0def}).

Applying a twist field which only depends on $x$, $\Phi(x,y)=x\phi$, in the 2D XY model, we have the hamiltonian
\begin{equation}
H(\phi)=-J\sum_{\langle i,j\rangle_x}\cos(\phi+\Theta_j-\Theta_i)-J\sum_{\langle i,j\rangle_y}\cos(\Theta_j-\Theta_i),
\end{equation}
where we can assume periodic boundary conditions in both directions (although in principle the $y$ boundary could be open). We can simplify the 
dependence on $\phi$ by using a standard trigonometric equality,
\begin{equation}
\cos(\phi+\Theta_j-\Theta_i)=\cos(\Theta_j-\Theta_i)\cos(\phi)-\sin(\Theta_j-\Theta_i)\sin(\phi),
\end{equation}
which we expand to second order in $\phi$:
\begin{equation}
\cos(\phi+\Theta_j-\Theta_i) \to \cos(\Theta_j-\Theta_i)(1-\phi^2)-\sin(\Theta_j-\Theta_i)\phi + {\cal O}(\phi^3).
\end{equation}
The hamiltonian in the presence of a small ($\phi \to 0$) twist can then be written as
\begin{equation}
H(\phi) \to H(0) + \hbox{$\frac{1}{2}$}\phi^2H_x -  \phi I_x,
\end{equation}
where $H_x$ is the $x$-bond part of the hamiltonian at $\phi=0$ and $I_x$ is the total spin ``current'' in the $x$ lattice direction:
\begin{equation}
I_x=J\hskip-0.5mm\sum_{\langle i,j\rangle_x}\sin(\Theta_j-\Theta_i).
\label{ixdef}
\end{equation}
The partition function in the presence of a small twist can then be written as
\begin{eqnarray}
Z(\phi) & =   & \int d[\Theta]{\rm e}^{-\beta H(\phi)} \\
        & \to & \int d[\Theta]{\rm e}^{-\beta H(0)} [1-\hbox{$\frac{1}{2}$}\beta\phi^2H_x+\ldots]
                 [1 + \beta \phi I_x + \hbox{$\frac{1}{2}$}(\beta\phi I_x)^2+\ldots],\nonumber
\end{eqnarray}
where the exponentials involving $H_x$ and $I_x$ have been Taylor expanded to the orders needed. We can now write this in a form with expectation values 
over the distribution for $\phi=0$. To second order in $\phi$:
\begin{equation}
Z(\phi) \to Z(0)[1  - \hbox{$\frac{1}{2}$}\beta\phi^2\langle H_x\rangle 
+ \beta \phi \langle I_x\rangle + \hbox{$\frac{1}{2}$}\beta^2\phi^2 \langle I_x^2\rangle].
\end{equation}
By symmetry $\langle I_x\rangle=0$ and the free energy (\ref{freeneergyphi}) is given by
\begin{equation}
F(\phi)=F(0)+ \half \phi^2 \bigl (\langle H_x\rangle -\beta\langle I_x^2\rangle \bigr ),
\end{equation}
and from this we can extract a simple expression for the spin stiffness (\ref{stifffdef}):
\begin{equation}
\rho_s=\frac{1}{N}( \langle H_x\rangle -\beta\langle I_x^2\rangle),
\label{rhosestimator1cl}
\end{equation}
which can be evaluated using Monte Carlo simulations. This result is for a twist field in the $x$ lattice direction. For a $d$-dimensional isotropic 
system, we can average over all the equivalent directions and write the spin stiffness as 
\begin{equation}
\rho_s=\frac{1}{Nd}\left ( \langle H\rangle -\beta \sum_{a=1}^d \langle I_a^2\rangle \right ),
\label{rhosestimator2cl}
\end{equation}
where the index $a$ corresponds to the current in lattice direction $a$. For an anisotropic $d$-dimensional system, the stiffness in general 
is different for all lattice directions, i.e., there are $d$ different stiffness constants, each of them given by a form  like 
(\ref{rhosestimator1cl}).

The twist-field definition of the stiffness can be easily generalized to systems with longer-range interactions. The forms (\ref{rhosestimator1cl})
and (\ref{rhosestimator2cl}) remain valid, with the current $I_x$ containing contributions from all interactions exactly as in the hamiltonian.

\paragraph{Relation to superfluidity and superconductivity}

A very interesting aspect of the spin stiffness of the XY model is that it can be directly related to the superfluid density of a superconductor
(or a superfluid such as $^4$He) \cite{fisher73,josephson66}. Like the magnetization of the XY model, the order parameter of a superconductor or a
superfluid is $U(1)$ symmetric (corresponding to the global phase of the wave function), and the twist field we discussed above is directly analogous 
to a magnetic flux (in the case of a superconductor). A non-zero stiffness corresponds to the Meissner effect exhibited by a superconductor. Monte 
Carlo simulations of the 2D XY model directly aimed at properties of thin superfluid films are discussed in Refs.~\cite{schultka94,schultka95}.

\paragraph{Spin stiffness scaling}

\begin{figure}
\includegraphics[width=12cm, clip]{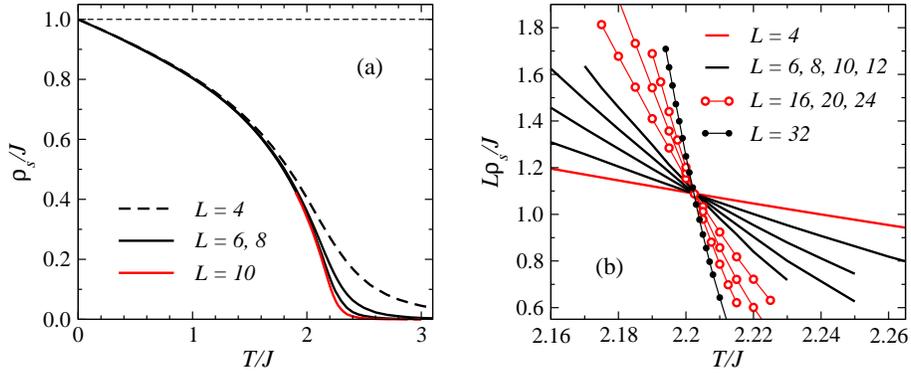}
\caption{(a) Temperature dependence of the spin stiffness of the 3D XY model for different lattices of size $N=L^3$. (b) Finite-size scaling according
to Eq.~(\ref{rhoscalelcl}). Curves of $L\rho_s$ for different $L$ cross each other (asymptotically for large $L$) at the critical temperature.
These results were obtained using a Monte Carlo cluster algorithm \cite{wolff}, and in all cases the error bars are too small to be visible.}
\label{wxy3d}
\end{figure}

The critical scaling properties of the spin (or superfluid) stiffness were first worked out in the context of superfluids \cite{fisher73,josephson66}. 
In the infinite $d$-dimensional system with $d>2$, it was shown that 
\begin{equation}
\rho_s \sim (T_c-T)^{(d-2)\nu},
\label{rhoscaletcl}
\end{equation}
 for $T \to T_{\rm c}$ from below. Here $\nu$ is the standard
correlation length exponent. According to the general finite-size scaling relation (\ref{atlscaling}), the size dependence of the stiffness 
exactly at $T_c$ is then given by
\begin{equation}
\rho_s \sim L^{2-d}.
\label{rhoscalelcl}
\end{equation}
The stiffness is therefore, like the correlation length and the Binder ratio, a useful quantity for locating the critical point without
having to adjust any unknown (or not precisely known) exponents. This is illustrated with Monte Carlo results for the 3D XY model in 
Fig.~\ref{wxy3d}, where $L\rho_s$ is size independent at the critical point.

\subsubsection{Kosterliz-Thouless transition in two dimensions}

In the 2D XY model there can be no transition into a phase with long-range magnetic order at $T>0$, according to the Mermin-Wagner theorem
\cite{mermin} (as discussed in Sec.~\ref{sec_heisenberg}). Remarkably, this system exhibits a different kind of phase transition, where
no long-range order develops but the spin correlations change from exponentially decaying to a power-law form \cite{kosterlitz73,kosterlitz74}. 
This {\it Kosterlitz-Thouless} (KT) transition is topological in nature, being a consequence of proliferation of unbound vortices (which are 
topological defects) in the spin configurations at temperatures $T>T_{\rm KT}$. For $T \le T_{\rm KT}$, vortices also exist (at a density which
vanishes at $T \to 0$) but they are all bound in vortex-antivortex pairs, which have no net vorticity (and therefore vanish upon course graining
of the spins). The power-law form of the spin correlations, $C(r) \sim r^{-\eta}$, applies for all $0 < T\le T_{\rm KT}$. The exponent $\eta$ is 
temperature dependent, with the value $\eta=1/4$ exactly at $T_{\rm KT}$ and $\eta \to 0$ as $T \to 0$, so that true long-range order exists at $T=0$.

Although there is no long rage order for $0 < T\le T_{\rm KT}$, the spin stiffness is actually non-zero in the KT phase---power-law correlations
with a sufficiently small exponent are enough to support an energy cost of a boundary twist. There is no power-law onset of the stiffness, but 
instead an even more prominent signal of the transition; a discontinuous jump at $T_{\rm KT}$ from $\rho_s=0$ to a non-zero value. 
Renormalization-group calculations for the continuum field theory corresponding to the 2D XY model have given very detailed information about 
the KT transition. One of the most important results is a rigorous relationship (due to Nelson and Kosterliz \cite{nelson77}) between the 
transition temperature $T_{\rm KT}$ and the spin stiffness exactly at this temperature (i.e., the size of the discontinuity);
\begin{equation}
\rho_s(T_{\rm KT}) = \frac{2T_{\rm KT}}{\pi}.
\label{rhosktrel}
\end{equation}
For finite lattices, the stiffness at $T_{\rm KT}$ approaches the infinite-size value with a logarithmic size correction \cite{weber87}:
\begin{equation}
\rho_s(T_{\rm KT},L) = \rho_s(T_{\rm KT},\infty)\left ( 1 + \frac{1}{2\ln(L) + c} \right ),
\label{rhoslogcorr}
\end{equation}
where $c$ is a system dependent parameter. Thus, one can say that the general finite-size scaling law (\ref{rhoscalelcl}) for $\rho_s$ at a critical 
point holds, but with a logarithmic correction to the leading-order size-independent form obtaining when $d=2$. 

\begin{figure}
\includegraphics[width=13.5cm, clip]{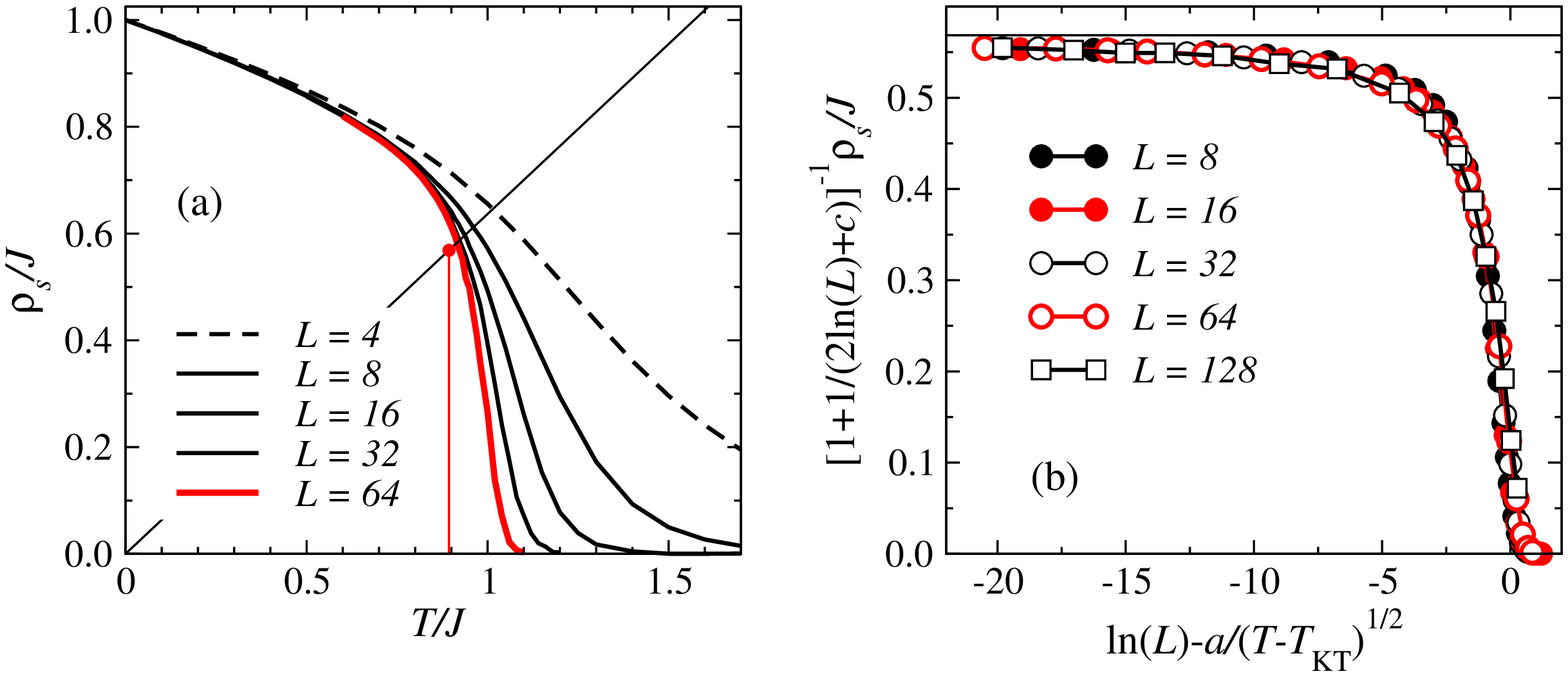}
\caption{(a) Monte Carlo results (obtained with a cluster algorithm \cite{wolff}) for the temperature dependence of the spin stiffness of the 
2D XY model for several lattices of size $L=2^n$. A discontinuity develops at $T_{\rm KT}$ as $L \to \infty$, as indicated with the vertical line 
at the known transition temperature ($T_{\rm KT}\approx 0.8933$ \cite{tomita02}). According to the Nelson-Kosterlitz relation (\ref{rhosktrel}), 
the stiffness exactly at $T_{\rm KT}$, for any system exhibiting a KT transition, must fall on the line shown; $\rho_s=2T/\pi$. (b) Finite-size 
data collapse according to the combined $(T,L)$ scaling hypothesis (\ref{rholtformkt}), using the known $T_{\rm KT}=0.8933$ and with the two
parameters, $a=1.5$ and $c=0.7$, chosen to obtain good data collapse. The vertical line shows the asymptotic $T\to T_{\rm KT}$, $L\to \infty$ value 
expected according to the Nelson-Kosterlitz relation.} 
\label{wxy2d}
\end{figure}

Fig.~\ref{wxy2d}(a) shows Monte Carlo results for the stiffness of the 2D XY model versus temperature. The jump expected in the thermodynamic limit is 
approached very slowly as a function of size. This can be related to the fact that the correlation length for $T > T_{\rm KT}$ does not diverge as a power-law,
but according to the exponential form
\begin{equation}
\xi \sim {\rm e}^{a/(T-T_{\rm KT})^{1/2}},
\label{xiktexpform}
\end{equation}
where $a$ depends on details of the lattice and the  interactions. Therefore, using the same arguments as we in the case of a standard critical point in 
Sec.~\ref{finitsizescaling}, the finite-size shift of $T_{\rm KT}$ [defined using, e.g., the temperature $T^*(L)$ at which $\rho_s$ drops
the most rapidly] is given by the form,
\begin{equation}
T^*(L) - T_{\rm KT} \sim \frac{1}{\ln^2(L)},
\end{equation}
which is much slower than the conventional power-law shift $T^*(L)- T_c \sim L^{1/\nu}$.

Using the standard finite-size scaling hypothesis (\ref{fsxilhypo}) and replacing $L^\sigma$ (the size dependence of the singular quantity exactly
at the critical point) by the logarithmic size correction in (\ref{rhoslogcorr}), we can write a hypothesis for the size and temperature dependence 
of the spin stiffness at the KT transition as
\begin{equation}
\rho_s(T,L)=\left ( 1 + \frac{1}{2\ln(L) + c} \right )f(e^{a/(T-T_{\rm KT})^{1/2}}/L),
\end{equation}
which, after taking the logarithm of the argument of the scaling function $f(x)$, can be written in terms of another function $g[\ln(x)]$ as
\begin{equation}
\rho_s(T,L)\left ( 1 + \frac{1}{2\ln(L) + c} \right )^{-1}=g[\ln(L)-a/(T-T_{\rm KT})^{1/2}].
\label{rholtformkt}
\end{equation}
Here $T \to T_{\rm KT}$ corresponds to the argument $z=\ln(L)-a/(T-T_{\rm KT})^{1/2} \to -\infty$.

The KT transition temperature of the 2D XY model has been extracted in different ways in many studies, e.g., in \cite{schultka94,tomita02}. 
Fig.~\ref{wxy2d}(b) shows a test of the scaling form (\ref{rholtformkt}), using $T_{\rm KT}=0.8933$, as obtained in \cite{tomita02}, and with
the constants $a$ and $c$ adjusted to obtain the (approximately) best data collapse onto a common scaling function for system sizes $L=8,16,\ldots,128$. 
The data are indeed well described by this form. This kind of plot confirms three different aspects of the $KT$ transition (which were
theoretically deduced at different stages of the history of the KT transition) at the same time; the exponentially divergent correlation length 
(\ref{xiktexpform}) \cite{kosterlitz73,kosterlitz74}, the logarithmic correction (\ref{rhoslogcorr}) \cite{weber87}, and the Nelson-Kosterlitz 
relation (\ref{rhosktrel}) \cite{nelson77}.

\subsection{Quantum Phase Transitions}
\label{qcpintro}

In the following sections of these notes, we will study several examples of quantum phase transitions, which take place in the ground state of a system 
as a function of some model parameter \cite{sachdevbook}. The scaling properties we have discussed above for classical continuous and first-order transitions 
still apply, with some important extensions and modifications. The nature of a classical thermal phase transition can be 
traced to singularities in the free energy. At a quantum phase transition, it is instead the ground state energy which exhibits singular behavior, 
which is manifested also in other quantities. This can be understood as arising from the $T>0$ free energy of the quantum system, which when $T \to 0$ 
becomes the ground state energy. That is of course also true classically, but the ground states of classical models normally do not evolve 
continuously as a function of the parameters (as we saw in the example of the frustrated Ising model in Sec.~\ref{j1j2ising}) and are therefore strongly 
first-order. In contrast, as already discussed in Sec.~\ref{qtrans}, quantum systems have non-trivial ground states, with zero-point 
fluctuations that evolve as the parameters are varied. Continuous phase transitions driven by divergent quantum fluctuations are common.

As we will see in connection with quantum Monte Carlo methods in Sec.~\ref{sec_sse}, a $d$-dimensional quantum system can formally be mapped, 
using path integrals, onto an equivalent classical statistical-mechanics problem in $d+1$ dimensions (albeit some times with a non-positive-definite 
distribution function). The size of the system in the new ``imaginary time'' dimension is the inverse temperature; $L_\tau=c/T$, where $c$ is a velocity. 
At $T>0$, this dimension is finite, while the spatial dimensions can be infinite. The system is then $d$-dimensional with a ``thickness'' 
$L_\tau$. The strict $(d+1)$-dimensionality applies when also $T\to 0$. In that case, a tunable parameter in the hamiltonian can play a role very similar 
to the temperature in a classical system. Interestingly, changing the temperature in this case is analogous to finite-size scaling in $L_\tau$ \cite{chn}, 
which can be used to deduce finite-temperature scaling properties close to quantum-critical points \cite{chubukov}.

In some cases, the $d$-dimensional quantum system has the low-energy properties of {\it the same kind of classical system} in $d+1$ dimensions. This 
is the case, e.g., for the dimerized Heisenberg models discussed in Sec.~\ref{qtrans}. The low-energy physics of these models can 
be mapped onto a 3D classical Heisenberg model, which is normally done via continuum field theories, such as the nonlinear $\sigma$-model 
\cite{chn,chubukov} (and, it, should be noted, the dimerization is only a means of tuning the strength of the quantum fluctuations, which in a 
course-grained uniform effective model is just represented by the coupling constant of a field theory, or the temperature in an effective 3D 
uniform classical model \cite{vojta04}). One can then presume that the quantum phase transition driven by tuning the dimerization strength should be 
in the universality class of the temperature-driven transition of the 3D classical Heisenberg model. In other cases, the low-energy mapping may give an 
effective system that does not correspond to any familiar classical model, but one can still say that the system corresponds to {\it some} $(d+1)$-dimensional 
effective model.

\paragraph{The dynamic critical exponent}

In many cases, such as the dimerized Heisenberg models mentioned above, the time dimension arising in the mapping to $d+1$ dimensions is equivalent (in 
an asymptotic sense) to the spatial dimensions. The scaling properties of such a system at a quantum phase transition are then obtained by just 
replacing $d$ by $d+1$ in the critical correlation function (\ref{cofreta}), in hyperscaling relations such as the second Eq.~(\ref{exprelations}),
and in the scalin forms (\ref{rhoscaletcl}) and (\ref{rhoscalelcl}) of the spin stiffness. In other cases, the correlations in the new dimension may be 
fundamentally different from those in the spatial dimensions. The {\it dynamic exponent} $z$ relates the power-laws associated with spatial and temporal 
correlations, e.g., if the spatial correlation length $\xi$ diverges when some parameter $g$ is tuned to its critical value as $|g-g_c|^{-1/\nu}$, then 
the temporal correlation length diverges as $|g-g_c|^{z/\nu}$, and in classical scaling relations one should replace $d$ by $d+z$ for the quantum
critical point. 

The dynamic exponent derives its name 
from the fact that it also governs the dispersion $\omega(q) \sim q^z$ of excitations of wave-number $q$. An important aspect of this is that the finite-size 
excitation gap is obtained by setting $q \propto 1/L$, giving the gap scaling $\Delta_L \sim 1/L^z$. This result can be used directly in numerical
calculations, and also has indirect consequences for the scaling properties of quantities that depend on the excitation spectrum (e.g., various 
susceptibilities). Extracting the dynamic exponent is an important aspect of computational studies of quantum phase transitions. Apart from this, 
quantum phase transitions, both continuous and first-order ones, can be analyzed with the same finite-size scaling methods as the classical transitions 
discussed above.

It should be noted that there is a dynamic exponent also in classical systems, but this does not come into play in equilibrium statistical mechanics
(because if a system has a kinetic-energy part of the hamiltonian, the associated phase-space probability distribution cancels out in expectation 
values). The classical dynamic exponent depends on how dynamics is introduced into the system \cite{halperin}. The Metropolis Monte Carlo algorithm 
for the Ising model is an example of classical dynamics (called Glauber dynamics), and the long-time decay of the autocorrelations at criticality is governed 
by an associated dynamic exponent. This dynamics does not, however, determine the equilibrium properties. In quantum mechanics, the dynamics cannot be 
separated out, but is an integral part of the equilibrium properties, because the full hamiltonian always enters and contains in it the static as well 
as dynamic properties \cite{sondhirpm}.

\section{Exact diagonalization methods}
\label{sec_diag}

By exactly diagonalizing its hamiltonian, complete knowledge of a quantum spin system can be obtained---with the eigenstates available, any static or dynamic 
quantity can be computed. In principle, all eigenstates can be computed exactly for a finite quantum system, by constructing the hamiltonian matrix and 
diagonalizing it numerically. In  practice, however, such exact diagonalization studies are limited to rather small lattices, a few tens of spins, because 
of the exponential increase of the basis 
size with the number of spins ($2^N$ states in the case of $S=1/2$). Great care therefore has to be taken in drawing conclusions about the thermodynamic limit, 
which may not even be possible (if the available lattices are too small to accommodate the infinite-size physics). Insights gained from exact diagonalization 
studies are nevertheless very useful, in their own right and as a complement to other calculations. Exact results for small lattices are also indispensable 
for testing the correctness of, e.g., quantum Monte Carlo programs. In addition, exact diagonalization methods provide a concrete path for learning many 
important aspects of quantum mechanics, in particular the symmetry properties of many-body states. 
 
\paragraph{Block diagonalization}

\begin{figure}[t]
\includegraphics[width=11cm, clip]{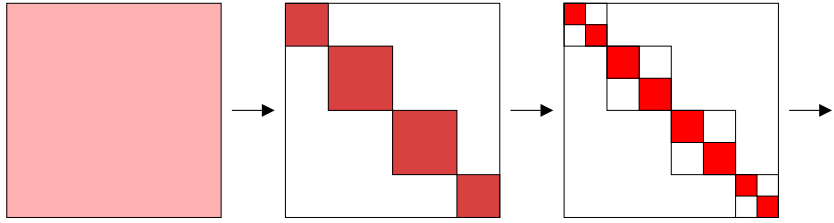}
\caption{Schematic illustration of block diagonalization. In the original basis, the hamiltonian has no apparent structure (left). By constructing 
states labeled by a conserved quantum number, the matrix breaks up into blocks (with all matrix elements zero outside the shaded squares) that 
can be diagonalized independently of each other (middle). Applying another symmetry (conservation law), the blocks can be further broken up 
into smaller blocks (right) labeled by two different quantum numbers, etc.}
\label{hamblocks}
\end{figure}
\vskip5mm

Given a hamiltonian $H$, the first step of an exact diagonalization calculation is to choose a basis in which it and other operators of interest will be expressed. 
The working basis for a spin-$1/2$ system normally consists of the single-spin states $\up_i$ and $\dn_i$, $i=1,\ldots,N$ (with the quantization direction normally taken 
as ${\bf z}$). However, because of 
the $2^N$ growth of the Hilbert space with the number of spins in the system, symmetries should be used whenever possible to first reduce the hamiltonian to 
a block-diagonal form, as illustrated in Fig.~\ref{hamblocks}. In such a scheme, the spin states are combined and ordered with the aid of applicable symmetry 
operations. The  blocks correspond to states with different conserved quantum numbers related the symmetries, e.g., crystal momentum conservation following from lattice 
translational symmetry or the conserved $z$-component of the total spin (the magnetization). The blocks can be diagonalized independently of each other, at a 
much reduced computational cost. In addition to the reduced computational effort, immediate access to the quantum numbers is very useful for classifying 
excitations. 

Some symmetries are relatively easy to take advantage of (e.g., the conservation of the magnetization) whereas others require some more work 
and lead to more complicated basis states (e.g., momentum states). Some symmetries that could be used in principle are normally not implemented, because the practical 
complications of the calculation may outweigh the benefits. Total spin conservation is an example of this.

The use of symmetries in exact diagonalization can be discussed using the language of group theory \cite{didier}. This formalism is not necessary (and often 
confusing), however, and here a more practical approach is taken, with no reference to group theory terminology. Group theory is actually very useful when dealing with 
complex lattices, but the power of its formalism can perhaps be better appreciated after a thorough understanding of symmetry operations and block-diagonalization 
has been gained through less formal methods for simple lattices. Here we consider 1D chains and 2D simple square lattices.

\paragraph{Outline of this Section}

Our discussion will first be framed around the $S=1/2$ Heisenberg chain as a concrete example in Sec.~\ref{hbchaindiag}. We first discuss its symmetries
and introduce a computer representation of the basis states using bits of integers in \ref{staterepresentation}. In \ref{makehamiltonian} we use this representation 
in pseudocodes to construct the full hamiltonian without any symmetries, and then block-diagonalize using magnetization conservation. Momentum states 
are discussed in \ref{momentum}, the use of parity (reflection symmetry) in \ref{reflection}, and spin-inversion symmetry in \ref{spininversion}. Complete 
diagonalization to obtain finite-temperature thermodynamic properties is illustrated with some results in \ref{expvalues}. To compute a small number of 
eigenstates for larger chains, the Lanczos method is developed in \ref{lanczos}. The utility of this technique for several 1D systems is illustrated in
Sec~\ref{sec_results1d}. The ground state and low-energy excitations of the Heisenberg chain are discussed in \ref{sec_hchain}. The dimerization transition taking 
place in the presence of a frustrating second-nearest-neighbor interaction is investigated in \ref{sec_j1j2chain}, and in \ref{sec_longrange} an extended 
model with long-range interactions is considered. Diagonalization of 2D systems is briefly described in \ref{sec_twodim}. Momentum states and other square-lattice 
symmetries are considered. Results pertaining to the N\'eel ground state of this system are discussed.

\subsection{Diagonalization of the Heisenberg chain}
\label{hbchaindiag}

We will study the $S=1/2$ antiferromagnetic Heisenberg chain with hamiltonian
\begin{equation}
H = J\sum_{i=0}^{N-1} {\bf S}_i\cdot {\bf S}_{i+1} =
J\sum_{i=0}^{N-1} [S^z_iS^z_{i+1}+\hbox{$\frac{1}{2}$}(S^+_iS^-_{i+1}+S^-_iS^+_{i+1})],
\label{heis0}
\end{equation}
where, for reasons that will become apparent below, it will here be practical to label the spins $i = 0,\ldots,N-1$. Periodic boundaries, 
${\bf S}_{N}={\bf S}_0$, will be assumed when we consider momentum states, but before that the boundary condition is arbitrary.

\subsubsection{Representations of states and symmetries}
\label{staterepresentation}

\paragraph{Lattice transformations}

We use the standard notation $|S^z_0,\ldots,S^z_{N-1}\rangle$ for the basis states, with $S^z_i$ from left to right always corresponding to the spin
states on lattice sites numbered $0,1,2,...$, irrespective of the ordering of the subscripts $i$. Thus, if we write a state $|S^z_1,S^z_0\rangle$, this is 
different from $|S^z_0,S^z_1\rangle$ unless $S^z_0=S^z_1$. The former could  refer to the state obtained from the latter when the two spins are switched 
by a permutation operator $P$, which can be operationally defined so as to affect the site indices; $P|S^z_0,S^z_1\rangle=|S^z_1,S^z_0\rangle$, e.g., 
$P|\up\dn\rangle=|\dn\up\rangle$. Generalizing this to $N$-spin lattice transformations, reflections, translations, and rotations (in two and three 
dimensions) are defined in terms of permutations of the spin indices. As an example, for a periodic chain we define the translation operator as
moving the spins one step cyclically to the ``right''; 
\begin{equation}
T|S^z_0,S^z_1,\ldots,S^z_{N-1}\rangle=|S^z_{N-1},S^z_0,\ldots,S^z_{N-2}\rangle. 
\label{tdef}
\end{equation}
This corresponds to decreasing the spin index by one (modulus the system size $N$) at each location $i$ in the ket: $S^z_i \to S^z_{i-1}$. When writing specific 
states with up and down spins denoted by $\up$ and $\dn$, e.g., $|\up\dn\up\dn\ldots\rangle$, indices are normally not needed (and, for, the sake of compactness 
of the notation, the arrows are also not be separated by commas). 

The Heisenberg hamiltonian (\ref{heis0}) with periodic boundary conditions is invariant with respect to translations, i.e., it commutes with $T$;
$[H,T]=0$. We can therefore construct momentum states $|\Psi(k)\rangle$, which by definition are eigenstates of the translation operator,
\begin{equation}
T|\Psi(k)\rangle = {\rm e}^{ik}|\Psi(k)\rangle.
\label{kstateproperty}
\end{equation}
Here the allowed momenta are $k=2n\pi/N$, with $n=0,\ldots,N-1$, following from the fact that $T^N=1$. States with different $k$ form their own individually
diagonalizable blocks of the hamiltonian. How the momentum states are constructed in practice and used in a computer program will be discussed in Sec.~\ref{momentum}.

The Heisenberg hamiltonian also commutes with the reflection (parity) operator, which we will define in a way generalizing the two-spin permutations already 
considered, in terms of the spin index transformation $i \to N-1-i$;
\begin{equation}
P|S^z_0,S^z_1,\ldots,S^z_{N-1}\rangle=|S^z_{N-1},\ldots,S^z_1,S^z_{0}\rangle.
\label{pdef}
\end{equation}
For an eigenstate of $P$, $T|\Psi(p)\rangle = p|\Psi(p)\rangle$, where $p=\pm 1$ since $P^2=1$. We will use $T$ and $P$ for block-diagonalization, 
although, as we will discuss in detail in Sec.~\ref{reflection}, they cannot always be used simultaneously because $[T,P]=0$ only in a sub-space 
of the Hilbert space. For a system with open boundaries, $T$ is not defined, but $P$ can be used. 

\paragraph{Spin quantum numbers}

Since the hamiltonian is spin-rotationally invariant, its eigenstates also have to be eigenstates of the square 
${\bf S}^2$ of the total spin, where
\begin{equation}
{\bf S} = \sum_{i=0}^{N-1} {\bf S}_i.
\label{totals}
\end{equation}
For an eigenstate we have ${\bf S}^2|\psi(S)\rangle= {\bf S} \cdot {\bf S}|\psi(S)\rangle =S(S+1)|\psi(S)\rangle$. Here there is potential for confusion, as 
the same symbol $S$ is used for both the spin magnitude of the individual spins (i.e., $S_i=S$) and the total spin of a many-body state. The context 
should always make the meaning clear (and we anyway only consider $S_i=1/2$ here). 

With the total spin conserved, we know that the states can form blocks labeled by the quantum numbers $(S,m_z)$, where $m_z \in \{-S,-S+1,\ldots,S\}$ is the 
total magnetization in the direction of the quantization axis,
\begin{equation}
m_z=\sum_{i=0}^{N-1} S^z_i.
\end{equation}
If we use momentum states, each $k$-block of course also splits into smaller blocks, $(k,S,m_z)$, because $[T,{\bf S}]=0$. 
It is easy to block diagonalize using $m_z$, but implementing the conservation of total $S$ is normally cumbersome (except for a very small number of spins) 
and therefore rarely used (although $S=0$ states in the valence-bond basis are some times used \cite{poilblanc}).
We will use $m_z$ conservation in combination with lattice symmetries. Note that $m_z$ conservation is more general than conservation
of the total $S$---even if we introduce some anisotropy in the hamiltonian by giving a different prefactor to, e.g., the Ising (diagonal) term in the
hamiltonian (\ref{heis0}), $m_z$ is still conserved although total $S$ is not. All the techniques discussed in this section can therefore be applied
directly also to such anisotropic models.

For the special (and most important) case $m_z=0$ (for even $N$), 
we can block-diagonalize using a discrete subset of all the possible rotations in spin-space; the spin-inversion 
symmetry, i.e., invariance with respect to flipping all the spins. This is defined formally by an operator we call $Z$;
\begin{equation}
Z|S^z_0,S^z_1,\ldots,S^z_{N-1}\rangle=|-S^z_0,-S^z_1,\ldots,-S^z_{N-1}\rangle .
\label{zdef}
\end{equation}
For this operator we again have $Z^2=1$ and the eigenvalues $z=\pm 1$. Since $Z$ commutes with both $P$ and $T$, it can be used together with these operators 
to further block-diagonalize $H$, which we will do in Sec.~\ref{spininversion}.

The total-spin operator ${\bf S}^2$ can be written in a form resembling the Heisenberg model with equal interactions among all the spins;
\begin{equation}
{\bf S}^2 = \sum_{i=0}^{N-1}\sum_{j=0}^{N-1} {\bf S}_i\cdot {\bf S}_{j} = 
2\sum_{i<j} {\bf S}_i\cdot {\bf S}_{j} + \frac{3}{4}N.
\label{ssquared}
\end{equation}
Constructing the matrix form of this operator, which we need for computing the quantum number S, is therefore very similar to constructing 
the hamiltonian matrix.

\paragraph{Bit representation of spin states}

$S=1/2$ models are special because $\dn$ and $\up$ spins can be represented directly in the computer by the bit values $0$ and $1$ of an integer. We will 
take advantage of this here. The bits are conventionally labeled starting from $0$, and that is why we also number the spins in that way. To refer 
to the bits $i=0,\ldots,31$ of an integer $s$ (or $i=0,\ldots,63$ for a ``long'' integer), we will use the notation $s[i]$. A basis state 
$|S^z_0,\ldots,S^z_{N-1}\rangle$ for $N$ spins is thus represented in the computer by an integer $s$ with $s[i]=S^z_{i}+1/2$ for $i=0,\ldots,N-1$ 
and $s[i]=0$ for $i>N-1$. 

\begin{figure}
\includegraphics[width=8cm, clip]{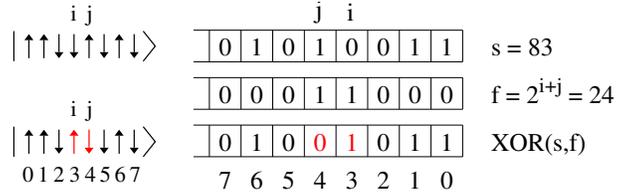}
\caption{The top line shows an $N=8$ spin state $|s\rangle$ and its representation as the eight first bits of an integer $s$. Note that we label spins
$i=0,\ldots,N-1$ from left to right, while the bits are conventionally labeled from right to left, as a binary number. To flip two spins $i$, $j$, 
a bitwise exclusive-or (XOR) operation with a mask $f$ (middle line) can be used. Bits $i$ and $j$ of $f$ are set to $1$, and all other bits are $0$
(i.e., $f=2^{i+j}$). The bottom line shows the outcome of the XOR operation [{\bf ieor}$(s,f)$ in Fortran 90].}
\label{bitstates1}
\end{figure}

Most computer languages have functions for examining and manipulating bits. The off-diagonal terms of the Heisenberg hamiltonian (\ref{heis0}) 
flip two spins. In pseudocodes we will accomplish this operation using a bit function {\bf flip}$(s,i,j)$, which flips ($0\leftrightarrow 1$) bits 
$i$ and $j$ of the integer $s$ representing the state. How this is implemented in practice depends on the language used. One possibility is to use a bitwise 
exclusive-or operation with a mask, as illustrated in Fig.~\ref{bitstates1}. In Fortran 90, this operation can be implemented as {\bf ieor}$(s,2^{i+j})$.
Later, we will also need functions that accomplish the various symmetry transformations of the states; translation, reflection, and spin-inversion. 

With standard 4-byte integers, the bit representation with a single integer works up to $N=32$. Using long integers, one can extend the scheme up to 
$N=64$. The latter is well beyond the maximum size for which exact diagonalization techniques can be used in practice, except for magnetization sectors 
with $2m_z=n_\up-n_\dn$ large enough for the block size $N!/(n_{\up}!n_{\dn}!)$ to be manageable. Magnetization $m_z=0$ and other low-$m_z$ sectors are 
typically of primary interest, however.

Discussing algorithms for constructing the basis set and the hamiltonian matrix, we will start from the simplest method, in which no symmetries at all 
are employed, and then implement the magnetization conservation. In Secs.~\ref{momentum}-\ref{spininversion} we include more symmetries. The actual 
diagonalization of the hamiltonian matrix and the use its eigenstates to calculate physical observables will be deferred to Sec.~\ref{expvalues}.

\subsubsection{Computer generation of the hamiltonian}
\label{makehamiltonian}

If we do not make use of any conservation laws, the hamiltonian consists of a single $2^N \times 2^N$ matrix. We then simply use the
numbers $a=0,1,\ldots,2^N-1$ to label the basis states. The bit-values of these integers correspond directly to the spin states. Determining the diagonal 
contributions $\langle a|S^z_iS^z_{1+1}|a\rangle=\pm 1/4$ to the hamiltonian matrix just involves examining the corresponding bit pairs
$a[i]$, $a[i+1]$ (same or different), whereas an off-diagonal operator $(S^+_iS^-_{i+1} + S^-_iS^+_{i+1})/2$ acting on a state $|a\rangle$ with $a[i] \not= a[i+1]$ 
generates the state $|b\rangle$ where the two bits have been flipped. The matrix element is then $\langle b|H|a\rangle = 1/2$. For a state with $a[i] = a[i+1]$ 
there is of course no off-diagonal matrix element. The following piece of pseudocode generates the full hamiltonian matrix $H$:

{\code
\cia {\bf do} $a=0,2^N-1$ \br
\cib    {\bf do} $i=0,N-1$ \br
\cic       $j=\mathbf{mod}(i+1,N)$                        \hfill \{2\}\break
\cic       {\bf if} $(a[i]=a[j])$ {\bf then} \br
\cid          $H(a,a)=H(a,a)+\frac{1}{4}$ \br
\cic       {\bf else} \br
\cid          $H(a,a)=H(a,a)-\frac{1}{4}$ \br
\cid          $b=\mathbf{flip}(a,i,j)$; $H(a,b)=\half$ \br
\cic       {\bf endif} \br
\cib    {\bf enddo} \br
\cia {\bf enddo}
\code}

\noindent
Here $j=\mathbf{mod}(i+1,N)$ is the ``right'' nearest neighbor of $i$, with the $\mathbf{mod}$ function taking care of the periodic boundary.
In Fortran 90, the test for $a[i]=a[j]$ can be implemented with the boolean function {\bf btest}$(a,i)$ to examine bit $i$ of $a$. Each bond operator 
in the hamiltonian corresponds to a single off-diagonal matrix element, while the diagonal elements have $N$ different contributions. 
Matrices corresponding to other operators of interest can of course be constructed in an analogous way, by examining and flipping bits according to whatever 
combinations of $S^z_i$ and $S^\pm_i$ operators that are involved.

\paragraph{Using fixed-magnetization blocks}

Moving up in sophistication, we next implement magnetization conservation. We want to construct the block hamiltonian acting on all states with given 
$m_z=(n_\up - n_\dn)/2$. There are $M=N!/(n_{\up}!n_{\dn}!)$ such states, and we need a list of them. The order within this list will be used as a label 
of the states of the block, i.e., we write $|a\rangle$ for the $a$:th state in the list. We then also need a list of integers $\{s_a\}$, the bits 
of which represent the spin configuration of the $a$:th state. Later, we will have to search the list $s_a$ to find the position label $a$ of a particular 
given state-integer $s$, and it is therefore practical to make the list ordered; $s_{a} < s_{a+1}$. We will some times use the notation $|s_a\rangle$ 
instead of $|a\rangle$ when referring explicitly to the spins, $|s_a\rangle = |s_a[0]-1/2,\ldots,s_a[N-1]-1/2\rangle$. The context will make it clear 
if a label inside $|\rangle$ refers to the position in the list of $M$ states or to the integer containing the spins.

To construct the state list, we loop over the integers $s=0,\ldots,2^N-1$ and check whether the number of set bits (the number 
$n_\up$ of spins $\up$ in the state) corresponds to the target sector; $n_\up = m_z+N/2$. After initializing a state counter 
$a=0$, we can use the following pseudocode to generate all the states with given magnetization:

{\code
\cia {\bf do} $s=0,2^N-1$ \br
\cib     {\bf if} $(\sum_i s[i]=n_\up)$ {\bf then} $a=a+1$; $s_a=s$ {\bf endif} \hfill \{3\}\break
\cia {\bf enddo}; $M=a$
\code}

\noindent
We now have a basis of size $M$ stored as the integers $s_a,a=1,\ldots,M$.

To construct the hamiltonian, we loop over the labels $a=1,...,M$ and use bit operations as before to act on the corresponding state-integers $s_a$. When  
an off-diagonal operation on $|s_a\rangle$ leads to another state $|s^*\rangle=|s_b\rangle$, and have to find the position $b$ of the integer $s^*$
in the state list. Since this list is ordered, we can do this by a bisectional search in $\sim \log_{\rm 2}(M)$ steps on average. Such a search proceeds through 
a series of bracketings, where in each step we can halve the possible range of $b$ by examining the state at the mid-point of a range 
$[b_{\rm min},b_{\rm max}]$, with the brackets $b_{\rm min}=1$ and $b_{\rm max}=M$ initially. The following subroutine finds the position $b$ of 
a state-integer $s^*$;

{\code
\cia {\bf subroutine findstate}$(s^*,b)$ \br
\cia $b_{\rm min}=1$; $b_{\rm max}=M$      \hfill \{4\}\break
\cia {\bf do} \br
\cib     $b=b_{\rm min}+(b_{\rm max}-b_{\rm min})/2$ \br
\cib     {\bf if} ($s^*<s_b$) {\bf then} \br
\cic         $b_{\rm max}=b-1$ \br
\cib     {\bf elseif} ($s^*>s_b$) {\bf then} \br
\cic          $b_{\rm min}=b+1$ \br
\cib     {\bf else} \br
\cic          {\bf exit} \br
\cib     {\bf endif} \br
\cia {\bf enddo}
\code}

\noindent
Division of an integer by $2$ should here be regarded in the standard way, i.e., $i/2$ for odd $i$ equals $(i-1)/2$. The {\bf exit} 
from the loop occurs when the basis state $s_b$ equals the target state $s^*$. One can also use more efficient search procedures, using so-called 
hash-tables \cite{linhash}. The bisection is much simpler, however, and sufficiently fast in most cases.

Using subroutine $\{4\}$, the part of the hamiltonian originating from operation on a spin pair $(i,j)$ in the state $|a\rangle$ can be 
constructed with the following modification of code $\{2\}$;

{\code
\cia       {\bf if} $(s_a[i]=s_a[j])$ {\bf then} \br
\cib          $H(a,a)=H(a,a)+\frac{1}{4}$                     \hfill \{5\} \break
\cia       {\bf else} \br
\cib          $H(a,a)=H(a,a)-\frac{1}{4}$ \br
\cib          $s^*=\mathbf{flip}(s_a,i,j)$;~ {\bf call findstate}$(s^*,b)$;~ $H(a,b)=\half$ \br
\cia       {\bf endif}
\code}

\noindent
If one is just interested in obtaining quick results for some very small lattice, or if the system is not periodic (an open chain or a system 
with random couplings, in which case the momentum is not conserved), it may be sufficient to construct the hamiltonian in this form and proceed to 
diagonalize it (successively for all the $m_z$-blocks desired). For serious work on translationally invariant (periodic) systems, it is worth implementing 
additional symmetries to further block-diagonalize the fixed $m_z$ blocks.

\subsubsection{Momentum states}
\label{momentum}

We now construct eigenstates of the translation operator $T$ defined Eq.~(\ref{tdef}). Translating $N$ steps brings the spins back to their original state. 
Thus, $T^N=1$, which implies eigenvalues ${\rm e}^{ik}$, where the set of $N$ non-equivalent momenta can be chosen as,
\begin{equation}
k=m\frac{2\pi}{N},~~~~ m=-N/2+1,\ldots,N/2,
\label{momentkm}
\end{equation}
with the lattice constant equal to $1$. A momentum state can be constructed using a {\it reference state} $|a\rangle$ (a single state in the $z$-component 
basis) and all its translations; 
\begin{equation}
|a(k)\rangle = \frac{1}{\sqrt{N_a}} \sum_{r=0}^{N-1} {\rm e}^{-ikr} T^r|a\rangle.
\label{akdef}
\end{equation}
It can easily be verified (by a shift of the summation index allowed due to the periodic boundaries)
that operating with the translation operator (\ref{tdef}) on this state gives $T|a(k)\rangle = {\rm e}^{ik}|a(k)\rangle$, 
which is the definition of a momentum state. 

To construct the momentum basis for given $k$ (and normally also given $m^z$) we have to find a set of representatives resulting in a complete set of 
normalizable orthogonal states. Clearly, for two states $|a(k)\rangle$ and $|b(k)\rangle$ to be orthogonal, the corresponding representatives  must 
obey $T^r|a\rangle \not= |b\rangle$ for all $r$. Therefore, among all the states of the set of translated states $|a(r)\rangle=T^r|a\rangle$, $r=0,\ldots,N-1$, 
only one should be used as a representative. With the labels referring to the bit representation, it will be practical to always choose the representative as the one 
for which the integer $a(r)$ is the smallest (as determined by carrying out all translations).

\paragraph{Normalization and excluded representatives}

If all the translated states $T^r|a\rangle$ are distinct, the normalization constant in (\ref{akdef}) is just $N_a=N$. Some reference states have 
periodicities less than $N$, however, and this affects the normalization. The periodicity of a state is defined as the smallest integer $R_a$ for which
\begin{equation}
T^{R_a}|a\rangle=|a\rangle,~~~~R_a \in \{1,\ldots,N\}.
\end{equation}
If $R < N$, then there are multiple copies of the same state in the sum in (\ref{akdef}), and the normalization constant must be modified 
accordingly. We could then also restrict the summation in (\ref{akdef}) to $r=0,\ldots,T_a-1$, but it is more practical in formal manipulations 
of the states to keep all $N$ terms regardless of $R_a$. 

An important aspect of the momentum basis is that the periodicity of the representative has to be compatible with the momentum in order for 
(\ref{akdef}) to be a viable state. The compatibility is related to normalizability. The sum of phase factors associated with the representative
state $|a\rangle$ in the sum in (\ref{akdef}) is
\begin{equation}
F(k,R_a)= \sum_{n=0}^{N/R_a-1} {\rm e}^{-iknR_a} = \left \lbrace
\begin{array}{ll}
N/R_a, & \hbox{if $kR_a$ is a multiple of $2\pi$,} \\
0, & \hbox{otherwise.}\end{array} \right.
\end{equation}
The normalization constant is then
\begin{equation}
N_a=\langle a(k)|a(k)\rangle = R_a |F(k,R_a)|^2,
\end{equation}
and, therefore, if $F(k,R_a)=0$, no state with momentum $k$ can be defined using the reference state $|a\rangle$. 
Thus, for given $|a\rangle$ the allowed momenta are those for which $kR_a$ is a multiple of $2\pi$, or
\begin{equation}
k = \frac{2\pi}{R_a}m,~~~m=0,1,\ldots,R_a-1.
\label{allowedk}
\end{equation}
For the allowed momenta, the normalization constant in (\ref{akdef}) is
\begin{equation}
N_a=\frac{N^2}{R_a}.
\label{na1}
\end{equation}
When the reference state is not equal to any non-trivial translation of itself, then $R_a=N$ and $N_a=N$. Note again that a given reference state can only appear in a 
single basis state. It is then clear that the momentum states are orthonormal; $\langle b(k')|a(k)\rangle = \delta_{ab}\delta_{kk'}$. Also note again that reference 
states that are not compatible with a given momentum will not appear in that block of states. An important aspect of constructing the momentum basis is to check 
the compatibility of a potential representative state with the momentum.

\paragraph{The hamiltonian matrix}

Next, we construct the hamiltonian matrix in the momentum basis. The periodic Heisenberg hamiltonian under consideration here is translationally 
invariant and consists of $N$ {\it bond operators} ${\bf S}_i\cdot {\bf S}_{i+1}$. It is convenient to lump all the diagonal terms together and consider 
the off-diagonal terms separately. To simplify the formalism to follow, we define operators accordingly;
\begin{eqnarray}
&&H_0 =  \sum_{j=1}^N S^z_jS^z_{j+1}, \label{hdia1}\\
&&H_j =  \half(S^+_{j-1}S^-_{j} + S^+_{j-1}S^-_{j}),~~~~j=1,\ldots,N,\label{hoff1}
\end{eqnarray}
so that $H=J\sum_{j=0}^N H_j$. We now set $J=1$. We need to find the state resulting when $H$ acts on the momentum state (\ref{akdef}).
Since $[H,T]=0$ we can write 
\begin{equation}
H|a(k)\rangle = \frac{1}{\sqrt{N_a}} \sum_{r=0}^{N-1} {\rm e}^{-ikr} T^rH|a\rangle
= \frac{1}{\sqrt{N_a}} \sum_{j=0}^N  \sum_{r=0}^{N-1} {\rm e}^{-ikr} T^rH_j|a\rangle,
\label{hoperak1}
\end{equation}
and we need to operate with the hamiltonian operators $H_j$ only on the reference state. For each operation we get a different state, or, in the diagonal
($j=0$) case, the same state. In either case we can write $H_j|a\rangle = h_j(a)|b'_j\rangle$,  where $h_j(a)$ is the matrix element coming from (\ref{hdia1}) or 
(\ref{hoff1}), and we do not, for simplicity of the notation, include any explicit indicator that $|b_j'\rangle$ also depends on $|a\rangle$. The prime 
in $|b_j'\rangle$ is there to indicate that this new state is not necessarily one of the reference states used to define the basis and, therefore, a 
momentum state should not be written directly based on it. Provided that $|b'_j\rangle$ is compatible with the momentum, there must be a reference 
state $|b_j\rangle$ which is related to it by some number of translations;
\begin{equation}
|b_j\rangle = T^{l_j} |b'_j\rangle,
\label{btransrep}
\end{equation}
and using this relation we have
\begin{equation}
H_j|a\rangle = h_j(a)T^{-l_j}|b_j\rangle,~~~~~l_j \in \{ 0,1,\ldots,N-1 \}.
\label{hatransrep}
\end{equation}
Here the notation is again simplified by not making explicit that $l_j$ depends on the actual state $|b_j\rangle$ (and therefore on $|a\rangle$). 
We can now write (\ref{hoperak1}) as 
\begin{equation}
H|a(k)\rangle = 
 \sum_{j=0}^N \frac{h_j(a)}{\sqrt{N_a}} \sum_{r=0}^{N-1} {\rm e}^{-ikr} T^{(r-{l_j})}|b_j\rangle,
\end{equation}
and by shifting indices in the summation, and also noting that $|b_j\rangle$ may have a different normalization factor (periodicity)
than $|a\rangle$, we obtain
\begin{equation}
H|a(k)\rangle = 
\sum_{j=0}^N h_j(a){\rm e}^{-ik{l_j}} \sqrt{\frac{N_{b_j}}{N_a}}|b_j(k)\rangle.
\end{equation}
We can now simply extract the matrix elements of the hamiltonian operators $H_j$;
\begin{equation}
\langle b_j(k)|H_j|a(k)\rangle = h_j(a){\rm e}^{-ik{l_j}} \sqrt{\frac{N_{b_j}}{N_a}}.
\label{hakbk}
\end{equation}
Strictly speaking, for the off-diagonal bond operators ($j>0$), this is not the only matrix element, because individually these are not translationally invariant 
operators. They therefore have matrix elements also between states with different momenta. Here we have in mind summing over all $j$, after which only the elements 
diagonal in $k$ survive.

There can be several terms in $H$ that contribute to the same matrix element (\ref{hakbk}), because it is possible (in fact very likely) that 
$H_j|a\rangle \propto T^{-l_i}|b_i\rangle$ and $H_j|a\rangle \propto T^{-l_j}|b_j\rangle$ with $|b_i\rangle=|b_j\rangle$ (but $l_i \not= l_j$). Note also 
that the matrix element $h_j(a)$ in (\ref{hatransrep}) is zero for an off-diagonal operator acting on two parallel spins. It should also be kept in 
mind that $|b_j(k)\rangle$ may not exist even if $h_j(a)>0$ in in (\ref{hatransrep}), if $|b_j\rangle$ is incompatible with the momentum.
The momentum-state matrix element (\ref{hakbk}) then does not exist.  With these caveats, which we will deal with in the implementations below, (\ref{hakbk}) 
specifies all the non-zero matrix elements of the hamiltonian. For the specific Heisenberg model considered here, we can substitute $h_j(a)$ by the actual 
values of the diagonal and off-diagonal matrix elements and obtain
\begin{eqnarray}
&&\langle a(k)|H_0|a(k)\rangle = \sum_{j=1}^N S^z_jS^z_j, \label{hmathchain1}\\
&&\langle b_j(k)|H_{j>0}|a(k)\rangle = {\rm e}^{-ik{l_j}}\frac{1}{2} \sqrt{\frac{R_a}{R_{b_j}}},~~~~~~~|b_j\rangle \propto T^{-l_j}H_j|a\rangle.
\label{hmathchain2}
\end{eqnarray}

Matrix elements of other translationally invariant operators (which is what should be used in the momentum basis) can be obtained in exactly the same way if the
momentum transfered by the operator is $0$ (i.e., a sum of identical local operators $O_i$, $i=0,\ldots,N-1$). For an operator $A_q$ which transfers momentum 
$q \not =0$ (such as $S^z_q$, the Fourier transform of $S^z_i$) the procedures differ only in that the basis states obtained when operating, $A_q|a(k)\rangle$, 
have momentum $k+q$ (and in that case the basis sets for both momenta involved have to be stored).

\paragraph{Constructing the momentum-state basis in a program}

We will now again use a label $a$ to refer to the position of the state in the basis, and store the corresponding spins in the form of the bits
of integers $s_a$, $a=1,\ldots,M$. A momentum state is defined in terms of its representative $|s_a\rangle$, from which the full 
state is generated using the translation operator $T$ according to (\ref{akdef}). In this case we do not know the basis size $M$ {\it a priori}, 
because the number of compatible representatives depends on the momentum. If all the translations of all states were unique, and there were no states 
incompatible with the momentum, then the number of states would equal $[N!/(n_\up!n_\dn!)]/N$. However, since many states have periodicities less than 
$N$ and some are incompatible with $k$ (unless $k=0$), this is only an approximate basis size. The $k=0$ momentum block is always the largest, because 
there are no states incompatible with this momentum. For large systems the fractions of disallowed states and periodicities $<N$ are small, 
and the above approximation to the basis size is then quite good. 

\begin{figure}
\includegraphics[width=5.5cm, clip]{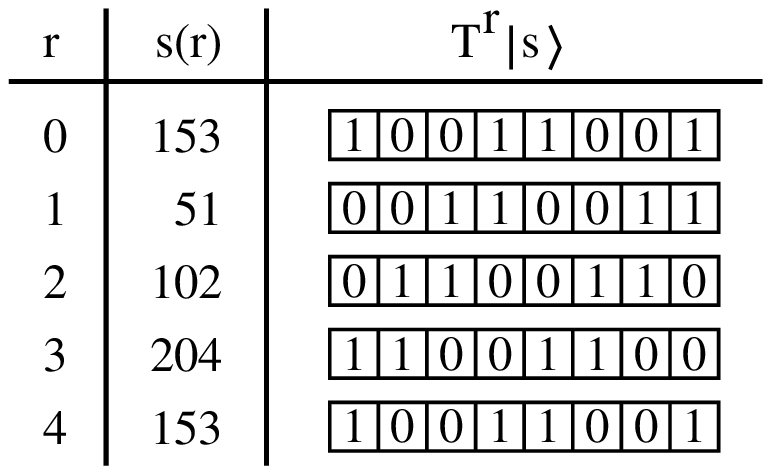}
\caption{Bit representation of the 8-spin state $|s\rangle = |\up\dn\dn\up\up\dn\dn\up\rangle$ (top row) along with its translations, 
$T^r|s\rangle$, with $r=1,\ldots,R$, where $R=4$ is the periodicity. The integers $s(r)$ correspond to $r$ successive cyclic permutations 
(to the left) of the bits. The representative is $s(1)=51$; the lowest integer in the set.}
\label{bitstates2}
\end{figure}

To implement the momentum basis in a computer program, we first need to decide how the representatives are chosen---in principle the representative of 
a state $|s\rangle$ could be any one of the members of the group of states related to it through translations; $|s(r)\rangle=T^r|s\rangle$, 
$r=0,\ldots,N-1$. Fig.~\ref{bitstates2} shows an example of all the unique translations of an 8-spin state in the bit representation.
We have to construct a list of the representatives and should be able to easily identify the representative corresponding to an arbitrary 
spin state. As already mentioned, when using the bit representation it is natural to pick as the representative the translated state $|s(r)\rangle$ 
for which the integer $s(r)$ is the smallest. 

To generate the basis, we loop over the integers $s=0,\ldots,2^N-1$, and, as in code $\{3\}$, process only those corresponding to a chosen 
magnetization. The operations needed to determine whether a state is a valid new representative are carried out by bit operations. When translating 
$|s\rangle$, if some $s(r>0) < s=s(0)$, then the representative of $|s\rangle$ is already in the basis and should 
not be used again [as in Fig.~\ref{bitstates2}, where the original integer $s=s(0)=153$ but the translation $s(1)=51<s(0)$]. If all $s(r) \ge s(0)$, 
then the representative of $|s\rangle$ is not yet in the list. However, $|s\rangle$ may still not be allowed, due to its periodicity potentially 
being incompatible with the momentum, according to (\ref{allowedk}).

In the pseudocode segments below, the momentum will be represented just by the integer $\in \{-N/2+1,\ldots, N/2\}$ multiplying $2\pi/N$ in (\ref{momentkm}), which we 
here call $k$. We will define a subroutine which checks whether or not a state-integer $s$ is a new valid representative and, for a valid representative, 
delivers its periodicity $R$. To carry out the translations of the state $|s\rangle$, we use cyclic permutations of its bits. We assume that this is accomplished 
with a function $\mathbf{cyclebits}(i,n)$, which performs a cyclic permutations to the ``left'' (as exemplified in Fig.~\ref{bitstates2}) of the $n$ first bits 
of the integer $i$. In Fortran 90, exactly this operation is available with the intrinsic function {\bf ishftc}$(s,d,n)$, which cyclically right-shifts the 
first $n$ bits of $s$ by $d$ steps. The definition of $T$ according to (\ref{tdef}) corresponds to $d=-1$. Our state checking subroutine can be implemented as:

{\code
\cia       {\bf subroutine checkstate}$(s,R)$ \br
\cia       $R=-1$;~ $t=s$                     \hfill \{6\}\break
\cia       {\bf do} $i=1,N$ \br
\cib          $t=\mathbf{cyclebits}(t,N)$ \br 
\cib          {\bf if} $(t<s)$ {\bf then} \br
\cic              {\bf return}  \br
\cib          {\bf elseif} $(t=s)$ {\bf then} \br
\cic              {\bf if} ($\mathbf{mod}(k,N/i)\not=0$) {\bf return} \br
\cic              $R=i$;~ {\bf return} \br
\cib          {\bf endif} \br
\cia       {\bf enddo} 
\code}

\noindent
Here the integer $R$ is first initialized to $-1$, and this value will be returned (and later used as an indicator of a disallowed representative) unless 
the input state-integer $s$ is the smallest among all the translated integers and the periodicity is compatible with the momentum (and then $R$ will equal 
the periodicity upon return). We can now easily construct the list of $M$ basis states;

{\code
\cia    {\bf do} $s=0,2^N-1$ \br
\cib       {\bf call checkstate}$(s,R)$                  \hfill \{7\}\break 
\cib       {\bf if} $R\ge 0$ {\bf then} $a=a+1$;~ $s_a=s$;~ $R_a=R$ {\bf endif} \br
\cia    {\bf enddo} \br
\cia    $M=a$ 
\code}

\noindent
The list $R_a$ of periodicities will be needed when constructing the hamiltonian.

\paragraph{Constructing the hamiltonian matrix}

To generate the hamiltonian, we loop over the representatives $|s_a\rangle$, $a=1,...,M$. For each of them, we check the bits corresponding to 
all nearest-neighbor pairs $(i,j)$. The diagonal matrix element can be handled exactly as in code $\{5\}$. For the off-diagonal part, we need 
a few minor modifications. After two spins have been flipped, the resulting state $|s'\rangle$ is typically not a representative, and we need to 
find it using Eq.~(\ref{btransrep}). The matrix element (\ref{hmathchain2}) also requires the number $l_j$ of translations used to bring $|s'\rangle$ 
to its representative $|s_b\rangle$. We implement these tasks as a subroutine, the contents of which are rather similar to the subroutine {\bf checkstate}, 
code $\{6\}$, that we used when constructing the basis. Given the state-integer $s$, we translate its bits in all possible ways and store the 
corresponding smallest integer $r$ (the potential representative) found so far, along with the corresponding number of translations, $l$;

{\code
\cia    {\bf subroutine representative}$(s',r,l)$ \br
\cia    $r=s'$;~ $t=s'$;~ $l=0$~                                \hfill \{8\}\break
\cia    {\bf do} $i=1,N-1$ \br
\cib       $t=\mathbf{cyclebits}(t,N)$ \br 
\cib       {\bf if} $(t<r)$ {\bf then} $r=t$;~ $l=i$ {\bf endif} \br
\cia    {\bf enddo} 
\code}

\noindent
Having found the representative $r$, we need to locate its position $b$ in the list $\{s_a\}$. This is done in the same way as before, with the subroutine 
{\bf findstate}, in code $\{4\}$. Note, however, that because of the periodicity constraint imposed by the momentum, it is now possible that the potential 
representative $r$ is actually not present in the list. Therefore {\bf findstate} is slightly modified, so that $b=-1$ is returned if there is no element 
$r$ in the list, using the following piece of code after the {\bf if}...{\bf endif} statements in code $\{4\}$;

{\code
\cia    {\bf if} $(b_{\rm min}>b_{\rm max})$ {\bf then} \br 
\cib       $b=-1$; {\bf return}                                        \hfill \{9\}\break 
\cia    {\bf endif} 
\code}

\noindent
For each spin flip, we can now add the contribution to the hamiltonian with the following code replacing the next-to-last statement in code $\{5\}$;

{\code
\cia    $s'={\bf flip}(s_a,i,j)$ \br
\cia    {\bf call representative}$(s',r,l)$       \hfill \{10\}\break  
\cia    {\bf call findstate}$(r,b)$  \br      
\cia    {\bf if} $(b\ge 0)$ $H(a,b)=H(a,b)+\half (R_a/R_b)^{1/2}{\rm exp}(i2\pi kl/N)$  
\code}

\noindent
The hamiltonian is complex, except when $k=0$ or $N/2$ (actual momentum $0$ or $\pi$). 

\subsubsection{Reflection symmetry and semi-momentum states}
\label{reflection}

We will next consider in addition to the translated states $T^r |a\rangle$ those that are generated by the reflection (parity) operator $P$ defined in
(\ref{pdef}). The operators $T$ and $P$ do not commute, and so it would at first sight appear that we cannot construct states that are eigenstates of
both operators simultaneously. These operators do, however, commute in the $k=0,\pi$ momentum blocks, as we will show explicitly below. In addition, we will construct 
{\it semi-momentum} states that are also parity eigenstates for any $k$. An advantage of such states is that they (and the hamiltonian) are real-valued, 
in contrast to the standard complex momentum states.

\paragraph{States with parity}

Consider the following extension of the momentum state (\ref{akdef});
\begin{equation}
|a(k,p)\rangle = \frac{1}{\sqrt{N_a}} \sum_{r=0}^{N-1} {\rm e}^{-ikr} T^r(1 + pP)|a\rangle,
\label{akpdef}
\end{equation}
where $p=\pm 1$. Clearly, this is a state with momentum $k$ [i.e., it satisfies Eq.~(\ref{kstateproperty})], but is it also an eigenstate of $P$ with 
parity $p$? We can check this by explicit operation with $P$, using $P^2=1$, $p^2=1$, and the relationship $PT=T^{-1}P$:
\begin{eqnarray}
&&P|a(k,p)\rangle = \frac{1}{\sqrt{N_a}} \sum_{r=0}^{N-1} {\rm e}^{-ikr} T^{-r}(P+p)|a\rangle \nonumber \\
      &&\hskip16mm = p\frac{1}{\sqrt{N_a}} \sum_{r=0}^{N-1} {\rm e}^{ikr} T^{r}(1 + pP)|a\rangle.
\end{eqnarray}
This is not exactly of the form (\ref{akpdef}), unless $k=0$ or $\pi$, for which ${\rm e}^{ikr}={\rm e}^{-ikr}$ (i.e., the momentum $k$ is equivalent to $-k$, and 
there is no directionality associated with the state). Thus, in these two special cases, parity and translational invariance can be used simultaneously for
block-diagonalization and $|a(k,p)\rangle$ is indeed a momentum state with parity $p$ (or, in other words, $[T,P]=0$ in the sub-spaces with momenta $k=0$ 
and $\pi$). 

\paragraph{Semi-momentum states and parity}

Except for the special cases $k=0,\pi$, parity cannot be used to further block diagonalize a momentum block of $H$. We can, however, use parity in 
combination with the momentum in a different way, by mixing momentum states with $\pm k$. We consider the sum and difference of these states;
\begin{equation}
|a^\sigma (k)\rangle = \frac{1}{\sqrt{N_a}}\sum_{r=0}^{N-1}C^\sigma_k(r)T^r|a\rangle ,
\label{akgdef}
\end{equation}
where $\sigma=\pm 1$ and we have, for convenience, introduced a function $C^\sigma_k(r)$;
\begin{equation}
C_k^\sigma(r) = 
\left \{ \begin{array}{ll}
\hbox{\hskip-1.5mm} \cos(kr), & \sigma=+1  \\
\hbox{\hskip-1.5mm} \sin(kr), & \sigma=-1. \end{array} \right.
\label{cfunc}
\end{equation}
We we will here refer to $k$ in (\ref{akgdef}) as the {\it semi-momentum}. Note that $\sigma$ is not a conserved quantum number, just an indicator
for how the momentum states have been combined into semi-momentum states. Strictly speaking, the special values $k = 0,\pi$ are still conventional 
(crystal) momenta (and the $\sigma=-1$ states do not exist for these $k$) and only $k$ in the range $0 < k < \pi$ should be referred to as a
semi-momenta. We will here consider all $0\le k\le \pi$ on the same footing (where it should be noted that only half of the first Brillouin zone is used, 
as the other half corresponds to the same states). The normalization constant is different for the special cases;
\begin{equation}
N_{a} = \left ( \frac{N}{R_a} \right )^{\hskip-1mm 2} \sum_{r=1}^{R_a} [C^\sigma_k(r)]^2 = \frac{N^2g_k}{2R_a},
\label{normsemi}
\end{equation}
where we have introduced the factor 
\begin{equation}
g_k = \left \lbrace \begin{array}{ll}
\hskip-1.2mm 1, & 0 < k < \pi, \\
\hskip-1.2mm 2, & k = 0,\pi .
\end{array} \right.
\label{gfactor}
\end{equation}
In practice, $g_k$ will not matter here, because we are only considering matrix elements of $H$, which are diagonal in $k$. The $g$-factors 
therefore cancel out in the ratios of normalization constants appearing in the matrix elements.

Checking overlaps, states with the same $k$ but different $\sigma$ are orthogonal,
\begin{equation}
\langle a^{-\sigma}(k)|a^\sigma(k)\rangle = \frac{1}{N_a} \sum_{r=1}^{R_a} \sin(kr)\cos(kr) = 0,
\label{olapsemi}
\end{equation}
and other requirements for orthonormality, $\langle a^{\tau}(k')|a^\sigma(k)\rangle = \delta_{\sigma\tau}\delta_{kk'}$, of the semi-momentum basis 
can also easily be verified. The advantage of semi-momentum states is that they are real-valued for all $k$, in contrast to the complex momentum states.

The following equalities---standard trigonometric identities---are useful when manipulating semi-momentum states; 
\begin{eqnarray}
C^\pm_k(-r) &=&  \pm C^\pm_k(r), \label{cgrel1} \\
C^\pm_k(r + d) &=& C^\pm_k(r)C^+_k(d) \mp C^{\mp}_k(r)C^{-}_k(d),\label{cgrel2}
\end{eqnarray}
where $\pm$ stands for $\sigma=\pm 1$. Using (\ref{cgrel2}) it is easy to see that the Hamiltonian acting on a semi-momentum state mixes
$\sigma=\pm 1$ states. With $H_j|a\rangle = h_j(a) T^{-l_j}|b_j\rangle$ we get
\begin{equation}
H|a^\pm(k)\rangle = \sum_{j=0}^N h_j(a)\sqrt{\frac{R_a}{R_{b_j}}} 
\Bigl ( C^+_k(l_j)|b^\pm_j(k)\rangle \mp C^{-}_k(l_j)|b^{\mp}_j(k)\rangle \Bigr ),
\end{equation}
from which the hamiltonian matrix elements can be extracted and written as
\begin{equation}
\langle b^{\tau}(k)|H_j|a^\sigma(k)\rangle = 
h_j(a)\tau^{(\sigma-\tau)/2}\sqrt{\frac{N_{b_j}}{N_a}} C^{\sigma\tau}_k(l_j).
\label{hamkg}
\end{equation}

\paragraph{Incorporating parity}

Since the hamiltonian (\ref{hamkg}) is not diagonal 
in $\sigma,\tau$, the number of states in a semi-momentum block is twice that in a conventional momentum block, and thus it would appear that there is
not much to be gained over complex momentum states by making hamiltonian real in this way (by making the states real). However, we can also incorporate 
parity in a semi-momentum state, by defining 
\begin{equation}
|a^\sigma(k,p)\rangle = \frac{1}{\sqrt{N^\sigma_a}}\sum_{r=0}^{N-1}C^\sigma_k(r)(1 + pP)T^r|a\rangle .
\label{akgpdef}
\end{equation}
This is also a semi-momentum state, because the reflected component $P|a^\sigma(k)\rangle=\sigma|Pa^\sigma(k)\rangle$ is the semi-momentum state obtained by 
using the reflected representative state $P|a\rangle$ instead of $|a\rangle$. It can also easily be verified that (\ref{akgpdef}) is an eigenstate of the 
parity operator for any semi-momentum; $P|a^\sigma(k,p)\rangle=p|a^\sigma(k,p)\rangle$ with $p=\pm 1$. This is simply due to the fact that in (\ref{akgpdef}) 
the operator $(1+pP)$ appears before the translation operators $T^r$, in contrast to the momentum state (\ref{akpdef}) where  $(1+pP)$ is written after $T^r$. 
The hamiltonian is thus diagonal in $p$, and the number of states in each $(k,p)$ block is roughly half of that in the original semi-momentum $k$ blocks. 
We are then back to the same block size as with the conventional momentum states, but with states with purely real coefficients. 

\paragraph{Orthogonality and normalization of semi-momentum states}

Examining the orthogonality of the states (\ref{akgpdef}), we have an apparent problem: When $T^m P|a\rangle = |a\rangle$ for some $m$, the states 
$|a^+(k,p)\rangle$ and $|a^-(k,p)\rangle$ are not orthogonal. We can then use Eqs.~(\ref{cgrel1}) and (\ref{cgrel2}) to write the 
parity-conserving semi-momentum state (\ref{akgpdef}) as a linear combination of non-parity semi-momentum states (\ref{akgdef});
\begin{equation}
|a^\pm (k,p)\rangle=\sqrt{\frac{N_a}{N^\pm_a}}
\Bigl (\bigl (1 \pm pC^+_k(m) \bigr )|a^\pm (k)\rangle - pC^-_k(m)|a^\mp(k)\rangle \Bigr ).
\label{akpnonortho}
\end{equation}
It is then clear that $\langle a^{-\sigma} (k,p)|a^\sigma (k,p)\rangle$ can be non-zero. To look at this more closely, we first determine the normalization
constant $N^\sigma_a$ in (\ref{akgpdef}). Note that we have attached $\sigma$ as a superscript to indicate that the normalization constant of the 
parity-conserving semi-momentum states can, unlike the normalization (\ref{normsemi}) of the plain semi-momentum states, depend on $\sigma$ (in addition 
to the implicit dependence on $k,p$). In the case $T^m P|a\rangle \not= |a\rangle$ for all $m$, the calculation of $N^\sigma_a$ is trivial. When 
$T^m P|a\rangle = |a\rangle$ for some $m$ we can use (\ref{akpnonortho}) and the orthonormality of the pure semi-momentum states $|a^\sigma(k)\rangle$, resulting in
\begin{equation}
N^{\sigma}_a = \frac{N^2g_k}{R_a} \times
\left \{ \begin{array}{ll} 
\hbox{\hskip-1.5mm} 1, & T^m P|a\rangle \not= |a\rangle~~ \forall m, \\
\hbox{\hskip-1.5mm} 1 + \sigma p\cos(km), & T^m P|a\rangle = |a\rangle .
\end{array}\right.
\label{normkgp}
\end{equation}
In the same way, we can calculate the overlap between the $\sigma=\pm 1$ states when $T^m P|a\rangle = |a\rangle$ for some $m$ and find 
\begin{equation}
\langle a^\mp (k,p)|a^\pm (k,p)\rangle = -p,~~~~(T^m P|a\rangle = |a\rangle {\rm ~for~some~}m).
\label{olapkgp}
\end{equation}
This is of course under the assumption that both the $\sigma = \pm 1$ states exist (i.e., $N^\sigma_a \not =0$). Thus, the $\sigma = \pm 1$ states in 
the case $T^m P|a\rangle = |a\rangle$ differ at most by a sign, and we should include only one of them in the block of states. For definiteness, we can 
choose $|a^+(k,p)\rangle$ if $N^+_a \not=0$ and $|a^-(k,p)\rangle$ else. We still have to pay attention to the non-zero overlap (\ref{olapkgp}) in some 
formal manipulations with the semi-momentum states, as we shall see shortly.

\paragraph{The semi-momentum hamiltonian}

To calculate the hamiltonian matrix elements, we first note that, instead of (\ref{btransrep}) for a plain momentum state, now when we act with a 
hamiltonian operator $H_j$ on $|a\rangle$ we get a state $|b_j'\rangle$ which is related to another representative state $|b_j\rangle$ by a number 
of translations and possibly a reflection:
\begin{equation}
|b_j\rangle =  T^{l_j}P^{q_j}|b_j'\rangle,~~~~ l_j \in \{0,\ldots,N-1\},~~~ q_j \in \{0,1\}.
\end{equation}
The result of $H$ acting on a semi-momentum state (or a $k=0,\pi$ momentum state with parity) can therefore be written in in the form
\begin{equation}
H|a^\sigma(k,p)\rangle = \sum_{j=0}^N \frac{h_j(a) (\sigma p)^{q_j}}{\sqrt{N^\sigma_a}}
\sum_{r=0}^{N-1}C^\sigma_k(r+l_j)(1 + pP)T^r|b_j\rangle .
\label{hactskp}
\end{equation}
where the representative $|b_j\rangle$ is related to $H_j|a\rangle$ 
\begin{equation}
H_j|a\rangle = h_j(a) P^{q_j} T^{-l_j}|b_j\rangle,~~~~ l_j \in \{0,\ldots,N-1\},~~~ q_j \in \{0,1\}.
\end{equation}
Using the relation (\ref{cgrel2}) we can write (\ref{hactskp}) as
\begin{eqnarray}
&&H|a^\sigma(k,p)\rangle = \sum_{j=0}^N h_j(a) (\sigma p)^{q_j}\sqrt{\frac{N^\sigma_{b_j}}{N^\sigma_a}} \times \\
&&~~~~~~\Bigl ( \cos(kl_j)|b_j^\sigma(k,p)\rangle  - \sigma \sqrt{\frac{N^{-\sigma}_{b_j}}{N^\sigma_{b_j}}}\sin(kl_j)|b^{-\sigma}(k,p)\rangle \Bigr ). \nonumber 
\end{eqnarray}
The ratio of the $\sigma= \pm 1$ normalization constants is $1$ if $T^mP|b_j\rangle \not= |b_j\rangle$ for all $m$, and otherwise it
can be written as
\begin{equation}
\sqrt{\frac{N^{-\sigma}_{b_j}}{N^\sigma_{b_j}}} = \sqrt{\frac{1-\sigma p \cos(km)}{1+\sigma p \cos(km)}} = 
\frac{|\sin(km)|}{1+\sigma p \cos(km)}. 
\end{equation}
Using the overlap (\ref{olapkgp}) we can extract the matrix elements. The ones diagonal in $\sigma$ are,
\begin{eqnarray}
&&\langle b_j^{\sigma}(k,p)|H_j|a^\sigma(k,p)\rangle = h_j(a) (\sigma p)^{q_j}\sqrt{\frac{N^\tau_{b_j}}{N^\sigma_a}} \times \nonumber \\
&& \hskip25mm\left\lbrace \begin{array}{ll}
\hskip-1.0mm \cos(kl_j), & P|b_j\rangle \not= T^m|b_j\rangle, \\
\hskip-1.0mm \frac{\cos(kl_j)+\sigma p\cos(k[l_j-m])}{1 +\sigma p\cos(km)}, & P|b_j\rangle = T^m|b_j\rangle,~~~~~~ 
\end{array}\right.  \label{hamkgp1}
\end{eqnarray}
whereas the off-diagonal ones are
\begin{eqnarray}
&&\langle b_j^{-\sigma}(k,p)|H_j|a^\sigma(k,p)\rangle = h_j(a) (\sigma p)^{q_j}\sqrt{\frac{N^\tau_{b_j}}{N^\sigma_a}} \times \nonumber \\
&& \hskip22mm\left\lbrace \begin{array}{ll}
\hskip-1.0mm -\sigma\sin(kl_j), & P|b_j\rangle \not= T^m|b_j\rangle, \\
\hskip-1.0mm \frac{-\sigma\sin(kl_j)+p\sin(k[l_j-m])|}{1-\sigma p\cos(km)}, & P|b_j\rangle = T^m|b_j\rangle.~~~~~~ 
\end{array}\right.  \label{hamkgp2}
\end{eqnarray}

\paragraph{Degeneracies}

In the conventional momentum basis, states with $\pm k$ are degenerate, but are orthogonal states (except when $k \not =0,\pi$). In 
the semi-momentum basis, $\pm k$ states are really the same state, differing at most (in the case $\sigma =-1$) by a sign (and that is why the semi-momentum 
Hilbert space only includes $0 \le k \le \pi$). The degeneracy has here been moved to the parity sector, with the $p \pm 1$ states being degenerate (but still orthogonal) 
for $k \not =0,\pi$. This can be seen by letting $(p,\sigma,\tau) \to -(p,\sigma,\tau)$ in the matrix elements (\ref{hamkgp2}), which leaves unchanged the set of matrix 
elements for $\tau,\sigma = \pm 1$. Thus, in calculations we need to consider only, e.g., $p=+1$ for $k \not =0,\pi$ (while for $k=0,\pi$ we only have $\sigma=\tau=1$
and the $p = \pm 1$ sectors are not degenerate).

\paragraph{Constructing the basis of semi-momentum states}

\begin{figure}
\includegraphics[width=8cm, clip]{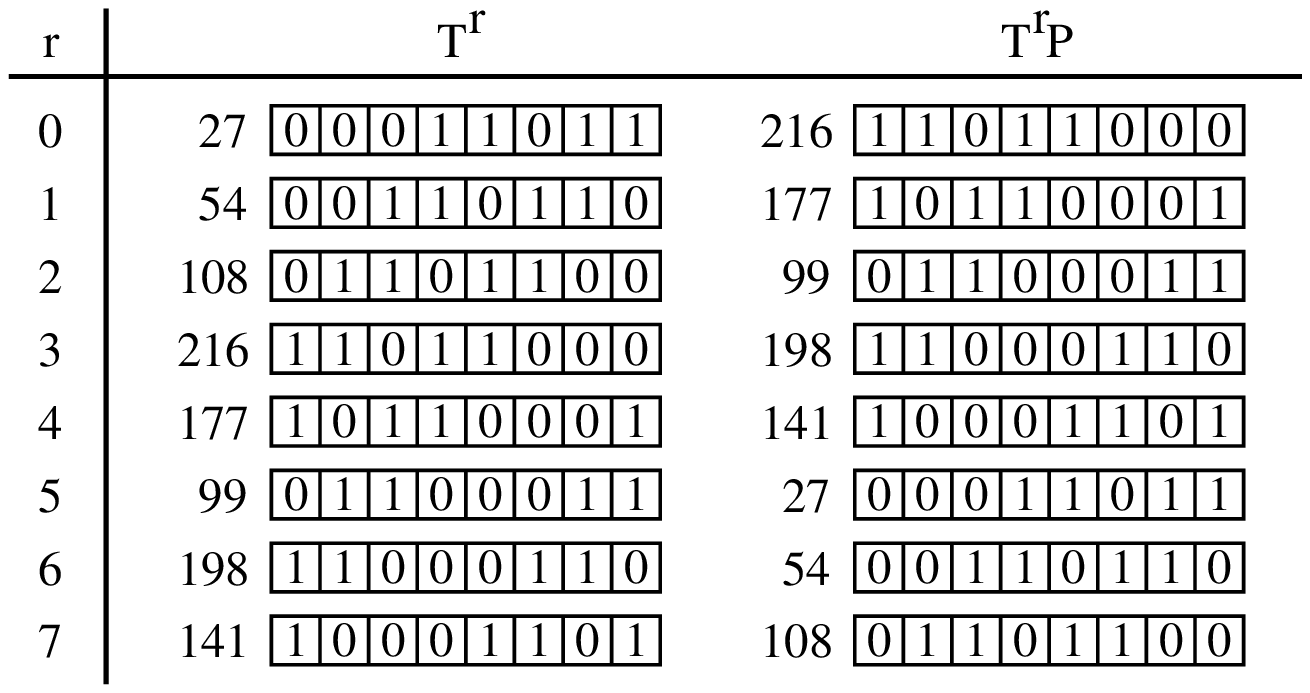}
\caption{A state $|s\rangle = |\up\up\dn\up\up\dn\dn\dn\rangle$ (upper left) along with all its transformations by the translation, $T$, and reflection, $P$, operator 
in the bit representation. The corresponding base-10 integer $s$ and its transformations are also shown--the smallest integer, $s=27$, corresponds to the 
representative state. For this particular state, $T^5P|s\rangle=|s\rangle$, i.e., $m=5$ in the normalization 
constant (\ref{normkgp}).}
\label{bitstates3}
\end{figure}

Generalizing the convention we used for momentum states, we now choose as the representative of a state $|s\rangle$ the state $|s(r,q)\rangle = T^rP^q|s\rangle$ 
for which the corresponding integer $s(r,q)$ is the smallest (where $r\in \{0,\ldots,N-1\}$, $q\in \{0,1\}$). States can now be disallowed in a given block not 
only due to incompatibility with the momentum, but also because of restrictions imposed by the reflection quantum number $p$ and the non-conserved state label 
$\sigma$. We thus have to determine which one, or both, of the $\sigma=\pm 1$ semi-momentum states (\ref{akgpdef}) should be included, according to the conditions 
discussed above. Fig.~\ref{bitstates3} illustrates all the combination of translations and reflections in the bit representation of a state 
which is related to its own reflection according to $|s\rangle = T^mP|s\rangle$, in which case only the $\sigma=-1$ or $\sigma=+1$ variant of the state should 
be included in the basis (and the state is compatible with all momenta, because its periodicity $R=N$).

When both $\sigma=\pm 1$ states are required, we will store two consecutive copies of the same representative, so that we can continue to use the location 
in the list $s_a$ as the label of each state within the block. We then also need to store the $\sigma$ labels for each representative, as well as the number 
of translations $m$ by which the reflection of a state is brought back onto itself, which is needed in the 
normalization and the hamiltonian.

As before, we will use a subroutine {\bf checkstate} to determine whether or not a state is a new representative to be added to the list. In addition to the 
periodicity $R$ of a new representative, this subroutine now also delivers the reflection-translation number $m$, when applicable. We can use $m=-1$ as a flag 
indicating that there is no $m$ for which $T^mP|s\rangle = |s\rangle$ (and otherwise $m\ge 0$). The operations on the reflected state are carried out after the 
instructions in $\{6\}$ have been performed, by first reflecting the spin bits in the state integer $s$ and storing it as $t$, $t[i]=s[N-1-i]$ for $i=0,\ldots,N-1$. 
This reflection is accomplished by a function {\bf reflectbits}$(s,N)$. Then the resulting integer is translated as before. We thus modify code $\{6\}$ 
according to:

{\code
\cia       {\bf subroutine checkstate}$(s,R,m)$ \br
\cia       {\bf ...}                                                 \hfill \{11\}\break
\cia       $t=\mathbf{reflectbits}(s,N)$;~ $m=-1$ \br
\cia       {\bf do} $i=0,R-1$ \br
\cib          {\bf if} $(t<s)$ {\bf then} \br
\cic             $R=-1$;~ {\bf return} \br
\cib          {\bf elseif} $(t=s)$ {\bf then} \br
\cic             $m=i$;~ {\bf return} \br        
\cib          {\bf endif} \br
\cib          $t=\mathbf{cyclebits}(t,N)$ \br 
\cia       {\bf enddo} 
\code}

\noindent
Here {\bf ...} represents all operations in code $\{6\}$. Note that the loop in the code above starts at $i=0$, and the translations are carried out after 
comparing $t$ with $s$, since a state can be its own reflection (in which case $m=0$). Also, since we have already determined the periodicity $R$ of the state, 
we need to go only up to $i=R-1$ in this loop. The function {\bf reflectbits} has to be implemented by hand; a corresponding internal function is normally
not available. In Fortran 90 one can use the functions {\bf btest}$(i,b)$, {\bf setbit}$(i,b)$, and {\bf clrbit}$(i,b)$, to examine, set, and clear 
individual bits $b$ of an integer $i$.

A state which has passed the above checks is still not necessarily a valid representative for both $\sigma=\pm 1$, and the state must therefore be further examined 
to determine which (one or both) of the $\sigma = \pm 1$ copies of the representative should be added to the basis. The following code segment stores the state 
information for $\sigma = \pm 1$ states that are to be included in the block based on the criteria discussed above;

{\code
\cia       {\bf call checkstate}$(s,R,m)$ \br      
\cia       {\bf do} $\sigma=\pm 1$~ (do only $\sigma=+1$ if $k=0$ or $k=N/2$) \hfill \{12\}\break
\cib           {\bf if} ($m\not=-1$) {\bf then} \br                       
\cic               {\bf if } $(1+\sigma p \cos(ikm2\pi/N)=0)$ $R=-1$ \br
\cic               {\bf if } $(\sigma=-1 {\bf ~and~} 1-\sigma p \cos(ikm2\pi/N)\not=0)$ $R=-1$ \br
\cib          {\bf endif} \br
\cib          {\bf if} $R>0$ {\bf then} $a=a+1$;~ $s_a=s$;~ $R_a=\sigma R$;~ $m_a=m$ {\bf endif} \br
\cia       {\bf enddo} 
\code}

\noindent
Recall that here $m=-1$ means that there is no $m$ such that $T^{m}P|s\rangle = |s\rangle$, whence both $\sigma=\pm$ states should be included. On the other
hand, if there is such an $m$, then only one of the states is included---we pick $\sigma=1$ if the normalization constant (\ref{normkgp}) for it is non-zero. 
This is accomplished above  by the two consecutive {\bf if} statements (where $R=-1$ is set for an invalid representative). We have to store $\sigma$, and since 
the periodicity $R$ is positive and non-zero, we can store it and $\sigma$ jointly as $\sigma R$ in a list $R_a$. A more dense packing of $R,\sigma,m$ into a single 
integer is of course also possible.

\paragraph{Constructing the semi-momentum hamiltonian}

For general semi-momentum, the same representative can appear once or twice in the list $s_a$, whereas for the special cases $k=0,\pi$ there is always only a 
single ($\sigma=1$) copy of each representative. We here construct code that can treat both cases, but simpler code can be written for the two special momenta.
When building the hamiltonian by looping over the state indices $a$, with the corresponding representatives $s_a$, we first check the previous and next items 
in the list. If the previous item $s_{a-1}=s_a$, then we skip over this representative, because it will already have been taken care of in the previous pass 
through the loop. If the next representative, $s_{a+1}$ equals $s_a$, then we store as $n=2$ the number of same representatives, and else set $n=1$. This part of 
the hamiltonian building is accomplished by;

{\code
\cia       {\bf do} $a=1,M$ \br 
\cib           {\bf if} ($a>1$ {\bf and} $s_a=s_{a-1}$) {\bf then}          \hfill \{13\}\break
\cic               {\bf skip } {\rm to next} $a$ \br
\cib           {\bf elseif} ($a<M$ {\bf and} $s_a=s_{a+1}$) {\bf then} \br
\cic               $n=2$ \br
\cib           {\bf else} \br
\cic               $n=1$ \br
\cib           {\bf endif} \br
\cib           {\bf ...}   \br
\cia       {\bf enddo} 
\code}

\noindent
Here the {\bf skip} command skips to the next iteration of the loop and {\bf ...} stands for the bulk of the loop, the main features of which 
we discuss next. 

Considering first the diagonal matrix elements $H(a,a)$, for a given $a$ we now first sum up all the diagonal contributions from the bit configurations
in $s_a$ and store the result as $E_z$. Taking into account the possibility of one or two entries of the representative ($n=1$ or $2$), we can use the following 
loop to assign the diagonal matrix elements accordingly;

{\code
\cia       {\bf do} $i=a,a+n-1$ \br 
\cib           $H(a,a)=H(a,a)+E_z$          \hfill \{14\}\break
\cia       {\bf enddo} 
\code}

\noindent
Turning next to the off-diagonal operations, looping over site pairs $(i,j)$, if the bits $s_a[i]\not=s_a[j]$, then the spins are flipped, resulting in 
a state-integer named $s'$ as in code $\{10\}$. We again have to identify the corresponding representative state $|r\rangle=T^lP^q|s'\rangle$ by carrying out 
all the symmetry operations to determine the transformation indices $l$ and $q$ appearing in the hamiltonian matrix elements (\ref{hamkgp1}) and 
(\ref{hamkgp2}). The subroutine accomplishing this is a simple modification of code $\{8\}$;

{\code
\cia    {\bf subroutine representative}$(s',r,l,q)$ \br
\cia    {\bf ... }                         \hfill \{15\}\break
\cia    $t=\mathbf{reflectbits}(s',N)$;~ $q=0$\br
\cia    {\bf do} $i=1,N-1$ \br
\cib       $t=\mathbf{cyclebits}(t,N)$ \br 
\cib       {\bf if} $(t<r)$ {\bf then} $r=t$;~ $l=i$;~ $q=1$ {\bf endif} \br
\cia    {\bf enddo} 
\code}

\noindent
Here {\bf ...} stands for the code after the header in $\{8\}$. After having called this subroutine, we can search for the representative $r$ in the list 
$\{s_a\}$. This is done in the same way as we did before, with the subroutine {\bf findstate}$(r,b)$. 

Since the hamiltonian is more complicated than in the standard
momentum basis, we also define a function {\bf helement}$(a,b,l,q)$, which returns the matrix element according to (\ref{hamkgp1}) or 
(\ref{hamkgp2}). The function takes as input the labels $a$ and $b$ (locations in the matrix) of the two representatives and the transformation indices 
$l,q$ delivered by {\bf representative}. The matrix element also of course depends on the momentum $k$ and the parity quantum number $p=\pm 1$, which we do 
not indicate explicitly here. The function is straight-forward to implement based on (\ref{hamkgp1}) and (\ref{hamkgp2}), and we do not list any code here.

As with the state-integer $s_a$ in code segment $\{13\}$, we also here need to take into account that the representative $s_b$ can appear once or twice 
(indicated below with $m=1$ or $2$) in the list of representatives. In case there are two copies, we do not know whether the subroutine {\bf findstate} has 
returned its first or second location. Thus, we again need to examine positions in the list of representatives adjacent to the one delivered. The following 
code segment does all that and then assigns the matrix elements:

{\code
\cia    $s'={\bf flip}(s_a,i,j)$ \br
\cia    {\bf call representative}$(s',r,l,q)$       \hfill \{16\}\break  
\cia    {\bf call findstate}$(r,b)$  \br      
\cia    {\bf if} $(b\ge 0)$ {\bf then} \br
\cib       {\bf if} ($b>1$ {\bf and} $s_b=s_{b-1}$) {\bf then}  \br
\cic          $m=2$; $b=b-1$ \br
\cib       {\bf elseif} ($b<M$ {\bf and} $s_b=s_{b+1}$) {\bf then} \br
\cic          $m=2$ \br
\cib       {\bf else} \br
\cic          $m=1$ \br
\cib       {\bf endif} \br
\cib       {\bf do} $j=b,b+m-1$ \br
\cib       {\bf do} $i=a,a+n-1$ \br
\cic            $H(i,j)=H(i,j)+{\bf helement}(i,j,l,q)$ \br
\cib       {\bf enddo} \br
\cib       {\bf enddo} \br
\cia    {\bf endif} 
\code}

\noindent
This piece of codes replaces the much simpler four-line code segment $\{10\}$ for the pure momentum basis. The advantage is that the matrix $H$
is now real-valued.

\subsubsection{Spin-inversion symmetry}
\label{spininversion}

Spin-inversion symmetry, with the operator $Z$ defined in (\ref{zdef}), can be used to reduce the hamiltonian block size in the magnetization sector $m^z=0$. 
The fact that the magnetization is conserved implies, however, that there is nothing to be gained by using this symmetry in the $|m_z|>0$ sectors (since such a 
basis would consist of mixed $\pm |m_z|$ states, and the blocks of symmetric and anti-symmetric combinations would be of the same size as the original blocks). 
In models that do not conserve $m_z$, e.g., the transverse-field Ising model, spin inversion can be exploited for all states. We denote by $z$ the eigenvalue 
of $Z$. Since $Z^2=1$, we again have $z=\pm 1$. 

Unlike the parity operation considered in the previous section, the spin-inversion operator $Z$ in (\ref{zdef}) commutes with the translation
operator, $[T,Z]=0$. The associated quantum number $z$ is therefore conserved together with $k$ in all momentum sectors. In the magnetization sector 
$m_z=0$, we can therefore always split a momentum block into two smaller ones by using states of the form
\begin{equation}
|a(k,z)\rangle = \frac{1}{\sqrt{N_a}} \sum_{r=0}^{N-1} {\rm e}^{-ikr} 
T^r \bigl (1+ zZ)|a\rangle,
\label{akzdef}
\end{equation}
where the normalization constant is easily obtained as
\begin{equation}
N_a= \frac{2N^2}{R_a} \times \left \lbrace \begin{array}{ll}
\hbox{\hskip-1.5mm} 1,~~            & T^mZ|a\rangle \not= |a\rangle ~~\forall~ m, \\
\hbox{\hskip-1.5mm} 1+z\cos(km),~~  & T^mZ|a\rangle = |a\rangle.
\end{array}\right.
\label{normkz}
\end{equation}
For a hamiltonian operation on $|\alpha\rangle$ resulting in
\begin{equation}
H_j|a\rangle = h_j(a) Z^{g_j}T^{-l_j}|b_j\rangle,~~~~ l_j \in \{0,\ldots,N-1\},~~~ g_j \in \{0,1\},
\end{equation}
we obtain the matrix element
\begin{equation}
\langle b_j(k,z)|H_j|a(k,z)\rangle = h_j(a)z^{g_j}{\rm e}^{-ik{l_j}} \sqrt{\frac{N_{b_j}}{N_a}},
\label{hamkz}
\end{equation}
which is valid for any $k$. As always, for $j=0$ this reduces to just $h_0(a)$.

\paragraph{Spin-inversion symmetry with semi-momentum states}

We can also consider semi-momentum states incorporating spin-inversion symmetry;
\begin{equation}
|a^\sigma(k,p,z)\rangle = \frac{1}{\sqrt{N^\sigma_a}}\sum_{r=0}^{N-1}C^\sigma_k(r)(1  + pP)(1 + zZ)T^r|a\rangle,
\label{akgpzdef}
\end{equation}
These states are eigenstates of $Z$ as well as $P$. 

When calculating the normalization and constructing the hamiltonian we now have to consider five different types of reference states, depending on 
which combinations of symmetry operations transform the reference state into itself;
\begin{equation}
\begin{array}{rrrl}
1)~~ & T^mP|a\rangle \not= |a\rangle,~~& T^{m}Z|a\rangle \not= |a\rangle~~&~~~~ T^{m}PZ|a\rangle \not= |a\rangle,\\
2)~~ & T^mP|a\rangle  = |a\rangle,~~   & T^{m}Z|a\rangle \not= |a\rangle~~&~~~~ T^{m}PZ|a\rangle \not= |a\rangle,\\
3)~~ & T^mP|a\rangle \not= |a\rangle,~~& T^{m}Z|a\rangle = |a\rangle~~      &~~~~ T^{m}PZ|a\rangle \not= |a\rangle,\\
4)~~ & T^mP|a\rangle \not= |a\rangle,~~& T^{m}Z|a\rangle \not= |a\rangle~~     &~~~~T^{m}PZ|a\rangle = |a\rangle,\\
5)~~ & T^mP|a\rangle = |a\rangle,~~    & T^nZ|a\rangle = |a\rangle~~        &\Rightarrow T^{m+n}PZ|a\rangle = |a\rangle.
\end{array} \label{tpzconditions}
\end{equation}
Here the inequalities should hold for all $m$ and the equalities for some $m,n$. The calculations follow the procedures of the previous section, and we just list 
the results. The normalization constants are:
\begin{equation}
N^{\sigma}_a = \frac{2N^2}{R_ag_k} \times
\left \{ \begin{array}{ll} 
\hbox{\hskip-1.5mm} 1,                       & 1) \\
\hbox{\hskip-1.5mm} 1+\sigma p\cos(km),          & 2) \\
\hbox{\hskip-1.5mm} 1+z\cos(km),                 & 3) \\
\hbox{\hskip-1.5mm} 1+\sigma pz\cos(km),        & 4) \\
\hbox{\hskip-1.5mm} [1+\sigma p\cos(km)][1+z\cos(kn)].   & 5)
\end{array}\right.
\label{normkgpz}
\end{equation}
For cases 2),4), and 5), only one state out of a pair with $\sigma = \pm 1$ should be included in the basis as the two states within such a pair differ merely 
by a factor. To be definite, we can again pick $\sigma=1$ if that makes $N^\sigma_a>0$ and $\sigma=-1$ else. Acting with a hamiltonian operator on a representative 
now leads to a new representative $|b_j\rangle$ according to
\begin{equation}
H_j|a\rangle = h_j(a)P^{q_j}Z^{g_j}T^{-l_j}|b_j\rangle.
\end{equation}
The $\sigma$-diagonal matrix elements of $H$ are
\begin{eqnarray}
&&\langle b_j^{\sigma}(k,p)|H_j|a^\sigma(k,p)\rangle = h_j(a) (\sigma p)^{q_j}z^{g_j}\sqrt{\frac{N^\tau_{b_j}}{N^\sigma_a}} 
\times \nonumber \hskip10mm \\
&& \hskip30mm \left\lbrace \begin{array}{ll}
\hskip-1.0mm \cos(kl_j), & 1), 3)\\
\hskip-1.0mm \frac{\cos(kl_j)+\sigma p\cos(k[l_j-m])}{1+\sigma p\cos(km)}, & 2), 5),\\
\hskip-1.0mm \frac{\cos(kl_j)+\sigma pz\cos(k[l_j-m])}{1+\sigma pz\cos(km)}, & 4),~~~~~~ 
\end{array}\right.  \label{hamkgpz1}
\end{eqnarray}
whereas the ones off-diagonal in $\sigma$ are,
\begin{eqnarray}
&&\langle b_j^{-\sigma}(k,p)|H_j|a^\sigma(k,p)\rangle = h_j(a) (\sigma p)^{q_j}z^{g_j}\sqrt{\frac{N^\tau_{b_j}}{N^\sigma_a}} 
\times \nonumber \hskip10mm \\
&& \hskip30mm\left\lbrace \begin{array}{ll}
\hskip-1.0mm -\sigma \sin(kl_j), & 1), 3), \\
\hskip-1.0mm \frac{-\sigma \sin(kl_j)+p\sin(k[l_j-m])}{1 -\sigma p\cos(km)}, & 2), 5),\\
\hskip-1.0mm \frac{-\sigma \sin(kl_j)+pz\sin(k[l_j-m])}{1- \sigma pz \cos(km)}, & 4),~~~~~~ 
\end{array}\right.  \label{hamkgpz2}
\end{eqnarray}
These expressions may seem rather complicated, but it should be noted that they are completely general for 1D systems, not just
for the Heisenberg chain considered here. Once they have been implemented and tested for some model, the code can easily be reused 
for other systems.

\paragraph{Program implementation; combining all symmetries}

Incorporating spin-inversion symmetry in the semi-momentum basis with $m_z=0$, with states $|a^\sigma(k,p,z)\rangle$ defined in (\ref{akgpzdef}), we have
to augment the subroutine {\bf checkstate} in code segment $\{11\}$. It should return the translation numbers corresponding not only to reflection $P$
of a representative, $T^{m_p}P|s\rangle = |s\rangle$, but also translation numbers in symmetry relationships involving $Z$ and $PZ$; $T^{m_z}Z|s\rangle = 
|s\rangle$ and $T^{m_{pz}}PZ|s\rangle = |s\rangle$. To accomplish the spin inversion in the bit representation, we define a function {\bf invertbits}$(s,N)$
which flips all the bits $s[i]\to 1-s[i]$, $i=0,N-1$. This can be easily accomplished without individual bit-level operations as $s=2^{N-1}-s$ (for $N$ less
than the number of bits in the integers used; otherwise a different formula has to be used). Codes $\{6\}$ and $\{11\}$ can be easily modified 
to also check the inverted and reflected-inverted states to determine $m_z$ and $m_{zp}$.

If the modified {\bf checkstate} returns a potential representative, $R\not=-1$, then again the state has to be examined further to determine whether it satisfies 
the further criteria for a representative for one or two states $\sigma = \pm 1$. An allowed representative now falls into one of five different classes 
depending on its symmetry properties as summarized in (\ref{tpzconditions}). The class, $c\in \{1,\ldots,5\}$ can be determined from the translation integers 
$m_p,m_z,m_{zp}$ delivered by {\bf checkstate}. After $c$ has been determined, the three translation integers can be reduced to two; the $m$ and $n$ in the 
normalization constants (\ref{normkgpz}). The class $c$ can be packed along with $m,n$ into a single integer, e.g., as $m+n(N+1)+c(N+1)^2$, which is stored 
along with the periodicity and $\sigma$ index packed as $\sigma R$ as before. 

For the construction of the hamiltonian, the subroutine {\bf representative}, the main parts of which are described in code segments $\{8\}$ and $\{15\}$, has 
to be amended further in order to also return an index $g$ corresponding to the number, $0$ or $1$, of spin-inversions needed, along with $q$ reflections and 
$l$ translations, to transform the state $|s\rangle $generated by a spin flip into the corresponding representative; $|r\rangle=T^{l_j}P^{q_j}Z^{g_j}|s\rangle$
The construction of the hamiltonian then proceeds exactly as before, with a modification only of the function {\bf helement} to include also $g$
as an argument in code segment (5.15). This is straight-forward, and there is no need to list any code here.

\paragraph{Examples of state block sizes}

\begin{table}
\begin{tabular}{rrrrr}
\hline
$N$ & $(+1,+1)$ & $(+1,-1)$ & $(-1,+1)$ & $(-1,-1)$ \\
\hline
 8 & 7 & 1 & 0 & 2 \\
12 & 35 & 15 & 9 & 21 \\
16 & 257 & 183 & 158 & 212 \\
20 & 2518 & 2234 & 2136 & 2364 \\
24 & 28968 & 27854 & 27482 & 28416 \\
28 & 361270 & 356876 & 355458 & 359256 \\
32 & 4707969 & 4690551 & 4685150 & 4700500 \\
\hline
\end{tabular}
\label{sizetab}
\caption{Size of the $k=0$ state blocks for magnetization $m_z=0$ and different parity and spin-inversion 
quantum numbers $(p,z)$.}
\end{table}

This is as far as we will go with applying symmetries for 1D systems. For the Heisenberg model, the total spin is also conserved, but incorporating 
that symmetry in the basis is much more complicated. In principle it can be done for total spin singlets, using the valence-bond basis \cite{poilblanc}, 
but it is rarely done in practice because the resulting hamiltonian is very dense, unlike the sparse matrices obtaining with the symmetries 
implemented here (the number of non-zero elements is proportional to $M\times N$ for an $M\times M$ matrix, and $N\ll N$). The sparseness 
will be very useful in the Lanczos calculations discussed in Sec.~\ref{lanczos}.

Table \ref{sizetab} shows the size of the Hilbert space blocks for several different chain sizes $N$ when all the symmetries are used. Here the momentum 
$k=0$ and the number of states is shown for all combinations of the parity and spin-inversion quantum numbers $(p=\pm 1,z=\pm 1)$. The largest block is always 
the fully symmetric one. As the system size grows, the 
relative variations in the block size diminish, and for large $N$ all sizes are rather well approximated by $N!/[4(\frac{N}{2}!)^2N]$. For $k=\pi$,
the blocks are approximately of the same size as in the table for $k=0$, and for other momenta, where both $\sigma=1$ and $-1$ states are allowed, 
the blocks are roughly twice as large.

\subsubsection{Expectation values and thermodynamics}
\label{expvalues}

We are now ready to diagonalize the hamiltonian and calculate physical observables. Here we will first consider complete diagonalization, meaning that we 
compute all the eigenvalues and eigenstates of $H$. In principle, the eigenvalues $\lambda_n$, $n=1,\ldots,M$, of a non-singular $M\times M$ matrix ${H}$ 
can be obtained by solving the secular equation,
\begin{equation}
{\rm det}[H-\lambda_n I]=0,
\end{equation}
where ${\rm det}[]$ denotes the determinant and $I$ is the unit matrix. The eigenvectors $v_n$ can 
subsequently be obtained by solving the linear system of equations
\begin{equation}
{Hv}_n = E_n{v}_n.
\end{equation}
However, since the secular equation is very complicated for a large matrix, this method is not used in practice. Most numerical matrix diagonalization methods 
are based on an iterative search for a unitary transformation $U$ such that
\begin{equation}
{U}^{-1}{H}{U}={E},
\label{dad}
\end{equation}
where ${E}$ is the diagonal eigenvalue matrix with $E_{nn}=E_n$ and, for a complex hermitian matrix $H$, the inverse ${U}^{-1}$ of the unitary matrix 
${U}$ is the transpose of its complex conjugate matrix; ${U}^{-1}={U}^{*T}$. If the  matrix is real and symmetric, we have ${U}^{-1}={U}^{T}$. 

The columns of the diagonalizing matrix $U$ contain the eigenvectors of $H$. This can be seen by multiplying (\ref{dad}) with $U$ from the left, giving 
${HU} = {UH}$. Since $E$ is diagonal, the $n$th column of $U$ on the right side of this equality is multiplied by the $n$th eigenvalue (i.e., the matrix 
element $E_{nn}=\lambda_n$). In the multiplication by $H$ on the left side, the $n$th column of $U$ gives the $n$th column $(HU)_n$ of $HU$, i.e., 
${H}{U}_n = E_{nn}{U}_n$, and thus the eigenvectors $v_n$ of $H$ are identified as $v_n = U_n$. 

The expectation value of some observable (operator) $A$ in an eigenstate of $H$ is given by the diagonal element of the corresponding matrix transformed
to the energy basis;
\begin{equation}
\langle n|A|n\rangle =[U^{-1}AU]_{nn}.
\label{anexp}
\end{equation}
One would typically be interested in the ground state, $n=1$ (assuming the eigenvalues to be ordered from the lowest to the highest), of the block 
in question (especially the actual ground state, in the block with the lowest energy of all). For a thermal average, all eigenvalues and/or 
eigenstates of all symmetry blocks are needed;
\begin{equation}
\langle A\rangle = 
\frac{1}{Z}\sum_j \sum_{n=1}^{M_j} {\rm e}^{-\beta E_{j,n}} [U_j^{-1}A_jU_j]_{nn},~~~~Z = \sum_j \sum_{n=1}^{M_j} {\rm e}^{-\beta E_{j,n}}.
\label{aexpect}
\end{equation}
Here $\beta=T^{-1}$ is the inverse temperature  (in units where $k_B=1$), and the index $j$ collectively denotes the different quantum numbers, 
$m_z,k,p,z$, of the blocks of size $M_j$. 

Matrix diagonalization is a standard linear algebra operation, and sophisticated subroutines are available in many software libraries. We will therefore not discuss 
the inner workings of matrix diagonalization procedures here. Almost always, a diagonalization routine delivers the eigenvalues in ascending order in a vector, 
along with the matrix $U$ with the corresponding eigenvectors. 

The number of operations needed to diagonalize an $M\times M$ matrix generally scales as $M^3$, and the memory required for storage is $\sim M^2$ (even for a 
sparse matrix, as intermediate steps normally do not maintain sparsity). This severely 
limits the size of the matrices that can be fully diagonalized in practice. Currently $M \approx 10^4$ can be handled without too much effort on a workstation, 
and a few times larger on a supercomputer. Looking at Table \ref{sizetab}, it is then clear that one cannot realistically carry out complete diagonalizations 
for Heisenberg chains larger than $N \approx 20$. Calculations aiming at just the ground state, and possibly some number of excited states, can be carried 
out in other ways for larger systems, using, e.g., the Lanczos method discussed in Sec.~\ref{lanczos}.

\paragraph{Total spin}

It is useful to know the total spin $S$ of the energy eigenstates. Since we do not use the conservation of $S$ when block-diagonalizing 
the hamiltonian, we have to calculate ${\bf S}^2$ using the states. As we have already noted, this operator is formally equivalent to a Heisenberg hamiltonian 
with long-range interactions, as written explicitly in (\ref{ssquared}). We can use a slightly modified version of the procedures we developed for 
constructing the hamiltonian to obtain the matrix for ${\bf S}^2$. We then transform it with $U$, after which we have $S(S+1)$ in the form of the 
diagonal matrix elements as in (\ref{anexp}). 

Let us look at some results obtained for a 16-site chain in the $k=0$ momentum sector. The eight lowest energies 
are listed for each of the symmetry sectors $(p,z)$ in table \ref{estab}, along with the calculated spin of the states. The lowest-energy state is a singlet 
with $(p=1,z=1)$. The first excited state of the system is actually not in the $k=0$ sector; it is a $k=\pi$ triplet with $(p=-1,z=-1)$, at energy 
$E=-6.87210668$.

\begin{table}
\begin{tabular}{rllllllll}
\hline
$n$ & $E(+1,+1)$  &$S$& $E(+1,-1)$  &$S$& $E(-1,+1)$  &$S$& $E(-1,-1)$ &$S$  \\
\hline
  1 & -7.1422964 & 0 & -4.9014133 & 1 & -4.1926153 & 2 & -5.7475957 & 1\\
  2 & -6.1223153 & 2 & -4.1067293 & 1 & -3.6537528 & 0 & -4.9014133 & 1\\
  3 & -5.5912905 & 0 & -3.9398439 & 1 & -3.6085498 & 2 & -4.6986358 & 1\\
  4 & -5.0981578 & 2 & -3.7756347 & 1 & -3.2192241 & 2 & -4.1067293 & 1\\         
  5 & -4.8142442 & 0 & -3.6808576 & 1 & -3.2129324 & 0 & -4.0007340 & 1\\
  6 & -4.5657878 & 0 & -3.5785191 & 3 & -3.1695648 & 2 & -3.9398439 & 1\\
  7 & -4.3243602 & 2 & -3.3678831 & 1 & -3.1652647 & 0 & -3.7756347 & 1\\
  8 & -4.1926153 & 2 & -3.3605397 & 3 & -3.1169772 & 2 & -3.6808576 & 1\\
\hline
\end{tabular}
\label{estab}
\caption{The eight lowest energies for a 16-site chain with momentum $k=0$ in the blocks with parity and spin-inversion 
quantum numbers $(p=\pm 1,z=\pm 1)$.}
\end{table}

One interesting feature to note in the table is that, in every symmetry sector, all the states have either even or odd spin. This is due to the spin-inversion 
symmetry. An $m_z=0$ state with $N$ sites and total spin $S$ can be written as a linear combination of states made up of $N/2-S$ singlets, 
$(|\up_i\dn_j\rangle - |\dn_i\up_j\rangle)/\sqrt{2}$, and $S$ triplets, $(|\up_i\dn_j\rangle + |\dn_i\up_j\rangle)/\sqrt{2}$ \cite{vbbasis}. 
Since the singlet is odd under spin inversion but the triplet is even, it follows that a spin $S$ state of an $N$-site system 
has spin-inversion quantum number $z=+1$ if $N/2$ and $S$ are both even or both odd, and has $z=-1$ else. Thus, even if we do not calculate $S$ we have 
some limited knowledge of it also from $z$. The low-energy states typically also have low spin, $S=0$ or $1$, and these can then be distinguished by 
$z$ alone. Note, however, that the lowest state in a given $(p,z)$ sector does not necessarily have the lowest (even or odd) spin, as exemplified 
in Table \ref{estab} by the lowest $(p=-1,z=1)$ level, which has $S=2$, not $S=0$. However, this is not a low-energy state, as there are 11 
states with lower energy in the $k=0$ sector (and numerous additional ones in other momentum sectors). The $k=0$ state with second-lowest energy also has 
$S=2$, but there are four $S=1$ states below it in other momentum sectors.

\paragraph{Magnetic susceptibility and specific heat}

As an example of thermodynamics, let us calculate two important properties of the Heisenberg chain; the specific heat $C$ and the magnetic 
susceptibility $\chi^z$. These can be evaluated according to the formulas,
\begin{eqnarray}
\label{cexpr} C      & = & \frac{d\langle H\rangle }{dt} = \frac{1}{T^2}\bigl ( \langle H^2\rangle - \langle H\rangle^2 \bigr ),  \\
\label{xexpr} \chi & = & \frac{d\langle m_z\rangle }{dh} = \frac{1}{T}\bigl ( \langle m_z^2\rangle - \langle m_z\rangle^2 \bigr ), 
\end{eqnarray}
which can both be easily derived using (\ref{aexpect}). In the definition of $\chi$, $h$ is an external magnetic field in the $z$-direction, which 
couples through a term $-hm_z$ added to the hamiltonian. We will here consider the zero-field case only, and thus $\langle m_z\rangle=0$.

Both $C$ and $\chi$ are special quantities in the sense that they do not depend on the states, just the energy spectrum. For the Heisenberg model
one can calculate these quantities by just considering the $m_z=0$ sector, because of the conservation of the total spin $S$. A spin $S$ state is 
$(2S+1)$-fold degenerate with magnetization $m_z=-S,\ldots S$. In the case of $C$, we therefore just have to sum \ref{cexpr} over the $m_z=0$ 
levels and weight them by this degeneracy factor to obtain the full average over all $m_z$. In the case of $\chi_z$, we can proceed in the same way, 
using also the fact that, in a spin-$S$ multiplet, the average $\langle m_z^2\rangle=\langle {\bf S}^2\rangle/3=S(S+1)/3$  in (\ref{xexpr}). 

The situation is slightly complicated by the fact that there can be ``accidental'' degeneracies not related to any apparent symmetry of the hamiltonian. 
If such degenerate states have different $S$, then the diagonalization procedure will give some random (depending on exactly how the diagonalization 
is done) mixed spins for the states. One should then in principle diagonalize ${\bf S}^2$ in the degenerate subspace to obtain the spin eigenvalues.
However, in practice these accidental degeneracies are very rare. For small Heisenberg chains they occur only in the $k=\pi,p=1,z=1$ sector. For $N=10$, 
there are two such states with $E=0.5$, for $N=12$ and $14$ there are no accidental mixed-$S$ degeneracies, and for $N=16$ there are four mixed levels at 
energy $2J$. With such a small number of degeneracies, and at such high energies, we can just ignore them. 

\begin{figure}
\includegraphics[width=11.5cm, clip]{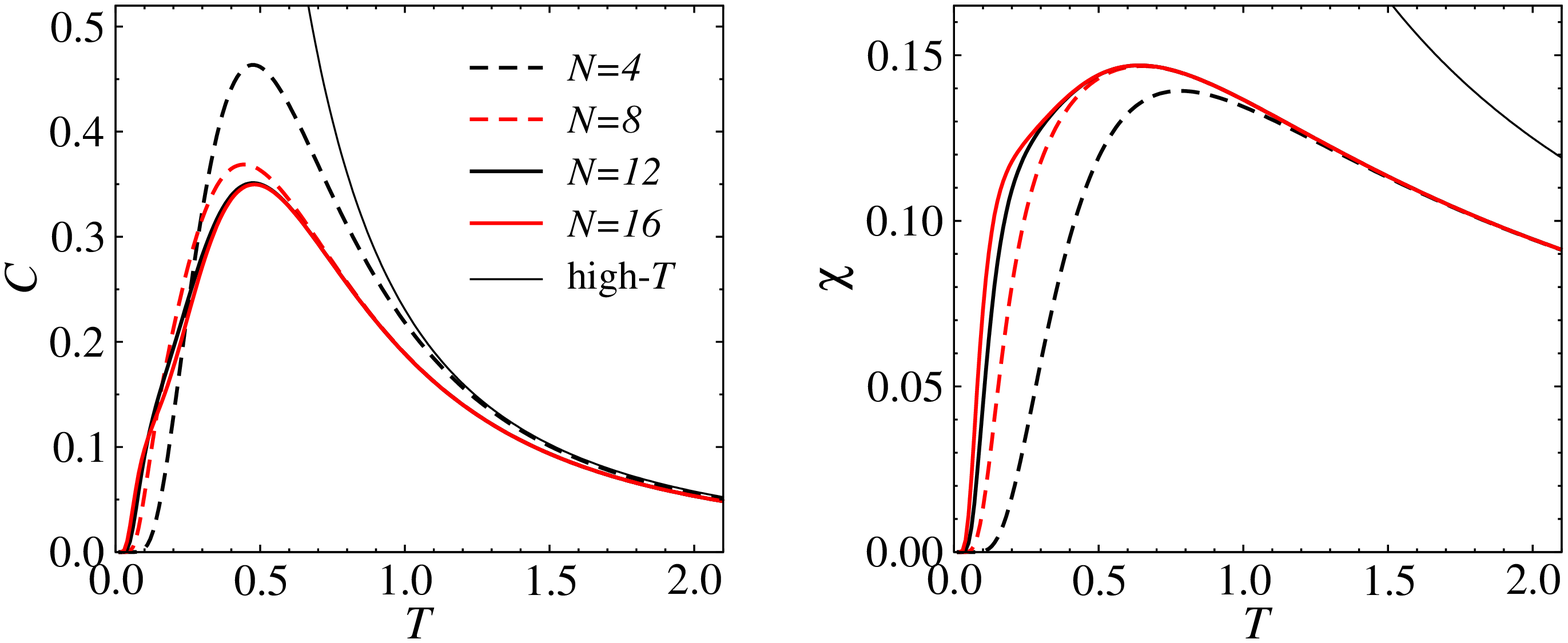}
\caption{Temperature dependence of the specific heat and the magnetic susceptibility for chains of
length $N=4,8,12$, and $16$. The thin curves show the leading high-temperature forms.} 
\label{cx}
\end{figure}

Fig.~\ref{cx} illustrates how the results for $C$ and $\chi$ converge to the thermodynamic limit with increasing system size. Going up to $N=16$, the results 
are well converged down to $T/J \approx 0.25$. It is natural that the convergence is more rapid at high temperatures, because in the limit $T \to \infty$ 
the spins are statistically decoupled and the system behaves as a set of independent spins. Thus the susceptibility per spin for $T \to \infty$ has the single-spin
Curie form; $\chi \to 1/4T$. For $C$, an analytic high-temperature form can be obtained by calculating the energy exactly for the $2$-site system 
(a single spin has constant energy, thus giving $C=0$), taking the derivative, and keeping only the leading term of an expansion in $T^{-1}$. This gives 
$C \to 3/13T^2$ when $T \to \infty$. The high-$T$ limits are also shown in Fig.~\ref{cx}. In the case of $C$, this form describes the behavior well 
down to $T/J \approx 1.5$, whereas in the case of $\chi$ the agreement is good only at higher temperatures.

When $T \to 0$, both $C$ and $\chi$ approach zero exponentially for a finite system, as a consequence of the finite gap between the ground state and excited 
states (due to which only the ground state contributes to thermal averages as $T \to 0$) and, in the case of $\chi$, the fact that the ground state of the 
Heisenberg chain for even $N$ is always a singlet (while for odd $N$ the ground state is an $S=1/2$ doublet and $\chi$ thus diverges as $T^{-1}$). In the 
infinite system, $C \to 0$ but $\chi$ approaches a constant non-zero value. 

Note that here, as well as in many cases we will encounter later on, when we talk about $T=0$ properties it is important to consider the order in which the 
limits $T \to 0$ and $N \to \infty$ are taken. The susceptibility of a chain with finite (even) $N$ always vanishes as $T\to 0$. The temperature at which the 
exponential drop to zero commences depends on the finite-size spin gap, which in the case of the Heisenberg model is $\propto N^{-1}$ (which we will also discuss
further and illustrate with data in Sec.~\ref{sec_hchain}). Thus, for given $N$, one can expect essentially thermodynamic-limit results down to some 
$T/J \propto N^{-1}$ (which can also be roughly seen in Fig.~\ref{cx}), and if one takes the limit $N \to \infty$ before $T \to 0$, then the susceptibility 
remains non-zero at $T=0$.
 
It is not possible to use exact diagonalization results to extrapolate reliably the properties of the Heisenberg chain to the thermodynamic limit at low 
temperatures. For the susceptibility, an extrapolation of exact diagonalization results \cite{bonnerfisher}, using a particular functional form \cite{estes}, 
was already attempted decades ago---the result is known as the Bonner-Fisher susceptibility. This extrapolation used also the known exact value of $\chi$ at 
$T \to 0$ ($N=\infty$), which was available from the exact Bethe ansatz solution of the ground state. The Bonner-Fisher curve has been commonly used in analyzing 
experimental results for quasi-one-dimensional antiferromagnets (weakly coupled chains). The most important feature here is the location and shape of the 
maximum at $T/J \approx 0.6$, which is already well converged for small systems, as seen in Fig.~\ref{cx}. It has for some time now been possible to use other 
computational methods to study various properties of the Heisenberg model on very long chains (as we will discuss in Sec.~\ref{sec_chainsse}), and thus 
the low-temperature limit can be accessed more reliably. There are also essentially exact results from field-theoretical studies combined with the Bethe 
ansatz \cite{eggert94}. These calculations show an anomalous feature; the approach to the $T \to 0$ limit is logarithmic, and therefore very 
difficult to reproduce with finite-$N$ calculations. The low-$T$ form of the Bonner-Fisher curve is therefore incorrect. Experimentally the most important
feature is the broad maximum. Often effects beyond the Heisenberg chain appears at lower temperatures
(e.g., due to inter-chain couplings, spin anisotropies, or disorder). The spin susceptibilities of some quasi-one-dimensional materials actually follows 
$\chi$ of the Heisenberg chain down to temperatures as low as $\approx J/50$, where the logarithmic behavior is prominent \cite{eggert96}.

\subsection{The Lanczos method}
\label{lanczos}

As we have seen in the preceding section, a complete diagonalization of the hamiltonian (or individual blocks) becomes prohibitively time consuming
for $S=1/2$ systems with more than $\approx 20$ spins. For higher $S$, the situation is of course even worse. If we restrict ourselves just to the ground
state, and possibly a number of low-lying excitations, we can reach systems roughly twice as large, by using a {\it Krylov-space technique}, such as the
Lanczos method. 

\subsubsection{The Krylov space}

The Krylov space is a sub-space of the full Hilbert space, constructed in such a way that the low-lying states of a hamiltonian $H$ of 
interest should be well approximated within it. Consider an arbitrary state $|\Psi\rangle$, e.g., a one with randomly generated vector elements
$\Psi(i)$, $i=1,\ldots,M$, in the $M$-dimensional Hilbert space in which $H$ is defined, and the expansion of this state in terms the hamiltonian 
eigenstates $|\psi_n\rangle$ (in order of increasing eigenvalues, which we here label $n=0,1,\ldots,M-1$). 
We operate with a power of the hamiltonian on this chosen state;
\begin{equation}
H^\Lambda|\Psi \rangle = \sum_{n=0}^{M-1} c_n E^\Lambda_n |\psi_n\rangle = c_{\rm max}E^\Lambda_{\rm max}\left [ 
|\psi_{\rm max}\rangle + \sum_{n \not= {\rm max}} \frac{c_n}{c_{\rm max}} \left (\frac{E_n}{E_{\rm max}} \right )^\Lambda |\psi_n\rangle \right ]. 
\label{hmexpansion}
\end{equation}
If the power $\Lambda$ is large, the state corresponding to the eigenvalue $|E_{\rm max}|$ with largest magnitude (i.e., ${\rm max}=0$ or ${\rm max}=M-1$) will 
dominate the sum, provided that the expansion coefficient $c_{\rm max} \not=0$. Hence, acting many times with the hamiltonian on the state will 
{\it project out} the eigenvector with the eigenvalue $E_{\rm max}$. If we want to make sure that the ground state $|\psi_0\rangle$ is obtained 
(exactly for $\Lambda \to \infty$ or approximatively for finite $\Lambda$), we can instead of $H^\Lambda$ use $(H-c)^\Lambda$, where $c$ is a positive number 
large enough to ensure that $|E_0-c| > |E_{M-1}-c|$. Here we will in the following assume that such a constant, if required, has already been absorbed 
into $H$.

While Eq.~(\ref{hmexpansion}) is guaranteed to produce the ground state when $\Lambda$ is sufficiently large,
a more efficient way to construct a state which approaches the ground state as $\Lambda\to \infty$ is to consider not only $H^\Lambda|\Psi\rangle$, but
the whole subspace of the Hilbert space spanned by the set of states $H^m|\Psi\rangle$, $m=0,\ldots, \Lambda$. These states can be constructed one-by-one by 
successive operations with $H$ on the initial state $|\Psi\rangle$. In this subspace, an optimal linear combination of vectors approximating the ground state 
(minimizing the energy expectation value) exists, and the way to find it is to diagonalize $H$ in the generated sub-space of $\Lambda+1$ vectors. In addition 
to projecting out the ground state for relatively small $\Lambda$ (often many orders of magnitude smaller than the size $M$ of the full Hilbert space), this 
approach can also accurately reproduce a number of low-lying excited states. 

The subspace of the Hilbert space obtained by acting multiple times with ${H}$ on an initial state is called the Krylov space. We are here discussing 
hamiltonians in quantum mechanics, but Krylov space methods of course apply to eigenvalue problems more generally as well, and are very widely used in 
many areas of science and engineering.

\subsubsection{The Lanczos basis and hamiltonian}
\label{sec_lancbasis}

In the Lanczos method \cite{lanczos52}, an orthogonal basis is constructed using linear combinations of the Krylov space states such that the hamiltonian written in this basis 
is tridiagonal. In a standard approach, which will be described below, a basis $\{ |f_m\rangle\}$, $m=0,\ldots,\Lambda$, is first constructed that is orthogonal 
but not normalized. These states will subsequently be normalized to yield an orthonormal set $\{ |\phi_m\rangle\}$, in which the hamiltonian takes a particularly 
simple tridiagonal form. We will also discuss a slightly different approach of directly generating the normalized basis $\{ |\phi_m\rangle\}$.

\paragraph{Generation of the Lanczos basis}

The construction of the basis $\{ |f_m\rangle\}$ starts from an arbitrary normalized state $|f_0\rangle$, of which it is required 
only that it is not orthogonal to the ground state of $H$ (if the goal is to find the ground state). This should be the case for a randomly generated
$|f_0\rangle$, but one can also start from some vector which is known to have a substantial overlap with the ground state. The next state is given by
\begin{equation}
|f_1\rangle = H|f_0\rangle - a_0|f_0\rangle ,
\label{f1}
\end{equation}
where the constant $a_0$ should be determined such that $|f_1\rangle$ is orthogonal to $|f_0\rangle$. To this end, we examine the overlap between the two states;
\begin{equation}
\langle f_1|f_0\rangle = \langle f_0|H|f_0\rangle - a_0\langle f_0|f_0\rangle
= H_{00} - a_0N_0,
\label{olap10}
\end{equation}
where we have introduced notation for the normalization constants and diagonal matrix elements of the hamiltonian that will also be used for subsequent states;
\begin{equation}
N_m  = \langle f_m|f_m\rangle,~~~~~~~~H_{mm} = \langle f_m|H|f_m\rangle. \label{nmhm} 
\end{equation}
For the overlap (\ref{olap10}) to vanish we must choose $a_0 = H_{00}/N_0$. 
The following, $m=2$ state is written in terms of the two preceding ones as
\begin{equation}
|f_2\rangle = H|f_1\rangle - a_1|f_1\rangle - b_0|f_0\rangle ,
\label{f2}
\end{equation}
where $a_1$ and $b_0$ can be chosen such that $|f_2\rangle$ is orthogonal to both $|f_0\rangle$ and $|f_1\rangle$. The overlaps are, 
using $H|f_0\rangle$ and $H|f_1\rangle$ obtained from Eqs.~(\ref{f1}) and (\ref{f2});
\begin{equation}
\langle f_2|f_1\rangle = H_{11} - a_1N_1,~~~~~~~~
\langle f_2|f_0\rangle = N_1 - b_0N_0.
\end{equation}
The appropriate coefficients are thus $a_1 = H_{11}/N_1$ and $b_0 =  N_1/N_0$.
For all subsequent iterations, Eq.~(\ref{f2}) generalizes to
\begin{equation}
|f_{m+1}\rangle = H|f_m\rangle - a_m|f_m\rangle - b_{m-1}|f_{m-1}\rangle ,
\label{fn1}
\end{equation}
and the coefficients rendering this state orthogonal to $|f_{m}\rangle$ and $|f_{m-1}\rangle$ are generalizations of the expressions we already
found for $a_0,a_1,$ and $b_0$;
\begin{equation}
a_m = H_{mm}/N_m,~~~~~~~~
b_{m-1} = N_m/N_{m-1}.
\label{abmlanczos}
\end{equation}
This can easily be checked by direct computation of the overlaps. It remains to be shown that with these coefficients the state $|f_{m+1}\rangle$ is orthogonal also 
to all previous states $|f_{k}\rangle$ with $k<m-1$. In an inductive proof, we can use the fact that all previously generated states are orthogonal to each other,
and obtain, 
\begin{eqnarray}
\langle f_{m+1}|f_{m-k}\rangle & = & 
\langle f_m|H|f_{m-k}\rangle -a_m\langle f_m|f_{m-k}\rangle 
-b_{m-1}\langle f_{m-1}|f_{m-k}\rangle  \\
& = & \langle f_m|f_{m-k+1}\rangle +a_{m-k}\langle f_m|f_{m-k}\rangle 
+ b_{m-k-1}\langle f_m|f_{m-k-1}\rangle = 0,~~ \nonumber
\end{eqnarray}
where we also used $H|f_{m-k}\rangle$ in the form given by Eq.~(\ref{fn1}). 

This iterative procedure (\ref{fn1}) is continued until $m=\Lambda$, where $\Lambda$ can be determined automatically, on the fly, 
according to some convergence criterion for computed quantities, as we will discuss below. First we need the hamiltonian matrix elements.

\paragraph{The hamiltonian in the Lanczos basis}

Having constructed a set of $\Lambda+1$  Lanczos vectors $|f_m\rangle$, $m=0,\ldots,\Lambda$, the hamiltonian in this basis can be constructed. 
We can use the expression for $H$ acting on one of the basis states, obtained from Eq.~(\ref{fn1});
\begin{equation}
H|f_m\rangle = |f_{m+1}\rangle  + a_m|f_m\rangle + b_{m-1}|f_{m-1}\rangle ,
\label{hfn}
\end{equation}
and thus the non-zero matrix elements are
\begin{eqnarray}
\langle f_{m-1}|H|f_m\rangle & = & b_{m-1}N_{m-1} = N_m, \nonumber \\
\langle f_{m}|H|f_m\rangle & = & a_{m}N_{m},  \\
\langle f_{m+1}|H|f_m\rangle & = &  N_{m+1}. \nonumber 
\end{eqnarray}
The normalized basis states are
\begin{equation}
|\phi_m\rangle = \frac{1}{\sqrt{N_m}}|f_m\rangle,
\end{equation}
and $b_m=N_{m+1}/N_m$ according to (\ref{abmlanczos}). The non-zero matrix elements are therefore
\begin{eqnarray}
\langle \phi_{m-1}|H|\phi_m\rangle & = & \sqrt{b_{m-1}}, \nonumber \\
\langle \phi_{m}|H|\phi_m\rangle & = & a_{m},  \\
\langle \phi_{m+1}|H|\phi_m\rangle & = &  \sqrt{b_{m}} . \nonumber 
\label{tridia}
\end{eqnarray}
This is a tridiagonal matrix, which can be diagonalized using special methods that are faster than generic diagonalization algorithms (available in many 
linear algebra subroutine libraries). The main advantage is not, however, that the matrix can be diagonalized more easily---it is anyway typically not 
very large and even if a generic diagonalization routine is used the time spent on that part of the calculation is negligible. The advantage
is that this matrix can be constructed relatively quickly, especially when the matrix $H$ is sparse and only its non-zero elements have to be considered
(stored or generated on the fly when acting with $H$ on a state). Some times, when pushing the method to very large matrices, it is also useful that only 
three of the large state vectors $|f_m\rangle$ have to be stored in the process. The tridiagonality is also very convenient when calculating spectral 
functions (dynamic correlations), as discussed, e.g., in Refs.~\cite{dagotto2,didier}.

Note that with the basis states labeled $m=0,\ldots,\Lambda$, the size of the tridiagonal matrix (the actual basis size) is $\Lambda$, not $\Lambda+1$, 
because for $m+1=\Lambda$ in (\ref{hfn}) the last diagonal matrix element generated for use in (\ref{tridia}) is $a_{\Lambda-1}$. The state $|f_{\Lambda}\rangle$ 
does not have to be constructed; only the term $H|f_{\Lambda -1}\rangle$ is needed to calculate the last coefficient $a_{\Lambda-1}$.

Since the cost of diagonalizing the Lanczos hamiltonian matrix is very low, one can do that after each new basis state has been generated, 
and follow the evolution of the eigenvalues. One can stop he procedure based on some suitable convergence criterion, e.g., the desired eigenvalues 
changing by less than some tolerance $\epsilon$ between iterations. The basis size required for convergence of course depends on the system studied, but 
typically, for quantum spin systems, $\Lambda$ in the range tens to hundreds should suffice.

\paragraph{Alternative formulation with normalized vectors}

The generation of  the Lanczos basis is normally discussed in terms of the un-normalized states $|f_m\rangle$, as we have done above. However, a direct computer
implementation of this procedure occasionally leads to numerical problems, because the normalization constants $N_m$ can become exceedingly large (if the eigenvalues 
of $H$ are large). It may then be better to work directly with the normalized states $|\phi_m\rangle$ (as an alternative to multiplying $H$ by a suitable factor). 
In fact, in practice this formulation is even simpler, and the additional computational cost is merely the normalizations at each step (which is normally small 
compared to other costs).

We start with a normalized state $|\phi_0\rangle$ and generate the second state according to
\begin{equation}
|\phi_1\rangle = \frac{1}{N_1}\bigl (H|\phi_0\rangle - a_0|\phi_0\rangle \bigr ).
\label{phi1def2}
\end{equation}
Here $N_1$ is a normalization constant, which is determined by direct computation of the scalar product of the constructed state 
$H|\phi_0\rangle - a_0|\phi_0\rangle$ with itself. Orthogonality with $|\phi_1\rangle$ again requires $a_0=H_{00}$. For each following state one can 
easily show that
\begin{equation}
|\phi_{m+1}\rangle = \frac{1}{\sqrt{N_{m+1}}}\bigl (H|\phi_m\rangle - a_m|\phi_m\rangle - N_m|\phi_{m-1}\rangle \bigr ) = \frac{|\gamma_{m+1}\rangle}{N_{m+1}} 
\label{phimdef2}
\end{equation}
is orthogonal to all previous states. Here the definitions of $a_m$ and $N_m$ differ from the previous ones in Eq.~(\ref{abmlanczos});
\begin{eqnarray}
a_m & = & \langle \phi_m|H|\phi_m\rangle, \nonumber \\
N_m & = & \langle \gamma_m|\gamma_m\rangle,
\end{eqnarray}
where $|\gamma_m\rangle$ is the generated state before normalization as in Eq.~(\ref{phimdef2}) above. The hamiltonian matrix elements are
\begin{eqnarray}
\langle \phi_{m-1}|H|\phi_m\rangle & = & \sqrt{N_{m}}, \nonumber \\
\langle \phi_{m}|H|\phi_m\rangle & = & a_{m},  \\
\langle \phi_{m+1}|H|\phi_m\rangle & = &  \sqrt{N_{m+1}} . \nonumber 
\label{tridia2}
\end{eqnarray}
In this formulation all the stored numbers are well behaved.

\paragraph{Degenerate states}

It should be noted that the Lanczos method cannot produce more than one member of a multiplet; out of a degenerate set of states, only a 
particular linear combination of them will be obtained (which depends on the initial state $|f_0\rangle$). To see the reason for this, we 
again look at the expansion (\ref{hmexpansion}) of a state $H^\Lambda|\Psi \rangle$, in which we assume that there are two degenerate states 
$|\psi_i\rangle$ and $|\psi_j\rangle$, $E_i=E_j$. In the expansion we can isolate these states from the rest of the terms;
\begin{equation}
H^\Lambda|\Psi \rangle = E^m_{i}(c_i|\psi_i\rangle +c_j|\psi_j\rangle) + \sum_{m\not=i,j} c_m E^\Lambda_m |\psi_m\rangle.
\end{equation}
For any $\Lambda$, the expansion contains the same linear combination of the states $|\psi_j\rangle$ and $|\psi_j\rangle$. 
Hence, in the subspace spanned by the set of states $H^m|\Psi \rangle$, $m=0,\ldots,\Lambda-1$, there is no freedom for obtaining 
different linear combinations of the two degenerate states. This of course generalizes also to degenerate multiplets with more than two states.

\paragraph{Loss and restoration of orthogonality}

When the basis size $\Lambda$ becomes large, the Lanczos procedure typically suffers from numerical instabilities. Round-off errors accumulated in the 
course of constructing the basis set will eventually introduce some non-orthogonality among the states. Such numerical errors can escalate and lead to 
successive sudden appearances (within some narrow ranges of $\Lambda$) of several identical eigenvalues (recall that the Lanczos scheme should never 
produce degenerate states). We will see an example of this further below.

Loss of orthogonality and the appearance of multiple copies of the same states is normally not a problem when the aim is to obtain only the ground state. 
It can complicate calculations of excited states, however. To remedy this, one can add to the basis construction a step where each new 
Lanczos vector constructed is explicitly orthogonalized with respect to all previous basis vectors. In the simplest implementation of such a stabilization procedure,
all the Lanczos vectors are stored (in primary memory or secondary storage). This can become problematic when dealing with very large basis sets, but the scheme 
is very simple. Working with the normalized states, each $|\phi_{m+1}\rangle$ constructed according to (\ref{phimdef2}) is orthogonalized with respect to 
all previous states, according to
\begin{equation}
|\phi_{m+1}\rangle \to \frac{|\phi_{m+1}\rangle-q|\phi_i\rangle}{1-q^2},~~~~q=\langle\phi_i|\phi_{m+1}\rangle,
\label{lancreort}
\end{equation}
successively for $i=0,\ldots,m$. This makes it possible to study a much larger number of excited states (in principle only limited by computer memory). 

Instead of converging several excited states in the same run, one can also target excited states one-by-one, starting each time from a vector which 
is orthogonal to all previous ones. Re-orthogonalization should then also be done with respect to those. 

\paragraph{Eigenstates and expectation values}

Diagonalizing the tridiagonal Lanczos hamiltonian results in eigenvalues $E_n$ and eigenvectors $v_n$, $n=0,\ldots,\Lambda-1$. We want these eigenvectors 
expressed in the original basis $\{|a\rangle\}$, in which we are able to evaluate the matrix elements $\langle b|O|a\rangle$ of operators $O$ of interest. 
First, the Lanczos basis states are
\begin{equation}
|\phi_m\rangle = \sum_{a=1}^{M}\phi_m(a)|a\rangle,~~~~m=0,\ldots,\Lambda-1,
\label{lanczosphi}
\end{equation}
and we denote the desired eigenvectors of the hamiltonian as
\begin{equation}
|\psi_n\rangle = \sum_{a=1}^{M} \psi_n(a)|a\rangle,~~~~n=0,\ldots,\Lambda-1.
\label{psisuma}
\end{equation}
The first few eigenvectors $v_n$ of the tridiagonal matrix accurately represent eigenstates of the hamiltonian (essentially exactly for sufficiently 
large $\Lambda$) in the Lanczos basis;
\begin{equation}
|\psi_n\rangle = 
\sum_{m=0}^{\Lambda-1} v_n(m)|\phi_m\rangle = \sum_{m=0}^{\Lambda-1} \sum_{a=1}^{M} v_n(m)\phi_m(a)|a\rangle,
\end{equation}
and thus the wave function coefficients we want to construct are given by
\begin{equation}
\psi_n(a) = \sum_{m=0}^{\Lambda-1} v_n(m)\phi_m(a),~~~~a=1,\ldots,M.
\label{psimake}
\end{equation}
If all the Lanczos vectors have been stored during the basis construction, this transformation can be carried out in a straight-forward manner.
If there is enough computer memory available, one should store all Lanczos vectors, but often calculations are pushing the limits of computer capacity, 
and then only the bare minimum of information can be stored, at the cost of longer calculation times. If we did not store the full set of Lanczos 
vectors during the basis construction, we need to generate them again, in the same iterative fashion as before, except that we already have the coefficients
$a_m$ and $N_m$ available (and $b_m$ if the scheme with un-normalized Lanczcos vectors $|f_m\rangle$ is used) and do not need to recompute them. We can 
transform the states according to (\ref{psimake}) on the run with the eigenvectors $v_m$, building up one or several of the eigenstates of the hamiltonian.

Having generated one or several eigenstates, now assumed to be stored in the form of the coefficients $\psi_m(a)$ in (\ref{psisuma}), an expectation value of 
some operator $O$ can be obtained by first acting on the state, giving an un-normalized state that we call $|\psi^O_n \rangle$,
\begin{eqnarray}
O|\psi_n \rangle = |\psi^O_n \rangle & = & \sum_{a=1}^M \psi_n(a) O | a \rangle \nonumber \\
& = & \sum_{a=1}^M\sum_{b=1}^M \psi_n(a) |b\rangle\langle b|O| a \rangle. \label{oexp1} \\
& = & \sum_{a=1}^M \psi^O_n(a) | a \rangle,~~~~~~ \psi^O_n(a) = \sum_{b=1}^M \psi_n(b) \langle b|O| a \rangle.\nonumber
\end{eqnarray}
We then evaluate the scalar product of this state with the original state $|\psi_n \rangle$;
\begin{equation}
\langle \psi_n|O|\psi_n \rangle =  \langle \psi_n|\psi^O_n \rangle = \sum_{a=1}^M \psi_n(a)  \psi^O_n(a) .
\label{oexp2}
\end{equation}
This somewhat cumbersome way of writing the expectation value corresponds to a typical computational procedure of first computing the state
vector $\psi^O_n$ and then computing its scalar product with $\psi_n$.

It is interesting to note that, in Lanczos calculations, we are often dealing with four different bases: First, we have the original ``computational'' 
basis of single states of $\up$ and $\dn$ spins. From these we generate a basis incorporating symmetries, e.g., momentum states (which normally would
be the states denoted as $|a\rangle$ above). We then construct the Lanczos vectors, which are particular linear combinations of those states. Finally, 
diagonalizing the tridiagonal matrix (which is an effective hamiltonian in the low-energy sector), we obtain the desired energy eigenstates. To do 
calculations with those states, we effectively do the basis transformations in reverse. The matrix elements in (\ref{oexp1}) and (\ref{oexp2}) are 
in the end carried out in the computational basis of $\up$ and $\dn$ spins (i.e., with the representative states used, e.g., to build the momentum states). 
We next discuss all these procedures in practical program implementations.

\subsubsection{Programming the Lanczos method}

While the Lanczos method can be applied to any symmetric (or hermitian) matrix, in the case of a spin hamiltonian there is an added advantage in that
the hamiltonian is a sparse matrix. Although the size of the hamiltonian (an individual symmetry-block) can be very large, the number of non-zero 
matrix elements is much smaller. For a model with short-range interactions on a lattice of $N$ sites, a hamiltonian block of $M$ states has on the order of 
$NM$ non-zero elements, which for large $M$ is much smaller than the total number of elements $M^2$. Since the most time consuming part of the construction 
of the Lanczos basis is the repeated operations with the hamiltonian matrix on a state vector, to generate the next basis state according to Eq.~(\ref{fn1}) or
(\ref{phimdef2}), the sparseness allows for enormous time savings. There are similarly significant memory savings advantages as well. The non-zero elements
of the hamiltonian should then be stored in a compact form, or generated on the fly as needed (which, when many symmetries are used, typically takes 
longer than even reading the elements from disk storage). We here implement the Lanczos method with the hamiltonian stored in a compact form in primary 
memory. 

We will generate the Lanczos basis of states $\{|\phi_m\rangle\}$ that are normalized at each step, using Eqs.~(\ref{phimdef2}) and (\ref{phi1def2}). 
In a computer program, these states are stored in the form of their vector components $\phi_m(a)$, in terms of which the Lanczos states are given by 
Eq.~(\ref{lanczosphi}). Here $a=1,\ldots,M$ labels the states of our working basis, which in the case of maximal use of symmetries would be momentum or 
semi-momentum states, e.g., $|a\rangle = |k,p,z,m_z=0,a\rangle$, but for simplicity we do not write out all the quantum numbers. The nature of the
basis states only come into play when acting with the hamiltonian (or some operator to be measured) on the basis states, and for this we need exactly
the matrix elements that have already been discussed for the various symmetry implementations in Secs.~\ref{momentum}, \ref{reflection}, and 
\ref{spininversion}. We will write pseudocodes assuming real-valued wave-function coefficients; the changes needed for complex states 
are self-evident. 

\paragraph{Lanczos basis construction and eigenstates}

To discuss the general structure of a Lanczos program, we begin by assuming that we have implemented a subroutine {\bf hoperation}$(\phi,\gamma)$ which acts with 
the hamiltonian on a state vector; $|\gamma\rangle=H|\phi\rangle$. Later, we will describe the implementation of this subroutine, which is the only part of the 
Lanczos procedure that requires a specification of the model and the symmetries implemented. We also use a subroutine {\bf normalize}$(\phi,n)$, which first 
computes $n=\langle \phi|\phi\rangle$ and then rescales $\phi$ so that $\langle \phi|\phi\rangle=1$ upon return. 

To start the basis generation, we first load the initial state $|\phi_0\rangle$ with a randomly generated normzalized state. We denote the elements 
$\phi_0(i)$, $i=1,\ldots,M$ and $\langle \phi_0|\phi_0\rangle = \sum_i\phi_0(i)^2=1$. The next Lanczos state is generated according to (\ref{phi1def2});

{\code
\cia    {\bf call hoperation}$(\phi_0,\phi_1)$ \br
\cia    $a_0=\langle \phi_0|\phi_1\rangle$;~ $\phi_1=\phi_0-a_0|\phi_1\rangle$ \hfill  \{17\}\break 
\cia    {\bf call normalize}$(\phi_1,n_1)$ 
\code}

\noindent
If we are not storing all the Lanczos vectors, we can cycle between three vectors, $\phi_0,\phi_1,\phi_2$, that contain the information we need at 
each step when implementing the iterative basis generation according to (\ref{phimdef2});

{\code
\cia    {\bf do} $m=1,\Lambda-1$ \br
\cib       {\bf call hoperation}$(\phi_1,\phi_2)$                \hfill  \{18\}\break 
\cib       $a_{m}=\langle \phi_1|\phi_2\rangle$;~ $\phi_2=\phi_2-a_{m}\phi_1-n_m\phi_0$ \br
\cib       {\bf call normalize}$(\phi_2,n_{m+1})$  \br 
\cib       $\phi_0=\phi_1$;~ $\phi_1=\phi_2$ \br
\cia    {\bf enddo} 
\code}

\noindent
If possible, we should store all the states, to avoid having to regenerate them at the later stage when computing expectation values (which doubles
the computation time). We have to store
all the states if we want to carry out additional re-orthogonalization, to ensure that numerical truncation errors do not eventually degrade the Lanczos 
basis. In that case we insert an additional loop to orthogonalize with respect to all previously generated states, according to Eq.~(\ref{lancreort});

{\code
\cia    {\bf do} $m=1,\Lambda-1$ \br
\cib       {\bf call hoperation}$(\phi_m,\phi_{m+1})$ \hfill  \{19\}\break 
\cib       $a_m=\langle \phi_m|\phi_{m+1}\rangle$;~ $\phi_{m+1}=\phi_{m+1}-a_m\phi_m-n_m\phi_{m-1}$ \br
\cib       {\bf call normalize}$(\phi_{m+1},n_{m+1})$  \br 
\cib       {\bf do} $i=0,m$ \br
\cic           $q=\langle \phi_{m+1}|\phi_{i}\rangle$;~ $\phi_{m+1}=(\phi_{m+1}-q\phi_i)/(1-q^2)$ \br
\cib       {\bf enddo} \br
\cia    {\bf enddo} 
\code}

\noindent
After completing either code segment $\{18\}$ or $\{19\}$, we have the contents of the tridiagonal matrix in Eq.~(\ref{tridia2}) and can proceed to feed 
this information into a diagonalization routine which delivers the eigenvalues $E_n$ and eigenvectors with elements $v_n(i)$, $n=0,\ldots,\Lambda-1$,
$i=1,\ldots,M$. It should be noted again that, with the conventions we have been using, the Lanczos basis size (the size of the matrix) is $\Lambda$; the last state
generated in $\{18\}$ and $\{19\}$, with $m+1=\Lambda$, is actually not needed (only the coefficient $a_{\Lambda-1}$ generated in the last iteration is needed).

The computational effort of the diagonalization is very small compared to the time spent on the basis construction. We may therefore as well diagonalize after each 
new Lanczos vector has been generated. We can then monitor how the energies evolve with the basis size. One can then stop when some convergence criterion is 
satisfied. One can, e.g., demand that the change in the ground state energy (or the highest excitation of interest) changes between steps $m-1$ and $m$ by less 
than some small number $\epsilon$. The Lanczos method is normally capable of converging energies to the numerical precision of the computer (and double 
precision should always be used). As we will see below, other properties converge slower than the energies.

\paragraph{Expectation values}

First, let us consider calculations of operator expectation values (besides the energy) in a program where we have not stored all the Lanczos basis vectors. 
Then, in order to be able to transform the ground state from the Lanczos basis to the original basis, where we can carry out ``measurements'' on the state,
we have to repeat the Lanczos basis construction {\it starting from the same initial state as in the first run}. If we had initialized with a random $\phi_0$, 
as in code $\{17\}$, and did not save this state, we can re-generate it by initializing the random number generator with the same seeds as in the first 
construction. In any case, with $\phi_0$ at hand, we can proceed as in code segment $\{18\}$, with the minor simplification that we now do not 
have to calculate the coefficients $a_m,n_m$, because we already have them. We can use $n_m$ for normalizing, instead of calling the subroutine {\bf normalize}
in $\{18\}$. During the basis re-construction, we transform the states with the ground state vector $v_0$ (or $v_n$ for an excited state $i$) that 
resulted from the diagonalization of the tridiagonal matrix, building up the ground state in a vector $\psi$;
 
{\code
\cia    {\bf call hoperation}$(\phi_0,\phi_1)$;~ $\phi_1=(\phi_1-a_{0}\phi_0)/\sqrt{n_1}$ \br
\cia    $\psi=v_0(0)\phi_0+v_0(1)\phi_1$ \hfill $\{20\}$\break 
\cia    {\bf do} $m=2,\Lambda-1$  \br                     
\cib       {\bf call hoperation}$(\phi_1,\phi_2)$ \br
\cib       $\phi_2=(\phi_2-a_{m-1}\phi_1-n_{m-1}\phi_0)/\sqrt{n_m}$ \br
\cib       $\psi=\psi+v_0(m)\phi_2$ \br
\cib       $\phi_0=\phi_1$;~ $\phi_1=\phi_2$ \br
\cia    {\bf enddo} 
\code}

\noindent
The elements of the vector $\psi$ now contain the ground-state wave-function coefficients in whatever basis is used and implemented through 
{\bf hoperation}. If we have stored all the Lanczos vectors, we of course do not need this step and can directly transform the states generated 
in code $\{19\}$.

It is particularly easy to calculate expectation values of operators that depend only on the $z$-components of the spins and is invariant under 
all the symmetry operations used in the basis. We then just have to weight the quantity calculated using the representative states with 
the corresponding wave-function coefficient squared (the state's probability). As an example, the spin-spin correlation function 
$\langle S^z_{i}S^z_{i+r}\rangle$ for all distances $r=0,\ldots,N/2$ can be obtained as:

{\code
\cia    {\bf do} $a=1,M$   \br 
\cib       {\bf do} $r=0,N/2$                       \hfill $\{21\}$\break 
\cic          {\bf do} $i=0,N-1$   \br 
\cid              $j=\mathbf{mod}(i+r,N)$ \br
\cid              {\bf if} $(s_a[i]=s_a[j])$ {\bf then} \br 
\cie                 $C(r)=C(r)+\psi^2(a)$ \br
\cid             {\bf else} \br
\cie                 $C(r)=C(r)-\psi^2(a)$ \br
\cid             {\bf endif}   \br 
\cic          {\bf enddo} \br 
\cib       {\bf enddo}  \br
\cia    {\bf enddo}  \br
\cia    $C=C/4N$
\code}

\noindent
Here $\psi^2(a)$ is the probability of the state $|a\rangle$, which when symmetries are incorporated is represented by a number $s_a$. As before, bit tests  
can be used to determine the relative spin orientation, now of spins separated by $r$ lattice spacings. Note that although the final spin correlation 
$\langle S^z_iS^z_j\rangle$ only depends on $r=|i-j|$, we still have to average over all $i$ above, because we are only using the representative states 
(which do not by themselves, without acting with the symmetry operators, obey any lattice symmetries). In principle we should also average over both 
reflections and spin-inversions of the representative, but after the translational averaging has been done the spin correlation function is also
invariant with respect to the other two symmetries (which is not the case for all operators). Note that since the Heisenberg hamiltonian is spin-rotation 
invariant, the correlation function $\langle S^z_{i}S^z_{i+r}\rangle$ calculated here equals $\langle {\bf S}_{i}\cdot {\bf S}_{i+r}\rangle/3$.

Calculating expectation values of $z$ off-diagonal operators that cannot be simply related to $z$ diagonal ones require explicit operations on the states, and 
thereafter evaluation of a scalar product, according to Eqs.~(\ref{oexp1}) and (\ref{oexp1}). This is in principle easy---essentially proceeding as in the 
construction of the hamiltonian in the preceding sections---but more time consuming than diagonal operators.

\paragraph{Compact storage of the hamiltonian}

We now discuss the inner workings of the subroutine {\bf hoperation}$(\phi,\gamma)$ that we employed in the Lanczos procedures. It implements the operation
\begin{equation}
H|\phi\rangle = |\gamma\rangle = \sum_{a=1}^M\sum_{b=1}^M \phi(a)\langle b|H|a\rangle |b\rangle.
\label{habbhab}
\end{equation}
This is of course where all the details of the symmetries employed will enter. The main difference with respect to the construction of the hamiltonian 
in complete diagonalization is that we do not want to store $H$ as a full $M\times M$ matrix, because we have in mind calculations were the number of basis states $M$ 
can be up to many millions (e.g., for $N=32$ in Table \ref{sizetab}). We therefore have to devise a convenient way of only storing its non-zero elements, of which 
there are of the order $NM$ in (\ref{habbhab}).  

We define the following data structures for the compact hamiltonian storage: For each $a=1,\ldots,M$ in (\ref{habbhab}) we store the number $e_a$ of non-zero matrix 
elements $\langle b|H|a\rangle$. We store the locations $b$ of these non-zero elements as consecutive integers in a vector with elements $B(i)$. In a program, it is 
convenient to use the elements $i=1,\ldots,e_1$ for $a=1$, followed by $i=e_1+1,\ldots,e_1+e_2$ for $a=2$, etc. The required size of the vector $B$ (the number of
non-zero elements) is initially not known, and, depending on the programming language used, may have to be allocated using an estimated size. We can store the 
corresponding non-zero matrix elements in a vector with floating-point values $H(i)$. We can also take advantage 
of the fact that the hamiltonian is a symmetric matrix and only store one of its ``triangles''.

With the above notation, carrying out the operation (\ref{habbhab}) is now as simple as:

{\code
\cia    {\bf subroutine hoperation}$(\phi,\gamma)$  \br
\cia    $\gamma=0$; $i=0$                \hfill $\{22\}$\break 
\cia    {\bf do} $a=1,M$  \br         
\cib       {\bf do} $j=1,e_a$ \br
\cic          $i=i+1$ \br             
\cic          $\gamma(B(i))=\gamma(B(i))+H(i)\phi(a)$ \br 
\cic          $\gamma(a)=\gamma(a)+H(i)\phi(B(i))$ \br 
\cib       {\bf enddo}  \br
\cia    {\bf enddo} 
\code}

\noindent
The counter $i$ keeps track of the position of the elements in the data structures $H$ and $B$. Diagonal matrix elements, $B(i)=a$, are double counted 
in this procedure since the contributions from the upper and lower triangle of the hamiltonian matrix are added. It is therefore assumed that their
stored values have been divided by $2$.

The matrix elements $H(i)$ and their locations $B(i)$ are generated in a way similar to what we did when constructing the complete hamiltonian, e.g., 
as in code segments $\{13\}$--$\{16\}$ in the semi-momentum basis with parity. We also discussed the simple extensions involving spin-inversion symmetry. 
Now we discuss the modifications needed when we wish to load only the non-zero elements into our compact storage. 

In the case of the semi-momentum basis, which we will consider here, the procedures are again somewhat complicated by the fact that the same 
representative can appear once or twice ($\sigma = \pm 1$) in the state list, and we want to take care of these at the same time in order to avoid 
repeating tasks unnecessarily. We therefore carry out the loop over the basis states $a$ and determine the number of same representatives $n$ as in 
code $\{13\}$. We cannot put the matrix elements directly into the storage vector $H$, because our scheme in code $\{22\}$ requires consecutive 
storage of all matrix elements for each column. With the way the individual bond operators in the hamiltonian are treated one-by-one in code 
$\{16\}$, when $n=2$ the two columns would be mixed up if we store each matrix element as it is generated. We therefore use temporary storage 
for the one or two columns being currently processed, and then later copy their contents into the appropriate positions of the final storage. 
Referring to the first ($a$) and potential second ($a+1$) column as $c=1$ and $c=2$, we store the data temporarily as $B_{c}(k)$ and 
$H_{c}(k)$, $k=1,\ldots,n_c$, where $n_c$ is the number of elements for the column in question. The diagonal elements (divided by $2$, as discussed 
above) are entered first into these temporary storage lists;

{\code
\cia    $n_1=0$;~ $n_2=0$  \br
\cia    {\bf do} $c=1,n$                           \hfill $\{23\}$  \break              
\cib       $B_{c}(1)=a+c-1$;~ $H_{c}(1)=E_z/2$; $n_c=1$  \br    
\cia    {\bf enddo} 
\code}

\noindent
For the off-diagonal matrix elements we proceed as in code segment $\{16\}$. Note, however, that several bond operations may lead to the same state $|b\rangle$ when 
acting on a given basis state $|a\rangle$, but in the compact storage we do not want to store individual contributions to the same matrix element separately. 
Everything is taken care of by replacing the loops over $i$ and $j$ in code $\{16\}$ by this extended version;

{\code
\cia   {\bf do} $i=a,a+n$   \br
\cia   {\bf do} $j={\bf max}(b,i),b+m$                         \hfill $\{24\}$\break              
\cib       $c=i-a+1$  \br
\cib       $E={\bf helement}(i,j,l,q)$ \br
\cib       {\bf do} $k=1,n_c$           \br
\cic          {\bf if} $B_c(k)=j$ {\bf exit}    \br
\cib       {\bf enddo}  \br
\cib       {\bf if} ($k>n_c$) {\bf then} $n_c=k$;~ $B_{c}(k)=j$ {\bf endif} \br
\cib       {\bf if} ($j=i$) {\bf then} $H_{c}(k)=H_{c}(k)+E/2$ {\bf else} $H_{c}(k)=H_{c}(k)+E$ {\bf endif} \br       
\cia   {\bf enddo} \br
\cia   {\bf enddo}
\code}

\noindent
Here {\bf max}$(i,b)$ on the second line ensures that only the elements in matrix triangle $j\ge i$ are constructed. The innermost loop over $k$ checks weather 
there is already a stored contribution to the matrix element $\langle j|H|i\rangle$. It has been assumed that when the loop over $k$ has been completed 
(without exiting before $k=n_c$), then $k$ takes the value $n_c+1$ (which is the case in many computer languages). Then $k=n_c+1$ if there is no 
prior location $j$ in the list, which means that it should be added to the list (and the size of the list is then $k$). We have also assumed that all $H_{c}(k)$ 
are initially set to zero, so that each new contribution $E$ can be added in the appropriate location of $H_c$. Note that although we are carrying out 
off-diagonal operations here, in the basis we are using, such operations can also lead to diagonal matrix elements, in which case we have to divide $E$ by $2$.

After having generated all the matrix elements originating from the current representative (i.e., completed the loop over all nearest-neighbor spin pairs), 
we copy the contents of the temporary storage vectors into the permanent full storage;

{\code
\cia   {\bf do} $c=1,2$     \br
\cib       {\bf do} $i=1,n_c$          \hfill $\{25\}$\break                      
\cic           $n_H=n_H+1$; $B(n_H)=B_c(i)$; $H(n_H)=H_{c}(i)$  \br               
\cib       {\bf enddo}    \br
\cia    {\bf enddo}     
\code}

\noindent
Here $n_H$ is a counter for the total number of non-zero matrix elements added so far. After this, the loop over state labels $a$ is closed, as in code
$\{13\}$. This completes the construction of the hamiltonian.

For other operators, it is not worth storing the matrix elements, because normally we do not reuse operators for observables many times. The 
hamiltonian is used repeatedly, however, up to a few hundred times, and so using code segment $\{21\}$ instead of having to carry out all the 
operations associated with extracting each individual hamiltonian matrix element from scratch every times can amount to a very significant speed-up. 

It is possible to further compactify the hamiltonian by not storing the matrix elements as double-precision numbers, but instead use a mapping to 
the actual numbers based on a table of integers. The number of unique values is often very small (tens or hundreds of values, so that one can
even use ``short'' integers as pointers to the actual values), and this can save some memory, at the cost of a somewhat more time consuming 
construction of the hamiltonian. The locations $B(i)$ have to be stored as four-byte integers, however.

When pushing the limits of the largest treatable system sizes, one may have to store the hamiltonian on disk (and reading successive portions
of it when executing code $\{22\}$), or generate it on the fly without storing it. The latter essentially amounts to executing code based on $\{24\}$ 
every time when acting with the hamiltonian.

\subsubsection{Convergence of Lanczos calculations}

\begin{figure}
\includegraphics[width=12.25cm, clip]{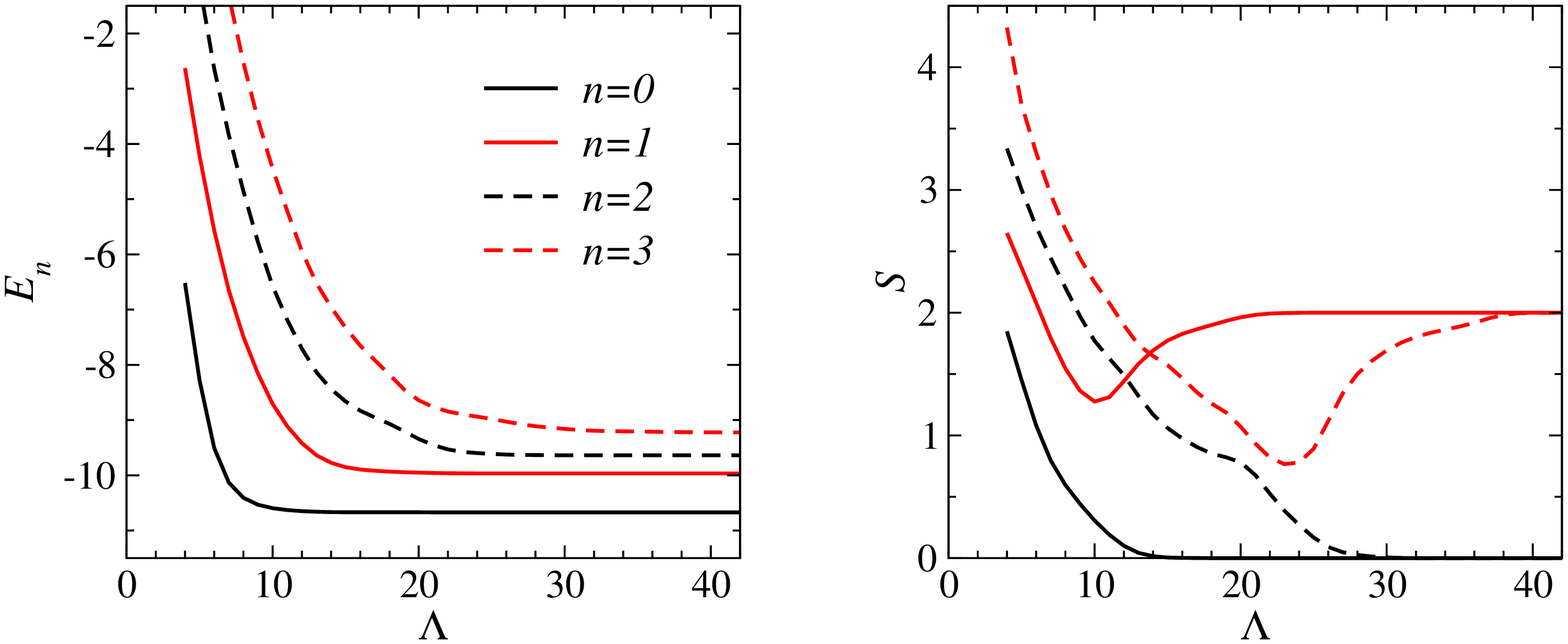}
\caption{Example of the convergence as a function of the Lanczos basis size of the energy (left) and the total spin (right) of the four 
lowest levels of the 24-site Heisenberg chain in the symmetry sector $k=0,p=1,z=1$. The spin $S$ is extracted using the assumption
that $\langle {\bf S}^2\rangle = S(S+1)$.}
\label{es24}
\end{figure}

The Lanczos method is essentially exact if a sufficiently large number of Lanczos vectors $\Lambda$ is used, and typically this number does not even 
have to be very large; on the order of a few tens to hundreds. The convergence should be checked by carrying out calculations for several $\Lambda$,
until no changes can be detected in the energies and expectation values of interest. The ground state converges the fastest, and energies converge 
faster than expectation values. As an example, Fig.~\ref{es24} shows results for an $N=24$ chain. The energy and the total spin of the four lowest 
levels in the symmetry sector of the ground state are shown versus $\Lambda$. The spin quantum number is calculated by acting with the squared total spin 
operator ${\bf S}^2$ on the states, according to Eqs.~(\ref{oexp1}) and (\ref{oexp2}), and extracting $S$ assuming ${\bf S}^2=S(S+1)$ (which is 
valid only when the states have converged to eigenstates of the operator). The energies are seen to converge monotonically, whereas this is not
necessarily the case for other quantities, as seen clearly for the $S=2$ states in this case. The details of the convergence of course depend on 
the initial state from which the Lanczos basis is constructed (which in this case was a random state). 
In this case all the four levels shown ($E$ as well as $S$) were converged to better 
than 10 decimal places at $\Lambda=60$, with the ground state having converged at that level already at $\Lambda=30$. Going to larger system sizes, the 
convergence becomes a little slower, but for this particular model there are no difficulties in converging several levels up to the largest system 
sizes that can feasibly be studied.

One can accelerate the convergence of a Lanczos calculation by starting from a state which is already close to the ground state. Such states may be
constructed in a number of ways, e.g., based on some approximate analytical method. But if there are no convergence problems this may not be worth
the additional effort. However, if a series of calculations are carried out as a function of some parameter in the hamiltonian, then subsequent 
calculations can be started from the ground state of the preceding parameter value, which is likely to have a significant overlap with the next
ground state. However, it should be noted that if the initial state is a good approximation to the ground state, it will have very small overlaps 
with the first few excited states, and hence only the ground state is likely to converge rapidly in such a calculation. If excited states are also
needed, this problem can be circumvented by starting the next calculations using a linear combination of eigenstates from prior calculations.

\begin{figure}
\includegraphics[width=12.25cm, clip]{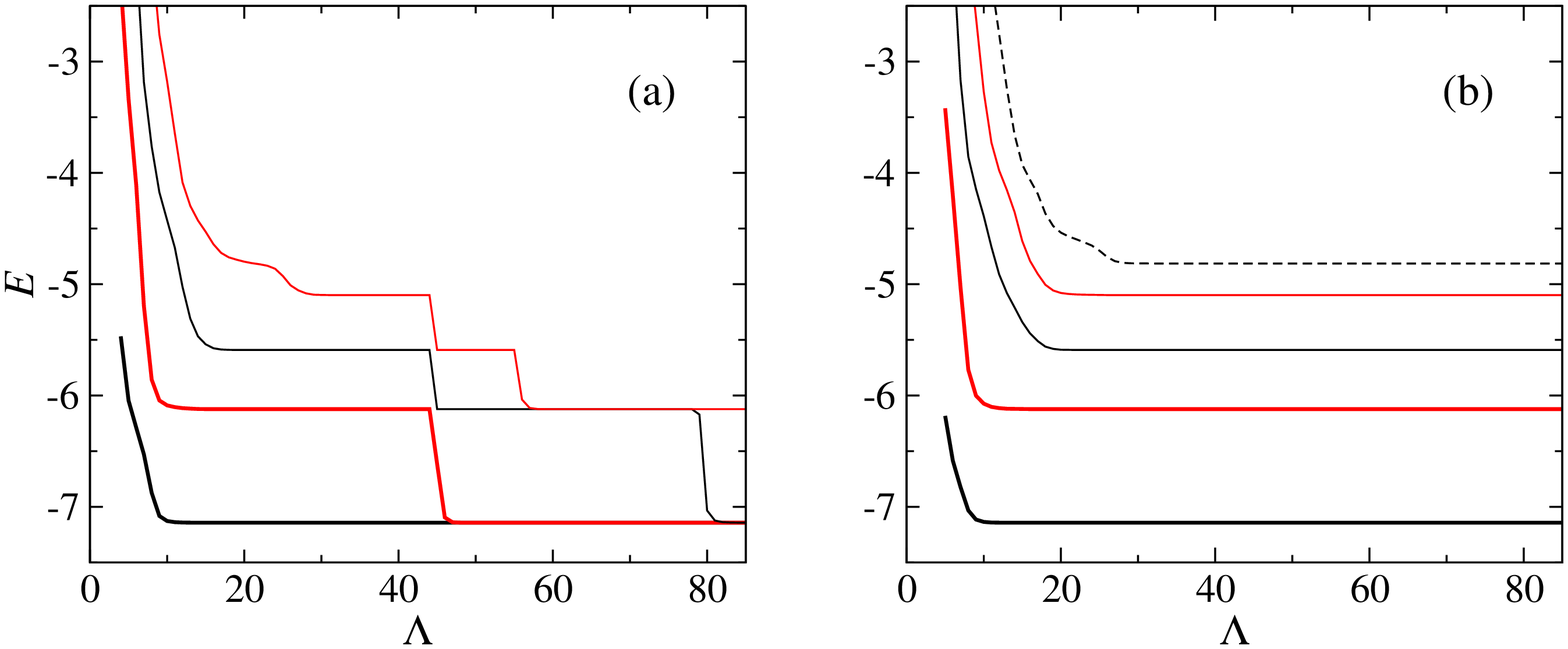}
\caption{(a) The four lowest energies as a function of the Lanczos basis size for a 16-site Heisenberg chain with quantum numbers ($k=0,p=1,z=1$).
Multiple copies of the same state appear successively due to loss of orthogonality. (b) The five lowest states of the same system
obtained with re-orthogonalization of the basis set.}
\label{lancbreak}
\end{figure}

\paragraph{Loss and recovery of orthogonality}

The Lanczos basis vectors should all be completely orthogonal to each other, but numerical truncation errors build up and eventually lead to escalating 
loss of orthogonality for some $\Lambda$. This manifests itself as artificial degeneracies, with excited states ``falling down'' onto lower states. An 
example of this is shown in Fig.~\ref{lancbreak}(a), where the four lowest energies of a 16-site chain in the ground-state symmetry sector are graphed 
versus $\Lambda$. The higher energies are seen to successively collapse onto the immediately lower energies, with only a few iterations taken for the 
levels to become completely degenerate after signs first appear of the imminent collapse. The basis then includes multiple copies of the same states. 

The loss of orthogonality may at first sight seem like a serious problem, but in practice this is no necessarily the case. Calculations for 
the ground state are not much affected by there being more than one copy of it, as long as one makes sure that it is properly normalized before using it 
to calculate expectation values, and the first excited states are normally (but not always) well converged before they fall down onto lower states.

Loss of orthogonality often occurs sooner for small systems than large ones. This is related to the fact that there are less low-energy states for small 
systems, whence the lowest states in the Lanczos basis can become ``overcomplete'' in this subspace (if the Lanczos basis is larger than the total number 
of states, it clearly is truly overcomplete and the calculation will not work). 

One can easily supplement the Lanczos method by an explicit re-ortho\-gonalization step, as in Eq.~(\ref{lancreort}) and implemented in code $\{19\}$.
The drawback is that this requires storage of all the Lanczos basis vectors, which may not be possible for large systems. The procedure may also become
time consuming if the number of states is large. As shown in Fig.~\ref{lancbreak}(b), re-orthogonalization enables convergence of many more eigenstates 
(limited only by memory and time constraints).

\subsection{1D states and quantum phase transitions}
\label{sec_results1d}

In this section we discuss several related 1D calculations using the Lanczos method and finite-size scaling methods for up to $N=32$ spins.
First we investigate the critical ground state of the standard Heisenberg chain, and also analyze some properties of excited states. Then we introduce 
frustration, studying the dimerization transition in the J$_{\rm 1}$-J$_{\rm 2}$ chain. Finally, we add long-range interactions in addition to frustration, 
in which case the continuous dimerization transition evolves into a first-order transition between a N\'eel state (which is possible even in a 1D system 
if the interactions are sufficiently long-ranged) and a dimerized VBS state.

\subsubsection{Ground state and excitations of the Heisenberg chain}
\label{sec_hchain}

Although the Heisenberg chain has an exact Bethe ansatz solution \cite{bethe}, the wave function is very complicated and in many cases numerical calculations 
for finite-size systems have to be used to extract information on physical properties from it \cite{caux,maillet}. The energy of the ground state \cite{hulthen} 
and the low-lying excitations \cite{cloizeaux} can be calculated exactly both for finite chains and in the thermodynamic limit, however.  Some other quantities 
can also be extracted for very large chains \cite{eckel}. In addition to using the exact solution, many properties of this class of system (i.e., with a wider
range of interactions maintaining the symmetries of the system) are known based on an asymptotically exact low-energy field-theory description; the 
Weiss-Zumino-Witten non-linear $\sigma$ model with topological coupling (or central charge) $c=1$ \cite{affleck1}. Equivalently, the Heisenberg chain also 
represents a special case (because of its spin-rotational invariance) of the Luttinger-liquid state, which describes a broad range of interacting 1D spin,
fermion, and boson systems \cite{schulz90,voit}. Numerical studies based on exact diagonalization and other unbiased finite-lattice techniques (such as QMC and 
DMRG) have played an important role in guiding and confirming these theories (see, e.g., \cite{affleck3} and \cite{eggert94}). Computational studies are also 
required for extracting non-universal (short-distance, higher-energy) properties that are not captured by universal continuum field theories 
(including quantities directly accessible to experiments, as discussed in, e.g, Refs.~\cite{lorenzana,takigawa}). Numerical results for the Heisenberg chain are 
also very important for enabling rigorous bench-mark tests of extrapolation techniques when studying other spin models, for which less is known from 
analytical calculations. 

\begin{figure}
\includegraphics[width=8.25cm, clip]{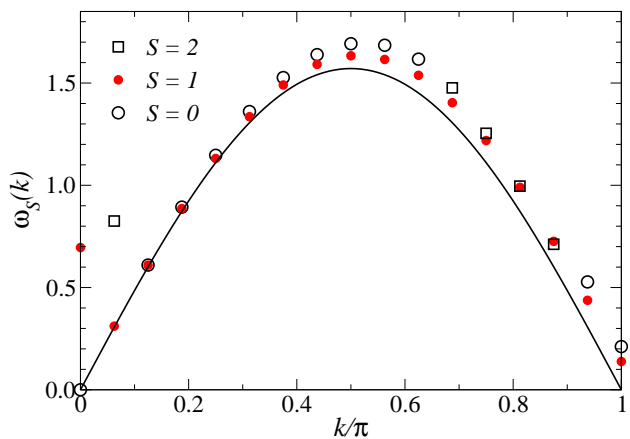}
\caption{Dispersion of the lowest excitations, relative to the ground state, in the sectors with spin-inversion quantum number $z=1$ (giving low-energy states 
with spin $S=0$ or $2$) and $z=-1$ (which always gives $S=1$ for the lowest state) for an $N=32$ chain. The solid curve below the data points is the exact 
infinite-size dispersion derived by des Cloizeaux and Person \cite{cloizeaux} using the Bethe ansatz.} 
\label{eq32}
\end{figure}

\paragraph{Ground state and low-energy excitations}

For a Heisenberg chain of size $N=4n$, the ground state has momentum $k=0$, parity $p=1$, and spin-inversion number $z=1$; it is a fully symmetric singlet state. 
For $N=4n+2$ the lowest-energy state is in instead in the completely antisymmetric sector; $k=\pi,p=-1,z=-1,S=0$. 

Let us look at the most important excitations of the Heisenberg chain. For $k$ not equal to the ground-state momentum, the lowest-energy state of a finite 
system is a triplet; hence $z=-1$ and $+1$ for chains of size $N=4n$ and $4n+2$, respectively. Fig.~\ref{eq32} shows the momentum dependence of the excitation 
energies $\omega(k)=E_S(k)-E_0$ for both the lowest $z=+1$ and $z=-1$ states of a $32$-site chain. The lowest $z=+1$ state always has $S=1$, while the lowest 
$z=-1$ state has either $S=0$ or $S=2$. The triplet energies are quite close to the exact $N=\infty$ triplet dispersion obtained from the Bethe ansatz 
\cite{cloizeaux}, $\omega_S(k)=\pi |\sin(k)|/2$, especially close to $k=0$. Close to $k=\pi$ the deviations are larger. The $S=0,2$ excitations are mostly 
slightly higher in energy, except that the $S=2$ state close to $ k/\pi=0.85$ is actually marginally lower than the triplet (likely a small-$N$ anomaly), 
and at $k/\pi=2/N$ it is much higher. 

For large $N \to \infty$, one would expect the singlet and triplet excitations to become degenerate with dispersion $\omega_{0,1}(k)=\pi |\sin(k)|/2$. This  
can be understood as originating from excitations of pairs of spin-$1/2$ soliton-like degrees of freedom called {\it spinons}. These spinons are very 
weakly interacting and in the infinite chain behave as independent particles, hence forming four degenerate levels (from which $S=0$ and $1$ states can be formed). 
The dispersion graphed in Fig.~\ref{eq32} is only the lowest edge of a continuum of spinon excitations, which can be calculated in detail using the Bethe 
ansatz \cite{caux} and has also been observed experimentally in quasi-1D antiferromagnets \cite{lake05}.

\begin{figure}
\includegraphics[width=8cm, clip]{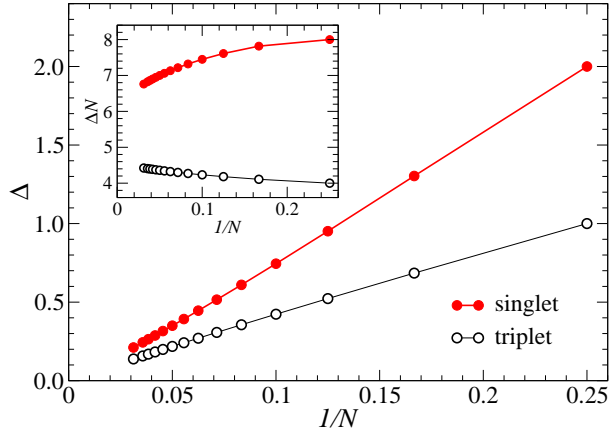}
\caption{Singlet and triplet gaps of the Heisenberg chain versus the inverse system size. The inset shows the gaps multiplied by $N$, 
illustrating the presence of multiplicative logarithmic corrections to the dominant $\sim 1/N$ scaling (dynamic exponent$z=1$).}
\label{gaps}
\end{figure}

The very lowest excitation for a $N=4n$ chain is a triplet at $k=\pi$, with $p=-1,z=-1$, whereas for $4n+2$ the lowest triplet has $k=0,p=1,z=1$ (i.e., the 
difference in momentum with respect to the ground state is always $\pi$, and $p,z$ are minus their ground-state values). Fig.~\ref{gaps} shows the finite-size 
scaling of these lowest triplet and singlet excitation energies (the singlet and triplet finite-size gaps) versus the inverse system size. They are
both seen to scale to zero as $1/N$, corresponding to a dynamic exponent $z=1$. There is a weak correction to this form originating from logarithmic corrections, 
which can be seen clearly when plotting the gaps multiplied by $N$ (in the inset of Fig.~\ref{gaps}). This gap scaling has been predicted in detail based on the 
the continuum field theory approaches \cite{affleck2,affleck3}. The prefactor of the gap scaling is directly related to the velocity of the spinon excitations,
but it is not easy to extract it reliably based on the finite-size data because of the log corrections. The velocity can also be extracted from the linear 
parts of the dispersion relation close to $k=0$ and $\pi$, $\omega_S(k)/k$  or $\omega_S(k)/(\pi-k)$, with the $k=0$ behavior being the easier to analyze, 
as seen clearly in Fig.~\ref{eq32}.

\paragraph{Spin correlation function}

The spin correlation function is one of the key characteristics of of any quantum spin system. The correlations in the ground state are of primary interest.
For a spin-isotropic and translationally invariant chain we can write it as
\begin{equation}
C(r) = \langle {\bf S}_i \cdot {\bf S}_{i+r}\rangle = 3\langle {S^z}_i{S^z}_{i+r}\rangle,
\label{cpincorr}
\end{equation}
for any reference site $i$. As we discussed in Sec.~\ref{sec_heisenberg}, the Mermin-Wagner theorem rules out antiferromagnetic long-range order in the Heisenberg 
chain. Instead, in the ground state $C(r)$ decays as $(-1)^r/r$ for large $r$, up to a multiplicative logarithmic correction (originating from a marginally 
operator in the field-theoretical description \cite{affleck2,singh1,giamarchi}).

\begin{figure}
\includegraphics[width=8.25cm, clip]{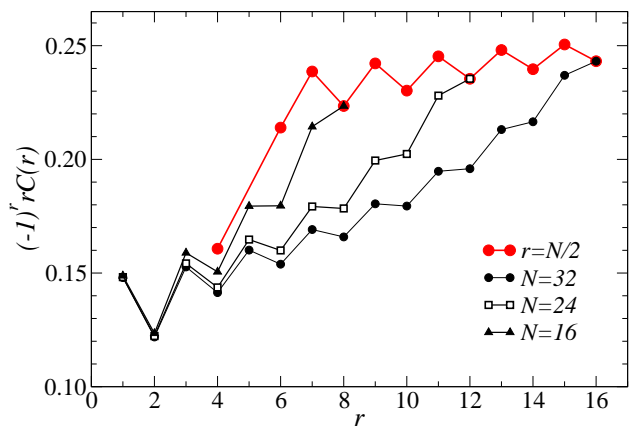}
\caption{Absolute value of the spin correlation function, multiplied by the separation $r$, in the ground state of the Heisenberg chain.
Results are shown as a function of distance for different system sizes, and also at $r=N/2$ versus the chain length $N$.}
\label{cor}
\end{figure}

Fig.~\ref{cor} shows ground-state results for $C(r)$ as a function of $r$ for system sizes $N=16,24$, and $32$. The results have been multiplied by $r$, so 
that an $r$-independent behavior should obtain for large $r$ if the asymptotic form is $\sim 1/r$. It is also useful to multiply by $(-1)^r$, in order to cancel out
the $\pm$ oscillations of the staggered phases. Note that there are remaining even-odd oscillations in $C(r)(-1)^r$. Such oscillations are quite 
common for various correlation functions of 1D systems. They diminish with increasing system size but are still strong for the small system studied 
here. The form of the oscillations is also predicted by the Luttinger liquid theory; they decay as $1/r^2$ \cite{voit}.

In Fig.~\ref{cor}, clear deviations from the $1/r$ form of $C(r)$ can be seen which are mainly due to the periodic boundary conditions, which enhance the 
correlations close to $r=N/2$. Open boundaries cause even more severe finite-size effects (in addition to not allowing the use of translational symmetry for 
block-diagonalization). The logarithmic corrections should play some role as well. As the chain length is increased, the correlation function at fixed 
$r$ converges to the infinite-$N$ limiting form. However, as can be seen in the figure, the maximum $r$ for which the results are approximately converged 
is a rather small fraction of $N$. For $N=32$, one could safely say that $C(r)$ is converged for $N$ only up to $r \approx 4$, which is not enough to say 
much about the long-distance behavior. An alternative, also shown in the figure, is to investigate the correlation function at fixed $r/N$ as a function 
of $N$, with $r/N=1/2$ the most natural choice for checking the long-distance behavior. Even though periodic boundaries enhance the correlations significantly 
at this point, the functional form of $C(r=N/2)$ is proportional to the infinite-size $C(r)$ [although the overall prefactor of function $C(r=N/2)$ 
versus $N$ will clearly be different from that in the actual $N\to \infty$ converged $C(r)$]. The leading $1/r$ form appears quite plausible based on the 
data in Fig.~\ref{cor}, and the remaining enhancement is consistent with the predicted logarithmic corrections. In Sec.~\ref{sec_chainsse} we will further 
investigate the long-distance correlations based on quantum Monte Carlo calculations for much longer chains.

\subsubsection{Frustration-driven quantum phase transition}
\label{sec_j1j2chain}

A very interesting aspect of the $S=1/2$ Heisenberg chain is that it exhibits several ground-state phase transitions (quantum phase transitions) when a 
next-nearest-neighbor interaction is added to the hamiltonian; \begin{equation}
H = J_1\sum_{i=1}^N {\bf S}_i \cdot {\bf S}_{i+1} + 
J_2\sum_{i=1}^N {\bf S}_i \cdot {\bf S}_{i+2}.
\label{j1j2ham}
\end{equation}
This model goes under the name of the $J_1$-$J_2$ chain, or the Majumdar-Ghosh model (after the authors of the first comprehensive study of the system 
\cite{majumdarghosh}). Here we will discuss the most important (and most well understood) transition, which occurs when both $J_1$ and $J_2$ are 
antiferromagnetic (positive). For convenience we define the ratio $g=J_2/J_1$. 

For $g < g_c$, $g_c \approx 0.2411$ \cite{nomura92}, the system is in the same phase as the Heisenberg chain with $J_2=0$ discussed in the previous section. 
The ground state is critical with antiferromagnetic correlations decaying as $1/r$ and the finite-size gaps scale as $1/N$ (up to log corrections in both 
cases). Only the prefactors (e.g., the velocity of the  excitations) and the strength of the log corrections change as a function of $g$. 

For $g>g_c$ the system is in a completely different kind of state, with exponentially decaying spin correlations and a triplet excitation gap which remains 
finite in the thermodynamic limit. The ground state has long-range dimer order (a simpler, 1D version of the 2D VBS states discussed in Sec.~\ref{vbsrvb}). 
The nearest-neighbor bond strengths (the $J_1$ energy 
contribution), $\langle B_i\rangle =\langle {\bf S}_i \cdot {\bf S}_{i+1} \rangle$, are of the alternating (period two) form $B_i = B +\delta(-1)^i$ in the 
symmetry-broken infinite-size state (of which there are two degenerate ones, with oscillations out-of-phase relative to each other). Unlike magnetic order, 
this kind of order is allowed at zero temperature in one dimension because the broken symmetry (the lattice translational symmetry) is discrete. The modulation 
$\delta$ becomes non-zero at $g=g_c$ and thereafter increases with $g$. The spin correlations are initially staggered (peaked at momentum $k=\pi$ in reciprocal 
space), but at $g\approx 0.52$ the change to $k=\pi/2$ \cite{bursill96}, and for $g>1$ the peak-value may change continuously with $g$ (a spiral state)
\cite{kumar}.

The model with ferromagnetic (negative) $J_1$ is also interesting. There are several transitions between states with different periodicities \cite{momoi}).
Here we will consider exclusively the antiferromagnetic case, in the regime $g<1$. We will use Lanczos results to investigate the phase transition into the 
VBS state, and also discuss the properties of this ordered state. 

\paragraph{The Majumdar-Ghosh point}

\begin{figure}
\includegraphics[width=6.5cm, clip]{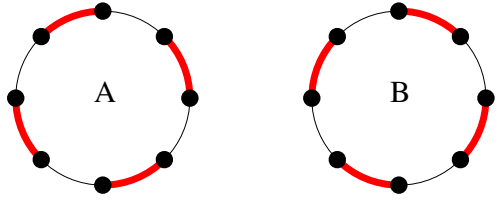}
\caption{Illustration of the degenerate ground states at the Majumdar-Ghosh point. The thick arcs illustrate singlets; the
product of these singlets on alternating bonds is an exact ground state at $g=1/2$.} 
\label{vbsring}
\end{figure}

Before we discuss the Lanczos results, it is wort noting that the existence of VBS order can be shown exactly at the special point $g=1/2$ (the Majumdar-Ghosh 
point), where the ground state is very simple \cite{majumdarghosh}. On a ring with even $N$, the ground state is a two-fold degenerate singlet product, as 
illustrated in Fig.~\ref{vbsring}. One can demonstrate this rather easily by just acting on the states with the hamiltonian, to show that they are eigenstates 
(while the proof of them being the lowest states is more involved \cite{broek80,shastry81,klein82}). Out of these degenerate states, $|\Psi_A\rangle$ and 
$|\Psi_B\rangle$, which break the translational symmetry, one can form symmetric and anti-symmetric states, which have momentum $k=0$ and $\pi$, respectively:
\begin{equation}
|\Psi(0)\rangle = (|\Psi_A\rangle + |\Psi_B\rangle)/\sqrt{2},~~~~~ |\Psi(\pi)\rangle = (|\Psi_A\rangle - |\Psi_B\rangle)/\sqrt{2}.
\label{mgpsiab}
\end{equation}
These are the states obtaining in Lanczos calculations with conserved momentum. Thus, the order parameter $\langle B_i\rangle=E_1$ is featureless when 
calculated on finite periodic systems. As always in exact finite-lattice calculations, the symmetry-breaking has to be observed by calculating correlation 
functions (unless we add some perturbation that breaks the symmetry between the two possible ordering patterns, which has its own complications as the
limit of vanishing perturbation has to be taken). The ground state away from the point $g=1/2$ is more complicated than than the simple singlet-products 
in Fig.~\ref{vbsring}, with fluctuations in the singlet pairings, but with remaining alternating higher and lower density of singlets on the nearest-neighbor bonds. 
The translationally invariant $k=0$ and $\pi$ states on finite rings correspond approximately to symmetric and anti-symmetric combinations as in Eq.~(\ref{mgpsiab}). 
It is, however, only at $g=1/2$ that these two states are exactly degenerate for finite $N$. In other cases, the states become degenerate only in the 
limit $N\to \infty$.

\paragraph{Dimer order parameter}

We already discussed VBS ground states in Sec.~\ref{vbsrvb} and an example of how to characterize this kind of order by dimer (or bond) correlations in 
Sec.~\ref{jqmodelintro}. We now investigate the dimer correlation function introduced in Eq.~(\ref{dxxdimorder}), written explicitly for a 1D system as
\begin{equation}
D(r) = \langle B_iB_{i+r}\rangle =\langle ({\bf S}_i \cdot {\bf S}_{i+1})({\bf S}_{i+r} \cdot {\bf S}_{i+1+r}) \rangle.
\label{drdef}
\end{equation}
This function should alternative between ``weak'' and ``strong'' values for large $r$ if there is VBS order (even if the symmetry is not broken). The 
calculation of this correlation function with the Lanczos method can be simplified by taking advantage of the rotational symmetry and compute 
$\langle ({\bf S}_i \cdot {\bf S}_{i+1}){S}^z_{i+r}{S}^z_{i+1+r} \rangle$, which is $1/3$ of $D(r)$. In principle, we could make it even simpler by 
defining the order parameter solely in terms of the $z$ components, $\langle {S}^z_i{S}^z_{i+1}{S}^z_{i+r}{S}^z_{i+1+r} \rangle$, which is not just 
a constant times (\ref{drdef}) but still a valid order parameter for a dimerized state. Here we consider the full $D(r)$.

\begin{figure}
\includegraphics[width=12.5cm, clip]{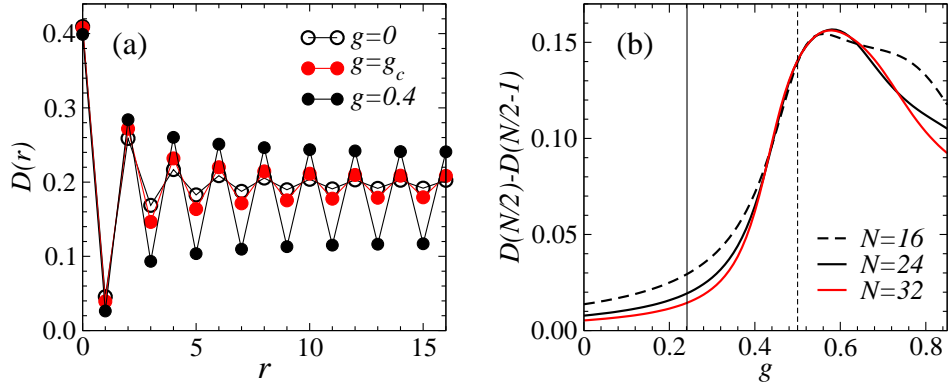}
\caption{(a) Bond correlations for $N=32$ chains at three different coupling ratios $g$. (b) Bond correlations versus $g$
for three different system sizes. The solid and dashed vertical lines indicate, respectively, the critical coupling $g_c$
and the exactly solvable point $g=1/2$.} 
\label{dcor}
\end{figure}

Fig.~\ref{dcor}(a) shows results for a 32-site system at three different couplings. At $g=0$, the standard Heisenberg model, the correlations decay rapidly
with $r$ and are very small at $r=N/2$. At the transition point, $g_c \approx 0.2411$ (which we will determine more precisely below), the correlations are 
clearly stronger. The Luttinger-liquid theory applied to spin chains predicts dimer correlations decaying as $1/r$ for $g \le g_c$ (with stronger logarithmic 
corrections when $g < g_c$) \cite{voit}. Going to larger coupling ratios, Fig.~\ref{dcor}(a) shows correlations clearly indicative of long-range order at 
$g=0.4$. To confirm the presence of long-range order, a finite-size scaling analysis has to be carried out. As we did with the spin correlation functions of 
the Heisenberg chain in the preceding section, it is then typically best to look at the correlations at the longest distance, $r=N/2$, versus the system size. 
In the case of the dimer correlations, where the ``bare'' correlation function is non-zero even in the disordered phase, one has to subtract off the constant 
corresponding to this background. One possibility is to subtract $\langle B_i\rangle^2$, which is the value to which $D(r)$ should converge for large $r$. 
Another option, which we will choose here, is to instead use the difference $D(N/2)-D(N/2-1)$. This quantity will be non-zero for $N\to \infty$ only if $D(r)$ 
oscillates, i.e., if there is long-range dimer order. Results for three different system sizes are shown versus the coupling ratio in Fig.~\ref{dcor}(b). 
Here the order parameter appears to be well converged in the range $0.4 < g < 0.6$ confirming that there is indeed long-range order. 

\paragraph{Finite-size extrapolations of spin and dimer correlations}

\begin{figure}
\includegraphics[width=12.5cm, clip]{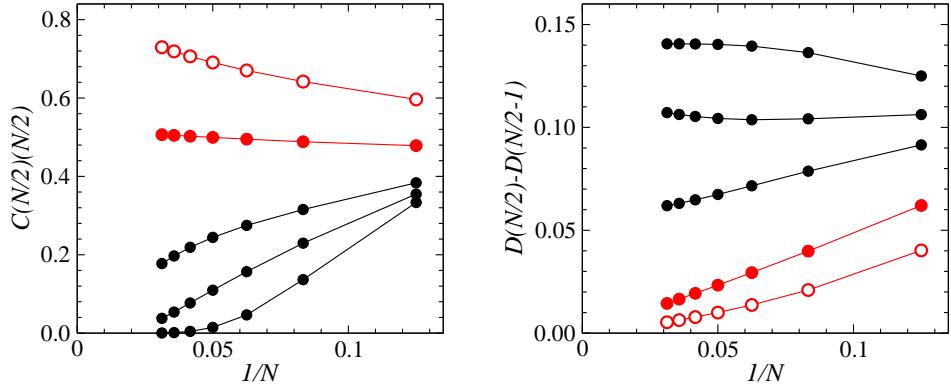}
\caption{Spin (left panel) and bond (right panel) correlations at the longest distance in periodic chains versus the inverse system size.
The couplings ratios for the five curves are $g=0,g_c,0.40,0.45,0.50$, in the order top to bottom (the left panel) and bottom to top (right panel), 
i.e., the three sets shown with black circles correspond to the VBS phase.}
\label{cor_n}
\end{figure}

To extract the infinite-size dimer order parameter quantitatively, it is in most cases necessary to perform an extrapolation, unless the system size is sufficiently 
large for there to be no remaining size-dependence of significance [which based on Fig.~\ref{dcor}(b) is the case close to $g=1/2$]. Fig.~\ref{cor_n} shows 
results for both dimer and spin correlations versus the inverse system size for representative $g$ values both inside and outside the VBS phase. 

The dimer correlations, shown in the right panel of Fig.~\ref{cor_n}, are of the $1/r$ form for $g<g_c$. There are logarithmic corrections, which are small
exactly at $g_c$. For $g > g_c$, inside the VBS phase, the correlations extrapolate to a non-zero value. The asymptotic $N \to \infty$ convergence should be 
exponential in this case (as can be demonstrated explicitly at $g=1/2$) but close to the transition it is in practice 
not possible to reach system sizes sufficiently large to observe this behavior. Instead, closer to $g_c$ the behavior appears to be essentially 
linear in $1/N$, as seen in the figure at $g=0.4$. However, at $g=0.45$ one can see that the behavior is actually non-monotonic, with the large-$N$ behavior 
consistent with an approach to the infinite-$N$ value from below, as at $g=1/2$. As seen in the right panel of Fig.~\ref{dcor}, for $g>0.6$ the dimer
order parameter decreases with $g$. Larger systems are required for proper extrapolations in this case, and other methods have to be used. It is believed that 
the system remains dimerized for all $g>g_c$ \cite{bursill96,kumar}.

The spin correlations should change from $1/r$ to exponentially decaying as the VBS phase is entered. To detect this change, the correlation function $C(r=N/2)$ 
is multiplied by $N/2$ in the left panel of Fig.~\ref{cor_n}. As we saw before, for $g=0$ there is a log-correction which enhances the long-distance spin
correlations. At $g=g_c$ these corrections are much smaller---it is known that the leading log corrections vanish at the dimerization transition 
\cite{eggert96b}---although the behavior is still not purely $1/r$ (which can be expected only in the limit $N \to \infty$). For $g$ inside the VBS 
phase, $rC(r)$ decays to zero, as expected. Close to $g_c$ one of course has to go to large system sizes---larger than the spin correlation length (which 
diverges at $g_c$)---in order to observe a pure exponential fall-off. 

For the $g$-values used in Fig.~\ref{cor_n}, the spin correlations are staggered, i.e.,peaked at $k=\pi$ in reciprocal space. For $g \approx 0.52$ 
the peak position changes rapidly to $\pi/2$ \cite{bursill96}, and for $g>1$ the spin structure most likely evolves into a spiral with continuously 
varying pitch \cite{kumar} (as is the case for the classical version of the model). The correlations always decay exponentially, however (unlike the
classical long-range spiral).

\paragraph{Determining the dimerization transition point}

The dimerization transition is known to be similar to the Kosterliz-Thouless transition of the classical 2D XY model. Unlike a conventional phase transition, 
the order parameter does not follow a power law at $g_c$, but is exponentially small close to $g_c$. It is therefore not possible to extract the infinite-$N$ 
order parameter close to $g_c$ based on the small systems accessible to Lanczos calculations---due to the cross-over behavior it is even difficult relatively 
deep inside the VBS phase. It would then appear to be very difficult to determine the location of the dimerization transition. There is, however, a very 
elegant way to extract the critical coupling in an indirect way, based on excited-state energies \cite{nomura92}.

As we saw in Sec.~\ref{sec_hchain}, the lowest excited state of the pure Heisenberg chain ($g=0$) is a triplet. On the other hand, we also know that in the VBS 
state the ground-state should be two-fold degenerate, and both these ground states must be singlets. At the exactly solvable point $g=1/2$, the degeneracy is 
exact for any $N$, but away from this special point the two states become degenerate only in the infinite-$N$ limits. The approach 
to degeneracy should be exponential in $N$ (as will be illustrated with data below), whereas in the case of the triplet excitation of the Heisenberg chain 
the gap closes as $1/N$, and the lowest singlet also approaches the ground state as $1/N$ (as shown in Fig.~\ref{gaps}). This being the case, for fixed $N$ there 
must be some coupling ratio $g_{\rm cross}(N)$ at which the singlet and triplet excitations cross each other. This point can be taken as a size-dependent definition 
of the transition point, and as $N \to \infty$ it should approach the actual critical coupling; $g_{\rm cross}(N \to \infty)=g_c$.

\begin{figure}
\includegraphics[width=13.25cm, clip]{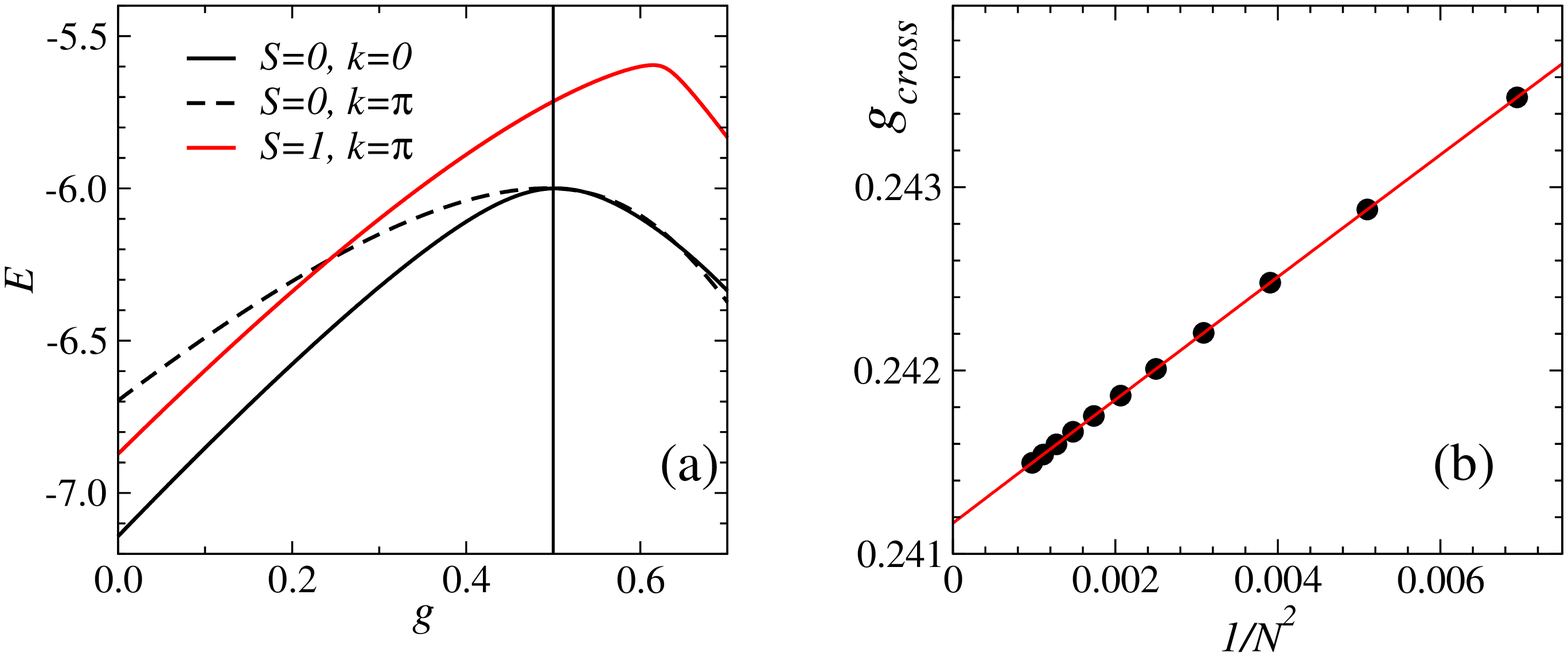}
\caption{(a) The three lowest energies for a 16-site chain as a function of the coupling ratio $g$. The crossing point $g_{cross}$ of 
the $k=\pi$ singlet and triplet levels can be used as a size-dependent definition for the critical (dimerization) coupling $g_c$. The vertical 
line indicates the exactly solvable Majumdar-Ghosh point, at which the singlet states are degenerate. (b) The crossing point versus $1/N^2$. 
The curve is a polynomial fit in $1/N$ (without a $\propto 1/N$ term, as the leading correction is $\propto 1/N^2$).}
\label{cross}
\end{figure}

Fig.~\ref{cross}(a) shows the three energy levels of interest for a 16-site chain, with the corresponding quantum numbers indicated
as well (these quantum numbers apply to all system sizes $N=4n$). Note how the two singlets become degenerate as $g \to 1/2$. As expected, the singlet and 
triplet excited levels cross; for this system size at $g\approx 0.242$. The crossing points converge very rapidly as a function of $N$, as shown in 
Fig.~\ref{cross}(b)---the leading corrections are known to be proportional to $1/N^2$ \cite{eggert96b}. Based on these results it is possible to determine 
$g_c$ very precisely. Fitting a high-order polynomial to crossing points for $8 \le N \le 32$ gives $g_c=0.2411674(2)$, where $(2)$ indicates the uncertainty 
in the last digit based on fluctuations in the extrapolated value when different ranges of system sizes are included in the fits and the order of the 
polynomial is varied. 

\paragraph{Finite-size gaps in the VBS phase}

\begin{figure}
\includegraphics[width=13cm, clip]{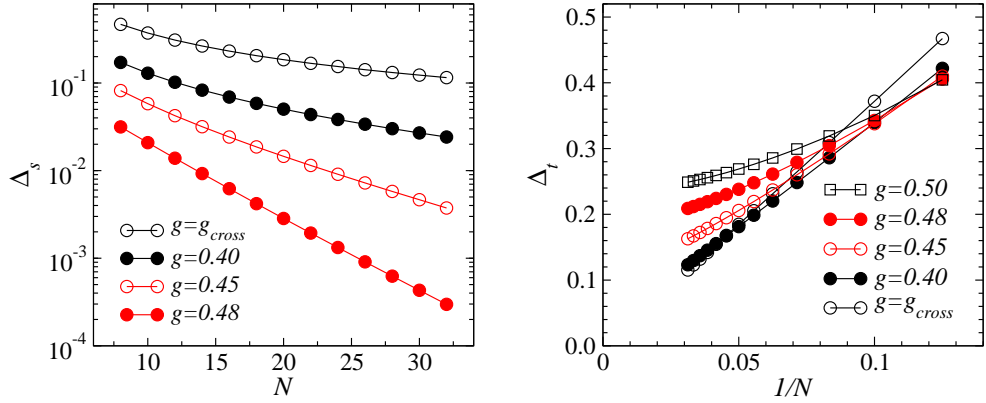}
\caption{Singlet (left panel) and triplet (right panel) gaps as a function of system size for different coupling ratios. Here $g_{\rm cross}$ is the
size-dependent coupling at which the lowest singlet and triplet excitations are degenerate (and thus $\Delta_s=\Delta_t$). The gap at $g_{\rm cross}$ decays as 
$1/N$ (as seen more clearly in the right panel). In the left panel a log scale for $\Delta_s$ versus $N$ is used to show the exponentially rapid closing of the 
gap between the two singlet states that become degenerate for $N\to \infty$ in the ordered VBS phase.  To the right, the triplet gap is plotted 
versus $1/N$ to show its convergence to a non-zero value in the VBS phase.}
\label{stgaps}
\end{figure}

As already mentioned, to accommodate symmetry breaking in the thermodynamic limit, one would expect the singlet-singlet finite-size gap to close exponentially 
fast with increasing system size inside the VBS phase (with the exception of the Majumdar-Ghosh point, where the degeneracy is exact for any size). The exponential 
gap scaling is demonstrated in the left panel of Fig.~\ref{stgaps} for some representative values of $g$. At $g_c$ the behavior is instead $\propto 1/N$, which 
is not clearly seen in this lin-log plot, but more clearly in the log-log plot in the right panel. In the right panel triplet gaps are also shown versus $1/N$. 
Here extrapolations to non-zero values inside the VBS phase are apparent. The finite gap essentially corresponds to the energy required to promote a singlet 
bond into a triplet. Note that only the ground state is exactly solvable at the Majumdar-Ghosh point, and the triplet gap is size dependent also here. At 
$g_c$ the gaps scale as $1/N$---in the figure the behavior at the size-dependent crossing points $g_{\rm cross}$ is shown and also scales as $1/N$.

\subsubsection{Chains with long-range interactions}
\label{sec_longrange}

While long-range spin ordering is not possible in Heisenberg chains with finite-range interactions, long-range interactions make magnetic order 
possible at $T=0$. A transition between a N\'eel state and a quasi-long-range-ordered (QLRO) state (the Heisenberg critical state with spin 
correlations decaying as $1/r$, discussed in the preceding sections) takes place in a system with distance-dependent couplings of the form 
$J_r \propto (-1)^{r-1}/r^\alpha$ \cite{laflorencie}. The signs here favor antiferromagnetic order, and there is no frustration. When the 
exponent $\alpha<\alpha_c \approx 2$, the ground state has true long-range N\'eel order, while for $\alpha>\alpha_c$ the system is in the
QLRO phase. The critical value of the long-range interaction parameter $\alpha_c$ depends on details of the couplings, e.g., on the nearest-neighbor 
coupling $J_1$ when all other $J_r$ are fixed, and the critical exponents at the transition are continuously varying (in contrast to the constant 
exponents throughout the QLRO phase). 

\begin{figure}
\includegraphics[width=7.5cm, clip]{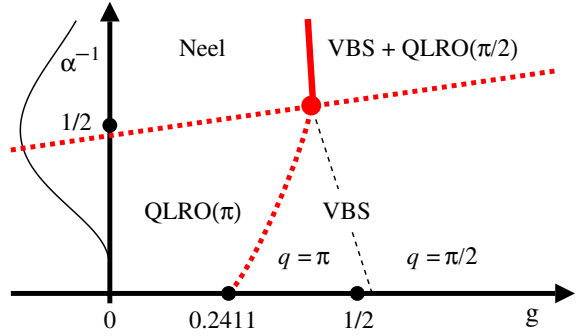}
\caption{Ground state phases of the chain with frustration parameter $g$ and exponent $\alpha$ in the long-range interaction, reproduced from
Ref.~\cite{awslr}. The dashed curves indicate continuous phase transitions, whereas the thick solid curve represents a first-order transition. 
The thin solid curve for $g<0$ corresponds to the system with unfrustrated interaction studied in \cite{laflorencie}. At $\alpha^{-1}=0$ 
(the J$_{\rm 1}$-J$_{\rm 2}$ chain) the dominant spin correlations in the VBS state change from $q=\pi$ to $\pi/2$ at $g\approx 0.52$ 
\cite{bursill96}. The curve where this change occurs for $\alpha^{-1}>0$ connects to the multi-critical point where all the phase boundaries 
come together.}
\label{lrphases}
\end{figure}

Another interesting example of a system with long-range interactions is the Haldane-Shastry chain \cite{haldane2,shastry88}, with frustrated interactions 
$J_r = 1/r^2$. It has a critical ground state similar to that of the standard Heisenberg chain, but, in field-theory language, the marginal operator causing the 
leading log-corrections vanishes \cite{essler}. The system is, thus, right at a dimerization transition such as the one discussed above for the 
J$_{\rm 1}$-J$_{\rm 2}$ chain. It can be noted that $J_2/J_1=1/4$ in the Haldane-Shastry model, which is quite close to the critical ratio $g_c=0.2411$ of 
the J$_{\rm 1}$-J$_{\rm 2}$ chain. Thus, the interactions beyond distance $r=2$ in the Haldane-Shastry chain only play a minor role (but, importantly, 
actually make the system exactly solvable \cite{haldane2}).

In the presence of long-range interactions one can also realize a direct 1D quantum phase transition between a N\'eel state and a VBS. Here we 
discuss the system introduced in Ref.~\cite{awslr}, which combines unfrustrated long-range interactions and short-range frustration (at separation 
$r=2$) according to the hamiltonian
\begin{equation}
H = \sum_{r=1}^{N/2} J_r \sum_{i=1}^N {\bf S}_i \cdot {\bf S}_{i+r},
\label{ham}
\end{equation}
where the distance dependence of the couplings is given by
\begin{equation}
J_1=\frac{1}{J_{\Sigma}},~~~~J_2=g,~~~~J_{r> 2} = \frac{1}{J_{\Sigma}}\frac{(-1)^{r-1}}{r^{\alpha}},~~~~~ J_{\Sigma}= 1+ \sum_{r=3}^{N/2}\frac{1}{r^\alpha}.
\label{lrfham}
\end{equation}
Here, with the normalization by $J_{\Sigma}$ of all couplings but $J_2$, the sum of all non-frustrated interactions $|J_r|$ equals $1$, and the limit 
$\alpha \to \infty$ therefore corresponds exactly to the J$_{\rm 1}$-J$_{\rm 2}$ chain with $g=J_2/J_1$. This will be useful when comparing the two 
models and also guarantees a finite energy per spin for $N\to \infty$, even for $\alpha<1$. Instead of summing $J_r$ up to $r=N/2$, one could also 
include $N/2 < r < N$. This should not affect the phase boundaries and critical exponents for $\alpha>1$, however. 

We will study the evolution if the dimerization transition occurring as a function of the frustration strength $g$ as the inverse $\alpha^{-1}$ 
of the long-distance interaction exponent is increased from $0$. As shown in the semi-quantitative phase diagram in Fig.~\ref{lrphases} (constructed 
on the basis of Lanczos results, as will be discussed below) this continuous transition persists until $\alpha \approx 2$, while for smaller $\alpha$ 
it evolves into a first-order transition between the N\'eel state and a state with coexisting VBS order and critical (or possibly long-ranged)
spin correlations at wave-number $q=\pi/2$. This state is denoted in the phase diagram as VBS+QLRO$(\pi/2)$. 

\begin{figure}
\includegraphics[width=12.5cm, clip]{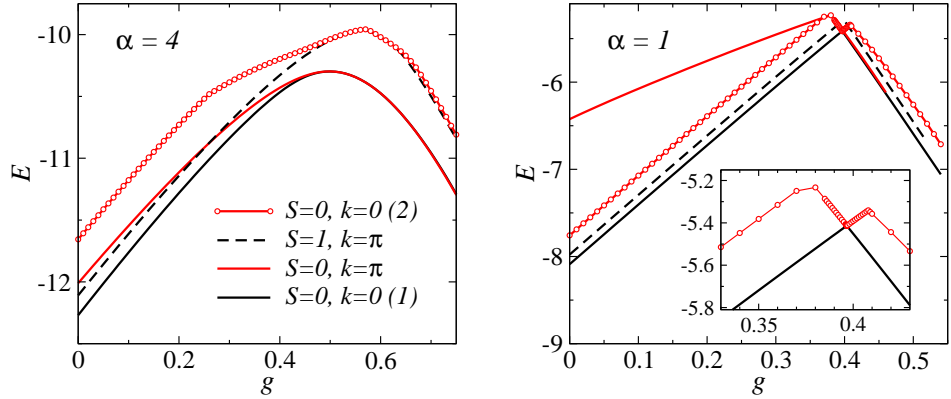}
\caption{Low-energy levels of a $16$-spin system at two values of the long-range exponent; $\alpha=4$ (left) and $\alpha=1$ (right). The spin $S$ 
and the momentum $k$ of the states are indicated in the left panel (and note that two levels are shown with $S=0,k=0$). The inset in the right panel shows 
the avoided level crossing of the two $k=0$ singlets in greater detail. There is a very small gap between these singlets where they come together, but 
this is much smaller than what can be resolved in this figure.}
\label{lrf_levels}
\end{figure}

\paragraph{Evolution of the dimerization transition}

\begin{figure}
\includegraphics[width=8.5cm, clip]{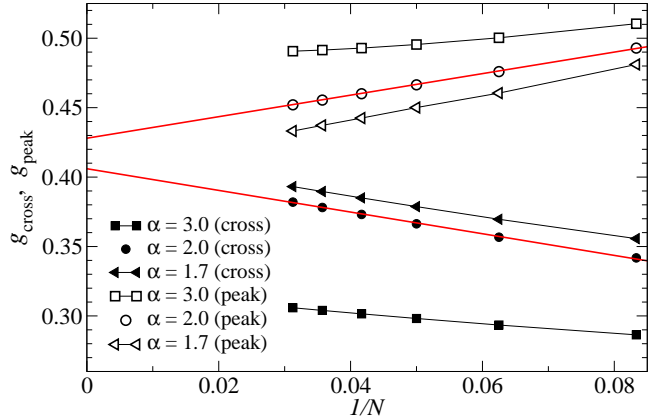}
\caption{Dependence on the inverse chain length of the singlet-triplet crossing point $g_{\rm cross}$ and the location $g_{\rm peak}$ of the ground state 
energy maximum for different long-range interaction exponents $\alpha$. The two lines show extrapolations of the $g=2.0$ data to $N=\infty$.}
\label{lrf_extrap}
\end{figure}

We have discussed how to study the dimerization transition based on crossings of excited-state energies for the J$_{\rm 1}$-J$_{\rm 2}$ chain, with an 
illustration in Fig.~\ref{cross}(a). The same physics applies in the presence of the long-range interaction as well, if $\alpha$ is sufficiently 
large. This is shown for a  $16$-spin chain at $\alpha=4$ in the left panel of Fig.~\ref{lrf_levels}. The lowest $k=0$ and $k=\pi$ singlets should 
become degenerate in the VBS phase for $N \to \infty$ (so that a symmetry-broken dimerized states can be formed). A region of very near degeneracy for 
$g>1/2$ can be seen in the figure. The region of approximate degeneracy, which is not easy to demarcate precisely, expands very slowly toward smaller 
$g$ with increasing $N$. As shown in Fig.~\ref{lrf_extrap}, the singlet-triplet crossing point converges with the system size [although slower than 
the $1/N^2$ convergence shown for J$_{\rm 1}$-J$_{\rm 2}$ the chain in Fig.~\ref{cross}(b)] and can reliably give the QLRO$(\pi)$--VBS phase 
boundary $g_c(\alpha)$ for $\alpha > 2$.

Another interesting feature of the energy levels is that, upon decreasing $\alpha$ below $\approx 2$, the broad maximum in the ground state energy 
versus $g$  becomes increasingly sharp. As seen in the right panel of Fig.~\ref{lrf_levels}, at $\alpha=1$ it has developed into a sharp tip due to an 
avoided level crossing with the second singlet at $k=0$. The real singlet-triplet crossing has moved to the same region. An avoided level crossing 
between two states with the same quantum numbers, leading to a discontinuity in the derivative of the ground state energy with respect to $g$ for 
$N\to \infty$, is the hall-mark of a first-order transition. It should be noted that it is not just the two lowest singlets that exhibit this kind of 
avoided level crossing. Other low-energy states as well come together in the neighborhood of the transition, in a complicated cascade of level 
crossings---all of these should converge to a single point at the first-order transition when $N\to \infty$. The nature of the phases at this transition 
will be discussed below. First, let us investigate in more detail how the dimerization transition evolves from continuous to first-order.

Fig.~\ref{lrf_extrap} shows the size dependence of the level crossing $g_{\rm cross}$ and the location $g_{\rm peak}$ of the maximum in the 
ground state energy. In the standard Majumdar-Ghosh frustrated chain the size correction to the crossing point is $\propto 1/N^2$ [as seen in 
Fig.~\ref{cross}(b)], which also can be seen for large $\alpha$. For smaller $\alpha$, the corrections instead seem to be  $\propto 1/N$, but a 
cross-over to $1/N^2$ for large $N$ seems likely as long as the transition remains continuous.  The peak location moves in the opposite direction. 
For some $\alpha$ and $N\to \infty$, $g_{\rm cross}$ and $g_{\rm peak}$ should coincide. The results indicate that both $g_{\rm cross}$ and $g_{\rm peak}$ 
have dominant $1/N$ corrections at this point. Line fits are shown in Fig.~\ref{lrf_extrap} at $\alpha=2$, were there is still a small gap between 
the two extrapolated values (and the extrapolations may not be completely accurate, if the linear scaling holds asymptotically only at the special 
$\alpha$-value for which $g_{\rm cross}=g_{\rm peak}$). At $\alpha=1.7$, where the transition is first-order, they should coincide (and then the asymptotic 
size correction should be exponential).

\paragraph{First-order N\'eel--VBS transition}

\begin{figure}
\includegraphics[width=8.5cm, clip]{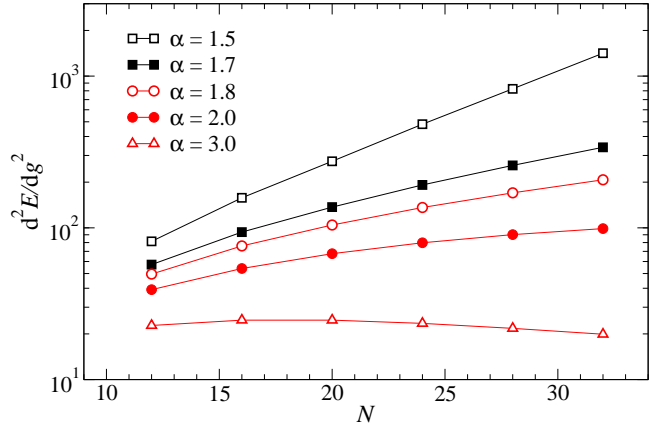}
\caption{Size dependence of the second derivative of the ground state energy with respect to the frustration parameter $g$ at the 
point $g_{\rm peak}$ where the ground state energy takes its maximum value.}
\label{lrf_deriv2}
\end{figure}

To confirm an avoided level crossing with a discontinuous energy derivative for $\alpha > 1.8$ (approximately), the second derivative of the ground 
state energy at its maximum is graphed on a lin-log scale in Fig.~\ref{lrf_deriv2}. These results were obtained based on calculations of the energy on a very 
dense grid of points close to the peak value, to which a polynomial could be reliably fitted. The second derivative extracted from this polynomial grows exponentially 
with $N$ for $\alpha=1.5$, showing that the slope of the energy curve indeed changes discontinuously for an infinite chain. In contrast, at $\alpha=3$ 
the second derivative decreases for large $N$. For $\alpha=2$ convergence to a finite value also seems plausible, whereas $\alpha=1.7$ and $1.8$ appear 
to be close to a separatrix (where the form of the divergence is consistent with a power law) between the two different behaviors. This analysis suggests 
that the continuous dimerization transition changes smoothly into a first-order transition at $(g_m \approx 0.41,\alpha_m \approx 1.8)$. The singlet-triplet 
crossing moves toward the ground-state energy maximum and coincides with it at the multi-critical point $(g_m,\alpha_m)$, beyond which it develops into 
a first-order singularity. 

\begin{figure}
\includegraphics[width=14cm, clip]{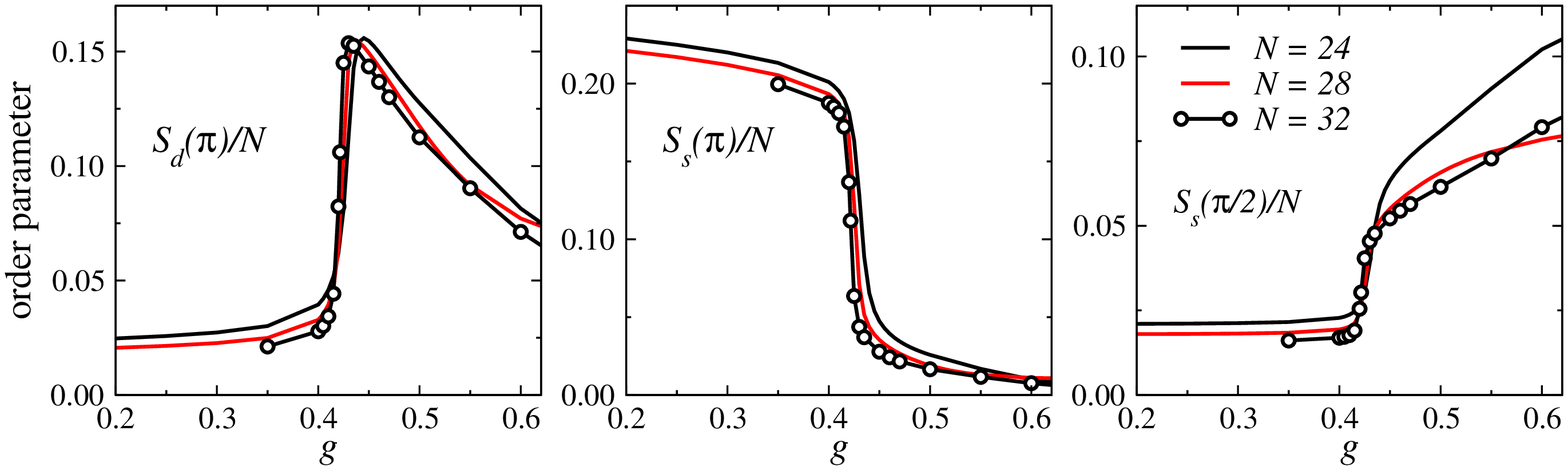}
\caption{VBS and magnetic order parameters defined in terms of the size-normalized dimer structure factor $S_d(\pi)$ (left) and spin structure 
factors $S_s(\pi)$ (middle) and $S_s(\pi/2)$ (right) versus the frustration strength $g$ in systems with long-range interaction exponent $\alpha=1.5$. 
Discontinuities develop with increasing system size at the first-order transition.}
\label{lrf_jumps}
\end{figure}

To analyze the states involved in the first-order transition, we next investigate the standard spin correlation function $C(r)$ and the dimer 
correlation function $D(r)$ defined in (\ref{drdef}). We will also study the Fourier transforms of these correlation functions; the static spin ($s$) 
and dimer ($d$) structure factors:
\begin{eqnarray}
&&S_s(q) =  C(0)+C(N/2)+2\sum_{r=1}^{N/2-1}\cos(qr)C(r), \\
&&S_d(q) =  D(0)+D(N/2)+2\sum_{r=1}^{N/2-1}\cos(qr)D(r).
\end{eqnarray}
If there is long-range order at some wave-vector $q=Q$, then the corresponding structure factor is proportional to $N$ (for large $N$), and order 
parameters can therefore be defined as $S_{s,d}(Q)/N$. Here $Q=\pi$ for both the N\'eel state and the dimerized VBS state. Fig.~\ref{lrf_jumps} shows 
the $g$-dependence of $S_{d}(\pi)/N$ and $S_{s}(\pi)/N$ at $\alpha=1.5$. Discontinuities are seen to develop at $g\approx 0.42$, in agreement with 
the first-order scenario (and at $g=1$, not shown here, the discontinuities are much sharper). Both the N\'eel and VBS order parameters are well 
converged to non-zero values in their respective phases, and small in the other phase (where they should vanish when $N\to \infty$, but it is not 
possible to observe this clearly because of the small system sizes). The spin structure factor at $q=\pi/2$ is also shown in the figure. Interestingly, 
it also becomes large in the VBS phase, while in the N\'eel phase it should decay to zero with increasing $N$ (which is plausible based don this data, 
but, again, not possible to see clearly for these small systems). Based on these results one might conclude that the VBS state obtaining at the 
first-order transition has co-existing long-range magnetic order at $k=\pi/2$ (i.e., a spiral with period four). Considering the small system sizes, 
this cannot be guaranteed, however. The behavior could also reflect a slow power-law decay of the spin correlations.

\begin{figure}
\includegraphics[width=12.25cm, clip]{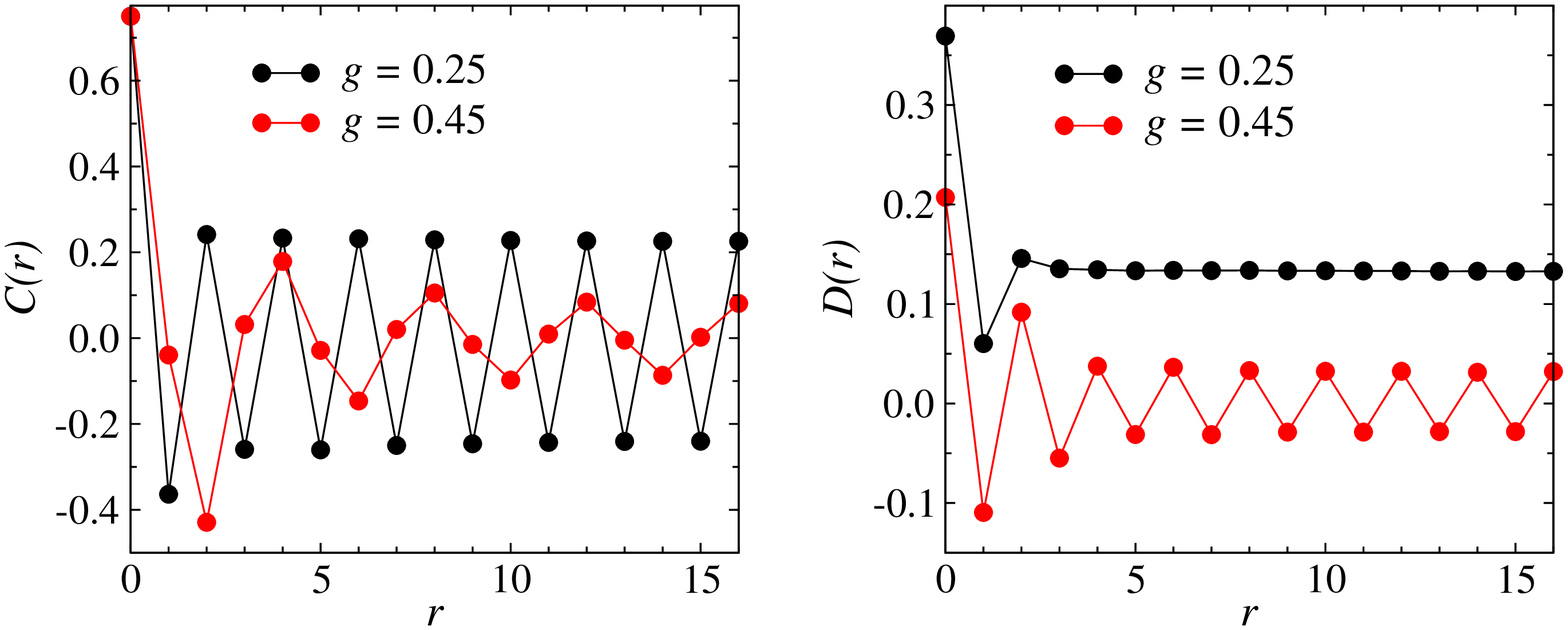}
\caption{Spin (left) and dimer (right) correlations in a 32-spin chain at $\alpha=1$ for two values of the frustration parameter. At $g=0.25$ 
and $0.45$ the system is in the N\'eel and VBS--QLRO$(\pi/2)$ phase, respectively. The first-order transition in this case occurs at $g\approx 0.39$.}
\label{lrf_corr}
\end{figure}

Fig.~\ref{lrf_corr} shows the real-space spin and dimer correlations at $\alpha=1$ for two $g$ values, at either side of the first-order transition. 
At $g<g_c$, in the N\'eel state, the spin correlations are staggered and clearly long-ranged (with almost no decay seen as a function of $r$ for 
$r>1$). At this coupling there is no structure in $D(r)$, i.e., there is no VBS order. For $g>g_c$ there is clearly VBS order, with $D(r)$ showing 
the characteristic staggered pattern. Strong period-four spin correlations are also observed, but it is not clear whether these are long-ranged or 
decay slowly to zero with increasing $r$. The behavior is consistent with an $1/r$ decay, but much longer chains would be needed to extract the behavior 
reliably. An interesting point to note here is that the dimer correlations oscillate around zero, even though no constant has been subtracted. This 
is in contrast to the oscillations around a positive value in the J$_{\rm 1}$-J$_{\rm 2}$ chain, as seen in Fig.~\ref{dcor}. Looking at the definition of 
the dimer correlation function in Eq.~(\ref{drdef}), this behavior is another manifestation of period-four spin structure, i.e., at odd separation of 
the bond operators ${\bf S}_i+{\bf S}_{i+1}$, a correlation between a ferromagnetic and an antiferromagnetic bond is measured. While the nature of 
the magnetic structure in this VBS is not completely clear, the correlation functions seem to indicate a tendency to bond-ordering of the form
$|ststst\cdots\rangle$, where $s$ and $t$ represent nearest-neighbor bonds with high singlet and triplet density, respectively. In Fig.~\ref{lrphases} the 
state has been denoted as magnetically quasi-long-range ordered, a VBS--QLRO$(\pi/2)$ coexistence state, but it could in fact also be a state with coexisting 
VBS and long-range magnetic order. It should also be noted that this phase could in principle be very sensitive to boundary conditions on small lattices, 
and it cannot be excluded that the actual magnetic structure is spiral-like, with continuously varying pitch (as in the J$_{\rm 1}$-J$_{\rm 2}$ chain 
with $g>1$ \cite{kumar}).

The VBS+QLRO$(\pi/2)$ state should have gapless spin excitations, regardless of the spin correlations as long as they are not exponentially decaying. 
The lowest triplet is at $k=\pi/2$. It is, however, difficult to demonstrate the gaplessness based on data for small systems, because the size dependence 
of the gaps (and other quantities) for system sizes $N=4n$ exhibit large even-odd $n$ oscillations (as well as other irregular size effects). In 
the conventional VBS phase (in the low-right part of the phase diagram in in Fig.~\ref{lrphases}) the lowest triplet 
is at $k=\pi$, even when the spin correlations (which are exponentially decaying in this phase) are peaked at $k=\pi/2$. The level crossing between the 
lowest $k=\pi$ and $k=\pi/2$ triplets can therefore in principle be used to extract the boundary between the VBS and VBS+QLRO$(\pi/2)$ phases. 
The size dependence of the crossing point is not smooth, however, and cannot be extrapolated very reliably. The boundary between dominant $k=\pi$ 
and $k=\pi/2$ spin correlations has also not been extracted accurately. This change in the spin correlations may be associated with a transition to 
a state with periodicity four \cite{bursill96}, although there are no sign of the VBS order changing.

\paragraph{QLRO--N\'eel transition}

Let us return to Fig.~\ref{lrf_levels} for another interesting feature of the level spectrum: The lowest singlet excitation for small $g$ has 
momentum $k=\pi$ for $\alpha=4$ but $k=0$ for $\alpha=1$. The switching of the order of these levels as a function of $\alpha$ for $g<g_c$ is associated 
with the N\'eel--QLRO$(\pi)$ transition. The level crossings can be used to extract this phase boundary very accurately up to $g \approx 0.25$ (while for 
higher $g$ the $N\to \infty$ extrapolations become difficult). While it is not completely clear why the $k=0$ and $k=\pi$ excited singlets cross each
other at this transition, it can again be noted that the QLRO$(\pi)$ phase the low-energy excitations arise from two deconfined spinons. The lowest 
(for $N=4n$) is a triplet at $k=\pi$. The lowest singlet is also at $k=\pi$, with a small finite-size gap to the triplet due to weak spinon-spinon 
interactions (and Fig.~\ref{gaps} shows and example of how these levels become degenerate as $N\to \infty$). In the N\'eel phase the spinons are no 
longer deconfined, and the structure of the low-energy spectrum changes. 

As we will discuss below in Sec.~\ref{lanc2d}, the 2D N\'eel state on finite
lattices has low-energy excitations analogous to quantum-rotor states, which have $S=1,2,3,\ldots$ and become degenerate with the ground state as 
the system size increases. They have momenta $k=(0,0)$ and $k=(\pi,\pi)$ for even and odd $S$, respectively. One might expect this kind of quantum 
rotor ``tower'' in the 1D N\'eel state as well. Apparently, the lowest $k=\pi$ singlet, which would not be part of such a rotor tower, is then pushed 
up much higher in energy, and the $k=0$ singlet takes over as the lowest singlet excitation.

It is useful to compare the level crossing approach with the QMC calculation in Ref.~\cite{laflorencie} for the 
unfrustrated model with $J_1=1$ and $J_{r>1}=\lambda (-1)^{r-1}/r^\alpha$. A reparametrization of this model for $\lambda=1$, to the convention used in 
(\ref{lrfham}), gives the curve for $g<0$ shown in Fig.~\ref{lrphases}. Finite-size scaling of QMC data for the N\'eel order parameter gave 
$\alpha_c=2.225 \pm 0.025$, for $N$ up to $4096$ \cite{laflorencie} . Extrapolating the $k=0,\pi$ singlet crossing points (which here have size 
corrections $\propto 1/N^{\beta}$, with $\beta\approx 1.50$) for 
$N\le 32$ gives a marginally higher (and probably more reliable) value; $\alpha_c=2.262 \pm 0.001$. Analyzing the singlet and triplet gaps at the crossings, 
assuming $\Delta \sim N^{-z}$, gives the dynamic exponent $z=0.764 \pm 0.005$, in very good agreement with Ref.~\cite{laflorencie}. For the frustrated model 
(\ref{lrfham}) on the QLRO$(\pi)$--N\'eel boundary, the gap scaling in $1/N$ give $z\approx 0.75$ for $g$ up $\approx 0.25$, while for larger $g$ it is not 
possible to reliably extract $\alpha_c$ and $z$ this way, because of large scaling corrections and the absence of level crossings for increasingly large 
systems as the multi-critical point $(g_m\approx 0.41,\alpha_m \approx 1.8)$ is approached (for reference $\alpha_c = 2.220 \pm 0.005$ for $g=0$ and 
$2.170 \pm 0.01$ for $g=0.2$). At the multi-critical point $z$ can be extracted using gaps at the singlet-triplet crossing point and the ground state energy 
maximum, which gives $z\approx 0.8$. It is, thus, possible that the dynamic exponent is constant, $z\approx 3/4$, on the whole QLRO($\pi$)--N\'eel 
boundary, including the multi-critical point where all phase boundaries come together.

\paragraph{Technical notes}

The model we have studied here is rather challenging for a 1D system, from a technical standpoint, because of the long-range interactions. 
The combination of long-range interactions and frustration makes it difficult to apply other computational techniques. While efficient QMC techniques can be 
applied to systems with unfrustrated long-range interactions \cite{awslrqmc,laflorencie}, this is no longer possible in the presence of the frustrating $J_2$ 
term, due to the sign problem (discussed in Sec.~\ref{sec_sse}). The DMRG method \cite{white1,schollwock2}, on the other hand, can handle frustration but not 
easily long-range interactions. It would still be interesting to try to apply DMRG, or related techniques based on matrix-product states \cite{verstraete1}, 
to this system, in particular to study further the spin structure in the putative VBS-QLRO($\pi/2$) state.

The Lanczos calculations presented above exploited all the symmetries discussed in Sec.~\ref{hbchaindiag}. The hamiltonian was stored in compact form on 
disk, with a separate file for each interaction distance $r$ in (\ref{lrfham}). These files were read and used one-by-one in each operation with the 
hamiltonian according to code $\{22\}$. For $N=32$, the number of Lanczos iterations needed was typically $60 \sim 80$ (and less for smaller systems). 
All basis states were saved in primary memory, and re-orthogonalization to all previous states was carried out after each iteration (as in code $\{19\}$).

\subsection{Two-dimensional systems}
\label{sec_twodim}

We now discuss exact diagonalization in two dimensions, using the important case of the Heisenberg model on the simple periodic square lattice as 
an example. As in one dimension, we can use momentum conservation to block diagonalize the hamiltonian within the sectors of fixed magnetization. It is a 
little more complicated to take maximum advantage of lattice symmetries to further split some of the blocks, because of the larger number of symmetry 
operations commuting with the hamiltonian. On the positive side, the reward for implementing all the lattice symmetries is that, for high-symmetry 
momenta, where lattice reflections and/or rotations can be used, the blocks are smaller for given $N$ than in one dimension. In practice, this 
may be of little help, since the maximum linear size $L$ of an $L\times L$ lattice that can be diagonalized is still pitifully small ($L=6$). 
Exact diagonalization studies are still of great value.

In one dimension, we used the reflection (parity) operation $P$ and constructed momentum eigenstates based on representatives $|a\rangle$ and their reflections, 
$(1 + pP)|a\rangle$ (and later we added also spin-inversion). Although $P$ does not commute with the translation operator, we found that for the two
special momenta $k=0,\pi$, both $k$ and $p$ are still good quantum numbers. For general $k$, $p$ is not a valid quantum number, but with semi-momentum 
states, mixing $\pm k$ states, we could use $P$ to accomplish a real-valued representation. In two dimensions one can in principle also 
construct a real-valued hamiltonian for any  momentum, but this is much more complicated in practice than in one dimension, and in the end it is not 
worth the effort (since it does not increase the system sizes that can be studied).
\footnote{A real maximally blocked basis exists because the hamiltonian is real 
and symmetric, which guarantees real eigenvalues and eigenvectors. If the hamiltonian is diagonalized numerically as a single block, the eigenvectors 
in degenerate multiplets will be mixed, and the translation operators will not be diagonal (because their eigenvalues are 
complex). The degenerate complex momentum states have been mixed in such a way as to make the linear combinations real. Such a superposition can in 
principle be applied to block-diagonalize the original hamiltonian matrix into semi-momentum like real-valued blocks.}  
 Here we will therefore work with standard complex momentum states.

\paragraph{Defining the lattice}

\begin{figure}
\includegraphics[width=3.25cm, clip]{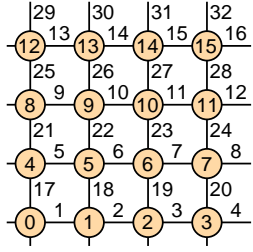}
\caption{Labeling of sites and bonds on a $4\times 4$ periodic square lattice. With the bit representation, it is practical to start the site 
labels from $0$, here according to $i=x+yL$ with $x,y \in \{0,\ldots,L-1\}$ with $L=4$. A bond is here labeled according to the ``left-down''
reference site $i$ to which it is connected; the horizontal and vertical bonds are $b_x=i$ and $b_y=L^2+i$, respectively.}
\label{sblabels2d}
\end{figure}

As in one dimension, we use the state notation 
$|S^z_0,\ldots,S^z_{N-1}\rangle$, which is practical with the bit representation of the spins. For a 1D system the spin indices correspond directly to 
the chain geometry, but for a 2D lattice we have to establish a labeling convention for the sites. We will consider rectangular lattices with 
$N=L_x \times L_y$ sites, at coordinates $(x,y)$ with $x=0,\ldots,L_x-1$ and $y=0,\ldots,L_y-1$. A natural choice is to number the sites 
$i=0,\ldots,N-1$ such that $i=x_i+y_iL_x$, as illustrated for a $4\times 4$ system in Fig.~\ref{sblabels2d}. In a notation not explicitly 
dependent on the lattice, we can write the Heisenberg hamiltonian as
\begin{equation}
H=J\sum_{b=1}^{N_b} {\bf S}_{i(b)} \cdot {\bf S}_{j(b)},
\label{hbsum}
\end{equation}
where $[i(b),j(b)]$ are the two nearest-neighbor sites connected by bond $b$. Specifying the lattice then just amounts to creating this list of site pairs.

\subsubsection{Momentum states in two dimensions}

\begin{figure}
\includegraphics[width=10cm, clip]{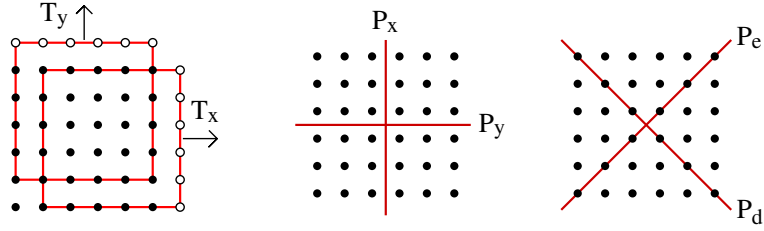}
\caption{Symmetries used for the two-dimensional square lattice. $T_x$ and $T_y$ translate the spins by one lattice spacing in the positive $x$ and 
$y$ directions, respectively. The lattice is periodic; the open circles represent sites of the opposite edges. $P_x$ and $P_y$ are reflections with respect 
to the $x$- and $y$-axis, and $P_d$ and $P_e$ reflect with respect to the two diagonal axes. For even $L$ (which normally should be used for an antiferromagnet), 
the $P_x$ and $P_y$ axes pass between lattice sites. The reflections $P_d$ and $P_e$ are about the lines connecting the far corners, which go through the 
sites on the diagonals (and hence leave the spins on those sites unchanged upon reflection).}
\label{squaretrans}
\end{figure}

We again define translations of the spin indices, now in both the $x$ and $y$ directions, as illustrated
in Fig.~\ref{squaretrans}, with corresponding operators $T_x$ and $T_y$ defined by
\begin{eqnarray}
&&T_x|S^z_0,\ldots,S^z_{N-1}\rangle = |S^z_{T_x(0)},\ldots,S^z_{T_x(N-1)}\rangle,\nonumber\\
&&T_y|S^z_0,\ldots,S^z_{N-1}\rangle = |S^z_{T_y(0)},\ldots,S^z_{T_y(N-1)}\rangle,
\label{tdef2d}
\end{eqnarray}
where the translated spin indices are
\begin{eqnarray}
&& T_x(i)=[x_i-1]_{L_x}+y_iL_x,\nonumber \\
&& T_y(i)=x_i+([y_i-1]_{L_y})L_x,
\end{eqnarray}
with $[\gamma_i-1]_{L_\gamma}$ denoting the modulus of $\gamma_i-1$ with respect to $L_\gamma$, i.e., $[-1]_{L_\gamma}=L_\gamma-1$. 

Using these translations, a momentum state based on a representative $|a\rangle$ is defined as
\begin{equation}
|a({\bf k})\rangle = 
|a(k_x,k_y)\rangle = \frac{1}{\sqrt{N_a}}\sum_{x=0}^{L_x-1}\sum_{y=0}^{L_y-1}{\rm e}^{-i(k_xx+k_yy)}T_y^yT_x^x|a\rangle,
\label{kstate2d}
\end{equation}
where the possible momenta are 
\begin{equation}
k_{\gamma}=\frac{2\pi}{L_{\gamma}}m_{\gamma},~~~~~ m_{\gamma}=0,1,\ldots,L_{\gamma}-1,~~~~~\gamma \in \{x,y\}.
\end{equation}
The normalization constant $N_a$ depends on the translational properties of the representative, i.e., the number of different states $D_a$ obtained 
among the group of $L_x\times L_y$ translations of the representative $|a\rangle$ (e.g., for a state with the spins in a checker-board pattern $D_a=2$). 
A representative is incompatible with the momentum if the sum of phases $F_a$ in (\ref{kstate2d}) over the translations bringing $|a\rangle$ onto 
itself vanishes. For a compatible state, the normalization constant $N_a=D_a|F_a|^2$. The easiest way to compute this in a program is simply to carry 
out all the translations and sum up $D_a$ and $F_a$ in the process, instead of using explicit formulas as we did for 1D systems (where one can of course 
also use the more brute-force approach).

The construction of the hamiltonian matrix proceeds as in the 1D case. We again split the hamiltonian into a diagonal
piece $H_0$ and off-diagonal bond operators $H_j$ as in (\ref{hdia1}) and (\ref{hoff1}), where now $j=1,\ldots N_b$. Acting with these operators on the 
representative $|a\rangle$, we again may have to translate the resulting state in order to obtain the new representative $|b_j\rangle$
corresponding to $H_j|a\rangle$, i.e.,
\begin{equation}
H_j|a\rangle = h_j(a)T_y^{-l_j^y}T_x^{-l_j^x}|b_j\rangle.
\end{equation}
Acting on a momentum state, we thus get
\begin{eqnarray}
&&H_j|a(k_x,k_y)\rangle = \frac{h_j(a)}{\sqrt{N_a}}\sum_{x=0}^{L_x-1}\sum_{y=0}^{L_y-1}{\rm e}^{-i(k_xx+k_yy)}
T_y^{(y-l_j^y)}T_x^{(x-l_j^x)}|b_j\rangle~~~~~~~~~~~~~~~~~ \\
&&~~~~~~~~~~~~~~~~~ =  \frac{h_j(a)}{\sqrt{N_a}}{\rm e}^{-i(k_xl_j^x+k_yl_j^y)}\sum_{x=0}^{L_x-1}\sum_{y=0}^{L_y-1}{\rm e}^{-i(k_xx+k_yy)}
T_y^yT_x^x|b_j\rangle,\nonumber
\end{eqnarray}
and taking the overlap with $\langle b_j(k_x,k_y)|$ gives the matrix element
\begin{equation}
\langle b(k_x,k_y)|H_j|a(k_x,k_y)\rangle = 
h_j(a) {\rm e}^{-i(k_xl_j^x+k_yl_j^y)}\sqrt{\frac{N_{b_j}}{N_a}}.
\label{hk2}
\end{equation}
The normalization constants (or some integer mapping to their possible numerical values) should be stored along with the representatives.

\paragraph{Incorporating other lattice symmetries}

We here consider a quadratic, $L\times L$, system. There are eight independent rotations and reflections of such a lattice, which can be 
taken as, e.g., the four $90^\circ$ rotations and a reflection about a horizontal or diagonal axis. It can be easily verified that additional reflections 
can be written as combinations of rotations and the first reflection. The easiest way to see this is to just consider all rotations and reflections 
of a $2\times 2$ array;
\begin{eqnarray}
&&\left (\begin{array}{ll}4 & 3
                       \\ 1 & 2\end{array}\right )
  \left (\begin{array}{ll}3 & 2
                       \\ 4 & 1\end{array}\right )
  \left (\begin{array}{ll}2 & 1
                       \\ 3 & 4\end{array}\right )
  \left (\begin{array}{ll}1 & 4
                       \\ 2 & 3\end{array}\right ) \nonumber \\
&&\left (\begin{array}{ll}3 & 4
                       \\ 2 & 1\end{array}\right )
  \left (\begin{array}{ll}2 & 3
                       \\ 1 & 4\end{array}\right )
  \left (\begin{array}{ll}1 & 2
                       \\ 4 & 3\end{array}\right )
  \left (\begin{array}{ll}4 & 1
                       \\ 3 & 2\end{array}\right ).
\end{eqnarray}
Here the first row contains all the rotations of the first array, and in the second row those arrays have been reflected by exchanging
the two columns. Any other reflection will just produce an array which is already in the above set of eight. Other permutations of the elements 
do not correspond to a combination of rotations and reflections (and hence do not correspond to symmetries of hamiltonians we are normally 
interested in). In practice, we can use any convenient set of reflections, or rotations and reflections, with which the eight unique transformations 
can be generated.

Using translations as well as a set of other symmetry operators, which we for now collectively add together into a single operator $Q$, with 
corresponding quantum numbers $\{q\}$, a momentum state can be defined as
\begin{equation}
|a^{{\bf \sigma}} ({\bf k},\{q\})\rangle = \frac{1}{\sqrt{N_a}}\sum_{r_x=0}^{L-1}\sum_{r_x=0}^{L-1}
{\rm e}^{-i(k_xx+k_yy)}T_y^{r_y}T_x^{r_x}Q|a\rangle .
\label{akpgdef2}
\end{equation}
As we have seen above, there are more symmetry operations than unique transformations. We can select the ones that are most convenient for given 
momentum. Instead of using the rotations of the square lattice, we will here use the reflections $P_x$ and $P_y$ about the $x$ and $y$-axis, respectively. 
These correspond to symmetries of the hamiltonian for any rectangular ($L_x\times L_y$) lattice (and note that the $180^\circ$ rotation is equivalent to 
$P_xP_y$). For generic momenta, the reflections do not commute with the translations, and there can be no further blocking using $P_x$ and $P_y$, but
there are high-symmetry momenta for which $P_x$ or $P_y$, or both, can be used. For $k_x = \pm k_y$ we will instead use the reflections $P_d$ and 
$P_e$ about the two perpendicular diagonal axes. All the reflections are defined in Fig.~\ref{squaretrans}. For the most special momenta, ${\bf k}=(0,0)$ 
and $(\pi,\pi)$, and $L_x,L_y$ arbitrary, we can also use $P_d$ (or $P_e$) in addition to $P_x,P_y$.  

To achieve maximal block-diagonalization for a 
given momentum $(k_x,k_y)$, the operator $Q$ in (\ref{akpgdef2}) can be chosen according to:
\begin{equation}
Q = \left\{
\begin{array}{lll}
1,  & {\rm general~} {\bf k}  &~~(a) \\
(1+p_xP_x),  & {\bf k}=(0,k_y),(\pi,k_y) &~~(b) \\
(1+p_yP_y),  & {\bf k}=(k_x,0),(k_x,\pi) &~~(c) \\
(1+p_dP_d),  & k_x= +k_y~ (L_x=L_y) &~~(d) \\
(1+p_eP_e), &  k_x= -k_y~ (L_x=L_y) &~~(e) \\
(1+p_yP_y)(1+p_xP_x), & {\bf k}=(0,0),(\pi,\pi),~p_x\not=p_y  &~~(f)\\
(1+p_dP_d)(1+p_yP_y)(1+p_xP_x), & {\bf k}=(0,0),(\pi,\pi),~p_x=p_y &~~(g)
\end{array}\right.
\label{qdef}
\end{equation}
where again all the reflection quantum numbers take the values $\pm 1$. Note the restriction $p_x=p_y$ in case (g), i.e., there are no such states 
for $p_x \not=p_y$. In that case, we can of course also use option (f), but using (g) splits the blocks further into two.

To prove that the symmetry operators $Q$ defined above indeed do have the corresponding conserved quantum numbers within the different
momentum sectors, the following relationships between the lattice transformations are useful:
\begin{equation}
\begin{array}{lll}
P_xT_x=T^{-1}_xP_x~  & ~~~~ & P_yT_y=T^{-1}_yP_y \\
P_dT_x=T_yP_d~      & ~~~~ &  P_dT_y=T_xP_d     \\    
P_eT_x=T^{-1}_yP_e~  & ~~~~ &  P_eT_y=T^{-1}_xP_e~ \\
P_dP_x = P_yP_d~    & ~~~~ &  P_dP_y = P_xP_d~   \\
P_eP_x = P_yP_e~    & ~~~~ &  P_eP_y = P_xP_e.
\end{array}
\label{oprels}
\end{equation}
The remaining pairs are commuting operators; $[T_x,T_y]=0$, $[T_x,P_y]=0$, $[T_y,P_x]=0$, $[P_x,P_y]=0$, $[P_e,P_d]=0$.
Let us just check the most complicated case, (g) in (\ref{qdef}). The permutations commute with the translations for these special momenta, 
for the same reasons as discussed for 1D systems in Sec.~\ref{reflection}. We then only have to investigate the properties of $Q$. It is 
clear that the state is an eigenstate of $P_d$, that being the left-most reflection operator. To check whether it is also an eigenstate of 
$P_x$ and $P_y$, we use some of the relationships in (\ref{oprels}) to verify that $P_{x}Q=p_{x}Q$ and $P_{y}Q=p_{y}Q$. After a little 
algebra we get
\begin{eqnarray}
&&P_{x}(1+p_dP_d)(1+p_yP_y)(1+p_xP_x)=\nonumber \\
&&~~~~~~~p_{x}(1+p_xp_yp_dP_d)(1+p_yP_y)(1+p_xP_x),
\end{eqnarray}
and an analogous result for $P_y$. Thus, if $p_x=p_y$, the state is indeed, for the special momenta $(0,0)$ and $(\pi,\pi)$, an eigenstate also 
of $P_x$, $P_y$, and $P_d$. 

If we want to study all momentum sectors, using the reflection symmetries does not buy us much, because in most of the blocks we have 
to use the generic option $Q=1$ in (\ref{qdef})(a). These symmetries are useful when investigating the ground state, which often is 
in the maximally blockable sector (g). The most useful aspect of the reflections may still be 
just the fact that they give insights into the symmetry-aspects of the states (e.g., to classify excitations). Apart from the lattice symmetries, 
spin-inversion symmetry can be used in the $m_z=0$ sector, exactly as in the 1D case.

We have considered combinations of reflection symmetries but could instead have used a combination of rotations and reflections. 
This would give us different quantum numbers, but the same size of the blocks. In group theory, there are {\it irreducible representations}, with 
standardized names, corresponding to various lattice symmetries. Instead of working out the applicable symmetries for a new case, one can look up 
the irreducible representations and their corresponding {\it character tables} and employ these to construct states with quantum numbers 
of a chosen irreducible representation. This is very useful when studying more complicated lattices. Ref.~\cite{didier} has a 
discussion of this more formal approach, with examples for chains and square lattices.

We have here used symmetries applicable only to $L_x\times L_y$ lattices with the edges along the $x$- and $y$ axis on the square lattice. With 
$L_x=L_y=L$, which is the most interesting if we are interested in approaching the infinite 2D lattices, we are then limited to $L=4$ and $L=6$ (since 
odd $L$ introduces undesirable frustration effects for an antiferromagnet). To increase the set of accessible lattice sizes $N$, one can also consider 
``tilted'' lattices, with edges that are not parallel to the square-lattice axes \cite{didier}.

\paragraph{Implementation}

To study a $6\times 6$ lattice we have to use long (8-byte) integers for the bit representation. To save time, it is ten better not to loop over all 
the $2^N$ possible state-integers and single out the ones corresponding to a given magnetization (as is done, e.g., in code $\{7\}$), but to use a scheme 
which from the outset only constructs states with a given $m_z$ (i.e., given number $n_\up$ of $\up$ spins). This can of course also be done for 1D 
systems. Such a more sophisticated generation of specific $m_z$ states starts with the integer $a=2^{n_\uparrow+1}-1$, in which the first $n_\up$ bits 
are $1$ and the rest are $0$, and then execute: 

{\code
\cia {\bf do} \br
\cib     {\bf call checkstate}$(a,{\rm pass})$ \hfill \{26\}\break
\cib     {\bf if} (pass) {\bf then} $M=M+1$;~{\rm store representative information} {\bf endif} \br
\cib     $c=0$ \br
\cib     {\bf do} $b=1,N$ \br
\cic         {\bf if} ($a[b-1]=1$) {\bf then} \br
\cid             {\bf if} ($a[b]=1$) {\bf then} \br
\cie                  $c=c+1$ \br
\cid             {\bf else} \br
\cie                  $a[0],\ldots,a[c-1]=1$;~ $a[c],\ldots,a[b-1]=0$;~ $a[b]=1$;~ {\bf exit} \br
\cid            {\bf endif} \br
\cic         {\bf endif} \br
\cib      {\bf enddo} \br
\cib      {\bf if} ($b=N$) {\bf exit} \br
\cia      {\bf enddo} 
\code}

\noindent
Here the inner loop is simply searching for the lowest bit position $b-1$, for which a set ($1$) bit of $a$ can be moved one step to the left (i.e., to a 
position $b$ where the bit currently is $0$). After such a position has been found, all the previously set bits below this position (the number of which is
kept track of with the counter by $c$) are moved to the lowest positions ($0,\ldots,c-1$). This bit evolution corresponds exactly to how the digits of a 
base-2 odometer advance from right to left. The contents of {\bf checkstate} depend on what symmetries are used, but it would be very similar to the 
implementations we discussed for the 1D case, apart from the fact that the translations [defined in (\ref{tdef2d})] are more complicated than just cyclic 
bit permutations and have to be implemented by hand. 

\begin{table}
\begin{tabular}{rrrrrr}
\hline
~~$p_x$ & ~~~~~~$p_y$ & ~~~~~~$p_d$ & ~~~~~~~~$z$ & ~~~~~~~~~~$M(L=4)$ & ~~~~~~~~~~~~~$M(L=6)$~ \\
\hline
$+1$ & $+1$ & $+1$ & $+1$ &  107 & 15,804,956~ \\
$+1$ & $+1$ & $+1$ & $-1$ &  46  & 15,761,166~ \\
$+1$ & $+1$ & $-1$ & $+1$ &  92  & 15,796,390~ \\
$+1$ & $+1$ & $-1$ & $-1$ &  38  & 15,752,772~ \\
\hline
$-1$ & $-1$ & $+1$ & $+1$ &   50 & 15,749,947~ \\
$-1$ & $-1$ & $+1$ & $-1$ &   45 & 15,739,069~ \\
$-1$ & $-1$ & $-1$ & $+1$ &   42 & 15,741,544~ \\
$-1$ & $-1$ & $-1$ & $-1$ &   36 & 15,730,582~ \\
\hline
$+1$ & $-1$ & ~ & $+1$ &      75 & 31,481,894~  \\
$+1$ & $-1$ & ~ & $-1$ &     108 & 31,525,574~ \\
\hline
\end{tabular}
\label{sizetab2d}
\caption{Sizes $M(L)$ of the $k=(0,0)$ state blocks for $L\times L$ lattices ($L=4,6$) with magnetization $m_z=0$ and different reflection 
and spin-inversion quantum numbers. For $p_x=-p_y$ there is no quantum number $p_d$. For $p_x=-1,p_y=1$ the block structure is the same
as for $p_x=1,p_y=-1$.}
\end{table}

The normalization constant, needed when constructing the hamiltonian matrix elements, should be delivered by {\bf checkstate} (if a 
representative has passed the tests). The simplest way to compute the normalization (instead of using formal expressions as we did in the 1D 
case) is again just to carry out all the symmetry operations of the representative state and add up the sum $F_a$ of factors (the complex momentum 
phases as well as the plus or minus signs from the reflection quantum numbers) in (\ref{akpgdef2}) for each symmetry operation bringing the representative 
onto itself. That number, along with the number $D_a$ of non-equivalent transformations of $|a\rangle$, gives the normalization $N_a=D_a|F_a|^2$
(and again $N_a=0$ if the representative is incompatible with the quantum numbers). We do not discuss further details of how to implement the 
basis generation and the construction of the hamiltonian matrix, as these tasks are straight-forward generalizations of the one-dimensional 
implementations discussed in Secs.~\ref{momentum} and \ref{reflection}. Spin-inversion symmetry in the $m_z=0$ sector can be implemented as 
discussed in Sec.~\ref{spininversion}.

\paragraph{Example of block sizes}

Table \ref{sizetab2d} lists block sizes for the square-lattice systems of interest in Lanczos calculations, for one of the special 
momenta, ${\bf k}=(0,0)$, where the largest number of symmetries can be exploited. The ground state of the Heisenberg model is in this block (in
the sub-block with all other quantum numbers equal to $1$). For $L=6$ even the smallest blocks have more than $15$ million states, and the largest 
blocks (for general ${\bf k}$) are about eight times larger. Blocks of this size can still be handled in Lanczos calculations (the smaller one 
rather easily on a standard workstation).

To test the basis construction, it is useful to check the sum of the sub-block sizes, which should equal the total number of states in the 
block. For $L=4$ there are 822 states with ${\bf k}=(0,0)$, which equals the sum of all the reflection-block sizes in Table~\ref{sizetab2d}.

\subsubsection{The N\'eel state and its quantum rotor excitations}
\label{lanc2d}

We now illustrate 2D Lanczos calculations with results for the Heisenberg model. Although the small lattices accessible with this method are not 
sufficient for quantitatively accurate extrapolations to the thermodynamic limit, the calculations do illustrate some important aspects of systems 
with N\'eel order (beyond what we discussed in the framework of spin-wave theory in Sec.~\ref{neel}). We introduce the {\it quantum rotor} mapping 
of the low-energy states of finite systems, and based on these discuss the magnetic susceptibility. We also calculate the sublattice magnetization.

\paragraph{Two-spin model of quantum-rotor states}

\begin{figure}
\includegraphics[width=5.75cm, clip]{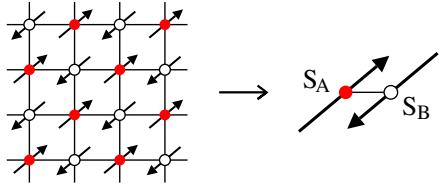}
\caption{Effective description of the rotationally invariant N\'eel vector ${\bf m}_s$ in terms of two large spins, ${\bf S}_A$, ${\bf S}_B$, corresponding to 
the sum of the spins on the two sublattices. There is an effective antiferromagnetic coupling between these spins, leading to a singlet ground state and a 
``tower'' of quantum rotor excitations of total spin $S=1,2,\ldots$ at energies $\Delta_S \sim S(S+1)/N$ above the ground state.}
\label{abspins}
\end{figure}

If there is antiferromagnetic order, the spins on sublattice A are predominantly oriented in the same direction, and the ones on sublattice 
B are predominantly in the direction opposite to those on A. If the number of spins $N$ is finite and the symmetry is not broken,
the over-all direction, defined, e.g., by the sublattice A spins, is not fixed, however. This situation can be captured by considering the sum
of the spins on the individual sublattices A and B \cite{anderson59},
\begin{equation}
{\bf S}_A = \sum_{i \in A} {\bf S}_i,~~~~~~~{\bf S}_B = \sum_{i \in B} {\bf S}_i,
\end{equation}
as two fixed-length spins $S_A=S_B= N/2$ (more precisely we would write, $S_A=S_B=m_sN$ for large $N$, but the constant is irrelevant), 
as illustrated in Fig.~\ref{abspins}. The two large spins are assumed to be antiferromagnetically coupled to each other in the 
simplest possible rotationally invariant way, which is through an effective Heisenberg interaction;
\begin{equation}
H_{AB} = J_{AB}{\bf S}_A \cdot {\bf S}_B = \half ({\bf S}^2 - {\bf S}^2_A - {\bf S}^2_B),
\label{habfull}
\end{equation}
where ${\bf S} = {\bf S}_A + {\bf S}_B$ is the total spin. Here ${\bf S}_A^2$ and ${\bf S}^2_B$ are just constants proportional to $N^2$, 
which can be neglected when we discuss excitation energies. However, these constants imply hat the coupling constant $J_{AB}$ should be 
$\propto 1/N$, in order for the total ground state energy to be $\propto N$. The ground state of (\ref{habfull}) has total spin $S=0$ and 
excitations with $S=1,2,\ldots$ at energies $J_{AB}S(S+1)/2$ above the ground state. 

We define a new $N$-independent coupling $J_{\rm eff}=NJ_{AB}/2$ and write the energies as
\begin{equation}
\Delta_S=J_{\rm eff}\frac{S(S+1)}{N}.
\label{estower1}
\end{equation}
These excitations are referred to as the tower of quantum rotor states. The states with $S\ll \sqrt{N}$ become degenerate as $N \to \infty$, and combinations 
of them can then be formed which are ground states with fixed direction of the N\'eel vector (in analogy with the infinite number of momentum states required
to localize a particle in quantum mechanics), thus allowing for the symmetry breaking that is the starting point for spin-wave theory. In the 
thermodynamic limit, the direction of the ordering vector is fixed (as the time scale associated with its rotations diverges \cite{anderson59}), and the 
quantum rotor-states are then in practice not accessed. They are neglected in standard spin-wave calculations (discussed in Sec.~\ref{neel}) from the outset 
because the order is by construction locked to the $z$ direction. One can still access the rotor energies in spin-wave theory, by considering systems in 
an external magnetic field, tuned to give a ground state with total magnetization $S^z=S$ \cite{oguchi,lavalle}. The rotor states are of great significance 
in finite clusters.

The effective coupling $J_{\rm eff}$ in (\ref{estower1}) for a given system can be determined if we can relate it to some physical quantity which 
depends on the rotor excitations. An obvious choice is the uniform magnetic susceptibility, $\chi=d\langle m_z\rangle/dh$. Calculating it for
the two-spin model when $T\to 0$ gives $\chi=3/J_{\rm eff}$. For the real Heisenberg model on a finite cluster in dimensions $d\ge 2$, $\chi$ 
should be dominated by the quantum rotor states when $T \ll 1/L$, because the lowest spin wave energy scales as $\propto 1/L$ (while the quantum 
rotor states scale as $1/L^d)$. Thus, we can write the effective quantum rotor tower for a Heisenberg model with N\'eel 
ground state as
\begin{equation}
\Delta_S=\frac{S(S+1)}{3\chi N},
\label{estower2}
\end{equation}
where $\chi$ should be evaluated in the limit $N\to \infty$ (first) and  $T\to 0$. Note that $I=(3/2)N\chi$ here plays the role of a moment of inertia, 
giving an analogy between (\ref{estower2}) and the energy spectrum of a rigid rotor in quantum mechanics. 

The relation (\ref{estower2}) can also be used as a way to compute the susceptibility of a Heisenberg models numerically; by extracting the lowest energies
as a function of $S$ (for small $S$, where the quantum-rotor mapping should apply). More precisely, the small-$S$ energies gives an estimate for $\chi$ 
as the $N\to \infty$, $S\to 0$ limit of the quantity $\chi(S,N)$:
\begin{equation}
\frac{1}{\chi(S,N)}=\frac{3N_S(E_S-E_0)}{S(S+1)}.
\label{estower3}
\end{equation}
Here $E_S$ denotes the lowest energy for total spin $S$. Note that we have to subtract the ground state energy ($S=0$) because in the two-spin effective 
model we only computed the excitation energies $\Delta_S$ with respect to the ground state energy (and the latter is not given accurately by the two-spin model). 
One would expect an $S$-independent behavior only for $S \ll L$, as the higher rotor states should be influenced by effects not taken into account 
in the two-spin model.

In analogy with the low-energy 1D quantum numbers discussed in Secs.~\ref{expvalues} and \ref{sec_hchain}, on the 2D square lattice (where $N=4n$ for all
even $L$) the quantum rotor states correspond to momentum $(\pi,\pi)$ and $(0,0)$ for odd and even $S$, respectively. For even $S$, the lowest states have 
reflection quantum numbers $p_x=p_y=p_d=1$ in (\ref{qdef}), while for odd $S$ the appropriate quantum numbers are $p_x=p_y=-1$, $p_d=1$. The lowest 
state for given $S$ can be obtained in the magnetization sector $m_z=S$ (and since $m_z\not=0$ we cannot use the spin-inversion symmetry here). 
The ($S=0$) ground state is in the fully symmetric sector; the momentum is $(0,0)$ and $p_x=p_y=p_d=z=1$.

\begin{figure}
\includegraphics[width=12cm, clip]{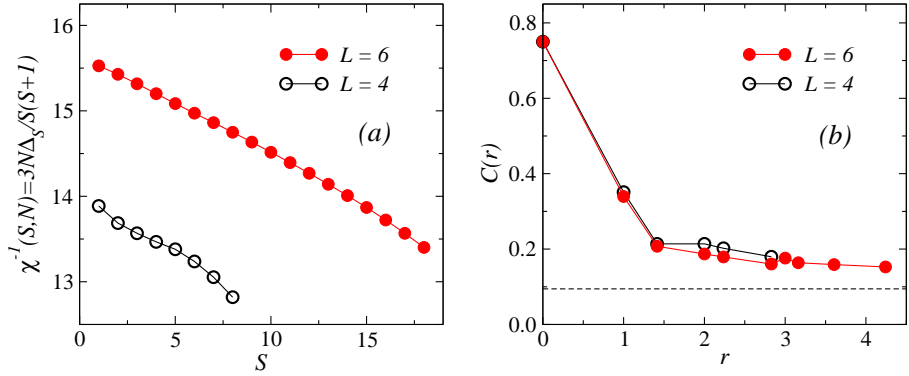}
\caption{Lanczos results for the Heisenberg model on $4\times 4$ and $6\times 6$ lattices. (a) The energy relative to the ground state  of the lowest 
state for each total spin sector, normalized by $NS(S+1)$, so that perfect quantum rotor excitations should produce $S$- and $N$-independent values.
(b) Spin correlation versus distance for all possible distances $r$ on the periodic lattices. The known $r\to \infty$ value (from the QMC results 
in Fig.~\ref{jjmag2}) is indicated by the dashed line.}
\label{tower2d}
\end{figure}

Lanczos results for $L=4$ and $L=6$ are shown in Fig.~\ref{tower2d}(a). There are clearly large corrections to (\ref{estower2}), as there is a 
significant decrease in $\chi^{-1}(S,N)$ with $S$ and an increase with $N$. For fixed $N$, the difference between $S=1$ and $S=N/2$ is roughly 10\%. 
Considering the fact that the limit $S \ll L$ cannot really be studied based on the small lattices, deviations of this order are not surprising. 
The rotor states have been studied using QMC calculations for much larger lattices \cite{lavalle,syljuasen02}. The most precise calculation for small 
$S$ and large $N$ gives $\chi^{-1}(S,N) \to 22.8$ \cite{syljuasen02}.

To understand the deviations from the rotor picture, one can use the analogy of a slightly non-rigid quantum rotor, which seems natural considering 
that the two-spin model is defined with fixed-length spins, while clearly in the real system the sublattice spins fluctuate (in a way which can 
depend on the total spin $S$). It may be possible to take these effects into account by adding higher-order terms $({\bf S}_A\cdot {\bf S}_B)^2$,
etc., in the interaction (\ref{habfull}). Details of this extended two-spin picture have not been worked out, however.

\paragraph{Transverse susceptibility}

For an infinite system at $T=0$ the spin-rotational symmetry is broken. One can then consider transverse and longitudinal components (with 
respect to the N\'eel vector), $\chi_\perp$ and $\chi_\parallel$, of the susceptibility. In the two-spin model it is clear that $\chi_\parallel=0$
and this is also true in the Heisenberg model. In a large system in which the symmetry is not broken, the components are rotationally averaged,
giving $\chi=(2/3)\chi_\perp$. The transverse susceptibility is relevant in practice in real N\'eel order quasi-2D systems (such as the 
high-$T_{\rm c}$ cuprates) at $T>0$, because of anisotropies and/or 3D couplings due to which the system can have a finite ordering temperature. 
The energy scales associated with the ordering are small, however, and the $T\to 0$ value of $\chi_\perp=(3/2)\chi$ of an isolated 2D
plane (modeled by the Heisenberg antiferromagnet) gives a good estimate of this quantity in the more complicated system.

\paragraph{Sublattice magnetization}

The first attempt to compute the sublattice magnetization of the 2D Heisenberg model was made using Lanczos calculations by Oitmaa and Betts
in 1978 \cite{oitmaa78}. At that time, the $6\times 6$ lattice was beyond reach, and extrapolations based on smaller lattices resulted in a value that 
is too large. Fig.~\ref{tower2d} shows the spin correlation function at all possible separations $r$ on $4\times 4$ and $6\times 6$ lattices. There 
is a slow decay with $r$, and it at least appears plausible that the results are approaching a value $\langle m_s^2\rangle >0$ as $r \to \infty$ 
(and $N \to \infty$). QMC calculations, results of which were already shown in Fig.~\ref{jjmag2}, give the $r\to \infty$ result indicated with
a dashed line in Fig.~\ref{tower2d}. The existence of long-range order in the 2D Heisenberg model was, in fact, under debate until the
first reliable QMC simulations were done in the late 1980s \cite{reger}.

One can attempt to extrapolate Lanczos results to $N\to \infty$ for, e.g., frustrated systems, were other calculations are very challenging 
\cite{dagotto3,schulz1}. The results must in general be viewed with caution, however, because it is assumed that the behavior for very small lattices are 
already exhibiting the ultimate asymptotic behavior. Fitting procedures based on  ``Betts clusters'' \cite{bettsclusters} of different shapes may then be 
misleading, even when the size dependence is smooth. There are examples where cross-overs occurs at larger distances \cite{mchains}, due to the presence
of some finite length-scale in the system, which has to be exceeded by $L$ before asymptotic behavior can be observed.

\section{Quantum Monte Carlo simulations and the Stochastic 
Series Expansion method}
\label{sec_sse}

Feynman's path integral formulation of quantum statistical mechanics \cite{feynman53} has played a major role in the development of QMC methods. In the case 
of spin systems and related lattice models, methods based on the path integral in imaginary time are commonly referred to as {\it world line methods} 
\cite{hirsch82}. These techniques were originally based on an approximate discretization of imaginary time---the Suzuki--Trotter decomposition of the 
Boltzmann operator ${\rm exp}({-\beta H})$ \cite{suzuki76,suzuki77}. Later, exact algorithms operating directly in the imaginary time continuum 
were developed \cite{beard,prokofev96}. The first practically useful QMC method did not, however, use the path integral. Already in the early 1960s, 
Handscomb developed an approximation-free method for the Heisenberg ferromagnet based on the power-series expansion of ${\rm exp}({-\beta H})$
and exactly computable traces of products of permutation operators (in terms of which the Heisenberg exchange can be written) \cite{handscomb62,handscomb64}. 
Although this scheme was also generalized to the Heisenberg antiferromagnet \cite{lyklema82,lee84,manousakis89} and some other systems \cite{chakravarty82}, 
world line methods were in general more efficient and dominated the field for a long time. The power-series approach to QMC calculations was revived with 
the introduction of a more generally applicable exact {\it stochastic series expansion} (SSE) formulation \cite{sandvik90,sandvik92}, in which 
also the traces are sampled (thus circumventing the previous reliance on permutation operator algebra). In addition to improved computational utility and efficiency, 
this generalization of Handscomb's approach also shows clearly how closely the discrete power series is related to the path integral in continuous imaginary 
time. Modern QMC algorithms based on the two formulations are also in the end rather similar \cite{syljuasen02,evertz1}. In particular, in both cases the 
spin configurations for some models (such as the Heisenberg model) can be sampled using highly efficient loop-cluster updates \cite{ying93,kawashima94,beard,sandvik99a}, 
which are generalizations \cite{evertz93} of  the classical Swendsen-Wang \cite{swendsenwang} cluster algorithm to quantum systems. Further generalizations of 
the loop concept to ``worms'' \cite{prokofev96} and ``directed loops'' \cite{sandvik99a,syljuasen02} have an even wider applicability (e.g., in the presence of 
external fields), and have enabled large-scale studies of a broad range of quantum spin and boson models. Some of these can now be studied at a level of detail 
approaching the state of the art for classical systems. 

In the case of frustrated spin systems, QMC methods are in general hampered by ``sign problems'', i.e., a non-positive definite path 
integral or series expansion \cite{henelius00}. While calculations can still be carried out in principle, and there has been some progress in controlling 
the sign fluctuations at high temperatures \cite{nyfeler08}, the statistical errors due to the mixed signs become uncontrollable at low temperatures. Apart 
from discussing the origins of the sign problem for frustrated models, we will here only consider unfrustrated (bipartite) systems. Despite constituting 
just a subset of the systems which we are in principle interested in, such models still exhibit a wealth of interesting physics and 
continue to provide important insights in cutting-edge research.

In these notes we will discuss implementations and applications only of the SSE approach, which for spin systems is normally more efficient and technically 
simpler than the continuous-time path integral. It is still useful to understand the relationships between the two schemes. In Sec.~\ref{sec_pathintegrals} 
we therefore first review path integrals in quantum statistical mechanics and world line QMC methods, in both discretized and continuous imaginary 
time. We introduce the general series expansion formulation of quantum statistical mechanics in Sec.~\ref{sec_sseformulation} and investigate its 
mathematical relationship to the path integral. An efficient implementation of the SSE method for the $S=1/2$ Heisenberg model is described in 
Sec.~\ref{sseheisenberg}. Illustrative calculations and results for several 1D and 2D systems are discussed in \ref{sec_sseapplications}; single
chains in \ref{sec_chainsse}, ladder systems in \ref{sec_ladders}, the uniform 2D lattice in \ref{sec_2dheis}, dimerized systems in \ref{sec_dimerized}
(focusing on the quantum phase transition between the N\'eel state and a non-magnetic state), and J-Q models in \ref{sec_jqresults} (with examples
of both continuous and first-order N\'eel--VBS transitions).

\subsection{Path integral and series expansion formulations \\ of quantum statistical mechanics}
\label{sec_pathintegrals}

The main technical problem in quantum statistical mechanics is how to deal with the Boltzmann operator ${\rm exp}({-\beta H})$. As we saw with exact
diagonalization methods in the previous section, a direct construction of the corresponding matrix becomes infeasible for systems with more than a 
few tens of spins. The path integral method
offers a way to transform the trace of this operator (the partition function) into a form that can be sampled using Monte Carlo methods. An alternative
is to start from a power-series expansion of the exponential. Here we introduce both these approaches as a foundation for implementing the SSE
method for Heisenberg models in Sec.~\ref{sseheisenberg}.

\subsubsection{The imaginary-time path integral}

The starting point of the path integral formulation is to write the exponential operator at inverse temperature $\beta$ as a product of $L$ operators 
with $\Delta_\tau=\beta/L$ in the exponent;
\begin{equation}
Z = {\rm Tr}\{ {\rm e}^{-\beta H} \} =  {\rm Tr}\left \{ \prod_{l=1}^L {\rm e}^{-\Delta_\tau H} \right \}.
\end{equation}
The trace can be expressed as a sum of diagonal matrix elements in any basis. We can also insert a complete set of states between each of the 
exponentials. The partition function then takes the form of an $L$-dimensional sum of products of matrix elements;
\begin{equation}
Z = \sum_{\alpha_0}\sum_{\alpha_1} \cdots \sum_{\alpha_L-1}
\langle \alpha_0|{\rm e}^{-\Delta_\tau H}|\alpha_{L-1}\rangle \cdots
\langle \alpha_2|{\rm e}^{-\Delta_\tau H}|\alpha_1\rangle 
\langle \alpha_{1}|{\rm e}^{-\Delta_\tau H}|\alpha_0\rangle.
\label{ztimeslices0}
\end{equation}
Formally, the exponential operator is equivalent to the Schr\"odinger time evolution operator ${\rm exp}{(-iHt)}$ (with $\hbar=1$) at imaginary (Euclidean)
time $t=-i\Delta_\tau$. One can therefore consider (\ref{ztimeslices0}) as a mapping of a $d$-dimensional quantum system to an equivalent system in $d+1$ 
dimensions, where the new dimension is imaginary time.\footnote{Often the equivalent ($d+1$)-dimensional system is referred to as an equivalent {\it classical} 
system. This terminology is, however, appropriate only in cases where all the path weights are positive, which we have not yet ascertained. It will be true for 
some classes of systems only, and those are the ones for which QMC calculations can be performed in practice. In some cases, e.g, the 
transverse-field Ising model, the effective model is an anisotropic version of the classical model in $d+1$ dimensions. In most cases, the equivalent 
classical system is not, however, the same kind of model as the original one with just one more dimension---typically the path integral corresponds to some 
completely different statistical mechanics problem in $d+1$ dimensions, with no apparent resemblance to the original $d$-dimensional system. In some cases, 
one can, however, show that the quantum system is equivalent, on large length and time scales, to the same classical system in $d+1$ dimensions, e.g., 
the 2D quantum Heisenberg antiferromagnet at low temperatures has the same properties as a 3D classical Heisenberg model \cite{chn}. This mapping is 
normally carried out using a basis of coherent spin states, as discussed, e.g., in the book by Auerbach \cite{auerbachbook}.}
The state index $l$ corresponds to a discrete set of imaginary time points $\tau_l=l\Delta_\tau$, with $0\le \tau_l \le \beta$ and periodic time-boundary 
conditions ($\alpha_L=\alpha_0$) for the states. Often the discrete times are referred to as ``time slices''. 

From a technical perspective, the purpose of writing $Z$ in the form (\ref{ztimeslices0}) is that, while the matrix elements of ${\rm exp}{(-\beta H)}$ are 
difficult to evaluate, we can use some approximation to evaluate the matrix elements of ${\rm exp}{(-\Delta_\tau H)}$ when $\Delta_\tau$ is small (the number of 
time slices is large) . We can then compute the weights for the different time-periodic ``paths'' $\alpha_0 \to \alpha_1 \cdots \to \alpha_{L-1} \to \alpha_0$ 
over which the system can evolve in the chosen basis. In QMC calculations, these paths are importance-sampled according to their weights 
in (\ref{ztimeslices0}).

To discuss the possible space-time paths and their weights, let us first consider a seemingly extreme method of approximating the exponential operator 
by just its Taylor expansions to linear order in $\Delta_\tau H$, i.e.,
\begin{equation}
Z \approx \sum_{\{ \alpha \}}
\langle \alpha_0|1-\Delta_\tau H|\alpha_{L-1}\rangle \cdots
\langle \alpha_2|1-\Delta_\tau H|\alpha_1\rangle 
\langle \alpha_{1}|1-\Delta_\tau H|\alpha_0\rangle,
\label{ztimeslices1}
\end{equation}
where $\{\alpha\}$ refers collectively to all the states $|\alpha_0\rangle,\ldots,|\alpha_{L-1}\rangle$. Since each factor now has an error of order $\Delta_\tau^2$ 
and there are $L=\beta/\Delta_\tau$ factors, the relative error in $Z$ at fixed $\beta$ is of the order $\Delta_\tau$. The leading neglected term in each expanded 
factor also contains $H^2$, and we might therefore suspect that the discretization error should also scale as $N^2$. However, to find exactly how the error scales 
with $\beta$ and $N$ requires a more careful analysis than just naively counting the neglected terms at the level of $\langle H^2\rangle \propto N^2$ 
and $L=\beta/\Delta_\tau$, because the paths contributing to the error are not the same as those contributing to the approximate $Z$ written as (\ref{ztimeslices1}). 
Based on arguments discussed later, to achieve an error which is independent of $\beta$ and $N$ we should expect to use at least of the order of $N\beta$ time 
slices in the linear approximation (\ref{ztimeslices1}), i.e., $\Delta_\tau \propto 1/N$. 

Far better approximations of the time slice operator can be used to reduce the number of slices (to independent of $N$ and proportional to $\beta$, with an 
error of $\Delta_\tau^2$ or smaller). However, for the purpose of illustrating many basic aspects of the path integral and QMC methods based on them, it is 
convenient to first use the linear approximation. In fact, as we will see shortly, it is even possible to take the limit $\Delta_\tau \to 0$ within this 
approximation and formulate QMC algorithms based on the exact continuous time path integral \cite{beard,prokofev96,syljuasen02}. We therefore discuss the 
linear version in some more detail, although in practice it is not recommended to implement an actual program using this scheme with $\Delta_\tau > 0$.  After 
understanding the properties of the paths in the linear approximation it will be easy to understand how to take the limit $\Delta_\tau \to 0$ or use a 
higher-order discrete approximation, such as the Suzuki-Trotter decomposition discussed in Sec.~\ref{sec_suzuki}.

\paragraph{Boson path integral and world lines}

At this stage it is better to continue the discussion with  a particular hamiltonian in mind, in order to have a concrete example of the paths and how 
one might go about sampling them.  Consider first a boson system with only kinetic energy. We here work on the lattice,\footnote{See \cite{ceperley95} 
for a review of boson path integrals in continuous space, and \cite{boninsegni06} for more recent progress on efficient QMC algorithms 
based on them.} and the purely kinetic-energy hamiltonian  can be written as (with an unimportant prefactor set to 1); 
\begin{equation}
H = K = -\sum_{\langle i,j\rangle} K_{ij} = -\sum_{\langle i,j\rangle} (a^+_ja_i + a^+_ia_j),
\label{hbosonkin}
\end{equation}
where $a^+_i$ and $a_i$ are, respectively, boson creation and destruction operators on the sites $i=1,\ldots,N$, and $\langle i,j\rangle$ is a pair of 
nearest-neighbor sites on an arbitrary lattice (in any number of dimensions, although for ease of visualization we will explicitly consider a 1D chain). 
For simplicity, we will consider hard-core bosons, for which the site occupation numbers are restricted to $n_i=0,1$ (which can be thought of as arising from a 
very strong on-site repulsion). Hard-core bosons also have a simple relationship to $S=1/2$ quantum spins. Occupied and empty sites correspond directly to $\up$ 
and $\dn$ spins, and with no interactions the boson hamiltonian is equivalent to the XY model, with $H_{ij}=-(S^+_iS^-_j+S^-_iS^+_j)$. There is no chemical 
potential in (\ref{hbosonkin})---we use the canonical ensemble with fixed number of bosons $N_B$ (density $\rho=N_B/N$ or, in the spin language, magnetization 
$m=\rho-1/2$). Later, we will see that it is easy to also incorporate interactions.

\begin{figure}
\includegraphics[width=11cm]{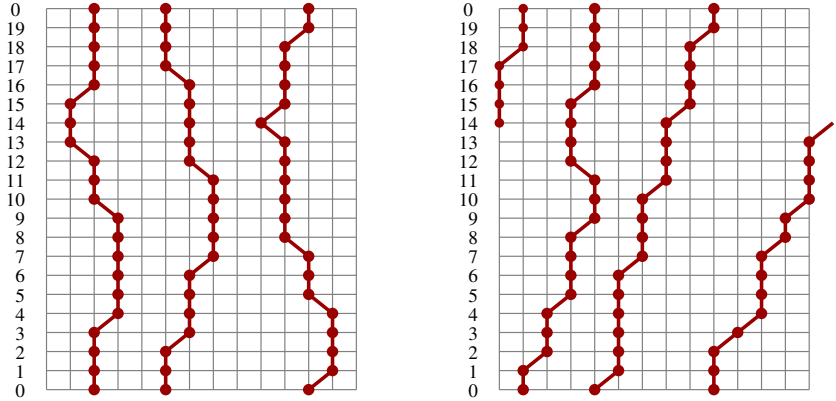}
\caption{Graphical representation of terms in the discrete path integral for a system of three bosons on a chain with 14 sites. There are 20 time 
slices, labeled by $l=0,\ldots,19$, which correspond to the states $|\alpha_l\rangle$ in Eq.~(\ref{ztimeslices1}). The bosons are represented by circles. They 
are connected to show more clearly the formation of world lines on the space-time lattice. Note the periodic boundary conditions in the time (vertical) direction, 
i.e., the first and last ($l=0$) states are the same. Note also that with the linear time slice operator $(1-\Delta H)$, there cannot be more than one particle 
hop (diagonal line segment) within a time slice. With periodic spatial boundary conditions, the identical particles can undergo cyclic permutations (winding), 
as in the configuration to the right.}
\label{pathintegral}
\end{figure}

We work in the basis of site occupation numbers $n_i$, $i=1,\ldots,N$. Consider a matrix element $\langle \alpha_{l+1}|1+\Delta_\tau K_{ij}|\alpha_{l}\rangle$, which 
appears in Eq.~(\ref{ztimeslices1}) when we also write each instance of $H$ as a sum over all the individual hopping terms $K_{ij}$ (and note the cancellation of the 
minus signs). There are then two possible relationships between the states $|\alpha_{l}\rangle$ and $|\alpha_{l+1}\rangle$ resulting in non-vanishing matrix elements; 
either $|\alpha_{l+1}\rangle=|\alpha_{l+1}\rangle$ or $|\alpha_{l+1}\rangle=K_{ij}|\alpha_{l}\rangle$. In the latter case, there must initially be a particle at 
site $i$, which is moved by $K_{ij}$ to a previously empty site $j$, or vice versa. Since these conditions must hold for all consecutive matrix elements, 
$l=0,\ldots,L-1,0$, each particle has to follow a ``world line'' (a term borrowed from the similar concept of a space-time path in relativity theory), which 
at each step of the imaginary time propagation either stays at the same spatial position or moves (jumps) by one lattice spacing. Two such world line 
configurations for a 1D system are illustrated in 
Fig.~\ref{pathintegral}. 

Note again the periodic boundary conditions in the imaginary time direction, which follow from the trace over $\alpha_0$ in (\ref{ztimeslices1}). These boundary 
conditions apply to the boson states (occupation numbers), but not strictly to the world lines. Since the bosons are identical particles, they can be permuted in 
the course of their evolution from $\tau=0$ to $\tau=\beta$ and still fulfill the required time-periodicity. The only possible permutation in a 1D hard-core 
system with nearest-neighbor hopping is a cyclic permutation involving all the particles on a periodic chain, resulting in a net particle current around the 
ring. An example of such a ``winding'' configuration is shown to the right in Fig.~\ref{pathintegral}. Other permutations can take place with soft-core bosons
(i.e., with no restriction on the occupation numbers $n_i$), and also hard core bosons in two or three dimensions (and even in one dimension if hopping beyond 
nearest neighbors is included), but only winding results in a net current. The net number of times world lines wrap around the system (the total current divided 
by the system length) is called the {\it winding number}. It is a topological quantity, characterizing a non-local property of the configuration. In higher 
dimensions, there can be winding in each of the spatial directions, with corresponding winding numbers. An interesting aspect of winding in bosons systems is 
that it (a properly normalized variance of the winding number) corresponds to superfluidity \cite{feynman53,ceperley86}. In spin systems, winding is related to 
the spin stiffness, or helicity modulus, as we will discuss further below.

We now denote by $W(\{\alpha\})$ the weight of a proper world line configuration $\{\alpha\}$ (i.e., one satisfying the constraints discussed above) 
in the partition function;
\begin{equation}
Z=\sum_{\{\alpha\}}W(\{\alpha\}).
\label{walphazsumboson}
\end{equation}
The weight has a very simple form in the case considered here: Each kinetic operation $K_{ij}$ is multiplied by the time step $\Delta_\tau$ in 
(\ref{ztimeslices1}), and the matrix elements of the boson operators is always $1$ in the hard-core case. The path weight is therefore
\begin{equation}
W(\{\alpha\})=\Delta_\tau^{n_K},
\label{walphabospaths}
\end{equation}
where $n_K$ is the total number of kinetic jumps, corresponding to diagonal world line segments in Fig.~\ref{pathintegral}. This weight can be simply generalized 
to the case of soft-core bosons, where the creation and annihilation operators are associated with factors $\sqrt{n_i}$ and $\sqrt{n_i-1}$. Below we will also 
consider interactions between the bosons, which lead to a more complicated factor multiplying the kinetic contribution (\ref{walphabospaths}).

\begin{figure}
\includegraphics[width=7.5cm]{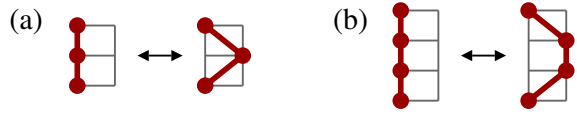}
\caption{Local updates of world lines, which change the number of kinetic jumps (diagonal world line segments) by $\pm 2$. In (a) the diagonal segments 
are directly following each other, whereas in (b) they are separated by one time slice. One can consider moves with arbitrary separation of the jumps.}
\label{wlmove}
\end{figure}

In a QMC simulation, one makes changes in the world lines in such a way that detailed balance is satisfied with the configurations distributed according to 
$W(\{\alpha\})$. Local updates of world lines are illustrated in Fig.~\ref{wlmove}. Note that while these local moves are ergodic within a sector of fixed 
winding number, they cannot change the topological winding number. In higher dimensions, local moves of individual world lines also cannot lead to any of the 
other permutations that need to be included. Local updates involving two world lines can be used to sample permutations, but are not enough to change the 
winding numbers associated with periodic boundaries. Algorithms based on simple local updates were historically successful in studies of some systems, but 
more sophisticated and powerful {\it loop updates} \cite{evertz1}, and related updating methods in an extended configuration space \cite{prokofev96,syljuasen02}, 
have been developed more recently. These are much more efficient in evolving the world lines in an ergodic way (instead of ``getting stuck'' or spending long 
times in some restricted part of the contributing configuration space) and can also lead to winding number changes. 

In this section we are primarily concerned with the path integral {\it representation}, and we we will not yet address the practical implementation of world line 
sampling (which we will do in detail only for the rather similar configurations arising in the series expansion formulation, although we will later in this 
section return to some more quantitative discussion of world line sampling as well). Next, we consider estimators for expectation values that we may want to 
compute. This will also give us some deeper insights into the properties of the world lines themselves. 

\paragraph{Expectation values and their world line estimators}

Consider an arbitrary operator $O$ and its thermal expectation value;
\begin{equation}
\langle O\rangle = \frac{1}{Z}{\rm Tr}\{O{\rm e}^{-\beta H}\}.
\end{equation}
Proceeding with the numerator as we did for $Z$ in 
Eq.~(\ref{ztimeslices0}), the exact time-sliced form of the expectation value can be written as
\begin{equation}
\langle O\rangle =
\frac{1}{Z} \sum_{\{\alpha\}}
\langle \alpha_0|{\rm e}^{-\Delta_\tau H}|\alpha_{L-1}\rangle \cdots
\langle \alpha_{2}|{\rm e}^{-\Delta_\tau H}|\alpha_1\rangle
\langle \alpha_{1}|{\rm e}^{-\Delta_\tau H}O|\alpha_0\rangle.
\label{otimeslices0}
\end{equation}
We would like to express this expectation value in the form appropriate for Monte Carlo importance sampling;
\begin{equation}
\langle O\rangle = \frac{\sum_{\{\alpha\}}O(\{\alpha\})W(\{\alpha\})} {\sum_{\{\alpha\}}W(\{\alpha\})}.
\label{pathintmcform}
\end{equation}
This is not always possible, however, because the paths contributing to the numerator in (\ref{otimeslices0}) may be different from those contributing 
to $Z$. If $W(\{\alpha\})=0$ for a configuration contributing to the expectation value, then no estimator $O(\{\alpha\})$ can be defined in (\ref{pathintmcform}). 
One then has to proceed with the calculation in a different way. We begin by discussing classes of expectation values for which (\ref{pathintmcform}) does apply.

\paragraph{Diagonal operators}

For quantities that are diagonal in the occupation numbers, e.g., a density correlator $\langle n_in_j\rangle$, the form (\ref{pathintmcform}) is trivially valid, 
because $O| \alpha_0\rangle = O(\alpha_0)|\alpha_0\rangle$ in (\ref{otimeslices0}), where $O(\alpha_0)$ denotes an eigenvalue of $O$. The estimator can then be taken 
as $O(\{\alpha\})=O(\alpha_0)$. This form of the estimator for a diagonal operator remains valid regardless of how the time-slice evolution operator 
${\rm e}^{-\Delta_\tau H}$ is approximated. Because of the cyclic property of the trace, by which the operator $O$ can be inserted anywhere in the product of time 
slice operators in (\ref{otimeslices0}), one can also average over all time slices and use
\begin{equation}
O(\{\alpha\})=\frac{1}{L}\sum_{l=0}^{L-1}O(\alpha_l),
\label{odiaaverage}
\end{equation}
which normally improves the statistics of a simulation. In practice, to save time without significant loss of statistics, one may only perform a partial summation 
in (\ref{odiaaverage}) over, e.g., every $N$th time slice (normally $L > N$), because eigenvalues $O(\alpha_l)$ and $O(\alpha_{l+1})$ of nearby states differ 
very little (at most corresponding to one single world line jump). States separated by $\propto N$ slices typically differ significantly and contribute independent 
statistics, at least to some degree (exactly how much can be statistically quantified in terms of an imaginary-time dependent correlation function 
$\langle O(\tau)O(0)\rangle$).

\paragraph{The kinetic energy}

Expectation values of off-diagonal operators are in general more complicated. The kinetic energy $\langle K\rangle$ is an exception. In addition to being
easy to evaluate, it is also a quantity of particular interest for understanding the properties of the path integral. At the level of the linear approximation 
of ${\rm e}^{-\Delta_\tau H}$ in $Z$, we can approximate $K{\rm e}^{-\Delta_\tau H}$ in the form (\ref{otimeslices0}) with $O=K$ by just $K$, because this is done 
only at a single time-slice and leads to a relative error of order $\Delta_\tau$. This is of the same order as the total error from the linear approximation made 
at all the other time slices in both (\ref{ztimeslices1}) and (\ref{otimeslices0}). Given a configuration $\{\alpha\}$ that contributes to $Z$, the estimator 
for a specific kinetic operator $K_{ij}$ is then
\begin{equation}
K_{ij}(\{\alpha\}) = \frac{\langle \alpha_1|K_{ij}|\alpha_0 \rangle}{\langle \alpha_1|1-\Delta_\tau K|\alpha_0 \rangle}.
\label{kijpathestim0}
\end{equation}
Here the numerator is non-zero only if there is a world line jump between sites $i$ and $j$ at the first time slice. In that case 
$K_{ij}(\{\alpha\})=1/\Delta_\tau=L/\beta$. In all other cases the estimator vanishes. We can again average over all time-slice locations of the operator
$K_{ij}$ in (\ref{otimeslices0}), which results in
\begin{equation}
\langle K_{ij}\rangle = \frac{\langle n_{ij}\rangle}{\beta},
\label{kijpathestim}
\end{equation}
where $n_{ij}$ denotes the number of kinetic jumps in the world line configuration between sites $i$ and $j$. Thus, the total kinetic energy is
given by the average of the total number of kinetic jumps $n_K$; $\langle K\rangle = -\langle n_K\rangle/\beta$. 

We could have derived the expression for the kinetic energy in a simpler way, by applying the thermodynamic formula for the internal energy 
(here just the kinetic energy); $E=\partial \ln(Z)/\partial \beta$, with $Z$ given by Eqs.~(\ref{walphazsumboson}) and (\ref{walphabospaths}). 
The more complicated derivation illustrates explicitly how the form (\ref{pathintmcform}) involves matching the configuration spaces of the numerator and 
denominator. This matching is not possible for generic off-diagonal operators; only for ones that are part of the hamiltonian. Before considering other cases, 
let us discuss another important aspect of the kinetic energy estimator.

The utility of the expression (\ref{kijpathestim}) is not just that it enables us to compute the kinetic energy. It also carries with it a fundamental message
about the path integral and the nature of the world lines. It is natural to ask how typical world lines will evolve as we increase the number of time slices. 
Specifically, how many kinetic jumps can we expect in a typical world line configuration? 

At low temperatures the kinetic energy should be almost temperature independent (approaching a constant as $\beta\to \infty$). It should be proportional to the 
lattice size $N$. Using Eq.~(\ref{kijpathestim}), we can therefore deduce the expected size and temperature scaling of the number of kinetic jumps as
\begin{equation}
\langle n_K\rangle = -\langle K\rangle \beta \sim N\beta.
\label{kijpathestim2}
\end{equation}
This tells us that there is a typical scale of roughness of the world lines; for large $L$, the typical time separation between jumps is $\sim 1/N$. 
As mentioned in the beginning of the section, we should expect to use of the order $N\beta$ time slices in a discrete path integral with the linear approximation 
of the time slice operator to avoid errors growing with $N$ and $\beta$. We now have the explanation for that, because it is clear from (\ref{kijpathestim2}) that 
a smaller number of time slices cannot accommodate the number of kinetic jumps necessary for the correct behavior of the kinetic energy (and then indirectly any 
other quantity).

The relationship between $\langle n_K\rangle$ and $\langle K\rangle$ is also important from another perspective; it indicates that it should be
feasible to formulate simulation algorithms directly in the continuum limit---even if we let the number of time slices $L \to \infty$, the number of kinetic 
jumps (``events'') in the contributing configurations stays finite. The continuous-time world lines can then be stored in the form of just a single state 
$|\alpha_0\rangle$ and the $n_K$ events; the times $\tau_i,i=1,\ldots,n_K$ at which they occur and the direction of each jump. With $\langle n_K\rangle$ scaling
as $N\beta$, it should be possible to construct Monte Carlo sampling algorithms with this scaling in system size and temperature of the time and memory requirements.

\begin{figure}
\includegraphics[width=11.75cm]{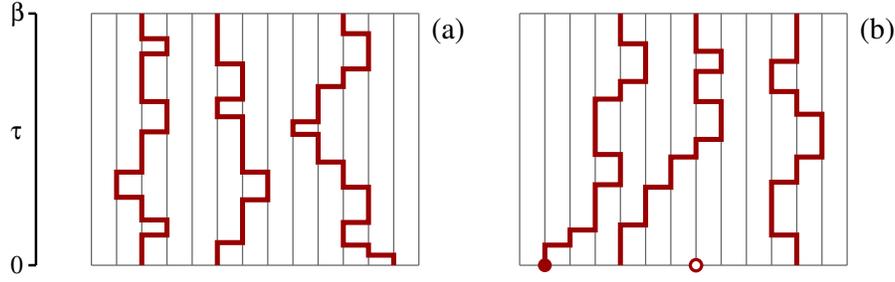}
\caption{Continuous-time world line configurations. Here the kinetic events (jumps) occur at arbitrary imaginary times $0 \le \tau < \beta$. Configuration (a) 
contributes to the partition function, whereas (b) includes a pair of creation (solid circle) and annihilation (open circle) operators separated by more
than one lattice spacing and does not contribute to $Z$ or diagonal expectation values. Such a configuration instead contributes to an off-diagonal 
expectation value $\langle a^+_ia_j\rangle$.}
\label{tpathintegral}
\end{figure}

Examples of world line configurations in continuous time is shown in Fig.~\ref{tpathintegral}. 
The configuration to the left contributes to the partition function, whereas the one to the right does not, because it does not satisfy all time-periodicity 
constraints. It instead contributes to the expectation value of an operator $a^+_ia_j$, the Fourier transform of which is the momentum 
distribution function $\rho(k)=\langle a^+_ka_k\rangle$. We briefly discuss such off-diagonal quantities next.

\paragraph{Off-diagonal operators and broken world lines}

If we consider a path $\{\alpha\}$ contributing to $Z$ and proceed to treat the expectation value of a general off-diagonal operator $a^+_ia_j$ in 
the same way as we did for the kinetic terms in (\ref{kijpathestim0}), it is clear that we always get zero, unless $i,j$ are nearest neighbors (in which case the 
operator is part of the kinetic energy). On the other hand, we can construct other paths, which do contribute to $\langle a^+_ia_j\rangle$ but not to $Z$, 
like the one in Fig.~\ref{tpathintegral}(b). If the sites $i,j$ are separated by more than one lattice spacing, the expectation value $\langle a^+_ia_j\rangle$ 
cannot be written in the standard Monte Carlo sampling form (\ref{pathintmcform}), because $W(\{\alpha\})=0$ for all $O(\{\alpha\})\not=0$. It is still possible to 
evaluate $\langle a^+_ka_k\rangle$, as well as the corresponding imaginary-time dependent function $\langle a^+_k(\tau)a_k(0)\rangle$ (the single-particle 
Green's function) by working in the combined space of the periodic world line configurations contributing to $Z$ and those with two defects corresponding to the 
presence of a creation and annihilation operator \cite{prokofev96}. 

\paragraph{The path integral including interactions}

It is not difficult to generalize the bosonic path integral discussed above to hamiltonians including potential-energy terms in addition to
the kinetic energy $K$ in (\ref{hbosonkin}). Let us denote by $V$ any interactions that are diagonal in the occupation number basis, e.g.,
attractive ($v_{ij}<0$) or repulsive ($v_{ij}>0$) terms of the form $v_{ij}n_in_j$. With $H=K+V$, we can decompose ${\rm exp}{(-\Delta_\tau H)}$ as
\begin{equation}
{\rm e}^{-\Delta_\tau H} = {\rm e}^{-\Delta_\tau K}{\rm e}^{-\Delta_\tau V} + {\cal O}(\Delta_\tau^2),
\label{kvtrotter}
\end{equation}
where the error is due to the commutator $[K,V]\not=0$. With $V$ diagonal, we can for each time slice in (\ref{ztimeslices0}) write
\begin{equation}
\langle \alpha_{l+1}|{\rm e}^{-\Delta_\tau H}|\alpha_l\rangle \approx {\rm e}^{-\Delta_\tau V_l}
\langle \alpha_{l+1}|{\rm e}^{-\Delta_\tau K}|\alpha_l\rangle,
\end{equation}
where $V_l$ denotes the potential energy evaluated at the $l$th time slice. The weight (\ref{walphabospaths}) of a world line configuration is 
therefore modified in the presence of interactions as
\begin{equation}
W(\{\alpha\})=\Delta_\tau^{n_K} {\rm exp}\left ( -\Delta_\tau \sum_{l=0}^{L-1}V_l \right ).
\label{walphabospaths2}
\end{equation}
The interaction part of the weight can be easily taken into account in elementary world line moves of the kind shown in Fig.~\ref{wlmove}, and also in
more sophisticated treatments in continuous imaginary time \cite{prokofev96}.

\paragraph{The continuum limit}

We have already discussed the fact that the continuum limit of the path integral can be used directly in QMC algorithms. However, examining the 
configuration weight, (\ref{walphabospaths}) for the purely kinetic hamiltonian or (\ref{walphabospaths2}) in the presence of interactions, we have an apparent 
problem; the weight vanishes as $\Delta_\tau \to 0$. This must clearly be compensated in some way by the number of configurations increasing, in such a way that the 
partition function remains finite. In Monte Carlo calculations we do not deal with the partition function directly and only need ratios of configuration weights to 
compute acceptance probabilities for world-line updates. Looking at such ratios for the simplest kinds of updates, the insertions and removals of opposite kinetic 
jumps illustrated in Fig.~\ref{wlmove}, they are also problematic in the continuum limit: Formally the probability for insertions and removals is zero and infinity, 
respectively. This is not just a problem in the continuum, but also for small $\Delta_\tau$, where the probability of accepting an insertion would be non-zero 
but very small. 

Proper ways to handle the continuum limit were introduced in the context of ``worm'' \cite{prokofev96}  and loop algorithms \cite{beard}, twenty years after 
simulations based on discrete path integrals were first introduced \cite{suzuki76,suzuki77}. Here we only consider local world line moves for a toy model, as an 
illustration of how Monte Carlo sampling can in fact rather easily be formulated in the continuum.

Fig.~\ref{wlmove} shows two different ways to introduce two events moving a world line to and from a neighboring site. Consider a line segment which is initially 
straight over $m$ time slices, between times $\tau_1$ and $\tau_2$, with $\tau_2=\tau_1+m\Delta_\tau$. Assuming that there is no world line on the neighboring site
within the range $[\tau_1,\tau_2]$, the weight change when introducing the two events is $\Delta_\tau^2$ (for a purely kinetic hamiltonian), regardless of where 
in this range the events occur. There are a total of $m(m-1)/2$ different ways of inserting events, which gives a total relative weight $\Delta_\tau^2m(m-1)/2$ for 
the subspace of such modified configurations, versus $1$ for the original configuration with two less events. In the continuum limit, the weight for all the 
possible updated configurations of this kind becomes $(\tau_2-\tau_1)^2/2$. As we have discussed above, when $\Delta_\tau \to 0$, the average total number of kinetic 
events remains finite. The typical length $\tau_2-\tau_1$ of a straight world line segments then also remains finite, and we should be able to construct a scheme 
with finite insertion probabilities. The key is that we do not just consider a single specific updated configuration, but a continuous range of possible 
updates.

\begin{figure}
\includegraphics[width=10.5cm]{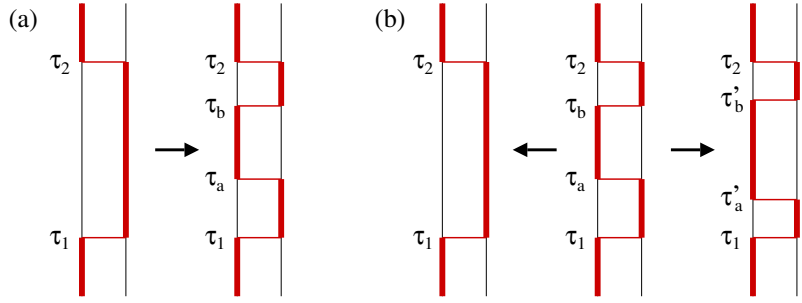}
\caption{Local world line moves in continuous time. In (a), two opposite kinetic jumps are inserted at randomly chosen times $\tau_a$ and $\tau_b$ between
two existing jumps at $\tau_1$ and $\tau_2$. In (b), two opposite jumps are either removed ($\leftarrow$) or the times of the two events are randomly changed 
to new arbitrary times $\tau_a'$ and $\tau_b'$ between $\tau_1$ and $\tau_2$ ($\rightarrow$). The acceptance probabilities are given in Eq.~(\ref{p2continuum}).}
\label{ctime}
\end{figure}

We also need to consider the removal of two events. In order to simplify the discussion, we here consider just a single boson on two sites. This avoids the 
complications of having to consider also constraints imposed by world lines on neighboring sites. The world line moves for which we will construct probabilities 
satisfying detailed balance are illustrated in Fig.~\ref{ctime}. The event insertion discussed above is illustrated in (a), where it should be noted that 
the times $\tau_1$ and $\tau_2$ correspond to two consecutive existing events (and in case there are no events, we take $\tau_1=0$ and $\tau_2=\beta$). We call 
the times of the two new events $\tau_a$ and $\tau_b$, and these are chosen at random anywhere between $\tau_1$ and $\tau_2$. In the opposite update of removing 
two events, we consider the total weight of two existing events $\tau_a$ and $\tau_b$ within two surrounding events at $\tau_1$ and $\tau_2$. Then we can again 
compute the total weight of the existing events, not only at the fixed current times $\tau_a$ and $\tau_b$ but at any times $\tau_a'$ and $\tau_b'$ within 
$[\tau_1,\tau_2]$. That relative weight is again $(\tau_2-\tau_1)^2/2$, versus $1$ for the single configuration with those two events absent. 

Thus, once $\tau_1$ and $\tau_2$ have been identified by inspecting the current configuration (which can be stored in the form of a list of events),
the acceptance probabilities for two-event insertion and removal are
\begin{equation}
P_{\rm insert} = \frac{(\tau_2-\tau_1)^2/2}{1+(\tau_2-\tau_1)^2/2},~~~~~~~
P_{\rm remove} = \frac{1}{1+(\tau_2-\tau_1)^2/2}.\label{p2continuum}
\end{equation}
For an accepted insertion, we generate $\tau_a$ and $\tau_b$ at random, while in a rejected removal we also generate new times $\tau'_a$ and $\tau'_b$
for the existing events.

In a larger system, with more than one boson, the above scheme would have to be modified to take into account the constraints of other world lines when a chosen 
world line is moved. The times $\tau_1$ and $\tau_2$ should then reflect those constraints. In the presence of interactions, the acceptance probabilities would 
be modified to take into account the potential-energy factor in (\ref{walphabospaths2}). In practice, one would not use these local updates, however, as loop 
and worm updates \cite{prokofev96,beard,syljuasen02} are not much more complicated to implement, but much more efficient. The purpose of the discussion here 
was to show how the apparent problems of the continuum limit can be overcome in principle.

\subsubsection{The Suzuki-Trotter decomposition}
\label{sec_suzuki}

While modern world line QMC algorithms for lattice models are normally based on the path integral in the continuum limit, historically
time-discretized variants based on a Suzuki-Trotter approximation of the time-slice operator ${\rm exp}{(-\Delta_\tau H)}$ were developed first. The discrete 
approach is still some times used and is the most practical option in some cases \cite{werner05}. We here discuss the main features of Suzuki-Trotter 
based methods and how they are applied to boson and spin systems.

One example of a Suzuki-Trotter (or split operator) approximation \cite{suzuki76,trotter59} was already written down in Eq.~(\ref{kvtrotter}). This is just one 
among many possible decompositions of an exponential of two (or more) non-commuting operators into a product of exponentials \cite{deraedt83}. The most commonly 
used approximants are, for any pair of operators $A$ and $B$ and a small factor $\Delta$;
\begin{equation}
{\rm e}^{\Delta (A+B)} = \left\lbrace
\begin{array}{l}
{\rm e}^{\Delta A}{\rm e}^{\Delta B} + {\cal O}(\Delta^2) \\
{\rm e}^{\Delta B/2}{\rm e}^{\Delta A}{\rm e}^{\Delta B/2} + {\cal O}(\Delta^3).
\end{array}\right.
\label{trotter23}
\end{equation}
Here the errors are also proportional to the commutator $[A,B]$. Using a larger number of judiciously chosen exponentials of functions of $A$ and $B$, the remaining 
error can be further reduced, in principle to an arbitrary high power of $\Delta$ \cite{hatano05}. High-order approximants are often too complicated to work 
with in practice, however. 

If we are interested in the trace of a product of exponentials, as in the path integral, it is easy to see that the two low-order approximants (\ref{trotter23})
are actually equivalent, because
\begin{equation}
{\rm Tr} \left \{ \prod_{l=1}^L {\rm e}^{\Delta B/2}{\rm e}^{\Delta A}{\rm e}^{\Delta B/2} \right \} =
{\rm Tr} \left \{ \prod_{l=1}^L {\rm e}^{\Delta A}{\rm e}^{\Delta B} \right \},
\label{traceprod23}
\end{equation}
due to the cyclic property of the trace. Thus, although the world line method is often discussed based on the first line of Eq.~(\ref{trotter23}), where the
error from each factor is $\propto \Delta_\tau^2$, the error is in effect smaller; $\propto \Delta_\tau^3$. However, since the number of exponential factors in the 
path integral is $L=\beta/\Delta_\tau$, the total error made in a product such as (\ref{traceprod23}) for fixed $N$ and $\beta$ is $\propto \Delta_\tau^2$. In
addition, one can show that for sufficiently large $L$, the error also does not grow with $N$ at fixed $\Delta_\tau$. In most cases, one can therefore keep
$\Delta_\tau$ independent of $N$ and $\beta$ (although for some classes of expectation values, care has to be taken to avoid error divergences when $\beta \to \infty$ 
\cite{fye86,fye87}).

In the context of path integrals, the utility of the Suzuki-Trotter approximation is that even though ${\rm exp}{(-\Delta_\tau H)}$ in the partition function 
(\ref{ztimeslices0}) cannot be easily evaluated, if we find a suitable decomposition of $H$ into two terms (or some small number of terms), $H=H_A+H_B$, 
we may be able to evaluate ${\rm exp}{(-\Delta_\tau H_A)}$ and ${\rm exp}{(-\Delta_\tau H_B)}$ exactly. The prototypical example is a one-dimensional system containing 
only nearest-neighbor kinetic energy and interactions, which we collectively denote $H_{i,i+1}$ for a site pair $i,i+1$. We then decompose $H=H_A+H_B$ according to
\begin{equation}
H_A=\sum_{{\rm odd~} i}H_{i,i+1},~~~~~~H_B=\sum_{{\rm even~} i}H_{i,i+1},
\label{hahbdecomp}
\end{equation}
With this arrangement, all terms in each of the sums $H_A$ and $H_B$ are mutually commuting; $[H_{i,i+1},H_{i+2m,i+2m+1}]=0$ for any $i$ and $m$. We can 
therefore split ${\rm exp}{(-\Delta_\tau H_A)}$ and ${\rm exp}{(-\Delta_\tau H_B)}$ into two-body operators without making any further errors, e.g.,
\begin{equation}
{\rm e}^{-\Delta_\tau H_A} = {\rm e}^{-\Delta_\tau H_{1,2}} {\rm e}^{-\Delta_\tau H_{3,4}} \cdots {\rm e}^{-\Delta_\tau H_{N-1,N}}.
\end{equation}
Using products like this in  the time-sliced partition function (\ref{ztimeslices0}), we can insert complete sets of states between all the exponentials.
Since each operator now involves only two sites, the matrix elements reduce to the form
\begin{equation}
\langle \beta|{\rm e}^{-\Delta_\tau H_{i,j}}|\alpha\rangle = \langle n_{\beta i},n_{\beta j}|{\rm e}^{-\Delta_\tau H_{ij}}|n_{\alpha i},n_{\alpha j}\rangle,
\label{sliceelements}
\end{equation}
which for hard-core bosons are the elements of a $4\times 4$ matrix. Since the bond operator $H_{i,j}$ conserves the number of bosons on the two sites involved, 
this matrix is bock-diagonal, with two single-element blocks (for $n_{\alpha i}=n_{\alpha j}=n_{\beta i}=n_{\beta j}$) and a $2\times 2$ matrix with both diagonal
and off-diagonal elements connecting the states with a single boson on the bond. It is thus easy to evaluate all the matrix elements, also if interactions are 
included (as long as they do not extend beyond nearest neighbors; otherwise a more complicated decomposition of $H$ has to be used).

\begin{figure}
\includegraphics[width=9.3cm]{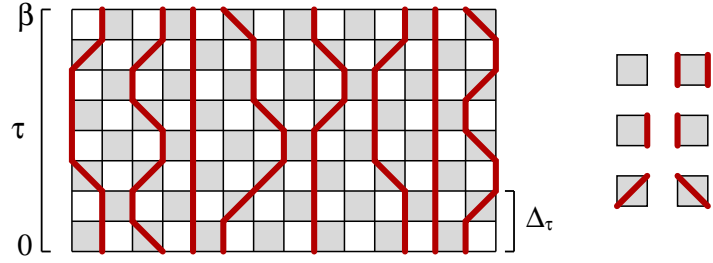}
\caption{A 1D world line configuration based on the checkerboard decomposition with the Suzuki-Trotter approximation. Kinetic jumps of the bosons (or flips
of a pair of $\up$ and $\dn$ spins) are allowed only across the shaded squares (plaquettes). A time slice of width $\Delta_\tau$ consists of two consecutive 
rows of plaquettes. The six isolated plaquettes shown to the right correspond to the non-zero matrix elements, which in the case of a spin model with Heisenberg 
interactions (for world lines and empty sites corresponding to $\up$ and $\dn$ spins, respectively) are given by Eq.~(\ref{trottermatelem}).}
\label{checkerboard}
\end{figure}

Pictorially, each matrix element (\ref{sliceelements}) corresponds to a four-site plaquette with zero, one, or two world line segments going through it, according 
to the occupation numbers. Considering all time slices and all site pairs $i,i+1$ in both ${\rm exp}{(-\Delta_\tau H_A)}$ and ${\rm exp}{(-\Delta_\tau H_B)}$, these 
plaquettes form a checkerboard pattern, with two adjacent rows corresponding to one time slice, as illustrated in Fig.~\ref{checkerboard}. For obvious reasons, the 
hamiltonian decomposition (\ref{hahbdecomp}) is also often called the checkerboard decomposition. 

For a hamiltonian conserving the total number of particles, the constraints on allowed world line configurations are similar to those in the linear approximation 
of the time-slice operator, with a few important modifications. Kinetic jumps are allowed only on the shaded plaquettes in Fig.~\ref{checkerboard}, but there is 
no further constraint on the number of jumps within a time slice, unlike in the linear time-slicing approximation illustrated in 
Fig.~\ref{pathintegral} (where there can be at most one jump in each time slice). The world line configurations still look very similar. They become equivalent 
when $\Delta_\tau \to 0$, in the physically relevant sense of their evolution from some time $\tau$ to some later $\tau'$, with $\tau'-\tau \gg  \Delta_\tau$.

In a Monte Carlo simulation, the world lines can be updated using simple moves of the kind illustrated in Fig.~\ref{wlmove}, with the constraint that the
diagonal loop segments (kinetic jumps) are allowed only on the shaded plaquettes in Fig.~\ref{checkerboard}. More complicated ``loop'' and ``directed loop''
updates, in which large segments of several world lines can be moved simultaneously, are used in modern algorithms \cite{evertz1,evertz93,syljuasen02} 
(which we will discuss in detail below in the context of the stochastic series expansion method).
\null\vskip4mm

\paragraph{Application to the Heisenberg model}

It is useful to consider a particular example of the path weights in the Suzuki-Trotter approach. Let us compute the plaquette matrix elements for the 
antiferromagnetic Heisenberg interaction; $H_{i,i+1}={\bf S}_i \cdot {\bf S}_{i+1}$. In this case the boson occupation numbers in (\ref{sliceelements}) are replaced 
by spin states $\up$ and $\dn$. We can consider the world lines forming between the $\up$ spins (and note that we could also draw world lines for the $\dn$ spins 
in pictures such as Fig.~\ref{checkerboard}; they occupy all sites not covered by $\up$ world lines and cross those lines at each diagonal segment). The calculation 
just involves straight-forward algebra and we just list the results for the six allowed (non-zero) matrix elements;
\begin{eqnarray}
&&\langle \up_i\up_j|{\rm e}^{-\Delta_\tau H_{ij}}|\up_i\up_j\rangle =
  \langle \dn_i\dn_j|{\rm e}^{-\Delta_\tau H_{ij}}|\dn_i\dn_j\rangle = +{\rm e}^{-\Delta_\tau/4} \nonumber \\
&&\langle \up_i\dn_j|{\rm e}^{-\Delta_\tau H_{ij}}|\up_i\dn_j\rangle =
  \langle \dn_i\up_j|{\rm e}^{-\Delta_\tau H_{ij}}|\dn_i\up_j\rangle = +{\rm e}^{\Delta_\tau/4}\cosh(\Delta_\tau/2)\label{trottermatelem}  \\
&&\langle \dn_i\up_j|{\rm e}^{-\Delta_\tau H_{ij}}|\up_i\dn_j\rangle =
  \langle \up_i\dn_j|{\rm e}^{-\Delta_\tau H_{ij}}|\dn_i\up_j\rangle = -{\rm e}^{\Delta_\tau/4}\sinh(\Delta_\tau/2)~~~~~~~~~~\nonumber
\end{eqnarray}
The weight of a world line configuration is a product of these matrix elements, all of which are pictorially represented in the right part of Fig.~\ref{checkerboard}.
Note the minus sign in front of the off-diagonal matrix elements. For an allowed world line configuration, all the signs cancel out due to the periodicity constraint 
on the world lines. This is true also for world line methods applied to bipartite lattices in higher dimensions, but for frustrated systems there is a ``sign problem''
because of the presence of both negative and positive weights (as we will discuss further in Sec.~\ref{sec_sseformulation}). In practice, world line methods and 
similar QMC approaches are therefore useful primarily for studies of bipartite spin systems and bosons models. For a fermion system, 
permutation of world lines also lead to sign problems, except in one dimension where only global cyclical permutations (winding) are possible 
(with associated signs that can be avoided by choosing periodic or anti-periodic boundary conditions \cite{hirsch82,sandvik92}).

In higher dimensions one can use the Suzuki-Trotter approach with various decompositions analogous to the one discussed above. Local updates such as those
we have discussed for 1D systems can be adapted to higher dimensions as well. It is worth reading Refs.~\cite{reger88,makivic91}, which are two 
pioneering world line studies that firmly established the basic properties of the 2D $S=1/2$ Heisenberg antiferromagnet.

Although the number of time slices used in the Suzuki-Trotter approach does not have to depend on the system size (because $\propto N$ events can take place within
a single time slice), the computational effort of Monte Carlo sampling the world-line configurations scales in the same way as in the continuous time formulation, 
as $N\beta$. This scaling is achieved in the exact continuum formulation by only storing and manipulating the times of the events, whereas in the Suzuki-Trotter 
approach one normally works with the full space-time lattice (although one could also in principle only store the events there). For this reason, the continuous-time 
methods actually normally run faster on the computer, in addition to not being affected by discretization errors. The only reason to use a discrete path integral 
would be in problems where it is difficult to take the continuum limit in practice, e.g., in effective models including dissipation, which leads to time-dependent 
interactions among the world lines \cite{werner05}.

\subsubsection{The series expansion representation}
\label{sec_sseformulation}

As an alternative to the discrete time-slicing approach or continuum limit of the path integral, one can also construct a configuration space suitable for 
Monte Carlo sampling by using the Taylor expansion of the Boltzmann operator;
\begin{equation}
{\rm e}^{-\beta H} = \sum_{n=0}^\infty \frac{(-\beta)^n}{n!}H^n.
\end{equation}
This approach was pioneered by Handscomb \cite{handscomb62,handscomb64}, who developed a method for studying the $S=1/2$ ferromagnet. The Taylor expansion was 
later considered more generally as a starting point for exact QMC algorithms for a wide range of models \cite{lyklema82,chakravarty82,sandvik90,sandvik92}. 
The power-series expansion of the exponential operator is convergent for a finite lattice at finite $\beta$ (which technically is due to the act that the 
spectrum of $H$ is bounded). In effect, as we will see below, the series expansion allows for a discrete representation of the imaginary time continuum, thus avoiding 
approximations but fully retaining the advantages of a finite enumerable configuration space.

Choosing a basis, the partition function can be written as
\begin{equation}
Z = \sum_{n=0}^\infty \frac{(-\beta)^n}{n!}
\sum_{\{\alpha\}_n} \langle \alpha_0|H|\alpha_{n-1}\rangle \cdots \langle \alpha_2 |H|\alpha_1\rangle\langle \alpha_1 |H|\alpha_0\rangle, 
\label{zssegeneral}
\end{equation}
where the subscript on $\{\alpha\}_n$ indicates that there are $n$ states to sum over. This expression can be compared with the linear-order discrete path 
integral (\ref{ztimeslices1}). Taking $H$ to be the bosonic kinetic energy in (\ref{hbosonkin}), we can clearly draw world line pictures very similar to those 
in Fig.~\ref{pathintegral} to represent the contributing terms. The main difference is that the number of ``slices'', the expansion power $n$, is varying and 
for given $n$ there are particle jumps at each slice. The new ``propagation'' dimension we have introduced in this representation is different from imaginary time, 
but it is clear that the label $p=0,\ldots,n-1$ of the states is closely related to imaginary time in the path integral. We will discuss the exact relationship 
further below.

In the case of only kinetic energy, all matrix elements in (\ref{zssegeneral}) equal $-1$ for an allowed configuration, the minus signs cancel and the weight is
\begin{equation}
W(\{\alpha\}_n)=\frac{\beta^n}{n!}.
\end{equation}
In more general cases, this weight will be modified by the product of matrix elements in (\ref{zssegeneral}). As in the path integral, the weights are 
positive definite (or can be made so with some simple tricks) for boson systems and quantum spins without frustration in the off-diagonal terms. We will here 
proceed under the assumption of positive definiteness and discuss the precise conditions for this further below.

It is useful to derive a general expression for the total internal energy $U=\langle H\rangle$. For any hamiltonian, we can write it as
\begin{equation}
E = \frac{1}{Z} \sum_{n=0}^\infty \frac{(-\beta)^n}{n!} \hskip-1mm \sum_{\{\alpha\}_{n+1}} \hskip-1mm
\langle \alpha_0|H|\alpha_{n}\rangle \cdots \langle \alpha_2 |H|\alpha_1\rangle\langle \alpha_1 |H|\alpha_0\rangle, 
\label{essegeneral}
\end{equation}
where the presence of an additional matrix element of $H$ in each term should be noted; for given $n$, the summation is over $n+1$ states instead of the $n$ states in 
the partition function (\ref{zssegeneral}). All configurations that contribute in (\ref{essegeneral}) contribute also to the partition function, 
but the weights differ. We can match the configurations explicitly by writing
\begin{equation}
E = -\frac{1}{Z} \sum_{n=1}^\infty \frac{(-\beta)^{n}}{n!}\frac{n}{\beta}
\sum_{\{\alpha\}_{n}} \langle \alpha_0|H|\alpha_{n}\rangle \cdots \langle \alpha_2 |H|\alpha_1\rangle\langle \alpha_1 |H|\alpha_0\rangle.
\label{essegeneral2}
\end{equation}
We can extend the sum over $n$ to include also $n=0$, because this term vanishes. The terms in (\ref{zssegeneral}) and (\ref{essegeneral2}) then match
exactly, and $n/\beta$ can be identified as the energy estimator. Thus, with the configurations sampled according to their weights in the partition 
function, the energy is simply given by
\begin{equation}
E = -\frac{\langle n\rangle}{\beta}.
\label{essenbeta}
\end{equation}
Writing $H$ as a sum, $H=-\sum_i H_i$, and using this for all instances of $H$ in (\ref{zssegeneral}) and (\ref{essegeneral}), we can derive a similar 
expression for the expectation value of an individual term $H_i$. The result for this estimator is the average number of times the operator appears in 
the operator string in the expansion of the partition function;
\begin{equation}
\langle H_i\rangle = \frac{\langle n_i\rangle}{\beta}.
\label{hiexpsse}
\end{equation}
In the case of a kinetic-energy term, this is identical to the path-integral expression (\ref{kijpathestim}). This shows that there is a one-to-one
correspondence between the paths in the two formulations, if we by path mean just the order in which kinetic events occur, without reference to the
times $\tau_l$ of the events in the discrete or continuous path integral. This correspondence holds quite generally. The difference is that in the
series expansion, the potential-energy terms are treated in the same way as the kinetic operators (i.e., they are not re-exponentiated, as they are in the 
path integral). Their presence corresponds to ``non-events'', which do not affect the world lines but are associated with diagonal matrix elements that 
modify the weight of the paths. In the path integral, there is instead the potential-energy factor in Eq.~(\ref{walphabospaths2}), which depends on the 
actual time-points of the kinetic events. 

Eq.~(\ref{essenbeta}) shows that although the sum over expansion orders $n$ in (\ref{zssegeneral}) extends up to $n=\infty$, it can in practice be truncated 
at some order $n_{\rm max} \propto N\beta$. To further motivate why this will only lead to an exponentially small and completely negligible error, one can make 
use of the estimator for the specific heat, obtained by taking the temperature derivative of (\ref{essenbeta});
\begin{equation}
C = \langle n^2\rangle - \langle n\rangle^2 - \langle n\rangle.
\label{cssenbeta}
\end{equation}
When the temperature $T\to 0$, $C$ should vanish, and (\ref{cssenbeta}) then shows that the variance of the distribution of $n$ equals $\langle n\rangle$. 
It is therefore clear that the distribution vanishes exponentially beyond some power $n \propto N\beta$. In practice, QMC (stochastic series expansion, 
SSE) algorithms based on this representation automatically sample the $n$-distribution according to the relative weights of the different $n$ sectors. 

By explicitly truncating the Taylor expansion, one can make the relationship between the series and path integral representations even clearer. Truncating 
at $n=L$, we can formally construct a fixed-size sampling space by augmenting all powers $H^n$ with $n<L$ by $L-n$ unit operators $I$. Allowing for all possible 
locations of the unit operators in the product of $L$ operators, we can define a modified operator string $S=S_1,S_2,\ldots, S_L$, in which $S_i \in \{H,I\}$, 
and do a summation over all these sequences. We then have to compensate the weight by the number $L\choose{n}$ of equivalent terms, giving
\begin{equation}
Z = \sum_{S} \frac{(-\beta)^n(L-n)!}{L!}
\sum_{\{\alpha\}_L} \sum_{\{S_i\}}
 \langle \alpha_0|S_L|\alpha_{L-1}\rangle \cdots \langle \alpha_2 |S_2|\alpha_1\rangle\langle \alpha_1 |S_1|\alpha_0\rangle, 
\label{zssegenera2}
\end{equation}
where $n$ now refers to the number of elements $S_i=H$ in the operator string (and we no longer need an explicit sum over $n$).
If we now consider the boson kinetic energy (\ref{hbosonkin}) and take the limit $L \to \infty$, the weight reduces to $(\beta/L)^n$, which is the same 
as the path integral weight (\ref{walphabospaths}) for $L$ time slices. In this limit, it is clear that the index $p=0,\ldots,L-1$ is related to
imaginary time according to $\tau=p\beta/L$. The difference is that the series expansion with the full weight, $\beta^n(L-n)!/L!$, is in practice exact 
for $L \propto \beta N$, whereas in the path integral we have to take the continuum limit to avoid a discretization error. For finite $L$, the series
index $p$ does not correspond exactly to imaginary time, but represents a distribution of imaginary times \cite{sandvik92}. Because of the close relationship 
between the propagation index $p$ and imaginary time, it is appropriate to refer to the $p$ space as the ``time'' dimension also in the series expansion.

The practical advantage of the series expansion is that it provides an exact but discrete representation of the imaginary time continuum. Algorithms 
based on it are normally easier to implement than continuous-time world lines and can be more efficient computationally. On the other hand, for a hamiltonian
with large potential-energy terms, the number of operators $\langle n\rangle$ in the expansion can be much larger than the number of kinetic events 
$\langle n_K\rangle$ in the path integral. The latter approach should then be more efficient, although an exact treatment necessitates manipulation of floating-point 
time variables. For quantum spin systems, there is in general a favorable balance between kinetic (off-diagonal) and potential (diagonal) energies, and 
the series expansion is then typically preferable. Next, we proceed to develop a QMC algorithm based on it.

\subsection{SSE method for the $S=1/2$ Heisenberg model}
\label{sseheisenberg}.

We now discuss in detail how to implement the SSE method for the $S=1/2$ Heisenberg model. Initially we do not have to specify the lattice and consider 
the hamiltonian written as a sum of bond operators,
\begin{equation}
H_b=J_{b}{\bf S}_{i(b)} \cdot {\bf S}_{j(b)},
\label{hbond}
\end{equation}
as in (\ref{hbsum}), with the lattice encoded as a list of sites $[i(b),j(b)]$ connected by the bonds, $b=1,\ldots,N_b$. An example of a 
labeling scheme in 2D was given in Fig.~\ref{sblabels2d}.

A positive-definite SSE can be constructed for any bipartite 
lattice, i.e., when all sites $i(b)$ and $j(b)$ belong to sublattice A and B, respectively, in any number of dimensions. This constitutes a large class
of interesting and important systems. Initially we will consider just a single antiferromagnetic coupling constant $J_{b}=J>0$, but the modifications
needed for non-uniform systems are very simple. 

\subsubsection{Configuration space}

It is useful to subdivide the Heisenberg interaction (\ref{hbond}) into its diagonal and off-diagonal parts in the standard basis of diagonal $z$ spin 
components. We then define operators with two indices, $H_{a,b}$, with $a=1,2$ referring to diagonal and off-diagonal, respectively, and $b=1,\ldots,N_b$ is
the bond index as before;
\begin{eqnarray}
H_{1,b}  & = & \hbox{$\frac{1}{4}$}-S^z_{i(b)}S^z_{j(b)}, \label{heishb12_1} \\
H_{2,b}  & = & \hbox{$\frac{1}{2}$}(S^+_{i(b)}S^-_{j(b)}+S^-_{i(b)}S^+_{j(b)}).\label{heishb12_2}
\end{eqnarray}
Here we have also introduced a minus sign and a constant in the diagonal operator, so that the full hamiltonian can be written as
\begin{equation}
H = -J\sum_{b=1}^{N_b}(H_{1,b}-H_{2,b})+\frac{JN_b}{4}.
\label{heisabsum}
\end{equation}
The reason for including the constant in $H_{1,b}$ is to make the series expansion positive-definite, as we will see shortly. The constant $JN_b/4$ 
is irrelevant in the algorithm but we will include it when calculating the energy. 

\paragraph{Series expansion of the partition function}

The general form of the partition function in the SSE approach can be written as (\ref{zssegeneral}), but here we will initially not write out all 
the complete sets of states inserted between the operators. We instead focus on the operators, expanding all instances of $H$ as sums over all the
bond operators in (\ref{heisabsum}). The starting point of the SSE algorithm for the Heisenberg hamiltonian is thus
\begin{equation}
Z=\sum_{\alpha}\sum_{n=0}^\infty (-1)^{n_2}\frac{\beta^n}{n!} \sum_{S_n}\left \langle \alpha \left | \prod_{p=0}^{n-1} H_{a(p),b(p)} \right |\alpha \right \rangle .
\label{zsseheis1}
\end{equation}
Here $J$ has been absorbed into $\beta=J/T$, $S_n$ refers to the products (strings) of the Heisenberg bond operators (\ref{heishb12_1}) and 
(\ref{heishb12_2}) originating from $H^n$,
\begin{equation}
S_n = [a(0),b(0)],[a(1),b(1)],\ldots,[a(n-1),b(n-1)],
\label{snindexlist}
\end{equation}
and $n_2$ is the total number of off-diagonal operators, i.e., the number of elements with $b(i)=2$ in the string. Note that the label $p$ referring to the position 
of an operator in the string here takes the values $0,\ldots,n-1$; this labeling will be more convenient than $p=1,\ldots,n$ in the program implementation. 

When the operator string acts on the state $|\alpha\rangle=|S^z_i,\ldots,S^z_N\rangle$ we get a succession of other basis states, with no branching into 
superpositions of more than one state. We will refer to these as propagated states $|\alpha(p)\rangle$,
\begin{equation}
|\alpha(p)\rangle \propto \prod_{i=0}^{p-1} H_{a(i),b(i)} \left |\alpha \right \rangle .
\label{propagatedstate}
\end{equation}
We have not written down the normalization of these states explicitly but will consider all $|\alpha(p)\rangle$ as properly normalized. These states of
course correspond to those in the summations over complete sets in (\ref{zssegeneral}), but in practice the SSE approach is framed around the summation
over the operator strings in (\ref{zsseheis1}), and we do not always have to write out the states explicitly. 

With the constant $1/4$ in (\ref{heishb12_1}), all operations on parallel spins destroy the states;
\begin{eqnarray}
H_{1,b}|\up_{i(b)}\up_{j(b)}\rangle = 0,&& H_{2,b}|\up_{i(b)}\up_{j(b)}\rangle = 0,~~~\nonumber\\
H_{1,b}|\dn_{i(b)}\dn_{j(b)}\rangle = 0,&& H_{2,b}|\dn_{i(b)}\dn_{j(b)}\rangle = 0.
\end{eqnarray}
An operator-state configuration $(\alpha,S_n)$ contributing to $Z$ thus has to involve only operations on anti-parallel spins, and the propagated states 
(\ref{propagatedstate}) are defined under this assumption. The corresponding matrix elements are 
\begin{eqnarray}
\langle \up_{i(b)}\dn_{j(b)}|H_{1,b}|\up_{i(b)}\dn_{j(b)}\rangle = \half,&&
\langle \dn_{i(b)}\up_{j(b)}|H_{2,b}|\up_{i(b)}\dn_{j(b)}\rangle = \half,~~~\nonumber\\
\langle \dn_{i(b)}\up_{j(b)}|H_{1,b}|\dn_{i(b)}\up_{j(b)}\rangle = \half,&&
\langle \up_{i(b)}\dn_{j(b)}|H_{2,b}|\dn_{i(b)}\up_{j(b)}\rangle = \half.~~~
\label{heismatelems}
\end{eqnarray}
The fact that all these are equal will be very useful in the sampling algorithm, and this was the reason for including the constant $1/4$ in the 
diagonal operators. 

In addition to the local constraints of only operations on anti-parallel spins, for the matrix element of the full operator product in (\ref{zsseheis1}) 
to be non-zero, the propagation also has to satisfy the periodicity $|\alpha(n)\rangle = |\alpha(0)\rangle$, where $|\alpha(0)\rangle=|\alpha\rangle$. 

Although one can formulate Monte Carlo sampling procedures in the space of operator strings of fluctuating length $n$ \cite{handscomb62,sandvik92,trebst07}, 
it is normally in practice somewhat easier to work within the fixed-size truncated space discussed in the previous section (and mathematically it is also easier 
to prove detailed balance in such a space).\footnote{A case where it is clearly better to work with sequences of fluctuating length $n$ is when the SSE approach 
is combined with a multi-canonical ensemble, in which a range of temperatures is sampled and therefore $n$ can fluctuate over a very wide range \cite{trebst07}.} 
In actual simulations, the cut-off $L$ will be determined automatically by the program, as we will discuss below, such that $L$ safely exceeds the largest $n$ ever 
sampled. Note again that {\it the truncation then does not cause any detectable errors and should not be considered as an approximation}. Defining the unit operators 
used for augmenting strings with $n<L$ as $H_{0,0}=I$ and including $[a(p),b(p)]=[0,0]$ as an allowed element in the index list (\ref{snindexlist}), the partition 
function can be written as
\begin{equation}
Z=\sum_{\alpha}\sum_{S_L}(-1)^{n_2}\frac{\beta^n(L-n)!}{L!}\left \langle \alpha \left | \prod_{p=0}^{L-1} H_{a(p),b(p)} \right |\alpha \right \rangle ,
\label{zsseheis2}
\end{equation}
where $n$ refers to the number of non-$[0,0]$ elements in the fixed-length operator string $S_L$ (and $n$ is summed implicitly by the sum over $S_L$). 
Since all matrix elements equal $1/2$, the weight of an allowed configuration is given by
\begin{equation}
W(\alpha,S_L)=\left ( \frac{\beta}{2}\right )^n\frac{(L-n)!}{L!}.
\label{wsseheis1}
\end{equation}
Here we have left out the sign factor $(-1)^{n_2}$, which, as we will discuss further below, always is positive for a bipartite (unfrustrated) system.
Note also that $L!$ is an irrelevant normalization factor and can be left out as well.

\begin{figure}
\includegraphics[width=9.5cm]{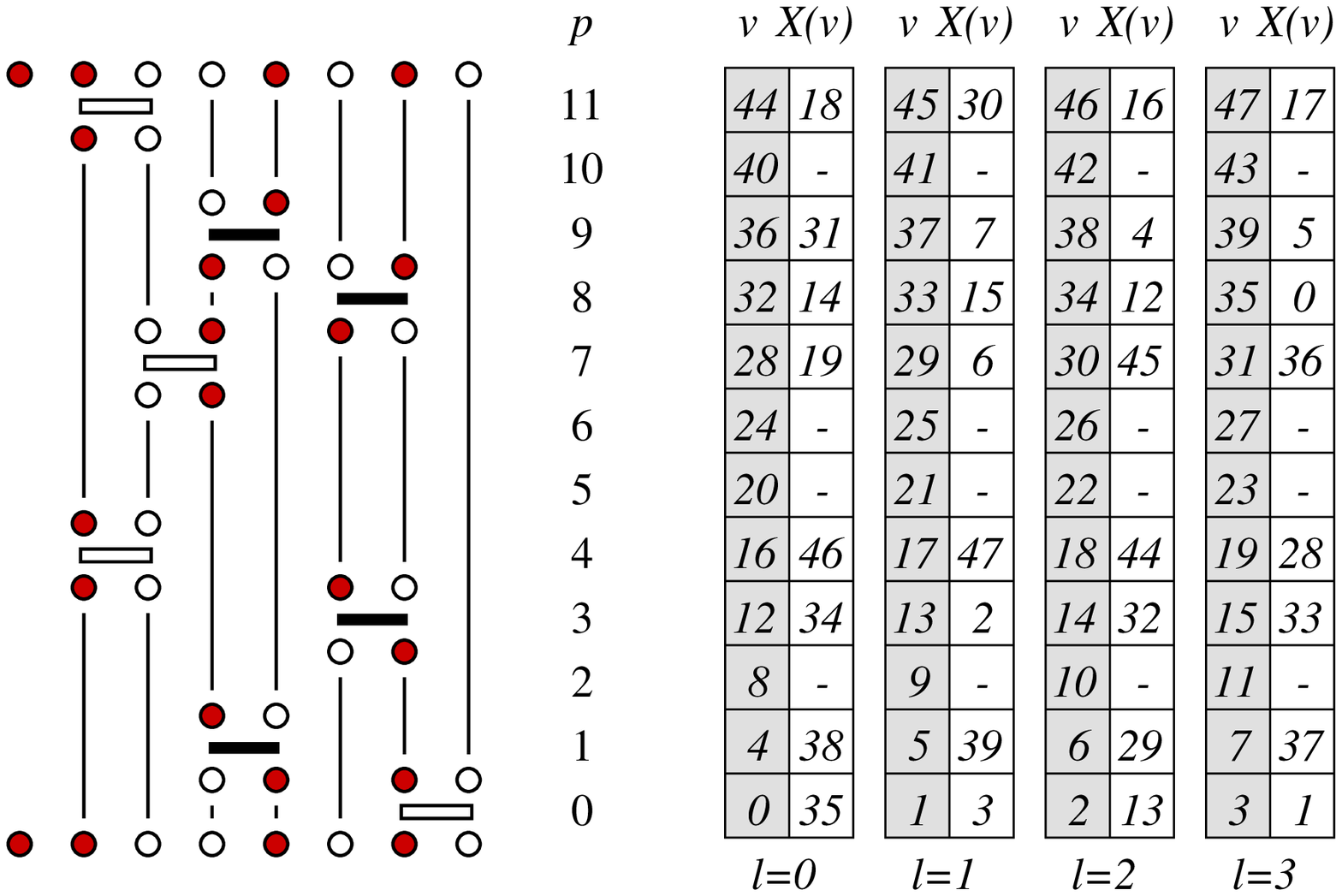}
\caption{An SSE configuration for an 8-spin chain, with all the propagated states shown. Open and solid bars indicate diagonal $H_{1,i}$ and off-diagonal 
$H_{2,i}$ operators respectively, while no bar between states corresponds to a ``fill-in'' unit operator $H_{0,0}$. The $\up$ and $\dn$ spins of the state 
$|\alpha\rangle$ are stored as $\sigma(i)=\pm 1$, and the operator string $S_L$ is encoded using even and odd integers for diagonal and off-diagonal
operators, respectively, according to $s(p)=2b(p)+a(p)-1$.}
\label{sseconfig1}
\end{figure}

With $\up$ and $\dn$ spins represented by the presence and absence of a hard-core boson, respectively, we could draw pictures like Fig.~\ref{pathintegral} to 
illustrate the rules for the contributing SSE configurations. However, we now also have diagonal operators that do not change the world lines but modify the path 
weight. In the path integral approach the diagonal operators are all brought together into a common exponential factor, Eq.~(\ref{walphabospaths2}), but there is 
no simple way to do this in the series expansion. As mentioned above in Sec.~\ref{sec_pathintegrals},  the SSE approach should not be used when the diagonal 
energy dominates, but in the case considered here we have $\langle H_{1,b}\rangle \approx \langle H_{2,b}\rangle$ and the diagonal operators do not dominate. As we 
will see below, the presence of a significant number of diagonal operators is in fact exploited in the SSE algorithm. When discussing the SSE method, it is useful 
to draw pictures showing explicitly the diagonal and off-diagonal operators. An example is given in Fig.~\ref{sseconfig1}, which includes also illustrations of the 
data structures that we will later  use in a computer implementation of the method. The states will be represented by integers $\sigma(i)=\pm 1$ corresponding 
to $S^z_i=\pm 1/2$. For compact storage of the operator string, it is also useful to combine the indices $[a(p),b(p)]$ into a single list of integers 
$s(p)=2b(p)+a(p)-1$. The full propagated states do not have to be stored simultaneously. They can be generated as needed from the single stored state 
$|\alpha(0)\rangle$ and the operator string. We will later introduce a different compact storage involving some spins of the propagated states as well.

\paragraph{Frustrated interactions and the ``sign problem''}

At first sight, it appears that we have a sign problem---a non-positive definite expansion---because of the factor $(-1)^{n_2}$ in (\ref{zsseheis1}). Actually, 
all the terms are positive for a bipartite lattice. This is because an even number $n_2$ of off-diagonal operators are required in every allowed configuration, 
in order to satisfy the ``time'' periodicity $|\alpha(L)\rangle=|\alpha(0)\rangle$. We already discussed this in the context of the world line method, where 
the off-diagonal matrix elements in (\ref{trottermatelem}) are negative, but the continuity of the world lines require an even number of these. This is yet
another example of the close relationship between the two approaches.

\begin{figure}
\includegraphics[width=10cm]{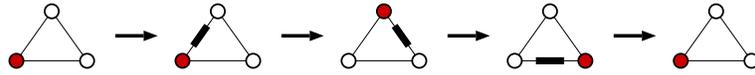}
\caption{An example of three off-diagonal operations (indicated by bars) bringing all spins on a triangle back to their original states. 
Each spin flip is associated with a minus sign, resulting in a negative path weight and a ``sign problem'' (due to cancellations of
configurations with different signs) in QMC simulations of this and other frustrated systems.}
\label{triangle}
\end{figure}

For frustrated systems, the series expansion is not positive-definite (and neither is the path integral, for exactly the same reason). This can be easily 
demonstrated for a system of three spins on a triangle. As shown in Fig.~\ref{triangle}, an allowed configuration can in this case contain three off-diagonal 
operators, resulting in an over-all minus sign. This is true for any system in which loops with an odd number of sites can be formed between antiferromagnetically 
interacting spins---this can be used as the definition of frustration. 

Positive-definiteness for a bipartite system can also be proved in a different way, by carrying out a unitary transformation of the spin operators on one of
the sublattices, say $B$, such that $S^+_j \to - S^+_j$ and $S^-_j \to - S^-_j$ (and no change in the diagonal operators $S^z_j$) for $j \in B$. This does not 
affect the spectrum of the model (since the commutation relations among all spin operators remain unchanged), but the sign in front of the off-diagonal terms 
in the hamiltonian (\ref{heisabsum}) changes to $+$. The factor $(-1)^{n_2}$ in (\ref{zsseheis1}) is then absent. For a frustrated system, no such 
transformation can remove all the signs.

Note that only the off-diagonal part of the interaction causes a sign problem. One can study systems with frustration in the diagonal part, but then the 
interaction is no longer spin-isotropic. Just neglecting the sign for a frustrated Heisenberg system corresponds to antiferromagnetic diagonal couplings but 
ferromagnetic off-diagonal couplings. Here we only consider isotropic interactions and restrict the discussion to unfrustrated systems.

For a bipartite system with periodic boundaries, the absence of sign problems holds strictly only if the lattice lengths $L_\gamma$ are even in all directions 
$\gamma=1,\ldots,d$. Winding configurations, like the one shown for the path integral in Fig.~\ref{pathintegral}(b), are negative if there is an odd number 
of windings around an odd number of sites. Clearly, in such cases there is also frustration, due to the boundaries. We will therefore consider only systems of 
even length (in all periodic lattice directions).

\paragraph{Linked-vertex storage of the configurations}

\begin{figure}
\includegraphics[width=10.5cm]{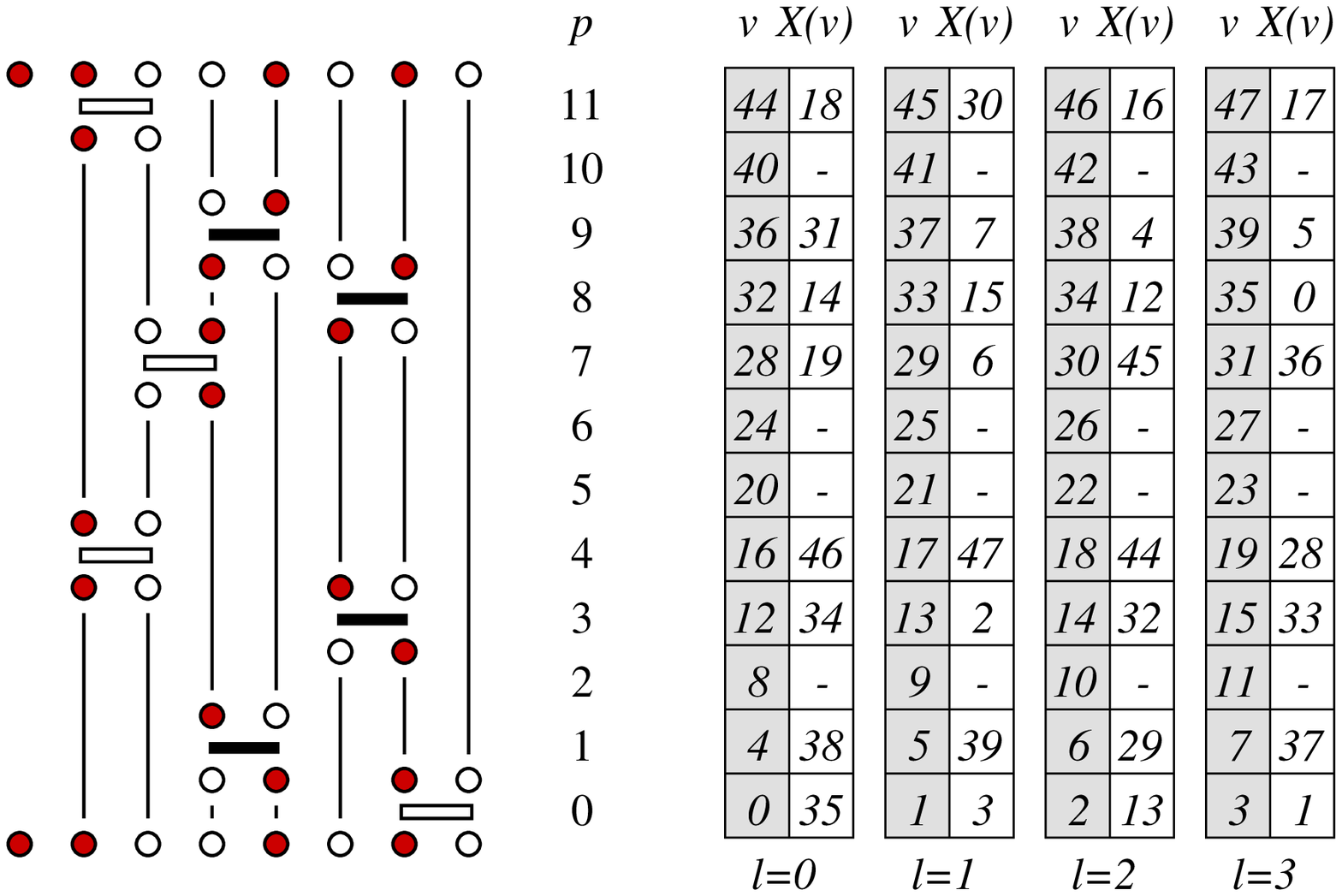}
\caption{Linked vertex storage of the configuration in Fig.~\ref{sseconfig1}. In the graphical representation to the left, 
constant spin states between operators have been replaced by lines (links) connecting the spins just before and after the operator acts. The links 
can be stored in a list $X(v)$, where the four elements $v=4p+l$, $l=0,1,2,3$, correspond to the legs (with the numbering convention shown in 
Fig.~\ref{vertexlegs}) of the vertex at position $p$ in the sequence $S_L$. For two linked legs $v$ and $v'$, $X(v)=v'$ and $X(v')=v$.}
\label{sseconfig2}
\end{figure}

We have discussed computer storage of the  SSE configuration as a state $|\alpha(0)\rangle$ and an operator string $S_L$. By acting sequentially with the 
operators, it is easy to generate all the propagated state $|\alpha(p)\rangle$ illustrated in Fig.~\ref{sseconfig1}. However, during the Monte Carlo sampling 
we will need to access operators and some properties of the propagated states also in non-sequential order---given an operator and the spins it acts on, we will 
need to to know which operators act on those spins next (when moving up as well as down in the operator sequence). It would be prohibitively time consuming to 
propagate a single state back and forth to extract this information, and also it would not be practical to store all the propagated states. We therefore use also 
another kind of data structure, in which the ``connectivity'' of the operators is explicit and represented as a network in a compact way. This {\it linked vertex} 
structure is illustrated in Fig.~\ref{sseconfig2}, using the same SSE configuration as in Fig.~\ref{sseconfig1}. Here, in the pictorial representation, the constant 
state of each spin between operators have been replaced by straight lines, which in a computer program will correspond to links (pointers). The operators are shown 
along with only the two spin states before and after each operator has acted. We will call these spin-operator objects {\it vertices} and refer to the four spins 
as the {\it legs} of the vertices. One stage of the Monte Carlo sampling procedures will involve making changes to the vertices (while their locations are changed
at another stage). The links will allow us to quickly move between the vertices and make a series of changes maintaining all the constraints.

\begin{figure}
\includegraphics[width=7cm]{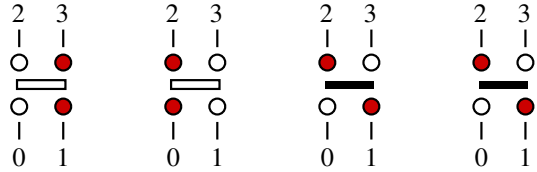}
\caption{Allowed vertices for the isotropic $S=1/2$ Heisenberg model. The numbering $l=0,1,2,3$ of the vertex legs corresponds to the position $v=4p+l$,
in linked-list storage illustrated in Fig.~\ref{sseconfig2}.}
\label{vertexlegs}
\end{figure}

The allowed vertices are dictated by the hamiltonian. In the case of the isotropic Heisenberg system considered here there are four of them, depicted in 
Fig.~\ref{vertexlegs}. They of course correspond to the non-zero matrix elements (\ref{heismatelems}), and, again, the constant $\frac{1}{4}$ in the diagonal
operator is the reason why there is no vertex with all four spins equal. Anisotropic interactions or an external magnetic field would necessitate inclusions 
of additional vertices \cite{sandvik99a,syljuasen02} (and the algorithm discussed here would then also have to be modified). Although the spin states at the four 
legs uniquely identify the vertices, we will continue to use also the open and solid bars in pictures, to indicate diagonal and off-diagonal vertices, respectively, 
for added clarity.

For a given position $p$ in the operator sequence $S_L$, the corresponding list element $s(p)$ tells us the operator type (diagonal or off-diagonal) and the 
bond $b$ on which it acts (as explained in Fig.~\ref{sseconfig1}). As will become clear below, along with this information, we only have to store the 
connectivity of the vertices, not their spin states. The links allowing us to jump between connected vertex legs are stored as a list $X(v)$, as explained 
in Fig.~\ref{sseconfig2}. For clarity of the illustration, the one dimensional list has here been arranged in four columns, with elements labeled $v=4p+l$, 
corresponding to each type of leg, $l=0,1,2,3$, with the labeling specified in Fig.~\ref{vertexlegs}. We will later describe an efficient way to construct 
this linked list, given the operator sequence. For now, it is sufficient to know that for a given operator at location $p$ in the sequence, the position of 
its $l$:th leg in the linked vertex list is $v=4p+l$. This leg is linked to another vertex leg with list address $v'=X(v)$. This kind of structure constitutes 
a doubly-linked (bi-directional) list, with $X[X(v)]=v$, in which we can move both ``up'' and ``down''. From a position $v$ in the list we can extract the 
corresponding operator location in $S_L$, $p=v/4$ (its integer part) and leg index $l={\rm mod}[v,4]$. We can move ``sideways'' to the other leg on 
the same vertex by changing the leg label $0 \leftrightarrow 1$ or $2 \leftrightarrow 3$, which in both cases can be expressed as a simple rule of an 
even$\leftrightarrow$odd change in the list address $v$. These movements will allow us to construct closed loops of changed spins and operators.

\subsubsection{Monte Carlo sampling procedures}

\begin{figure}
\includegraphics[width=14cm]{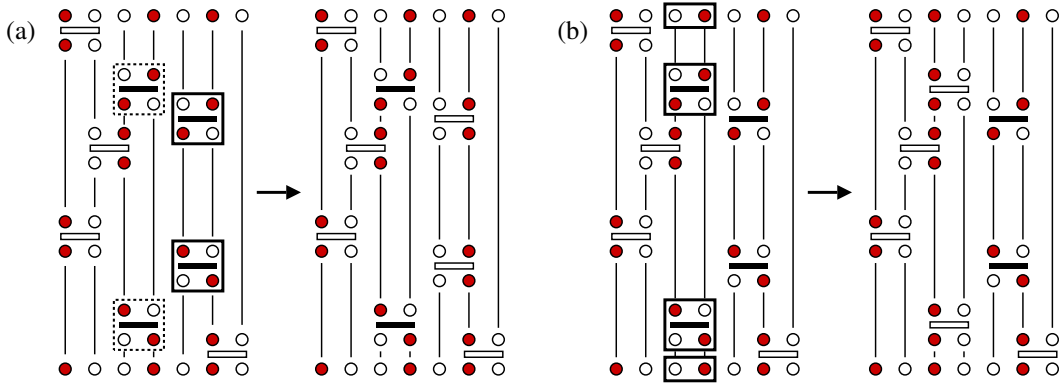}
\caption{(a) A pair of off-diagonal operators, indicated by solid-line boxes, which can be replaced by diagonal ones if the spins between the two 
operators are flipped as well. Such an update cannot be done with the operators enclosed by dashed boxes, because of the illegal spin configuration 
(a vertex with all four legs in the same spin state) that would result at the operator acting between the two boxed vertices on their left spin.
However, as shown in (b), this operator pair can be changed if instead the spins on the opposite sides of the operators are flipped. This also forces
a change in the stored state $|\alpha\rangle$.}
\label{localsseupdate}
\end{figure}

When Monte Carlo sampling the partition function (\ref{zsseheis1}), we have to make changes in the operator sequence $S_L$ as well as in the stored state 
$|\alpha\rangle$. It is clear from Figs.~\ref{sseconfig1} and \ref{sseconfig2}, however, that updates in these two data structures are not necessarily 
independent. Allowed changes (i.e., ones maintaining all configuration constraints) in operators at locations $p \in \{p_1,\ldots,p_2\}$ lead to corresponding 
changes in the propagated states in the range $|\alpha(p_1+1)\rangle, \ldots, |\alpha(p_2)\rangle$. Fig.~\ref{localsseupdate} illustrates such an update of 
two operators, where only the types of the operators are changed, diagonal$\leftrightarrow$off-diagonal (which leads to changes in the states similar to the 
world line update in Fig.~\ref{wlmove}). As the changed operator at $p_1$ acts on $|\alpha (p_1)\rangle$, the state $|\alpha (p_1+1)\rangle$ and subsequent states 
also change (by two flipped spins in the example), but when the last updated operator at $p_2$ has acted the modifications are ``healed'' and the resulting 
state $|\alpha(p_2+1)\rangle$ is the same as before the update. Since the propagated states form a cyclically permutable periodic structure, there is nothing 
special with the stored state $|\alpha\rangle=|\alpha(0)\rangle$ and it can change as well, depending on the locations of the updated operators.

Because of the time ($p$) periodicity of $|\alpha(p)\rangle$, the way state changes are forced by operator updates is not unique. In the above example, where we assumed 
$p_2>p_1$, we could also have acted first on $|\alpha (p_2)\rangle$ with the new operator at $p_2$. Then the states $|\alpha(p_2+1)\rangle, \ldots, |\alpha(L-1)\rangle$ 
as well as $|\alpha(0)\rangle, \ldots, |\alpha(p_1)\rangle$ would be affected. There are always two ways of changing the states in this way, but only one may be 
allowed in any given case because of the constraints, as illustrated in Fig.~\ref{localsseupdate}. In most cases, such updates done at random locations would not at all 
satisfy the constraints, and one has to specifically look for two (or more) operators that can be updated \cite{sandvik97}.

As seen in Figs.~\ref{sseconfig1} and \ref{sseconfig2}, the state $|\alpha\rangle$ can some times be updated without any changes to the operator sequence---since 
there  is no operator acting on the spin $\sigma(1)$, its state is arbitrary. Such free spins are very rare at low temperatures, when the average number of 
operators on each spin is large.

Focusing on the operator sequence, we will sample the number $n$ of non-unit operators $[a,b]_p\not=[0,0]$, their positions $p\in \{0,\ldots,L-1\}$ in $S_L$, 
bond indices $b \in \{1,\ldots,N_b\}$, and types induces $a\in \{1,2\}$. The general strategy is to let the expansion order $n$ change only in exchanges of 
diagonal operators with the fill-in unit operator; $[0,0]_p \leftrightarrow [1,b]_p$ (where $n \to n \pm 1$). Then, keeping all lattice locations $b$ of the 
operators fixed, we change the type index, $[1,b]_p \leftrightarrow [2,b]_p$ for some set of operators (as in the example with two operators above, but we 
will do it in a much more efficient way involving an arbitrary number of operators). With these off-diagonal replacements carried out properly, the sampling 
is ergodic in combination with the diagonal $n \to n \pm 1$ updates. We next consider the details of these updating procedures. We first assume that the 
expansion cut-off $L$ has been properly determined and then discuss  how this can be ensured in practice.

\paragraph{Diagonal updates}

\begin{figure}
\includegraphics[width=11cm, clip]{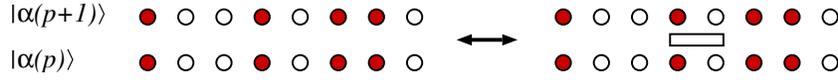}
\caption{Diagonal update. An operator is either inserted, $[0,0]_p\rightarrow [1,b]_p$, at a randomly selected bond or removed, 
$[1,b]_p\rightarrow [0,0]_p$.  In the former case, the update is canceled if the spins at the chosen bond are parallel, and therefore the state 
$|\alpha(p-1)\rangle$ must be known. The acceptance probabilities depend on the number of operators $n$ in the sequence before the update attempt, 
according to Eq.~(\ref{paccsseheis1}).}
\label{diagupdate}
\end{figure}

Updates of single diagonal operators, illustrated in Fig.~\ref{diagupdate}, can be carried out sequentially at the locations $p=0,\ldots,L-1$ in $S_L$. 
There are no constraints involved in such an update in the direction $[1,b]_p \to [0,0]_p$ (removal of a diagonal operator), whereas an insertion of a
diagonal operator, $[0,0]_p \to [1,b]_p$, is allowed only if the spins on bond $b$ are antiparallel, $\sigma(i(b)) \not= \sigma(j(b))$, in the propagated 
state $|\alpha(p)\rangle$ on which the operator acts. We therefore have to generate these states (storing only the one currently needed), which is simply 
done by flipping the two spins $\sigma(i(b))$ and $\sigma(j(b))$ each time an off-diagonal operator $[2,b]_p$ is encountered (in which case no 
single-operator update can be carried out at $p$). 

The weight ratio to use in the Metropolis acceptance probability is easily obtained from (\ref{wsseheis1}). Note, however, that we also need to correct for
the inherent imbalance of these update attempts---there is only a single unique way of carrying out the update in the direction $[1,b]_p \to [0,0]_p$, whereas 
when changing $[0,0]_p \to [1,b]_p$ the bond $b$ should be generated at random among the $N_b$ possible locations [and if, for the chosen $b$, the spins
$\sigma(i(b)) = \sigma(j(b))$, the update is immediately rejected and we proceed to the next $p$]. Therefore, following the criteria for detailed 
balance discussed in Sec.~\ref{montecarlo}, specifically the transition probability (\ref{pselpacc}) written as a product of selection and acceptance 
probabilities, we have to include the ratio $N_b$ of the selection probabilities for updates involving a specific bond $b$. We then obtain the following 
acceptance probabilities;
\begin{eqnarray}
P_{\rm accept}([0,0] \to [1,b]) & = & {\rm min}\left [ \frac{\beta N_b}{2(L-n)},1\right ], \label{paccsseheis1} \\
P_{\rm accept}([1,b] \to [0,0]) & = & {\rm min}\left [ \frac{2(L-n+1)}{\beta N_b},1\right ], \label{paccsseheis2}
\end{eqnarray}
where $n$ is the number of operators before the update (and after $n\to n+1$ or $n\to n-1$).  

To prove detailed balance for the diagonal updates, it may seem necessary to carry them out at random positions $p$, not sequentially.
Otherwise, after having carried out an update at some position $p$, there is zero probability of carrying out the reverse update as the next step. However, if 
we consider the whole sequence (sweep) of updates from $p=0$ to $P=L-1$ and follow this by a reverse updating sequence, starting from $P=L-1$ and 
ending at $p=0$, then detailed balance holds for the sweeps. In practice, one does not have to switch between the two directions of updating the 
operator sequence. 

\paragraph{Off-diagonal updates}

Updates involving off-diagonal operators clearly have to involve at least two operators in order to maintain the periodicity constraint on the
propagated sates. The simplest kind of pair update was already discussed and illustrated in Fig.~\ref{localsseupdate}. For a 1D system, such 
updates are ergodic with open boundary conditions, but with periodic boundaries local updates cannot change the topological winding number (which
in clear from Fig~\ref{pathintegral}, where no local deformations can affect cyclic permutations). 
This may still not be a very serious problem in principle, because for any $T>0$ only the sector with zero winding number contributes when the 
system size $N \to \infty$ (in practice for some large $N$, larger for lower $T$). If we are interested in ground state properties, we also can obtain 
the correct result when $T\to 0$ for fixed $N$, because in this limit an effective winding number defined within a finite range $p,\ldots,p+D_p$, with
$D_p\gg N$, of the propagated states $|\alpha(p)\rangle$ (corresponding to a finite range of imaginary times) can still fluctuate freely, irrespective 
of the constraint of zero global winding number. Physical quantities are then insensitive to the winding number \cite{henelius98}. To converge to the 
ground state, much lower $T$ has to be used, however.

In addition to their inability to change the winding number in one dimension, in higher dimensions the kind of operator substitutions shown in 
Fig.~\ref{localsseupdate} cannot lead to any of the permutations of same spins (permutation of particles in the boson language) that should be included 
(while in one dimension, cyclical permutations of all spins are the only possible permutations in the kind of system we consider here). To remedy this 
problem, one can construct other kinds of local updates, e.g., involving two operators acting on the corners of a plaquette of $2\times 2$ sites
on a 2D square lattice \cite{sandvik97}, which can sample among all such permutations within a given sector of winding numbers. Updates changing the 
topological winding number has to involve lines of sites wrapping around the boundaries, but for lattices of length larger than $\approx 20$ these 
targeted updates become very unlikely (i.e., it becomes too difficult to satisfy the constraints when a large number of operators are changed simultaneously) 
\cite{sandvik90,sandvik97}.

\begin{figure}
\includegraphics[width=12.75cm]{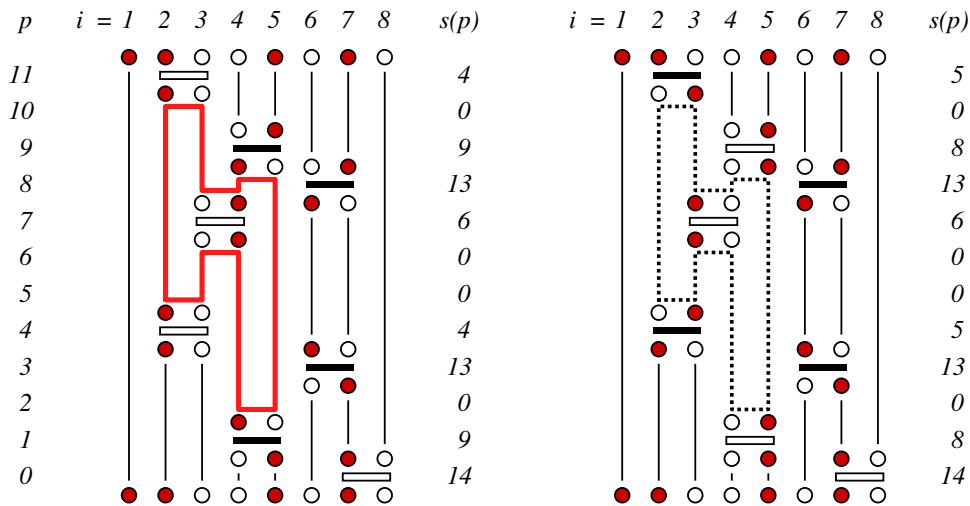}
\caption{A linked-vertex SSE configuration with one loop traced out and shown in both of its ``orientations'', along with the corresponding 
operator-index sequences. All spins covered by the loop are flipped, and operators are changed, diagonal $\leftrightarrow$ off-diagonal, each 
time the loop passes by (with no net change of an operator visited twice). Every vertex leg (spin) belongs uniquely to one loop, and spins not 
acted upon by any operator (here the one at $i=1$) can also be regarded as forming their own loops.}
\label{sseloop}
\end{figure}

Fortunately, instead of dealing with updates specifically targeting permutations and winding number sectors, there is a very efficient and simple type of 
{\it loop update} that accomplishes all these things automatically. This class of updates was initially introduced as a generalization of a cluster 
algorithm for the Ising model to a model where the flipped clusters take the form of loops; the classical six-vertex model \cite{evertz93}. The effective 
world line system for the $S=1/2$ Heisenberg model constructed using the discrete Suzuki-Trotter decomposition is exactly equivalent to an anisotropic 
six-vertex model, and the loop update for it was therefore at the same time a generalization of the classical cluster update to a quantum mechanical system. 
These ideas were subsequently applied also to continuous-time world lines \cite{beard} as well as to the off-diagonal updates in the SSE method \cite{sandvik99a}. 
The improvements in performance relative to local updates are enormous (as in the classical case, leading to a much reduced dynamic exponent) and brought 
simulations of quantum spin systems to an entirely new level. Like classical cluster algorithms, the loop updates are in practice limited to certain classes 
of models, of which the isotropic Heisenberg systems is one. Generalizations of the loop concept to {\it worms} \cite{prokofev96} and {\it directed loops} 
\cite{syljuasen02} (both of which can be regarded as loops that are allowed to self-intersect during their construction, unlike the original loop updates
where no self-intersection is allowed) are applicable to a wider range of systems.

For the $S=1/2$ model considered here, there is no reason to even discuss local off-diagonal updates in any greater detail, and we will just focus on how to 
implement the much more powerful loop updates. In the case of the SSE method, the operator string is again the main focus, and the loop update corresponds 
to constructing a loop of operators (vertices) connected by the links in the linked-list representation. 

\paragraph{Operator-loop updates}

An example of an operator-loop and how it is flipped is shown in Fig.~\ref{sseloop}. Here ``flipping'' refers to the spins along the loop (explicitly
those on the vertex legs and implicitly in all propagated states covered by the loop) as well as the operators themselves when we map the changes back 
to the sequence $S_L$; a diagonal vertex with two legs attached to a loop changes to off-diagonal, and vice versa. In the case of all four legs of a vertex 
belonging to the same loop (and example of which is present in the figure), the operator type does not change. 

The key aspect of the loop update is that the weight (\ref{wsseheis1}) only depends on the number of operators, which does not change when the operator-loop
is flipped. Therefore, according to the detailed balance rules, such a flip can always be accepted. It is also clear that loops can be large, and a loop 
flip can therefore lead to changes in the global spin permutations. The previously discussed pair substitutions correspond to flipping loops of two operators on 
the same bond. Any operator substitution specifically constructed to sample permutations can also be formulated as an operator-loop, including ones that change 
a winding number. 

Note that the loops are completely deterministic once the operator locations have been specified and each spin in the full space-time configuration belongs 
uniquely to one loop. The diagonal update is the {\it de facto} mechanism by which the loops are changed, and the purpose of the loop update is just to identify 
and flip some of the loops. It is then best to construct all the loops and to flip each of them with probability $1/2$ (which also maintains detailed balance), 
instead of constructing loops at random and always flipping them (in which case some loops would be constructed more than once and flipped unnecessarily). Free 
spins, on which there are no vertices (such as the first spin in Fig.~\ref{sseloop}), can also be considered as loops, since flipping such a spin implicitly 
flips a whole line of spins in the full configuration of time-periodic propagated state.

\subsubsection{Computer implementation}

A convenient definition of a Monte Carlo sweep in the SSE method is a full sequence of diagonal updates, followed by construction (and flip with probability $1/2$)
of all loops. From the loop illustration  in Fig.~\ref{sseloop}, it is clear why the linked-vertex storage $X()$ of the SSE configuration is useful; constructing 
a loop corresponds to moving in this list using very simple rules (following a link or moving laterally to the adjacent spin on the same vertex). On the other hand, 
in the diagonal update the original storage illustrated in Fig.~\ref{sseconfig1}, using a single spin state $\sigma()$ and the operator string $s()$, is more convenient. 
The configuration will always be stored in this way, and the linked vertices will be constructed before each set of loop updates. The loop updates are carried out 
by moving in the linked list $X()$, updating the corresponding operators in $s()$ at the same time (as indicated in Fig.~\ref{sseconfig2}). The stored spins $\sigma()$ 
can also be affected by the loop updates, and this can be taken care of after all loop flips have been carried out. We now discuss pseudocode implementations of 
all the main steps involved in a Monte Carlo sweep.

\paragraph{Diagonal update}

We assume that we currently have a valid configuration stored (which in the beginning of the simulation can be a random spin state and an operator
sequence with only $[0,0]$ elements), using $\sigma(i)=\pm 1$ for the individual spins and the compact storage of $[a,b]_p$ as single integers $s(p)$, 
with even and odd numbers $2b$ and $2b+1$ corresponding to diagonal and off-diagonal operators, respectively (as illustrated in Fig.~\ref{sseconfig1}). 
A sweep of diagonal updates can then be implemented in the following way;

{\code
\cia       {\bf do} $p=0$ {\bf to} $L-1$ \br
\cib          {\bf if} ($s(p)=0$) {\bf then}     \hfill \{27\} \break
\cic              $b={\bf random}[1,\ldots,N_b]$;~ {\bf if } $\sigma(i(b))=\sigma(j(b))$ {\bf skip} {\rm to next} $p$ \br
\cic              {\bf if} (${\bf random}[0-1]<P_{\rm insert}(n)$) {\bf then} $s(p)=2b$;~ $n=n+1$ {\bf endif}  \br
\cib          {\bf elseif} (${\bf mod}[s(p),2]=0$) {\bf then} \br
\cic              {\bf if} (${\bf random}[0-1]<P_{\rm remove}(n)$) {\bf then} $s(p)=0$;~ $n=n-1$ {\bf endif}  \br 
\cib          {\bf else} \br
\cic              $b=s(p)/2$; $\sigma(i(b))=-\sigma(i(b))$;~ $\sigma(j(b))=-\sigma(j(b))$ \br
\cib          {\bf endif} \br
\cia       {\bf enddo} 
\code}

\noindent
Here $P_{\rm insert}(n)$ and $P_{\rm remove}(n)$ are the acceptance probabilities (\ref{paccsseheis1}) and (\ref{paccsseheis2}) for inserting and removing a diagonal 
operator when the current expansion order is $n$. These probabilities can be precalculated or evaluated on the fly. One can also just precalculate the $n$-independent 
constant $\beta N_b/2$ and accept a removal if ${\bf random}[0-1]\times (L-n)<\beta N_b/2$ and an insertion if ${\bf random}[0-1]\times \beta N_b/2<L-n+1$. 
This avoids the more time consuming divisions required with the full acceptance probabilities  (\ref{paccsseheis1}) and (\ref{paccsseheis2}), or storage of 
a large number of the $n$-dependent probabilities. 

The sites belonging to bond $b$ can be stored in a list $[i(b),j(b)]$. With the exception of this list, the SSE sampling algorithm is lattice independent. 
It is also easy to modify the above code for non-uniform ($b$-dependent) Heisenberg coupling strengths. The diagonal update is the only stage at which the 
interactions appear explicitly, and no modifications at all are required in the loop update.

\paragraph{Construction of the linked vertex list}

We will construct the linked list by traversing the operators $s(p)$, starting from $p=0$. Recall that for vertex leg number $l=0,1,2,3$ of an operator 
at $p$, the corresponding links are stored as $X(v)$ with $v=4p+l$, as illustrated in Figs.~\ref{sseconfig2} and \ref{vertexlegs}. To construct the list, 
we will make use of two other data structures. The position $v$ in $X()$ of the first vertex leg for spin $i$ will be stored as $V_{\rm first}(i)$, while 
$V_{\rm last}(i)$ will be the position of the last leg on $i$ encountered so far. Note that the first leg on a given spin is always below the operator (before 
the operator has acted, $l=0,1$), whereas the last leg is above the operator ($l=2,3$). Initially all $V_{\rm first}(i)$ and $V_{\rm last}(i)$ are set to $-1$. 
Later, when encountering an operator $s(p)$ acting on spin $i$ and $V_{\rm last}(i)\ge 0$, then $V_{\rm last}(i)$ is the previous list element corresponding
to an operation on $i$. We then use the corresponding lower leg ($l=0$ or $1$, depending on which of the two spins acted on equals $i$) of the new operator 
and set the links between $v=4p+l$ and $V_{\rm last}(i)$; $X(v)=V_{\rm last}(i)$ and $X(V_{\rm last}(i))=v$. On the other hand, if $V_{\rm last}(i)=-1$, then we 
have found the first operation on $i$ and set $V_{\rm first}(i)=v$. In both cases, we set the element for the last operation on $i$ as $V_{\rm last}(i)=v+2$, 
where the addition of $2$ corresponds to the vertex leg on spin $i$ after the operation (leg index $l=2$ or $3$). For each operator, we have to examine both 
the spins $i_1$ and $i_2$ on which it acts. In the following pseudocode, the vertex list positions to be filled for the operator at $p$ are $v_0+l$, with $v_0=4p$, 
where the leg pairs $l=0,2$ and $l=1,3$ correspond to spins $i_1$ and $i_2$, respectively. We use $v_1$ and $v_2$ to denote the list elements of the last (previous) 
operation on the two spins. All links, except those across the boundary $p=L-1,0$, can be constructed according to;

{\code
\cia    {\bf do} $p=0$ {\bf to} $L-1$ \br
\cib       {\bf if} ($s(p)=0$) {\bf skip} {\rm to next} $p$  \hfill \{28\} \break
\cib       $v_0=4p$;~ $b=s(p)/2$;~ $i_1=i(b)$;~ $i_2=j(b)$;~ $v_1=V_{\rm last}(i_1)$;~ $v_2=V_{\rm last}(i_2)$ \br
\cib       {\bf if} ($v_1\not=-1$) {\bf then} $X(v_1)=v_0$;~ $X(v_0)=v_1$ {\bf else}  $V_{\rm first}(i_1)=v_0$ {\bf endif} \br
\cib       {\bf if} ($v_2\not=-1$) {\bf then} $X(v_2)=v_0$;~ $X(v_0)=v_2$ {\bf else}  $V_{\rm first}(i_2)=v_0+1$ {\bf endif} \br
\cib       $V_{\rm last}(i_1)=v_0+2$;~ $V_{\rm last}(i_2)=v_0+3$ \br 
\cia    {\bf enddo} 
\code}

\noindent
To connect the links across the time boundary we can use the arrays containing the first and last list positions;

{\code
\cia    {\bf do} $i=1$ {\bf to} $N$ \br
\cib       $f=V_{\rm first}(i)$ \hfill \{29\}\break
\cib       {\bf if} ($f\not=-1$) {\bf then} $l=V_{\rm last}(i)$; $X(f)=l$; $X(l)=f$ {\bf endif} \br
\cia    {\bf enddo} 
\code}

\noindent
Since there are also positions in $X()$ containing no links, corresponding to fill-in operators $s(p)=0$, we should have some way to distinguish these.
We can set all $X(v)$ to a negative number, e.g., $X(v)=-1$, before constructing the links. Then any $v$ for which $X(v)\ge 0$ can be used as a starting 
point to trace a loop.

\paragraph{Implementation of the operator-loop update}

We want to trace all the loops and flip each of them with probability $1/2$. In order to make sure that no loop is considered more than once, we should
always start a new loop from a position in $X()$ not previously visited. We then need some flag to indicate whether a position has been visited or not.
Instead of allocating separate storage for such flags, we can actually use the list $X()$ itself---the links stored in it only have to be used once, and
as soon as a link at $v$ has been used we can set $X(v)$ to a negative number, to indicate that this position should not be used again. To start a new loop, 
we can look for the first position $v_0$ for which $X(v_0)\ge 0$. Before traversing the loop, we make the random decision of whether or not to flip it. If we do 
not flip, the loop tracing still has to be carried out, in order to flag all the corresponding $X(v)$ as visited. For a flipped loop, we should also change the 
operators in $s()$ to reflect the change diagonal$\leftrightarrow$off-diagonal for each visited vertex. We will take care of possible flips of free spins in 
the stored state $|\alpha\rangle$ later. The structure of a code for tracing and flipping the loops is;

{\code
\cia       {\bf do} $v_0=0$ {\bf to} $4L-1$ {\bf step} $2$ \br
\cib          {\bf if} ($X(v_0)<0$) {\bf skip} {\rm to next} $v_0$     \hfill \{30\} \break
\cib          $v=v_0$ \br
\cib          {\bf if} (${\bf random}[0-1]<1/2$) {\bf then}  \br
\cic              $\bullet$ {\rm traverse the loop};~ {for all $v$ in loop, set $X(v)=-1$} \br
\cib          {\bf else}  \br
\cic              $\bullet$ {\rm traverse the loop};~ {\rm for all $v$ in loop, set $X(v)=-2$}\br
\cic              $\bullet$ {\rm flip the loop (change operator types $s(p=v/4)$ while loop is traversed)} \br
\cib          {\bf endif}  \br
\cia       {\bf enddo} 
\code}

\noindent
Here $v_0$ is a tentative starting point for a new loop, and we actually start a new loop only if $X(v_0)$ is a non-negative number. Note that we only have to 
consider even starting points (the {\bf step} $2$ on the first line of $\{30\}$), because of the structure of the vertex list and the loops, i.e., two consecutive 
(even, odd) elements in $X(v)$ always correspond to the same loop. To not visit the same loop more than once, we set $X(v)=-1$ for loops that are only visited, and $X(v)=-2$ 
for loops that are also flipped. We will need the distinction between flipped and only visited loops later, when updating the stored state $|\alpha\rangle$ to 
reflect the flipped loops. 

To trace a loop starting at some $v_0$, we can use the leg index $l_0={\bf mod}[v_0,4]$ and move to the adjacent leg $l'_0$ on the same vertex. With the 
leg labeling convention in Fig.~\ref{vertexlegs}, the rule for the leg adjacent to a leg $l$ can be summarized as $l=(0,1,2,3) \to l'=(1,0,3,2)$. This 
rule can be very efficiently implemented using a bit-level operation on $l$, as $l'$ corresponds to flipping 
($0\leftrightarrow 1$) the lowest bit of $l$. Here we just assume that such functionality is available in the programming language used [e.g., in Fortran 90 
one can use the exclusive-or operation of the integer $l$ with $1$;  $l'={\bf ieor}(l,1)$] and assign $l'_0={\bf flipbit}(l_0,0)$, where $0$ is the bit flipped. 
We actually do not need to extract the leg index itself, but can do the corresponding move to the adjacent leg just by manipulating its location $v_0=4p_0+l_0$ 
in $X()$. We can use the same bit flip method to find it; $v_0'={\bf flipbit}(v_0,0)$. To move to the next vertex, we use the link; $v_1=X(v_0')$. This completes 
one step of the loop tracing procedure. We next proceed in the same way to find the location $v_1'$ of to the adjacent leg, and from there we move to the leg 
$v_2$ linked to it. This continues until at some step $k$ we find $v_k=v_0$. We have then completed a full loop. In pseudocode form the traversal of a loop and 
flipping it can be accomplished by (where no subscripts or primes are needed on the visited list locations $v$);

{\code
\cia       $v=v_0$ \br
\cia       {\bf do}                         \hfill \{31\} \break             
\cib            $p=v/4$;~~ $s(p)={\bf flipbit}(s(p),0)$;~~ $X(v)=-2$  \br
\cib            $v={\bf flipbit}(v,0)$;~~ $X(v)=-2$ \br
\cib            $v=X(v)$;~~ {\bf if} ($v=v_0$) {\bf exit} \br
\cia        {\bf enddo}
\code}

\noindent
Here the flip of the operator type diagonal$\leftrightarrow$off-diagonal is also accomplished using the {\bf flipbit} procedure, which corresponds
exactly to the changing the operator code $s(p)$ from $2b$ (diagonal operator) to $2b+1$ (off-diagonal operator) or vice versa. In the case in code
$\{30\}$ where the loop is not to be flipped, the only difference is that the two statements involving $p$ and $s(p)$ in $\{31\}$ are absent, and we mark 
all visited locations as $X(v)-1$ instead if $-2$ (information which will be used to update the state $|\alpha\rangle$).

After all loops have been traced, we also have to update the spins in the stored state $|\alpha\rangle$. For each site $i$ which has a flipped loop passing 
through it, the stored spin $\sigma(i)$ should be flipped. We should also flip with probability $1/2$ any spin that has no loop connected to it (i.e., spins
on which no operator acts). Both these tasks can be taken care of using the information stored in $V_{\rm first}(i)$ during the construction of the linked vertex 
list. If $v=V_{\rm first}(i)=-1$, that implies that there is no operator acting on spin $i$. Otherwise, we can use the flag $-1$ or $-2$ now stored in $X(v)$ to 
determine whether the loop passing through $i$ has been flipped [with $X(v)=-2$ for flipped loops]. In pseudocode form:

{\code
\cia       {\bf do} $i=1$ {\bf to} $N$ \br
\cib            $v=V_{\rm first}(i)$           \hfill   \{32\} \break 
\cib            {\bf if} ($v=-1$) {\bf then} \br 
\cic                {\bf if} ({\bf random}[0-1]$<1/2$) $\sigma(i)=-\sigma(i)$ \br             
\cib            {\bf else} \br
\cic                {\bf if} ($X(v)=-2$) $\sigma(i)=-\sigma(i)$ \br             
\cib            {\bf endif} \br
\cia        {\bf enddo}
\code}

\noindent
This completes the loop update, and we can repeat the cycle consisting of: (i) performing a sweep of diagonal updates, (ii) constructing the linked 
vertex list, and (iii) tracing/flipping all the loops and updating $|\alpha\rangle$ accordingly. These procedures constitute one Monte Carlo sweep 
in the SSE method.

\paragraph{Equilibration and expansion truncation}

We now discuss the expansion cut-off $L$ (when the fixed-length scheme is used). We should make sure that $L$ exceeds the maximum expansion 
order $n$ that will be sampled, so that the truncation in practice is not an approximation. Since new diagonal operators are inserted through exchanges 
with the fill-in unit operators, we also want there to be a reasonably large number of these $s(p)=0$ elements present. But since the memory and CPU 
time scales with $L$, we should not make $L$ excessively large. One could in principle define an optimal $L$, for which the acceptance rate of the diagonal updates
is maximized. In practice, it does not matter much, however, whether $L$ is really optimal, since the off-diagonal updates are anyway more important in determining 
the autocorrelation times. We will therefore just make sure  that some significant fraction of the elements $s(p)=0$, so that the sampled $n$ never reaches 
$L$ and that many insertions of operators can be attempted in the diagonal update.

\begin{figure}
\includegraphics[width=9.5cm, clip]{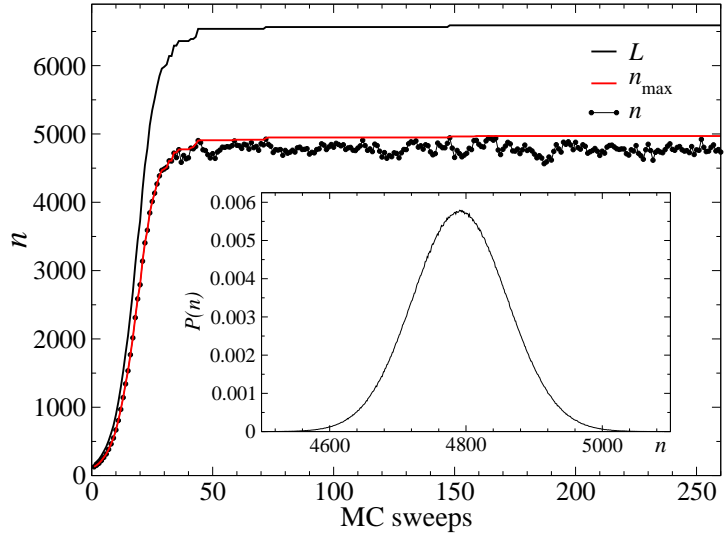}
\caption{Evolution of the expansion cut-off $L$ at the initial stage of an equilibration run of a $16\times 16$ Heisenberg system at $\beta=16$. The number 
of operators $n$ in the string after each Monte Carlo sweep is also shown, along with the maximum $n$ reached so far. The final cut-off after $5000$ sweeps 
was $L=6764$. The inset shows the distribution of $n$ in a subsequent run of $10^{7}$ MC sweeps.}
\label{ncut}
\end{figure}

We can aim for $L \approx (1+a) \langle n\rangle$ with, e.g., $a=1/3$. Since the relative fluctuations of $n$ are small, $\sim \sqrt{n}$, we can 
do this by letting $L \to a\times n$ and augmenting $S_L$ with additional $s(p)=0$ elements after each Monte Carlo sweep when $a\times n>L$ [setting
$s(p)=0$ for $L_{\rm old}+1<p<L_{\rm new}$ or distributing the zeros randomly. Fig.~\ref{ncut} shows an example of how this procedure works in practice, 
during equilibration of a $16\times 16$ system at inverse temperature $\beta=16$. Initially $L$ was set to $N/2=128$. The adjustment of $L$ quickly 
converges to an acceptable value, and, as shown in the inset, in a subsequent long simulation $n$ never comes close to $L$.

\subsubsection{Estimators for physical observables}
\label{estimators}

The autocorrelation time in SSE simulations with loop updates is typically very short; often just a few Monte Carlo sweeps or less \cite{syljuasen02,evertz1}.
Calculations of physical observables (``measurements'') should therefore normally be performed after every (or every few) Monte Carlo sweep. As discussed in
Sec.~\ref{montecarlo}, the autocorrelation times do not dictate how often measurements {\it may} be done, only with what frequency independent data are
generated. Autocorrelations, thus, affect the error bars but have no effect on the correctness of computed expectation values (i.e., it is not wrong to measure 
correlated data, although it may be wasteful if the measurements take a long time to evaluate) provided that the total simulation time is much longer than the 
autocorrelation time. Data binning should be used to compute error bars reliably, as also discussed in Sec.~\ref{montecarlo}. We here discuss several types of 
observables of interest in SSE calculations. More details and derivations of the expressions can be found in Refs.~\cite{sandvik90,sandvik92,sandvik97b}.

\paragraph{Energy and specific heat}

We already discussed some observables in Sec.~\ref{sec_sseformulation}; the internal energy (\ref{essenbeta}) and the related expectation value (\ref{hiexpsse})
of an individual operator in the hamiltonian, as well as the specific heat (\ref{cssenbeta}). Given that $\langle H\rangle=-\langle n\rangle/\beta$, it 
may seem surprising that the specific heat $C=(\langle H^2\rangle-\langle H\rangle^2)/T^2$ is not given just by the fluctuation in $n$, but there is also a 
term $-\langle n\rangle$. This is because $\langle H^2\rangle=\langle n(n-1)\rangle/\beta^2$, which can be easily shown using the same procedures leading to 
$\langle H\rangle=-\langle n\rangle/\beta$ in Sec.~\ref{sec_sseformulation}. Note that this expression includes the constant $1/4$ subtracted from each 
bond operator in (\ref{heishb12_2}), but that this constant cancels out in the specific heat expression (\ref{cssenbeta}). The specific heat in practice
becomes difficult to compute reliably (i.e., its statistical error is large) at low temperatures (where it becomes small), because it is the difference 
of two large numbers ($\sim N^2\beta^2$).

\paragraph{Diagonal operators}

Expectation values of operators diagonal in the $z$-component basis are also easy to evaluate, using averages over the propagated states;
\begin{equation}
\langle O_z\rangle = \frac{1}{L}\sum_{p=0}^{L-1} \langle \alpha(p)|O_z|\alpha(p)\rangle 
= \frac{1}{L}\sum_{p=0}^{L-1} O_z(p).
\label{ozsseaverage}
\end{equation}
It is not necessary to evaluate this average using all $p$, as successive propagated states differ by at most two flipped spins. One can instead use a partial 
summation over $p$, e.g., $O_z(0)+O_z(N)+O_z(2N)+\ldots$, to save time. In some cases, however, one may as well just propagate $O_z(p)$ from $p=0$ to $p=L-1$ 
by first computing $O_z(0)$ and then update the value $O_z(p)$ according to the spin flips occurring when propagating $|\alpha(p)\rangle$. For example, 
the fully averaged contribution to the expectation value of the squared staggered magnetization from an SSE configuration can be computed as follows:

{\code
\cia       $m=(1/2)\sum_i \phi_i\sigma(i)$;~ $m_{s2}=0$ \br       
\cia       {\bf do} $p=0$ {\bf to} $L-1$ \hfill   \{33\} \break 
\cib            {\bf if} ({\bf mod}(s(p),2)=1) {\bf then} \br
\cic                 $b=s(p)/2$;~ $\sigma(i(b))=-\sigma(i(b))$;~ $\sigma(j(b))=-\sigma(j(b))$ \br
\cic                 $m=m+2\phi_i\sigma(i(b))$  \br
\cib            {\bf endif} \br
\cib            {\bf if} ($s(p)\not=0$)~  $m_{s2}=m_{s2}+m^2$ \br
\cia        {\bf enddo} \br
\cia        $m_{s2}=m_{s2}/(nN^2)$
\code}

\noindent
Here $\phi_i = \pm 1$ is the staggered phase factor for site $i$ and $m$ contains the staggered magnetization, which 
evolves as the operator string is traversed. The change in $m$ when two antiparallel spins are flipped can be expressed as $2\phi_i\sigma(i(b))$, because 
$\phi_j\sigma(j(b))=\phi_i\sigma(i(b))$. The average of $m$ squared is accumulated as $m_{s2}$, in this example only using the $n$ propagated states generated 
by the original index sequence without the fill-in unit operators [i.e., skipping the steps where the operator $s(p)=0$]. One can also sum over all $L$ 
instances of the states, in which case the result would be divided by $L$ instead of $n$ on the last line. As written above, the special case $n=0$ is not 
treated correctly, but the code is easy to modify by just using $m_{s2}=m^2/N^2$ in that case.

The computational effort of the measurement in $\{33\}$ scales as $\beta N$, i.e., the same as the SSE sampling algorithm.  To compute the Fourier transform 
of the spin correlations (the static structure factor) at some arbitrary momentum [noting that $m_s^2$ corresponds to $S({\bf \pi})$, with ${\bf \pi}$ denoting
the staggered wave-vector, e.g., ${\bf \pi}=(\pi,\pi)$ in two dimensions], e.g., for use in the correlation-length definition (\ref{xiadef}) or (\ref{xibdef}), 
one can use a similar procedure for the real and imaginary parts of $m({\bf q})$,
\begin{equation}
m({\bf q}) = \sum_{\bf r}S^z_{\bf r}\cos({\bf q}\cdot {\bf r}) + i\sum_{\bf r}S^z_{\bf r}\sin({\bf q}\cdot {\bf r}),
\label{magqdef}
\end{equation}
with $\phi_i$ in $\{33\}$ replaced by the corresponding sine and cosine factors [and explicitly considering the changes from both the flipped spins 
at $i(b)$ and $j(b)$]. The structure factor is then accumulated according to $S({\bf q})=S({\bf q})+{\rm Re}^2\{m({\bf q})\}+{\rm Im}^2\{m({\bf q})\}$.
 
For computing the full correlation function,
\begin{equation}
C_z({\bf r}_{ij})=\langle S^z_iS^z_j\rangle,
\label{czrheis}
\end{equation}
averaged over all $i,j$ [or the Fourier transforms for all ${\bf q}$, which can be written as $\langle m({\bf q})m({\bf -q})\rangle$], it would not be 
practical to update all these $\propto N$ different functions after each spin flip as in $\{33\}$. It is then better to compute all the correlations from 
scratch in, e.g., every $N$:th propagated state. This still leads to a rather expensive scaling $\propto \beta N^2$ of the effort to measure all the 
correlations---a factor $N$ worse than the sampling. One should then judge whether it is better to compute only a subset of the correlations [e.g., along 
some lines in the $(x,y)$ plane] or just measure them less frequently, in order for the measurements not to completely dominate the calculation.

\paragraph{Susceptibilities}

Another important class of observables are generalized susceptibilities, i.e., linear response functions of the form
\begin{equation}
\chi_{AB}=\frac{\partial \langle A(b)\rangle}{\partial b},
\end{equation}
where $b$ is the prefactor in a field term $bB$ added to the hamiltonian and $A$ is the operator whose response to this perturbation we want to compute. 
For example, we may be interested in the response $\chi_{ij}$ at site $j$ when a magnetic field acts only on site $i$, in which case $A=S^z_j$ and $B=S^z_i$. 
Such a susceptibility is given by the Kubo formula \cite{sandvik90}
\begin{equation}
\chi_{AB}=\int\limits_0^\beta d\tau \langle A(\tau)B(0)\rangle - \beta \langle A\rangle\langle B\rangle,
\label{kuboformula}
\end{equation}
where $A(\tau)={\rm e}^{-\tau H}A{\rm e}^{\tau H}$. If both $A$ and $B$ are diagonal, this Kubo integral can be evaluated in SSE simulations using the 
generic formula
\begin{equation}
\int\limits_0^\beta d\tau \langle A(\tau)B(0)\rangle=\left \langle \frac{\beta}{n(n+1)}\left ( \sum_{p=0}^{n-1}A(p) \right )\left ( \sum_{p=0}^{n-1}B(p) 
\right )\right \rangle + \left \langle \frac{\beta}{n+1}\sum_{p=0}^{n-1}A(p)B(p) \right \rangle.
\label{chiabssedef}
\end{equation}
The sums over $A(p)$ and $B(p)$ can be computed using code similar to $\{33\}$. 

We are often interested in susceptibilities for which $A=B$, e.g., the magnetic response $\chi({\bf q})$ at wave-vector ${\bf q}$, in which case 
$A=B=m({\bf q})$, with $m({\bf q})$ the Fourier transform of the spin configuration given by (\ref{magqdef}). Since the exact value of $\chi({\bf q})$ 
must be real-valued, one only has to compute the real part of (\ref{chiabssedef}). If all local response functions $\chi_{ij}=\chi({\bf r}_{ij})$ are 
computed in a simulation, the Fourier transform $\chi({\bf q})$ can later be evaluated using these. These response functions are directly accessible in 
NMR experiments; see Ref.~\cite{sandvik95} for an example.

Note that if $A$ and $B$ commute with the hamiltonian, then $A(\tau)=A(0)=A$ [and $A(p)=A$ and $B(p)=B$ are independent of $p$ in (\ref{chiabssedef})] 
and the susceptibility reduces to the classical expression $\chi_{AB}=\beta(\langle AB\rangle-\langle A\rangle\langle B\rangle)$. In practice, for the
Heisenberg model the only case where this form applies is the uniform magnetic susceptibility;
\begin{equation}
\chi=\chi(0)=\frac{\beta}{N}\langle M_z^2\rangle,~~~~~~M_z=\sum_{i=1}^N S^z_i.
\label{chi0sse}
\end{equation}
Here the term $\beta\langle M_z\rangle^2$ in (\ref{kuboformula}) vanishes, since $H$ does not include a magnetic field.

If the operators $A$ and $B$ in (\ref{kuboformula}) are not diagonal, the susceptibility is more complicated in general. Here we will only discuss an important 
special case, in which the estimator actually is very simple. If the operators involved are two terms of the hamiltonian, as defined in Eqs.~(\ref{heishb12_1})
and (\ref{heishb12_2}) and here referred to just as $H_A$ and $H_B$ for any two instances of those (any two diagonal or off-diagonal bond operators), then 
the susceptibility measurement just involves counting the numbers $N(A)$ and $N(B)$ of those operators in the sampled SSE operator sequences;
\begin{equation}
\chi_{H_aH_B} = \frac{1}{\beta}\left [ \langle N(A)N(B)\rangle -\delta_{A,B}\langle N(A)\rangle \right ].
\label{chioddssedef}
\end{equation}
The most important example of this type is the current susceptibility $\chi_{I_xI_x}$, where the spin current operator $I_x$ (here in the lattice 
$x$ direction, for definiteness) is defined by
\begin{equation}
I_x = \sum_{i=1}^N[ S^-({\bf r}_i)S^+({\bf r}_i+\hat {\bf x}) - S^+({\bf r}_i)S^-({\bf r}_i+\hat {\bf x}) ],
\label{ixdefsse1}
\end{equation}
where we assume, for simplicity, a hamiltonian with only nearest-neighbor interactions. For longer-range interactions, there would be corresponding current
terms between the same site pairs as in the hamiltonian. Although (\ref{ixdefsse1}) is not exactly a sum of the off-diagonal operators (\ref{heishb12_2}) 
used in the SSE sampling, those operators can be written as sums of two parts,
\begin{equation}
H_{2,b}=H^+_{b}+H^-_{b},~~~~~~~H^+_{b}=S^+_{j(b)}S^-_{i(b)},~~~~H^-_{b}=S^-_{j(b)}S^+_{i(b)},
\label{h12plusminus}
\end{equation}
where the site pairs $[i(b),j(b)]$ in each case are assumed to be ordered in such a way that the $+$ and $-$ terms transport one unit of spin in the 
positive and negative direction ${\bf r}_j-{\bf r}_i$ of the bond, respectively. A key point here is that, although the SSE configurations contain the 
full off-diagonal operators $H_{2,b}$, only the $+$ or $-$ part in (\ref{h12plusminus}) contributes in each instance (with the other part destroying the 
state it acts on). One can therefore use (\ref{chioddssedef}) to evaluate local current-current response functions of the form
\begin{equation}
\Lambda_{b_1b_2} = \int\limits_0^\beta d\tau \langle I_{b_1}(\tau)I_{b_2}(0)\rangle,
\label{lambdab1b2}
\end{equation}
where $I_{b}$ here denotes the current operator at bond $b$. Using (\ref{chioddssedef}) one obtains
\begin{equation}
\Lambda_{b_1b_2} = \langle [N^+(b_1)-N^-(b_1)][N^+(b_2)-N^-(b_2)]\rangle - \delta_{b_1b_2}\langle N^+(b_1)-N^-(b_1)\rangle,
\label{lambdab1b2res}
\end{equation}
where $N^+(b)$ and $N^-(b)$ denote the number of operators in the SSE operator sequence transporting spin in the positive and negative direction, 
respectively, across the bond $b$. The meaning of this becomes clear when looking at a graphical representation of an SSE configuration, such as 
Fig.~\ref{sseconfig1}. There one just has to sum the number of off-diagonal events in which an $\uparrow$ spin is moved to the right 
(the positive direction) minus the number of left-moving events. The term $N^+(b)-N^-(b)$ is zero, unless the winding number in the lattice
direction defined by bond $b$ is not zero, which can be seen in the illustration of winding numbers in Fig.~(\ref{pathintegral}). It can be noted
that the form (\ref{lambdab1b2res}) of the current response is identical in SSE and path integral (world line) methods.

\paragraph{Spin stiffness}

We discussed the spin stiffness in the context of the classical  XY model in Sec.~\ref{stiffkt}. The basic definitions of this quantity at $T=0$, 
Eq.~(\ref{stiffe0def}), and at $T>0$, Eq.~(\ref{stifffdef}), are identical for quantum spin systems (XY or Heisenberg models) as well as bosonic systems 
more broadly (where the analogous quantity is the superfluid phase stiffness, which is proportional to the superfluid density). Deriving Monte Carlo 
estimators for SSE or world line methods, one finds that the term $\beta \langle I_x^2\rangle$ in (\ref{rhosestimator2cl}), which is a classical 
response function, should be replaced by the corresponding quantum mechanical Kubo formula, giving the spin stiffness in the form
\begin{equation}
\rho_s = \frac{1}{N}\langle H_x\rangle - \frac{1}{N}\int\limits_0^\beta d\tau \langle I_x(\tau)I_x(0)\rangle,
\label{rhosdefsse}
\end{equation}
where we assume that the phase twist is imposed in the lattice $x$ direction. The Kubo integral here consists of a sum of bond-current response functions 
of the form (\ref{lambdab1b2}), and using the result (\ref{lambdab1b2res}) one finds that the energy term $\langle H_x\rangle$ is exactly canceled 
by the $\delta$-function terms coming from (\ref{lambdab1b2res}). This leads to a very simple expression for the spin stiffness in terms of the SSE
(or world line) configurations:
\begin{equation}
\rho_s = \frac{1}{\beta N}\langle (N^+_x-N^-_x)^2\rangle .
\label{rhosdefsseestim}
\end{equation}
Here $N^+_x$ and $N^-_x$ denote the total number of operators transporting spin in the positive and negative $x$ direction, respectively. For a spatially 
isotropic system in two or three dimensions, this can of course be averaged over the dimensions, whereas for an anisotropic system the stiffness depends 
on the direction. 

The spin stiffness is often expressed as the fluctuation of the winding number (here in the $x$ direction, with analogous expressions for other 
directions),
\begin{equation}
W_x = \frac{1}{L_x} (N^+_x-N^-_x),
\label{wxdef}
\end{equation}
i.e., the size normalized current ($W_x=0,\pm1,\pm2,\ldots$). For a $d$-dimensional system with $N=L^d$ the stiffness is just 
$\rho_s=\langle W^2_a\rangle/\beta$ for any direction $a$, but for other shapes (e.g., $N=L_x\times L_y$ with $L_x\not=L_y$) the lengths also 
enter and (\ref{rhosdefsseestim}) looks simpler.

In derivations of the spin stiffness, it is normally for simplicity assumed that the spins order in the XY plane (in spin space), whereas in a Heisenberg 
model the direction of the order parameter in spin space is not restricted to this plane (the symmetry is not broken) in an SSE simulation of a finite 
system. A simple rotational averaging argument shows that the expression (\ref{rhosdefsseestim}) should be multiplied by $3/2$ in order to obtain the 
correct stiffness of a symmetry-broken state (i.e., when considering the thermodynamic limit). The expression (\ref{rhosdefsseestim}) applies also to the 
quantum XY model \cite{harada97}, where the result should not be multiplied by $3/2$.

A winding number estimator completely analogous to (\ref{rhosdefsseestim}) was originally derived in the context of the superfluid phase stiffness of bosons 
in continuous space \cite{pollock87}. The role of global cyclical permutations of particles in superfluidity dates back to Feynman's pioneering work on path 
integrals \cite{feynman53}.

\subsubsection{Improved estimators}
\label{improvedestimators}

The operator-loop update in the SSE method (as well as loop updates more broadly \cite{evertz1,evertz93}) is an example of a cluster update. Such
non-local updates were first developed for classical Monte Carlo simulations of the Ising model \cite{swendsenwang}. One aspect of cluster methods is that 
it is possible to take averages of estimators for physical quantities over all orientations of the clusters, because the configuration weight does not change 
upon flipping a cluster. This is immediately clear in SSE simulations of $S=1/2$ Heisenberg models, because the weight (\ref{wsseheis1}) only depends on the 
number of operators $n$ in the sequence, which does not change when a loop is flipped. If the number of clusters (here operator-loops) is $m$, then the total 
number of equal-weight configurations is $2^m$, and the average over all of these configurations can provide a much less noisy estimator than one depending on 
just a single configuration. The crucial point here is that, for many important quantities, this average can be computed analytically, and the resulting 
{\it improved estimator} is of a simple form that can be evaluated rapidly in simulations. Here we only discuss the rather simple cases of the the static 
(equal-time) structure factor and the uniform magnetic susceptibility. For improved estimators for some other quantities, see the review article by 
Evertz \cite{evertz1}.

\begin{figure}
\includegraphics[width=5.5cm, clip]{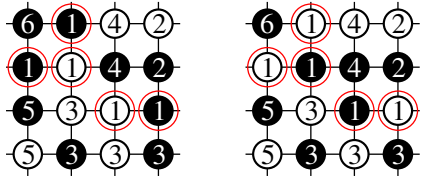}
\caption{Example of clusters formed by space-time loops passing through a propagated state $|\alpha(p)\rangle$ (for arbitrary fixed $p$). Here there are
six such clusters, labeled $1,\ldots,6$. Open and solid circles correspond to $\uparrow$ and $\downarrow$ spins, respectively, in $|\alpha(p)\rangle$.
When a loop is flipped, all spins in the corresponding cluster are also flipped, as indicated here with two different configurations corresponding to
the two states of cluster $1$ (the sites enclosed by larger circles). Note that the spins within each cluster are always in one of the two staggered 
configurations.}
\label{impclusters}
\end{figure}

Consider a propagated state $|\alpha(p)\rangle$, e.g., the stored $|\alpha(0)\rangle$. In the linked vertex representation of the SSE
configuration, illustrated in Fig.~\ref{sseloop}, there is a loop passing through each of the spins in this state (with spins without operators 
acting on them also considered as individual loops). The same loop can go through many spins in $|\alpha(0)\rangle$, and all spins belonging to the
same loop form a cluster, in the sense that if the loop is flipped all the spins in the clusters are flipped simultaneously. Note that the loops
are objects in space-time, while the clusters discussed here are defined on a cut at fixed time (here propagation index $p$). A cluster 
can consist of several parts that appear disconnected in space, since such pieces can be connected in the larger space-time volume where the loops 
exists. An example of clusters on a 2D lattice is shown in Fig.~\ref{impclusters}

Since we are dealing with a bipartite lattice, and because the loop structure is such that the spin on a vertical loop segment (referring to pictures
such as Fig.~\ref{sseloop}) changes each time one changes direction when moving along a loop, the spins within a cluster formed at a given state
$|\alpha(p)\rangle$ always have a staggered structure. The staggered magnetization $m_s(j)$ of a cluster labeled $j$, with $j=1,\ldots,C$, where $C$ is 
the total number of clusters, is then $m_s(j)=\pm n_j$, where $n_j$ is the number of spins in cluster $j$. For a given configuration, the total staggered 
magnetization $M_s=\sum_{j=1}^{C} m_s(j)$. When averaging the square of this sum over all the different realizations of cluster orientations, the cross 
terms $\langle m_s(i)m_s(j)\rangle=0$ (for $i \not =j$). One is then left with just the $i=j$ contributions, and the staggered structure 
factor is simply given by
\begin{equation}
S({\bf \pi}) = \frac{1}{4N}\left \langle \sum_{j=1}^{C} n_j^2 \right \rangle.
\end{equation}
Structure factors at other wave-vectors ${\bf q}$ are only marginally more complicated, demanding in place of the cluster sizes $n_j$ a summation 
over each cluster of the phases $\phi_{\bf r} {\rm exp}(i{\bf q}\cdot {\bf r})$, where ${\bf r}$ refers to sites on a given cluster and $\phi_{\bf r}=\pm 1$ 
is the staggered phase factor, which takes care of the staggered spin structure within the clusters (and the denominator $4$ corrects for the fact that 
the spin values are $\pm 1/2$). One can here also use the fact that the true structure factor must be real-valued for any ${\bf q}$.

In principle, equal-time correlation functions such as the structure factor can also be averaged (fully or partially) over the propagation index $p$, 
as in Eq.~(\ref{ozsseaverage}). This, however, requires more work for the improved estimator than in code $\{33\}$ for the simple estimator, because 
it takes some book keeping during the loop update to construct the clusters for several fixed $p$, and doing so may not always pay off.
Without this averaging, however, a simple $p$-averaged estimator, such as the one implemented in code $\{33\}$, may actually give better results at 
low-temperatures, where the gain due to averaging can be very significant. The case ${\bf q}=0$ is special in this regard, because this corresponds to 
the total squared magnetization, which is a conserved quantity (i.e., independent on the SSE propagation index $p$), and no further averaging 
over $p$ can then be done to improve the statistics further. The optimal estimator for the uniform susceptibility (\ref{chi0sse}) is therefore
\begin{equation}
\chi = \frac{\beta}{4N}\left \langle \sum_{j=1}^C \left ( \sum_{i=1}^{n_j}\phi_i \right )^2 \right \rangle.
\label{chiloopestim}
\end{equation}
Susceptibilities at other wave-vectors involve the full space-time loop structure, not just the clusters (cut through the loops) formed at
a fixed state. For example, the staggered susceptibility is given by the sum of the squares of all the loop sizes \cite{evertz1}.

\subsubsection{Program verification}

QMC programs should always be verified by comparing results for small systems with exact diagonalization data. When correctly implemented,
the SSE method should be exact, which means that the deviation of a computed quantity from its true value should be purely statistical, due to the 
finite number of sampled configurations. We have discussed how to quantify the statistical fluctuations in terms of ``error bars'' in Sec.~\ref{montecarlo}. 
Deviations beyond the error bars are due either to programming errors of flaws in the random number generator used. While most programming errors would lead to 
obviously wrong results, there are also possible subtle errors that may only lead to minute deviations from exact results for small lattices. Such systematical 
errors may grow with the system size, with potentially grave consequences. The same may be said about random number generators; bias effects due to imperfect random 
numbers may be very small for small lattices, but can become more significant for larger systems. It is therefore important to make comparisons with 
exact diagonalization results based on very long SSE runs, to detect possible small deviations. While it is impossible to strictly prove that a program is correct 
in all respects, agreement to within very small error bars with exact data makes this very likely.

\paragraph{Test results}

\begin{figure}
\includegraphics[width=10cm, clip]{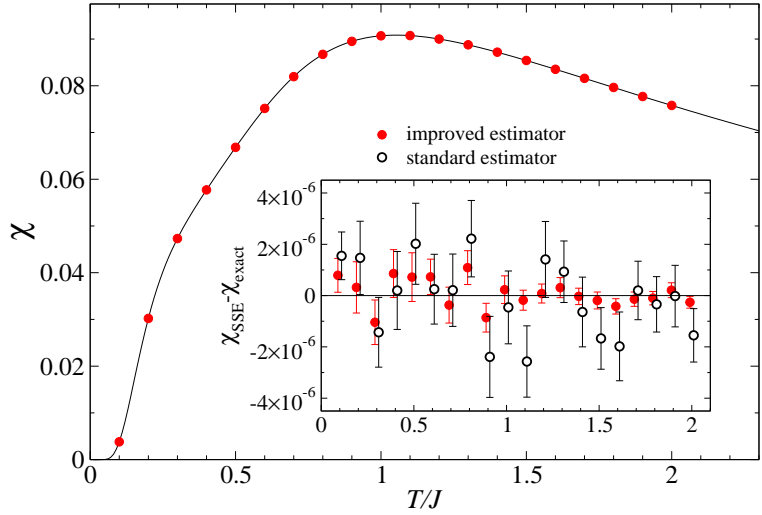}
\caption{The uniform susceptibility of a $4\times 4$ Heisenberg system versus the temperature. The curve is the exact result from a full diagonalization.
The points are SSE results based on $10^{10}$ updating sweeps for each $T$. The error bars cannot be resolved on this scale. The inset shows the deviation
of $\chi$ from the exact result with error bars, for both simple and improved (loop) estimators. The data points have been slightly shifted off their actual 
$T$ values in order for the error bars of the two estimators not to overlap.}
\label{l4comp}
\end{figure}

We now discuss some test results for 1D and 2D Heisenberg models. The random number generator used in these calculations (and most other SSE calculations 
discussed  in these lecture notes) was a simple $64$-bit linear congruential generator with multiplier $2862933555777941757$ \cite{knuth88} and addition of
$1013904243$.

Fig.~\ref{l4comp} shows the susceptibility per spin of a $4\times 4$ system. For each temperature $10^{10}$ updating sweeps and measurements were carried
out, with the data subdivided into $100$ bins for computing the statistical errors. For $T=J/10$, this required approximately 100 CPU hours on a mid-range PC
workstation. The error bar (defined as one standard deviation of the estimated fluctuation of the average, as discussed in Sec.~\ref{montecarlo}) is typically 
$\approx 10^{-6}$ when the standard estimator for $\chi$ is used, and even smaller with the improved estimator (\ref{chiloopestim}). The relative error (the 
error bar divided by $\chi$) is $\approx 5\times 10^{-6}$ for $T/J\approx 1$. The deviations of the averages are completely consistent with the size of the 
error bars. Recall that for correctly computed statistical errors, one should expect about $2/3$ of the data points to bracket the true data within one 
error bar. The gain in precision with the improved estimator can be much larger than in Fig.~\ref{l4comp} for larger systems.

\begin{figure}
\includegraphics[width=8.75cm, clip]{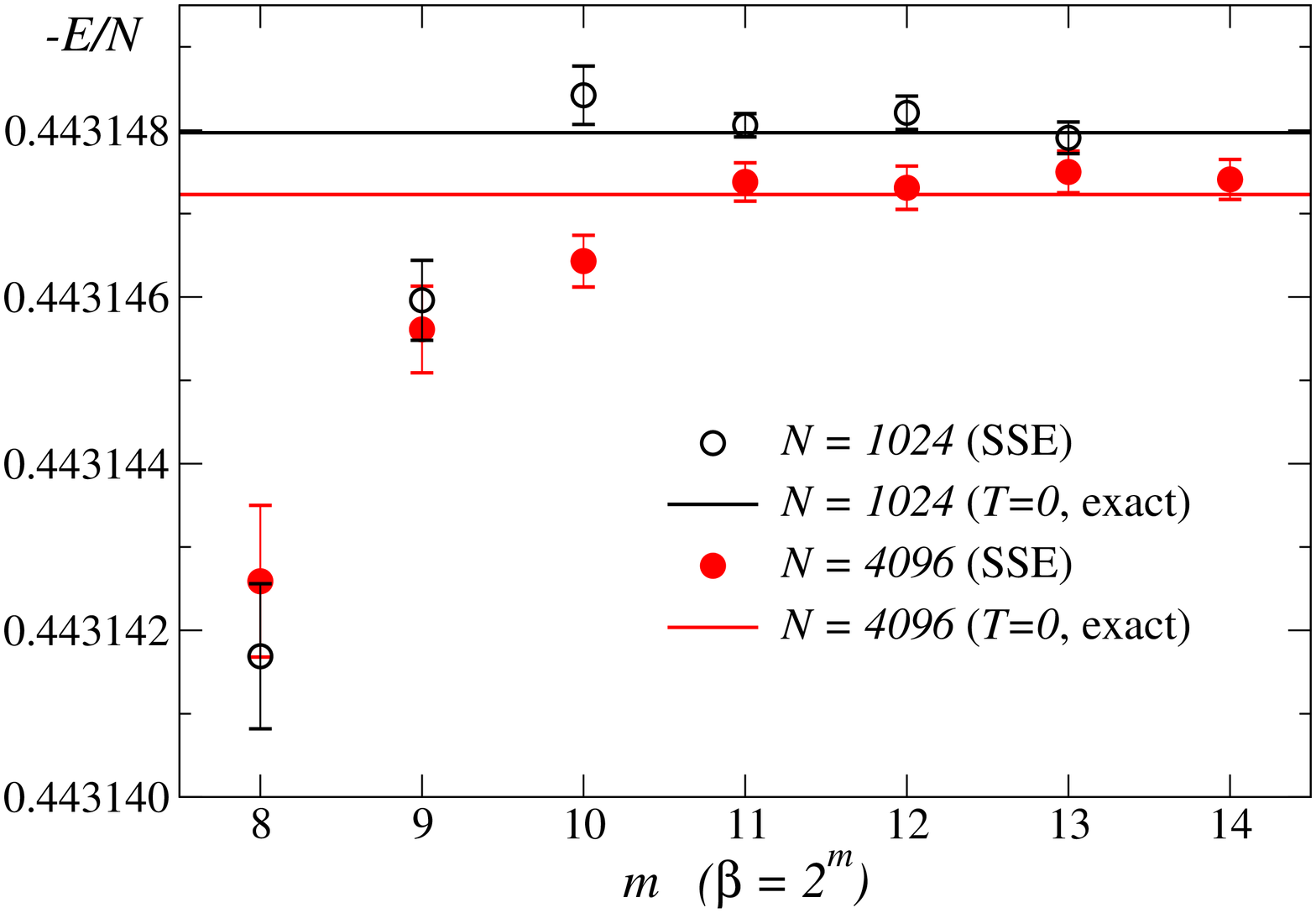}
\caption{SSE results for the internal energy per spin at several inverse temperatures $\beta=2^m$ for Heisenberg chains of length $N=1024$ and $4096$ compared 
with the corresponding exact (Bethe ansatz) ground state energies. The SSE results were obtained using $1-2 \times 10^6$ updating sweeps for each $m$, which
required several hundred CPU hours for the largest $N,\beta$.}
\label{ecomp}
\end{figure}

While it is essential to confirm the unbiased nature of the SSE calculations on small lattices, one might still worry about potential problems with the random number
generator for larger systems. In one dimension, we can also test SSE calculations against exact Bethe ansatz results for very long chains in the ground state. 
To approach the ground state in SSE calculations, it is convenient to use inverse temperatures of the form $\beta=2^m$ and go to sufficiently large $m$ for 
calculated quantities to become $m$ independent. This approach is illustrated in Fig.~\ref{ecomp}, which shows the internal energy, $E=\langle H\rangle$, computed 
using the simple expansion-order estimator (\ref{essenbeta}) for chains of length $N=1024$ and $4096$. Numerically computed Bethe ansatz results for these chain 
lengths are listed in \cite{karbach98}. The agreement between the calculations is perfect within error bars, with no detectable temperature dependence for the 
last three or four points for each $N$. This good agreement, to within relative statistical errors as low as $5\times 10^{-7}$, shows quite convincingly that
the calculation is for practical purposes completely unbiased and that ground states of even quite large systems can be studied.

\subsection{Applications of SSE to 1D and 2D systems}
\label{sec_sseapplications}

We next discuss several illustrative results for 1D and 2D Heisenberg models obtained with the SSE method. Results showing clearly the logarithmic corrections 
to critical behavior in the 1D chain are presented in \ref{sec_chainsse}. The qualitatively different ground states of $n$-chain ladder systems with even and
odd $n$ are discussed in \ref{sec_ladders}. The sublattice magnetization and some other low-energy parameters associated with the N\'eel state are computed for 
the standard 2D Heisenberg model in \ref{sec_2dheis}, and in \ref{sec_dimerized} dimerization is introduced in this model, to drive a quantum phase transition into 
a plain non-magnetic state. Quantum critical finite-size scaling behaviors of various quantities are analyzed. The more complex case of a N\'eel--VBS transition 
is investigated in \ref{sec_jqresults}, using two types of J-Q models that exhibit, respectively, continuous and first-order transitions.

\subsubsection{The Heisenberg chain}
\label{sec_chainsse}

\paragraph{Spin correlations at $T=0$}

\begin{figure}
\includegraphics[width=8.5cm, clip]{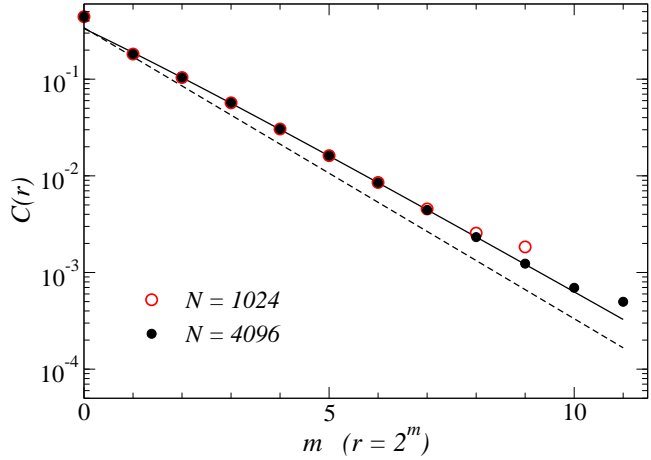}
\caption{Spin correlation function at distances $r=2^m$ for chains of length $N=1024$ and $4096$. The error bars are smaller than the symbols. The solid curve 
is of the expected form $Ar^{-1}\ln(r/r_0)^{1/2}$, with $A=0.21$ and $r_0=0.08$. The dashed curve shows the form $\propto r^{-1}$ for comparison. These results 
were obtained using inverse temperatures $\beta=2^{13}$ and $2^{14}$ for $N=1024$ and $4096$, respectively, which is sufficient for $T\to 0$ convergence.}
\label{crhchainsse}
\end{figure}

In Sec.~\ref{sec_hchain} we discussed Lanczos results for the spin correlation function of the Heisenberg chain and saw some hints of the expected logarithmic
correction to the $\sim 1/r$ critical behavior (Fig.~\ref{cor}). The system sizes accessible with the Lanczos method are not sufficient for studying these 
scaling corrections quantitatively, however. As we saw above, with the SSE method unbiased studies of the ground state is possible for chains of several 
thousand spins (with careful checks of the convergence  to the $T \to 0$ limit). Fig.~\ref{crhchainsse} shows the spin correlations for $N=1024$ and $4096$ 
at distances $r=2^m$, graphed on a log-log scale. To save time, only the correlations at these distances were computed [for a scaling $N\log(N)$ of the time
to carry out spatially averaged measurements]. The results for the two system sizes coincide closely for $r$ up to $2^7$, indicating convergence to the infinite 
size values up to this distance for $N=1024$ (and therefore up to $r \approx 2^9$ for $N=4096$, since the convergence behavior should scale approximately 
linearly with $N$). 

The expected form $|C(r)|=A\ln^{1/2}(r/r_0)r^{-1}$ \cite{affleck2,singh1,giamarchi} is very well reproduced up to $r=2^9$ for $N=4096$. The parameters 
$A$ and $r_0$ obtained from a fit are listed in the figure caption. If one leaves the exponent $\sigma=1/2$ of the logarithm as a free parameter to
be obtained from the data based on a fit, the exponent indeed comes out close to $0.5$, but with a rather large error bar of, roughly, $\pm 0.1$. To 
really investigate the exponent carefully, one should further increase the chain length (which is possible in principle).

If one did not know about the existence of a log correction and tried to extract the form of the spin correlations on the basis of numerical calculations alone,
one might at first sight conclude that the decay is $\propto 1/r^\alpha$ with $\alpha \approx 0.85$, based on the data in Fig.~\ref{crhchainsse}. There 
are, however, small but significant deviations from a pure power-law, which can be detected only if the relative statistical errors are sufficiently small. In 
the data shown in the figure, the error bars are too small to discern. While the absolute errors are small, typically $\approx 2\times 10^{-6}$ for $N=4096$ 
(based on approximately $10^{6}$ Monte Carlo sweeps at $\beta=2^{14}$ and using an improved estimator), the relative error is much larger, about $0.002$ 
for $r=2^7$, growing to $0.02$ for the longest distance $r=2^{11}$. 

\paragraph{Low-temperature magnetic susceptibility}

\begin{figure}
\includegraphics[width=9.5cm, clip]{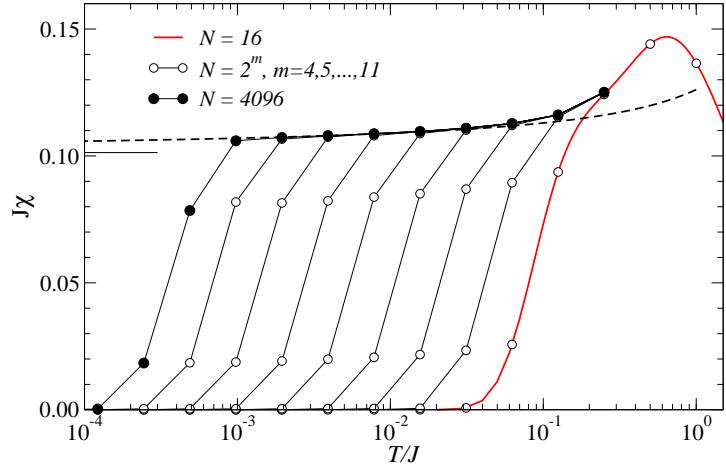}
\caption{The circles show SSE results for the uniform susceptibility of the Heisenberg chain at temperatures $T$ of the form $2^{-m}$, $m=0,1,2,...$ for 
different power-of-2 chain lengths up to $N=4096$. Error bars are much smaller than the circles. The exact $N=16$ result is shown as the solid curve and 
the dashed curve shows the low-$T$ form (\ref{chithchain}). The asymptotic $T \to 0$ value, $J\chi=1/\pi^2$, is indicated by the short horizontal line 
segment on the left side of the graph.}
\label{xchainsse}
\end{figure}

Logarithmic corrections are also important in the Heisenberg chain at $T>0$. The most prominent example is the uniform magnetic susceptibility, for which a 
renormalization-group study of the low-energy field-theory predicted the form \cite{eggert94,nomura93}
\begin{equation}
\chi(T) = \frac{1}{2\pi c} + \frac{1}{4\pi c \ln(T_0/T)},
\label{chithchain}
\end{equation}
where the $T\to 0$ value $x(0)=1/{2\pi c}$, where $c=J\pi/2$ is the spinon velocity, agrees with the Bethe ansatz solution of the ground state of the 
Heisenberg chain. The Bethe ansatz can also be extended to $T>0$. Good agreement with the asymptotic expression (\ref{chithchain}), with the parameter 
$T_0=7.7$ (adjusted to fit the Bethe ansatz results), was found at low temperatures ($T/J < 0.02$) \cite{eggert94}. Note that the logarithmic correction 
in (\ref{chithchain}) implies a very slow convergence to the $T=0$ limit, and it would therefore be difficult to extrapolate numerical results without 
knowing about the presence of this kind of correction. A higher-order logarithmic correction to (\ref{chithchain}) is also known \cite{nomura93}, but 
here we will just consider the leading-order correction.

Fig.~\ref{xchainsse} shows SSE results for the susceptibility obtained using chain lengths of the form $N=2^n$ and temperatures $T/J=2^{-m}$. Due to
the excitation gap in chains with finite $N$ (discussed in Sec.~\ref{sec_hchain}), $\chi$ decays exponentially to zero below a temperature $T \sim 1/N$.
The log-lin scale used in the figure makes these finite-size effects very clear and also shows how the results converge rapidly to the thermodynamic
limit form once the finite-size gap becomes smaller than $T$. The agreement with the form (\ref{chithchain}), using the value $T_0 =7.7$ determined in
Ref.~\cite{eggert94}, is very good for $T/J$ below $0.05$ (in fact, for some unknown reason, the agreement appears to be even better than in 
Ref.~\cite{eggert94}).

The main utility of the form (\ref{chithchain}), beyond its role in establishing the correctness of the low-energy field-theory, is that it is valid 
not only for the simple Heisenberg chain considered here, but for any 1D spin system (at sufficiently low temperatures) which is in the same phase as the 
Heisenberg chain. This includes the frustrated J$_{\rm 1}$-J$_{\rm 2}$ chain discussed in Sec.~\ref{sec_j1j2chain}, for $J_2/J_1$ less than the dimerization 
point $(J_2/J_1)_c \approx 0.241$. Exactly at the dimerization point the leading logarithmic correction should vanish \cite{eggert96b} (but there are still 
other, higher-order corrections), while above the transition point a spin gap opens and $\chi \to 0$ even for $N=\infty$. The asymptotic form (\ref{chithchain})
should be valid throughout the QLRO($\pi$) phase of the chain including longer-range interactions [the model discussed in Sec.~\ref{sec_longrange}], as well as 
in many other systems. The parameters $c$ and $T_0$ depend on the model parameters and fitting of numerical data provides a way to extract, in particular, the 
velocity $c$. As we will see in the next section, ladder systems consisting of an odd number of chains also are in the same critical phase as the single chain, 
and (\ref{chithchain}) applies also there.

\subsubsection{Ladder systems}
\label{sec_ladders}

A ladder lattice consists of a fixed number $L_y$ of coupled 1D chains (often referred to as the legs of the ladder) with the chain length $L_x$ taken to 
infinity (or, in practice, $L_x/L_y$ sufficiently large to give results converged to this limit). 
Strongly-correlated quantum systems in this geometry \cite{dagotto1,dagotto92,sheng09} play an important role as a means of 
interpolating between one and two dimensions. Many materials exhibit a structure of weakly coupled ladders with small $L_y$, which motivates studies of 
$L_y=2,3$, etc. On a more fundamental level, it is very interesting to see how the special properties of 1D systems evolve with increasing $L_y$ and approach 
the 2D limit \cite{frischmuth96}. Here we will investigate some of the most essential properties of Heisenberg ladders with $L_y$ up to $6$. 

\begin{figure}
\includegraphics[width=11cm, clip]{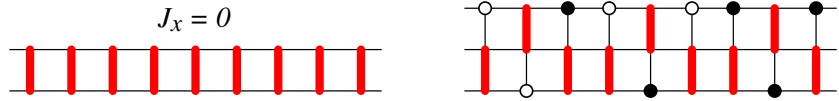}
\caption{Illustration of the ground states of the 2-leg and 3-leg Heisenberg ladders at inter-rung coupling $J_x=0$. The vertical bars indicate valence 
bonds (singlets), and the circles show unpaired $\up$ and $\dn$ spins. The ground state of the 2-leg ladder is a unique singlet-product, whereas in the 
3-leg case each rung-state is two-fold degenerate, with $S^z_{\rm rung} = \pm 1/2$. Each of these states is a symmetric combination of the two states with 
the unpaired spin on the upper and lower chain. Here a random configuration of the location and $S^z$ of the unpaired spin is shown for each rung.}
\label{laddervb}
\end{figure}

The most important aspect of the physics of $S=1/2$ Heisenberg ladders is that they have completely different low-energy properties for even and odd $L_y$ (the
number of sites on each rung of the ladder). For even $L_y$ there is always a spin gap (which vanishes as $L_y \to \infty$), while odd-$L_y$ systems are gapless
and have properties similar to a single 1D chain (below an energy or temperature scale which vanishes as $L_y \to \infty$). This difference can be understood roughly 
based on a simple picture of valence bonds \cite{white94}, illustrated in Fig.~\ref{laddervb}. If there is no coupling between the rungs ($J_x=0$, $J_y>0$), 
the ground state of the 2-leg ladder is just the product of rung singlets---a unique (non-degenerate) state with a gap $J_y$ to the first excited state (in
which one of the singlets is promoted to a triplet). For the 3-leg ladder, on the other hand, the ground state of an individual rung is two-fold degenerate, 
with $S_z=\pm 1/2$. When coupling the rungs ($J_x>0$), this degeneracy is lifted, but, for any ratio $J_x/J_y$, there is a remnant of the degeneracy, which 
can be understood as arising from $L_x$ interacting $S=1/2$ degrees of freedom. This leads to low-energy properties similar to those of a single $S=1/2$ chain. 
In the 2-leg ladder, the singlet-triplet excitation gap remains non-zero for any $J_x/J_y$, and the low-energy excitations above this gap form a band of 
propagating rung triplets \cite{troyer94}. The spin correlations are exponentially decaying with distance, with the correlation length diverging as $L_y \to \infty$. 

In the valence bond basis (discussed in Sec.~\ref{vbsrvb}), the 2-leg ladder with $J_x>0$ is still dominated by short bonds (one can say that it is an RVB spin 
liquid, but, due to the constraints of the ladder geometry, it is also appropriate to call it a VBS), whereas the critical state of the 3-leg ladder 
requires bond probabilities that decay with the bond length as a power-law \cite{white94} (and such a state may be called a critical RVB state). The same 
pictures remain valid also for the ground states of ladders of larger width, but the energy scale associated with ladder behavior vanishes as $L_y \to \infty$, 
with cross-overs into 2D behavior at higher energies. The relevant energy scale is the spin gap for even $L_y$ and the spin stiffness for odd $L_x$---as we 
discussed in connection with the KT transition in Sec.~\ref{stiffkt}, power-law correlations can be sufficient to sustain an non-zero spin stiffness, and
this is the case with the $\sim 1/r$ correlations in the single chain and odd-leg ladders.

Here we will discuss results only for spatially isotropic couplings; $J_x = J_y$. For even $L_y$, the behavior is similar for periodic and open boundary
conditions in the $y$ direction (and for $L_y=2$ periodic boundaries only corresponds to doubling $J_y$), while for odd $L_y$ the gapless nature of the
system for all $J_x/J_y$ applies only for open $y$ boundaries. In the ``tube'' geometry (periodic $y$ boundaries), odd $L_y$ leads to frustration, and the
behavior is then much more complex, with several possible ground state phases \cite{sakai08}. In the SSE calculations discussed below, open $y$ boundaries 
were used in all cases, while the $x$ boundaries were periodic. 

\paragraph{Spin correlations}

\begin{figure}
\includegraphics[width=10cm, clip]{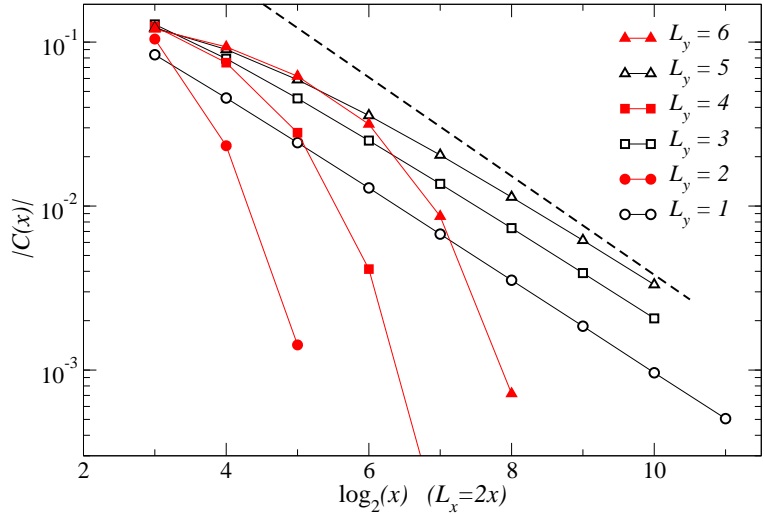}
\caption{Spin correlation function at the longest $x$ distance (with a certain averaging over $y$ separations, as discussed in the text) in $L_x\times L_y$ 
Heisenberg ladders with different $L_y$ as a function of the length $L_x$. The dashed line has the form $C(x) \propto 1/x$.}
\label{laddcorr}
\end{figure}

In the preceding section we studied the spin correlations of the single chain and confirmed the presence of a logarithmic correction to the $\propto 1/r$ 
critical behavior. We now test this behavior for odd $L_y>1$, and also look at the exponentially decaying correlations for even $L_y$. 

With open boundary conditions in the $y$ direction, the correlator $\langle {\bf S}_i \cdot {\bf S}_j\rangle$ is not a function just of the separation 
${\bf r}_{ij}$ between the two spins, but depends on the $y$ coordinates of both spins. Here we use the maximal reflection-symmetric (in the $y$-direction) distance 
between two spins, i.e., for a given spin at $(x,y)$, with $x=1,\ldots,L_x$ and $y=1,\ldots,L_y$, the correlation is computed with the spin at $(x+L_x/2,L_y-y+1)$, 
with the periodic boundary condition taken into account in the $x$ direction. Averages are then taken over all $(x,y)$. This is just one convenient choice, 
and it is not important exactly how the separation in the $y$ direction is treated, because the correlations anyway depend only weakly on it when the $x$ separation 
is large.

Fig.~\ref{laddcorr} shows the long-distance correlation function versus $x=L_x/2$ for ladders of width $L_y=1-6$. The qualitative difference between even
and odd $L_y$ is clear, and in the case of odd $L_x>1$ the behavior for large $L_x$ is very similar to the $L_x=1$ system, with deviations from the
form $C(x) \propto 1/x$ that can be explained by a logarithmic correction (which we will not analyze in more detail here). For even $L_x$, there is
an exponential decay for large $x$, but for $L_y=6$ a different short-distance behavior can already be seen emerging. For very large $L_y$ and $x$ up to $\sim L_y$,
one should expect the correlation function to be of the 2D form, for both even and odd $L_y$, i.e., $C(x)$ should approach the value $m_s^2 \approx 0.095$ 
of the 2D squared staggered magnetization (illustrated with data in Fig.~\ref{jjmag2}). For $x\gg L_y$, there should be a cross-over into either an exponential 
fall-off (for even $L_y$) or the $1/x$ form with a logarithmic correction (for odd $L_y$). In Fig.~\ref{laddcorr} one can see that the correlation at
$x=8$ is almost the same for $L_y=4-6$, but for larger $x$ there is still an increase with $L_y$. To clearly observe 2D behavior followed by cross-overs 
into either even- or odd-leg ladder asymptotic behavior, much larger $L_y$ would be required.

\paragraph{Susceptibility}

\begin{figure}
\includegraphics[width=12.5cm, clip]{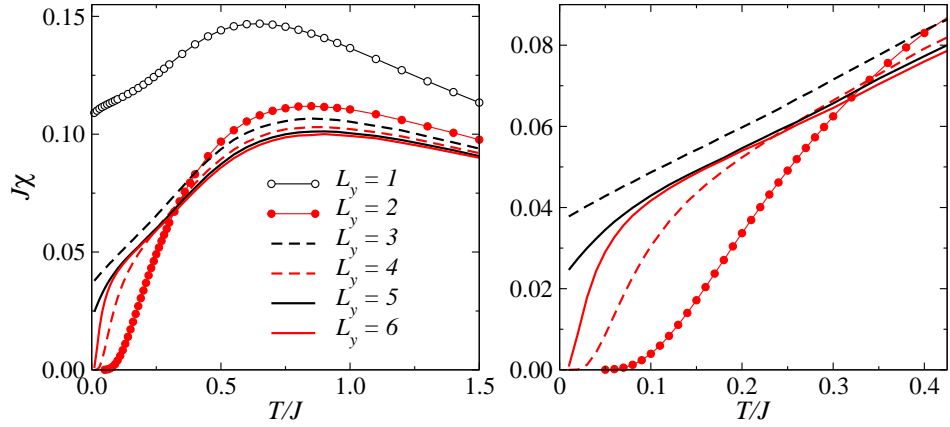}
\caption{Temperature dependence of the susceptibility of Heisenberg ladders of different width $L_y$. The length $L_x$ was sufficiently large
for each $T$ to represent accurately the $L_x\to \infty$ limit. The right panel shows the low-temperature behavior for $L_y=2-6$ on a more
detailed scale. The statistical errors are not discernible, thanks to the use of the improved susceptibility estimator (\ref{chiloopestim}).}
\label{laddsusc}
\end{figure}

The temperature dependence of the susceptibility provides a convenient way to extract the spin gap $\Delta$ for even-$L_y$ ladders, and it is also 
interesting to investigate the applicability of the expected asymptotic form (\ref{chithchain}) for odd $L_y$. Fig.~\ref{laddsusc} shows SSE results for $L_y=1-6$, 
with $L_x$ sufficiently large for each temperature to converge accurately to the thermodynamic limit. For the lowest temperatures $L_x=1024$ was used. For fixed $T$, 
convergence as a function of $L_y$ can be seen for $T/J>0.2$, where the results for $L_y=4,5$, and $6$ are almost the same. The converged curve corresponds 
to the 2D limit (for which results for large $L\times L$ lattices are shown in Fig.~\ref{hbsus2d}). The qualitative difference between ladders of 
even and odd width is clear at low temperatures, with $\chi$ for even $L_x$ decaying to zero exponentially below an $L_y$ dependent temperature. For odd
$L_y$, it can be noted that the $T\to 0$ value decreases with $L_y$. One can argue that the $T\to 0$ susceptibility per rung should be roughly independent of 
$L_y$ for small $L_y$ \cite{frischmuth96}, and, thus, the susceptibility per spin should scale approximately as $1/L_y$. We will not do any fitting to the asymptotic 
form (\ref{chithchain}) here, because it is valid only for very low temperatures for $L_y\ge 3$ (and a higher-order logarithmic correction may be necessary 
unless extremely low temperatures are used \cite{frischmuth96}). Note that the $\chi(0) \sim 1/L_y$ form is not inconsistent with the 2D behavior
$\chi_{2D}(0)>0$, because the ladder behavior (for both even and off $L_y$) applies only below some $T^*(L_y)$ which goes to zero when $L_y \to \infty$.

\begin{figure}
\includegraphics[width=9.75cm, clip]{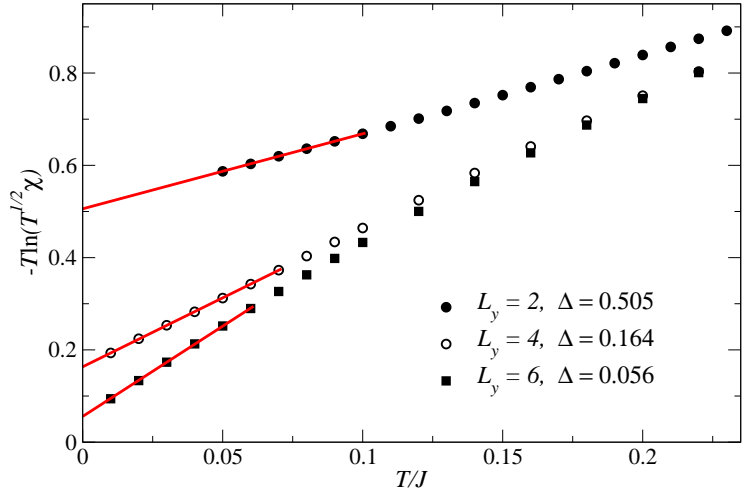}
\caption{The susceptibility of even-width ladders analyzed according to Eq.~(\ref{chitgapeven}). Extrapolations in the linear low-$T$ regime, shown as 
lines extending through the points used in line fits, give the spin gaps indicated in the figure. The statistical precision is roughly the number of
digits shown.}
\label{laddgap}
\end{figure}

Let us now analyze the $L_y$ dependence of the spin gap $\Delta$ for the even-width ladders. One expects roughly an exponential low-temperature
susceptibility, $\chi \sim {\rm exp}(-\Delta/T)$, due to the fact that the ground state is a singlet (non-magnetic, with $\langle M^2_z\rangle =0$) and the 
first excited state is a triplet (magnetic, with $\langle M^2_z\rangle =2/3$). This behavior is, however, modified by the fact that there is a whole continuum 
of magnetic states (for $L_x \to \infty$) above $\Delta$. A low-temperature form of the uniform susceptibility of the 2-leg ladder,
\begin{equation}
\chi(T) = \frac{a}{\sqrt{T}} {\rm e}^{-\Delta/T},
\label{chitevenleg}
\end{equation}
has been obtained by analyzing the limit of weak inter-rung couplings perturbatively \cite{troyer94}. This form can be expected to hold asymptotically
for $T\to 0$ for any $J_y/J_x$ and even $L_y$ (for $L_x \to \infty$). The gap can be extracted from numerical $\chi(T)$ data by considering the 
logarithm of $\sqrt{T}\chi$ in (\ref{chitevenleg}), giving
\begin{equation}
-T\ln(\sqrt{T}\chi) = \Delta - T\ln(a).
\label{chitgapeven}
\end{equation}
Thus, the intercept of a line fitted to $-T\ln(\sqrt{T}\chi)$ versus $T$ equals the gap $\Delta$. Fig.~\ref{laddgap} shows data analyzed in this way. In all cases, 
a linear behavior obtains at low temperatures, and line fits give the gaps listed in the figure. A field theoretical treatment shows that the gap for a spin-$S$ 
ladder system should decrease with $L_y$ as $\Delta \sim {\rm exp}(-\pi S L_y)$ \cite{dellaringa97}, which is not in good agreement with the results obtained here 
(and less precise earlier calculations \cite{frischmuth96}). Calculations for larger $L_y$ would have to be carried out (which can certainly be done) to investigate 
this issue in more detail.

It would also be interesting to study the odd-leg ladders for larger $L_y$ and lower temperatures than what has been done until now, to extract the $L_y$ dependence
of the parameters $c$ and $T_0$ in (\ref{chithchain}). An alternative approach is to map the low-energy properties of a ladder onto a single chain with longer-range 
interactions \cite{frischmuth97}, which arise because the $S=1/2$ degrees of freedom of the isolated rungs are not completely localized to individual rungs once they are
coupled. The localization length increases with $L_y$, and eventually, for $L_y \to \infty$, an effective model with sufficiently long-ranged interactions 
to produce N\'eel order (as discussed in Sec.~\ref{sec_longrange}) should obtain. The mapping procedure has not yet been tested for ladders with large $L_y$, however, 
and one should therefore also study the full ladder systems at lower temperatures and for larger $L_y$.

\subsubsection{Long-range order in two dimensions}
\label{sec_2dheis}

We already discussed the nature of the antiferromagnetically ordered (N\'eel) ground state of the 2D Heisenberg model in Secs.~\ref{neel} and \ref{lanc2d}. 
Extrapolation of the best currently available QMC results \cite{awshg} for the squared sublattice magnetization, shown in Fig.~\ref{jjmag2} gives $m_s=0.30743(1)$ 
for the infinite system, where $(1)$ denotes the statistical error in the preceding digit. This result deviates only by about $1\%$ from the linear spin wave result 
$m_s=0.3034$. Higher-order spin wave calculations \cite{canali,hamer,igarashi} give $m_s\approx 0.3070$. As discussed in Sec.~\ref{neel}, this good agreement with the 
actual value can be traced to the fact that the quantum fluctuations are not that strong, reducing the sublattice magnetization by only about $40\%$ from the classical 
value. Other important ground state quantities, such as the spin stiffness \cite{hamer94}, the transverse magnetic susceptibility, and the spin wave velocity, are in 
similar good agreement with spin wave theory (but we will not make any detailed comparisons of values here). 

After discussing some general data fitting issues, we will here compute the energy and the spin stiffness using finite-size extrapolations of $T\to 0$ converged
SSE results (with proper convergence confirmed using checks such as those shown in Fig.~\ref{ecomp} for all quantities of interest). In addition to the $L\times L$ 
lattices normally used, we will also consider rectangular $L_x\times L_y$ lattices with $L_x=2L_y$. This enables a consistency check of the extrapolations. We also 
discuss the finite-temperature susceptibility and extrapolate it to zero temperature, using $L\times L$ lattices sufficiently large to completely eliminate 
finite-effects for the range of temperatures considered. We finally discuss the divergence of the correlation length as $T\to 0$, which is another important 
manifestation of the  N\'eel ordered ground state \cite{chn}. 

Extrapolations of the ground state parameters of the 2D Heisenberg model have been presented in a large number of papers previously, e.g., 
Refs.~\cite{runge92,wiese92,sandvik97,kim98,beard98}. 
For most practical purposes, the precision already achieved is quite sufficient. Since the model is one of the most important 
prototypical systems in quantum magnetism, it is, however, useful to continue to establish more precise benchmark calculations (which can be useful, e.g., for testing
other methods \cite{chernyshev}). Some of the results presented below, in particular the ground state energy, represent the most precise calculations to date. 
In addition to the extrapolated $N=\infty$, $T=0$ values of the various physical quantities, their finite-size and temperature corrections are also of interest, 
because they have been predicted in great detail based on field-theoretical methods \cite{hasenfratz93}, and numerical test are useful to establish the range of
validity of the low-energy theories. The corrections are statistically much noisier and require very long simulations to establish precisely. The discussion here is
only intended to give a flavor of what can be done.

\paragraph{Data fitting issues}

When extrapolating finite-size result to the thermodynamic limit, it is useful to know the expected form of the size corrections based on analytical calculations.
Away from  a critical point, the size corrections for systems of dimensionality $d>1$ normally take the form of a polynomial in the inverse system length $1/L$, 
but some times the leading correction is $\propto 1/L^a$ with $a$ an integer larger than $1$. As discussed in Sec.~\ref{dimsystems}, the leading size correction to the 
sublattice magnetization of the 2D Heisenberg model can be obtained from the spin wave theory of the N\'eel state and is $\propto 1/L$. This result is also a more 
general consequence of the fact that the order parameter is a vector, as discussed in Sec.~\ref{scaling}. The ground state energy $E/N$ per site has a 
leading correction $\propto 1/L^3$, and thanks to this high power it is relatively easy to obtain the energy to high precision even based on rather small lattices 
\cite{sandvik97}. The leading correction to the spin stiffness is $\propto 1/L$ \cite{einarsson}. Even if the leading power is not known, one can normally find it 
empirically based on data, provided that the numerical precision is sufficiently high. When fitting data to a polynomial one may find that the coefficient of the linear 
term (and possibly higher-order terms) is very small, which then makes it likely that it actually should be exactly zero. It is best to use a fitting program which 
allows one to specify exactly what powers in $1/L$ should be included. 

For a given set of data (here a quantity computed for a set of lattice sizes, but we will use the same technique also for fitting temperature dependent 
quantities as powers in $T$) one should include high enough powers of $1/L$ for the fit to be statistically sound, i.e., $\chi^2$ per degree of freedom 
($\chi^2/{\rm dof}$) should be close to $1$ for a large data set (as discussed in standard texts on data analysis). The statistical errors of the fitting 
parameters can be computed based on the error bars of the data. The safest way, which does not rely on any assumptions of the errors being small (although 
normally they should be), is to carry out a large number of fits with Gaussian noise added to the data (with the standard deviation of the noise equal 
to the corresponding error bar) and compute the standard deviation of the distribution of the resulting parameter values. Note that when increasing the 
number of fitting parameters (here powers of $1/L$), the error bars of the best-fit parameters normally increase (and $\chi/{\rm dof}$ may also get worse). 
One should therefore not include many more powers than needed to bring $\chi^2/{\rm dof}$ close to $1$. To safeguard against possible remaining effects of 
higher-order corrections, one may nevertheless want to include one more parameter than strictly needed for and acceptable $\chi^2/{\rm dof}$.

\begin{figure}
\includegraphics[width=13.5cm, clip]{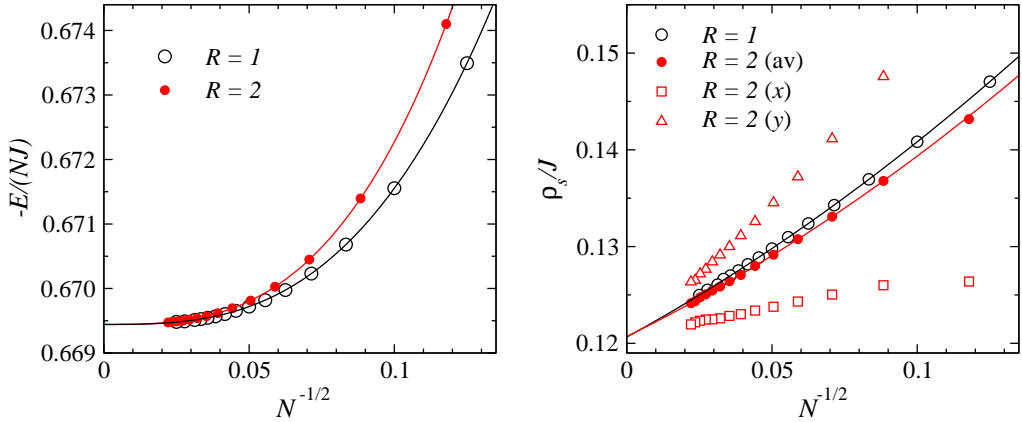}
\caption{Size dependence of the ground state energy (left) and spin stiffness (right) of 2D Heisenberg lattices with aspect ratio $R=1$ (where $N=L^2$) and 
$2$ (where $N=2L^2$). For $R=2$, the spin stiffness on a finite lattice is different in the $x$ and $y$ direction. Both values are shown here along with the 
average. The curves are polynomial fits discussed in the text.}
\label{ewhb2d}
\end{figure}

\paragraph{Ground state energy and spin stiffness}

The energy per site $E/N$ and the spin stiffness $\rho_s$ were evaluated using the estimators (\ref{essenbeta}) and (\ref{rhosdefsseestim}), respectively, with
the energy adjusted for the constant $1/4$ subtracted from each bond operator in the SSE simulations. As discussed in Sec.~\ref{estimators}, the value of the spin 
stiffness extrapolated to the thermodynamic limit should be multiplied by a factor $3/2$, to account for the fact that the spin-rotational symmetry is not 
broken in the SSE simulations.

Comparing extrapolations for lattices of different shapes is a good way to check for detectable consequences of finite-size corrections beyond those included
in the data fits \cite{mchains,chernyshev}. Fig.~\ref{ewhb2d} shows results for both $N=L\times L$ (aspect ratio $R=1$) and $N=2L\times L$ ($R=2$) lattices, graphed
versus $1/\sqrt{N}$ ($=1/L$ for $R=1$ and $\propto 1/L$ for $R=2$). These results were computed using inverse temperatures $\beta$ as high as $32\times L$ for 
the largest systems ($L$ up to $40$ for $R=1$ and up to $32$ for $R=2$). For the largest systems several hundred CPU hours were used. The error bars are too 
small to be visible in the figure. An example of the results produced, for the $32\times 32$ system $E/N=-0.6695115(8)$ and $\rho_s=0.12606(5)$, based
$5\times 10^6$ Monte Carlo sweeps. The error bars are of similar magnitude for the other systems as well.

For $R=1$ using the systems with $L\ge 6$ and a $5th$-order polynomial (without the linear and quadratic terms, which are predicted not to be present, as
discussed above) gives $E/N=-0.6694421(5)$, while a fit of the same order to the $R=2$ data with $L_y\ge 6$ delivers $E/N=-0.6694422(6)$. These results are in 
perfect statistical agreement with each other, and it is then permissible to use their statistically weighted average, $E/N=-0.6694421(4)$, as a final estimate of the
ground state energy. It should be noted that the numerical values depend slightly on what lattices are included in the fit and the order of the 
polynomial used. Once the fit is statistically sound, these fluctuations should be consistent with the statistical errors.

Turning now to the stiffness, note first that for a finite lattice with  aspect ratio $R\not=1$, there are two stiffness constants, for phase twists imposed
in the $x$  and $y$ direction. Both of them are graphed for $R=2$ in the right panel of Fig.~\ref{ewhb2d}, along with the arithmetic average of the two and
the $R=1$ values. It can be seen here that the $R=1$ stiffness and the average of the $R=2$ stiffnesses are better behaved for extrapolations than the individual 
$x$ and $y$ values for $R=2$, and no fits are therefore included for the latter. Quadratic fits for $R=1$ and $R=2$ (using $L\ge 6$ and $L_x \ge 8$ data) gives 
$\rho_s=0.12065(4)$ and $\rho_s(\infty)=0.12070(6)$, respectively. Taking the average of these statistically consistent values, and including the
rotational factor $3/2$, gives the final estimate $\rho_s=0.18100(5)$ for the infinite system.

\paragraph{Properties at $T>0$} 

Since the Mermin-Wagner theorem rules out magnetic order in the 2D Heisenberg model for any $T>0$, the correlation length must diverge as $T\to 0$ 
for the behavior to be consistent with the ordered ground state. The behavior is similar in the classical 2D Heisenberg model, where the correlation
length diverges exponentially. The long-distance correlations and fluctuations of the quantum system can in fact formally be mapped onto a classical
system with renormalized couplings \cite{chn}, and the low-temperature regime in which such a mapping holds is referred to as the {\it renormalized
classical} regime. Many predictions in this regime have resulted from field-theoretical treatments \cite{chn,hasenfratz93,chubukov}. Remarkably detailed 
results have been derived for the temperature dependence of, e.g., the correlation length and the susceptibility. The predicted forms, which are believed 
to be asymptotically ($T \to 0$) exact, depend only on the ground state parameters (e.g., the spin stiffness and the spin wave velocity $c$). 

\begin{figure}
\includegraphics[width=13cm, clip]{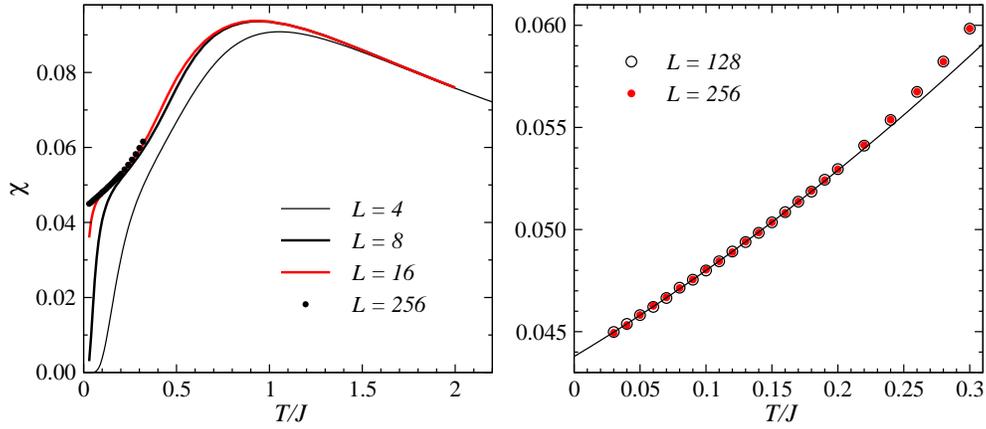}
\caption{Uniform susceptibility of the 2D Heisenberg model computed using several $L \times L$ lattices. The right panel shows low-$T$ results for 
$L=128$ and $256$ on a more detailed scale (with the two data sets coinciding within statistical errors, indicating convergence to the thermodynamic limit).
The solid curve in the right panel is a fit to the $L=256$ data (using only the $T/J \le 0.18$ data points) of the form $\chi=a+bT+cT^2$ with the parameters 
discussed in the text. Error bars are too small to discern but are typically $\approx 2\times 10^{-5}$ for the low-temperature $L=256$ results.}
\label{hbsus2d}
\end{figure}

Let us first investigate the uniform magnetic susceptibility, which should have the following form for $T\to 0$ (in an
infinite system) \cite{hasenfratz93} :
\begin{equation}
\chi(T)=\chi(0)\left [1+ \frac{T}{2\pi\rho_s} + \left (\frac{T}{2\pi\rho_s}\right )^2\right ].
\label{chit2dform}
\end{equation}
The only model dependent parameter here is the ground state spin stiffness, which we determined above. To compare numerical data with this result,
we first have to make sure that we can achieve the thermodynamic limit. For fixed temperature, we can study the behavior as a function of the system 
size, where we here use quadratic lattices with $N=L^2$. The left panel of Fig.~\ref{hbsus2d} shows results for several lattice sizes. For any finite 
system the susceptibility vanishes as $T\to 0$ due to the singlet ground state and finite-size gap in the spectrum. As we discussed in Sec.~\ref{lanc2d}, 
the lowest excitations are the quantum rotor states, with gaps scaling as $1/N$. The $T/J < 1/L$ susceptibility for a finite system is dominated by 
these excitations. Once $T>1/L$ spin waves can also be excited, and one may therefore suspect that a system size of at least $L>1/T$ will be required 
for convergence to the thermodynamic limit. Here we will not study the convergence in detail, but just conclude based on the data in Fig.~\ref{hbsus2d} 
that the $L=256$ results are safely converged down to $T/J=0.03$ and can be used to test the form (\ref{chit2dform}).

Tests of analytical predictions can be carried out in different ways. We could here use the value of $\rho_s$ extracted from the finite-size extrapolations 
of the $T=0$ data above and check the agreement between the SSE results for $\chi(T)$ and Eq.~(\ref{hbsus2d}), with only $\chi(0)$ adjusted as a fitting 
parameter to obtain the best agreement. Another way would be to also adjust $\rho_s$, the best-fit value of which can be compared with the 
result of the ground state extrapolations. Here we will proceed in a different way, which tests both the $T$ and $T^2$ corrections in (\ref{chit2dform}). 
Fitting SSE data to a form $\chi=a_0+a_1T+a_2T^2$, we can use the form of the coefficients in (\ref{chit2dform}) to extract corresponding estimates 
$\rho_s(1)$ and $\rho_s(2)$ for the spin stiffness from the linear and quadratic coefficients $a_1$ and $a_2$. These estimates should agree with each 
other and with the result of the ground state calculation if the temperatures used in the fit are sufficiently low. At higher temperatures, corrections 
including higher powers of $T$ will be important and should lead to a bad fit to the quadratic form and disagreements between the different stiffness 
estimates. 

The quadratic fit is good and consistent values of the exponents are obtained if only data for $T/J \le 0.18$ are included. The fit is shown in the 
right panel of Fig.~\ref{hbsus2d}, and the stiffness constants extracted from the parameters are $\rho_s(1)=0.180(3)$ from slope and $\rho_s(1)=0.179(6)$ 
from the quadratic correction. These values are in good agreement with (but noisier than) the ground state result $\rho_s=0.18100(5)$. Fixing the form 
(\ref{chit2dform}) and extracting the best-fit values of $\chi(0)$ and $\rho_s$ gives a less noisy estimate \cite{kim98}, but since we are still dealing 
with corrections to a $T=0$ quantity it is not easy to achieve the same precision as we did by extrapolating the finite-size values of $\rho_s$ at $T=0$.

The $T\to 0$ susceptibility extracted from the above fit is $a_0=\chi(0)=0.04378(3)$. As we discussed in Sec.~\ref{lanc2d}, the transverse susceptibility 
$\chi_\perp$ in the symmetry-broken state at $T=0$ is this number multiplied by $3/2$, giving $\chi_\perp=0.06567(4)$. In combination with the spin stiffness, 
this value gives us access to another ground state parameter; the spin wave velocity. We discussed the spin stiffness constant as an elastic modulus in 
Sec.~\ref{stiffkt}. This elastic-medium approach to the long-wavelength properties of quantum magnets is also referred to as the {\it hydrodynamic description} 
\cite{halperin69}. The transverse susceptibility is there analogous to a mass density, and one can relate the spin wave velocity to the spin 
stiffness and the susceptibility according to 
\begin{equation}
c=\sqrt{\frac{\rho_s}{\chi_\perp}}.
\end{equation}
Using the values extracted above for $\rho_s$ and $\chi_\perp$ gives the velocity $c=1.6602(6)$. Recall that linear spin wave theory 
(Sec.~\ref{spinwave}) gives $c=\sqrt{2}$. The difference between these two values is captured very well by spin wave theory including 
$1/S$ corrections, and the ``renormalization factor'' of $c$ can also be computed quite precisely using a series expansion around 
the Ising model \cite{singh89}.

\begin{figure}
\includegraphics[width=9.5cm, clip]{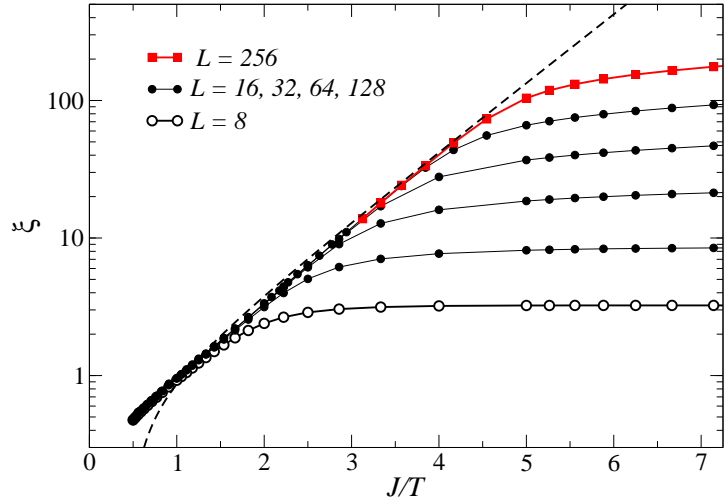}
\caption{The temperature dependent correlation length (graped versus the inverse temperature) of the 2D Heisenberg model computed for different 
lattice sizes using the spin structure factor definition (\ref{xiadef}), divided by the factor $\sqrt{15/16}$ coming from Eq.~(\ref{xiaxi}). 
The dashed curve is the expected asymptotic ($L=\infty$, $T\to 0$) form (\ref{corrlen2dform}) with the parameters $\rho_s=0.1810$ and $c=1.660$ 
determined previously.}
\label{hb2dxi}
\end{figure}

The temperature dependence of the correlation length of the infinite system is also known, including a correction to the leading exponential
divergence \cite{chn,hasenfratz93}:
\begin{equation}
\xi(T)=\frac{e}{8}\frac{c}{2\pi\rho_s}\left (1 - \frac{4\pi\rho_s}{T} \right ){\rm exp}\left (\frac{2\pi\rho_s}{T} \right ).
\label{corrlen2dform}
\end{equation}
We have already determined the two parameters involved here, the ground state spin stiffness and velocity, and we will use their values quoted 
above to directly test the analytical prediction against SSE results based on the ``second moment'' structure factor definition (\ref{xiadef}) of the 
correlation length.

Since the correlation length is a rapidly divergent quantity, very large lattices are required to converge the calculations to the thermodynamic limit at 
low temperatures. The lattice length $L$ should be several times larger than $\xi$ to achieve complete convergence. Fig.~\ref{hb2dxi} shows results 
for several lattices sizes up to $L=256$. While larger systems have been studied in the past \cite{kim98,beard98}, already these results, which are well 
converged for $T/J \ge 0.25$, show that the form (\ref{corrlen2dform}) describes the numerical data very well. The small deviations have been discussed
in the literature \cite{kim98,beard98} and can be understood as arising from higher-order corrections to the form (\ref{corrlen2dform}) .

\subsubsection{Quantum phase transition in a dimerized system}
\label{sec_dimerized}

We now study the quantum phase transition in the 2D columnar dimerized Heisenberg model illustrated in Fig.~\ref{dlattices}(b). The phase transition 
of the ground state takes place as a function of the coupling ratio $g=J_2/J_1>1$ and is caused by quantum fluctuations, which here correspond to an
increasing density of singlets on the dimers and at some point lead to the loss of the N\'eel order existing when $g\approx 1$. We already examined some 
SSE results for the sublattice magnetization of this system in Fig.~\ref{jjmag2}, and the behavior indicated a phase transition between a N\'eel-ordered 
and a nonmagnetic ground state at coupling ratio $g\approx 1.9$. In this section we apply the machinery of finite-size scaling at criticality, which we 
discussed in the context of classical systems in Sec.~\ref{scaling} and generalized to quantum systems in Sec.~\ref{qcpintro}. Here we first analyze the 
critical behavior of several quantities calculated at sufficiently low temperatures to access the ground state critical behavior, and then discuss 
consequences of the $T=0$ quantum critical point at non-zero temperatures. Before that, let us spend a few words on the implementation of the SSE algorithm 
for a model with non-uniform couplings.

\paragraph{SSE method}

An appealing feature of the SSE algorithm for the Heisenberg model is that the coupling strengths only enter in the acceptance probabilities
(\ref{paccsseheis1}) and (\ref{paccsseheis2}) in the diagonal updates, where the coupling $J$ is absorbed in the inverse temperature $\beta=J/T$. 
For non-constant couplings, we only have to define $\beta_b=J_b/T$ for each bond $b$. In the implementation in pseudocode $\{27\}$, some scheme has 
to be used to identify the coupling corresponding to a generated bond $b$ in an insertion attempt. In a system with a large number of different 
couplings (which could even be random, different for every bond) the couplings (in the form of $\beta_b$) should be stored as a list. For systems 
with just two different couplings, such as the dimerized lattice considered here, it is better to just order the bonds in such a way that the coupling 
can be quickly determined, e.g., with $b\le N/2$ and $b> N/2$ corresponding to $J_2$ and $J_1$, respectively. Note that for an operator removal, 
unlike the case of uniform couplings in code $\{27\}$, we now also need to extract the current bond, $b=S(p)/2$, in order to determine the 
probability.

\paragraph{Finite-size scaling for $T \to 0$}

\begin{figure}
\includegraphics[width=13.5cm, clip]{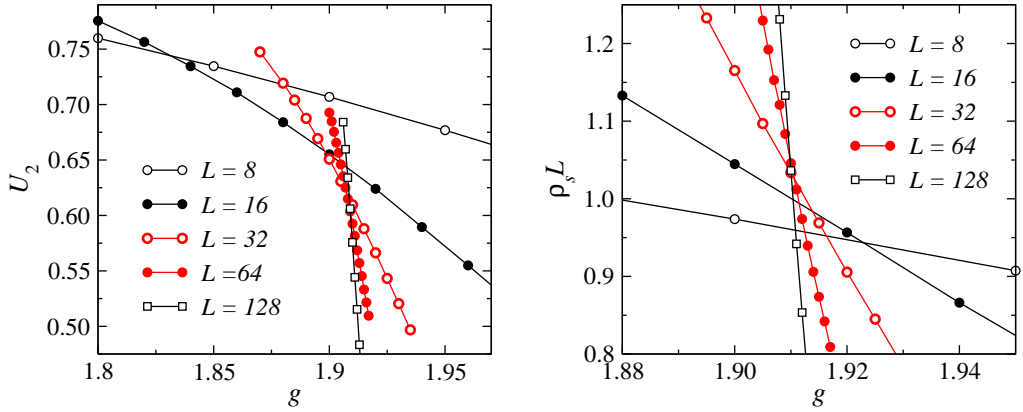}
\caption{Binder cumulant (left) and spin stiffness (in the $x$ direction) multiplied by the system length (right) of the dimerized Heisenberg 
model. The crossing points of these curves for different $L$ tend toward the critical value of the coupling ratio $g$. Error bars are much smaller 
than the symbols.}
\label{dimbindrho}
\end{figure}

When analyzing the ground state, we can proceed as in the case of the standard uniform 2D Heisenberg model in the previous section, simulating systems
at sufficiently low temperatures to achieve $T=0$ convergence of all the quantities of interest. This approach is discussed for various dimerized
systems in, e.g., Refs.~\cite{wang,wenzel1} (as well as in many older works). Another approach is to study systems at inverse temperature $\beta=L^z$, 
where $z$ is the dynamic critical exponent (which we discussed in Sec.~\ref{qcpintro}) \cite{krauth}. This is motivated in the following way, by a generalization 
of the finite-size scaling hypothesis (\ref{fsxilhypo}): In a quantum system the scaling function $f(\xi/L)$ should be replaced by a function with 
two arguments, $f(\xi/L,\xi_\tau/L_\tau)$, where the correlation length in the imaginary time dimension depends on the spatial correlation length 
$\xi$ according to $\xi_\tau \sim \xi^z$ (which defines the dynamic exponent) and the length of the system in the imaginary time direction is 
$L_\tau=c/T\sim \beta$ (where $c$ is a velocity). If we choose $\beta \propto L^z$, then the scaling function can be written as $f[\xi/L,(\xi/L)^z]$, 
which is a function of the single argument $\xi/L$. Thus, the finite-size scaling procedures can be used exactly as in the classical systems discussed 
in Sec.~(\ref{finitsizescaling}). This is the case also if we take the limit $\beta \to \infty$ for each $L$ (in practice finite $\beta$ large enough 
for convergence to this limit), because then $\xi_\tau/L_\tau \to 0$, and there is again only one argument $\xi/L$ left in the scaling function.

There is plenty of evidence already that $z=1$ in dimerized Heisenberg models, and we will here use systems with $\beta=L$. This allows for studies 
of larger systems than in the $\beta\to \infty$ limit, although it is not {\it a priori} clear which approach is in the end better, since the corrections 
to the leading finite-size scaling behavior can be different. Here we use $L$ up to $L=128$. We will also test explicitly that systems with $\beta=L$ 
exhibit behavior consistent with $z=1$, by studying quantities which depend on $z$.

We first locate the critical coupling by examining quantities that should be size independent at $g_c$. Fig.~\ref{dimbindrho} shows the $g$ dependence of 
both the Binder cumulant and the spin stiffness, with the latter multiplied by $L$ to compensate for the expected quantum critical scaling form $\rho_s \sim 1/L$, 
obtained the classical form (\ref{rhoscalelcl}) with $d \to d+z=3$. 

The Binder cumulant is defined according to (\ref{u2defn}), with the number of components $n=3$. Note, however, that (\ref{u2defn}) is defined with the full 
scalar product $m^2={\bf m}\cdot {\bf m}$ in (\ref{r2def}), whereas with the SSE method we here only compute the $z$ component expectation values $\langle m_z^2\rangle$ 
and $\langle m_z^4\rangle$ (the off-diagonal components being more difficult to evaluate \cite{dorneich01}). One can easily find the geometrical factors relating 
these by integrating the $z$ component $\cos(\Theta)$ of a classical 3D unit vector over the the angles, giving $\langle m^2\rangle=3\langle m_z^2\rangle$ and 
$\langle m^4\rangle=5\langle m_z^2\rangle$. For locating the critical point, these factors play no role, and we could also use the plain Binder ratio defined 
as $R_{2z}=\langle m_z^4\rangle/\langle m_z^2\rangle^2$. 

Since the dimerized lattice does not have $90^\circ$ rotational symmetry, the stiffness constants in the $x$ and $y$ directions are different. Although the 
numerical values are indeed quite different, their scaling behaviors close to the critical point is very similar, however [the $x$ stiffness is approximately 
a factor $2$ larger---the dimers are oriented in the $x$ direction as in Fig.~\ref{dlattices}(b)]. Only the $x$ stiffness is shown in Fig.~\ref{dimbindrho}(b). 

\begin{figure}
\includegraphics[width=9.75cm, clip]{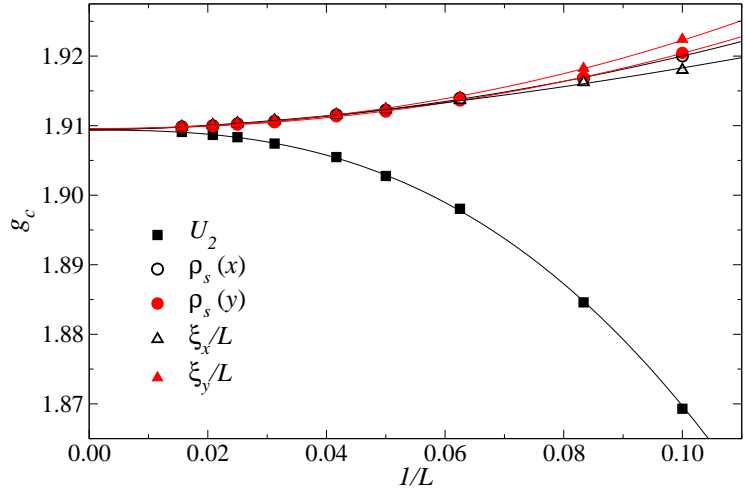}
\caption{Size dependent critical coupling for the dimerized Heisenberg model extracted from $(L,2L)$ crossing points of the Binder cumulant, 
the spin stiffness constants, and the correlation lengths. The curves show fits to the form $g_c(L)=g_c(\infty)+a/L^\omega$.}
\label{dimcrossgc}
\end{figure}

Curve crossings are indeed seen in Fig.~\ref{dimbindrho} for both $U_2$ and $\rho_sL$, and after some significant drift of the crossing points (e.g., for systems 
of size $L$ and $2L$) for small $L$, they seem to converge to roughly the same value in both cases. Note that the crossing points for $U_2$ and $\rho_s$
approach $g_c$ from opposite directions, which can be useful for bracketing the critical value \cite{wang,wenzel1}. Crossing points can be located numerically by 
fitting a polynomial of suitable order to some of the data points, repeating the procedure several times with added Gaussian noise to compute error bars. 
Fig.~\ref{dimcrossgc} shows results of such procedures for the Binder ratio, the $x$ and $y$ stiffness constants, as well as the correlation lengths [computed 
using the definition (\ref{xiadef})] in both the $x$ and the $y$ direction. Fits to the data points of the form $g_c(L)=g_c(\infty)+a/L^\omega$ are also shown. 
This form describes well all the data for $L\ge 10$ (the sizes shown in the figure). All the extrapolated values of $g_c$ fall within the range $[1.9094,1.9096]$, 
and if the $L=10$ data are excluded the range narrows even further. The exponent $\omega$ is in the range $2 \sim 2.5$ for all quantities (being largest
for $U_2$). Treating all five values obtained in these extrapolations as independent statistical data gives $g_c=1.90948(4)$ as a  final estimate for the critical 
point. This is in good agreement with (but with smaller error bar than) a recent estimate $g_c=1.9096(2)$ obtained using $T\to 0$ data for the same quantities 
on lattices with $L$ up to $64$. The crossing point shift exponents $\omega$ are also in good agreement.

The existence of a limiting value of the location $g_c$ at which the $L\rho_s$ curves cross does not prove that $z=1$. One also should check that the values 
of $L\rho_s$ are well behaved, i.e., that a crossing point in the plane $(g,\rho_sL)$ really forms. A plot such as Fig.~\ref{dimcrossgc} for $\rho_sL$ at the 
crossing points for lattices of size $L$ and $2L$ confirms that this is the case. Examining the Binder cumulant in a larger window of couplings, one can not 
find any indications of negative values, which would be a sign of phase coexistence at a first-order transition (as discussed in Sec.~\ref{firstordertrans}). 
Thus, the scaling behavior so far supports at continuous quantum critical point with $z=1$.

We could now proceed to perform data collapse fits in order to find the correlation length exponent $\nu$, as we discussed for classical systems in
Sec.~\ref{scaling}. This has been done for dimerized Heisenberg model, including also scaling corrections \cite{wang,wenzel1}, and the results are in
good agreement with the expected 3D classical Heisenberg universality class. Here we instead just discuss the exponent $\eta$ appearing in the critical 
correlation function (\ref{cofreta}), where we should again replace $d$ by $d+z=3$. The staggered structure factor $S(\pi,\pi)$ is the spatial integral 
(sum on the 2D lattice) of the correlation function (\ref{czrheis}), while the Kubo integral (\ref{kuboformula}) for the staggered susceptibility $\chi(\pi,\pi)$ 
corresponds to a 3D space-time integral (a 2D lattice sum and an imaginary time integral). Performing these integrals with the above critical form of the 
correlation function, with a cut-off equal to the system size $L$, gives $S(\pi,\pi) \sim L^{d-z-\eta}$ and $\chi(\pi,\pi) \sim L^{d-\eta}$. We will test this 
behavior of the critical system and extract the exponent $\eta$, using the value of $g_c$ found above. Instead of performing new simulations at this point 
(which is known only to within a statistical error), one can perform polynomial interpolations within the existing data sets. One can then also easily check 
the behavior for points slightly off the best estimate of $g_c$ (plus and minus one error bar), to check the sensitivity of the fitted $\eta$ to the location 
of the critical point. The simplest way to analyze the critical behavior is to fit a straight line to $\ln(S)$ and $\ln(\chi)$ versus $\ln(L)$ (as was done for 
$S(\pi,\pi)$ of this model in \cite{wenzel1}). Some corrections to scaling are always expected, and if the data have small error bars a straight line can only 
be fitted to large lattices. With the data sets used here, statistically acceptable linear fits are only obtained when using system sizes $L\ge 48$. We will 
therefore also include subleading corrections and assume the following forms
\begin{equation}
S(\pi,\pi) = aL^{1-\eta}+bL^{\omega},~~~~~~~\chi(\pi,\pi) = aL^{2-\eta}+bL^{\omega},
\label{sxfitforms}
\end{equation}
where one would expect the subleading exponent $\omega$ to be much smaller (possibly even negative) than the leading exponents. One could in principle 
perform a combined fit with $\eta$ fixed to be the same for the two quantities (with $\omega$ and the constants different), but here we will fit the 
data sets independently.

\begin{figure}
\includegraphics[width=14cm, clip]{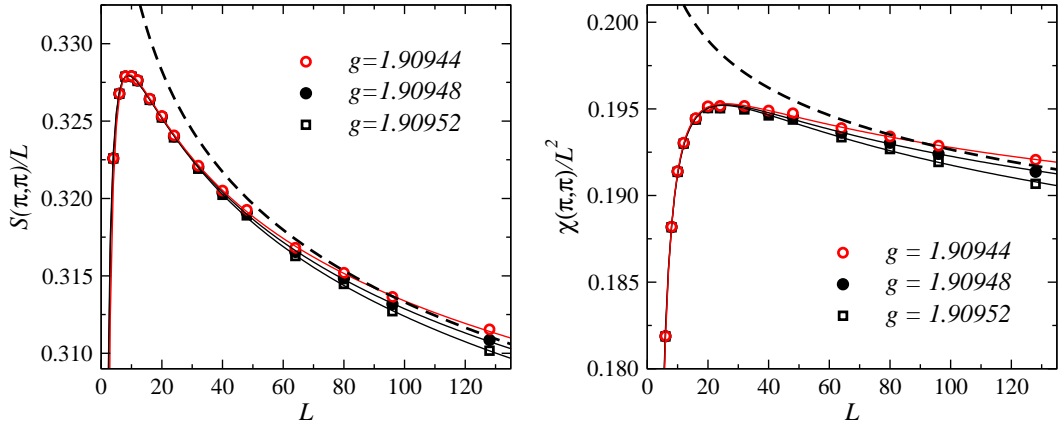}
\caption{Finite-size scaling of the staggered structure factor (left) and susceptibility (right) in close proximity of the estimated critical coupling 
ratio $g_c=1.90948(4)$. The powers of $L$ corresponding to $z=1$ and $\eta=0$ have been divided out. The remaining asymptotic size dependence should then 
be governed by the actual value of $\eta$. The fits to the forms in (\ref{sxfitforms}) give $\eta=0.029(3)$ from $S(\pi,\pi)$ and $\eta=0.020(4)$ from 
$\chi(\pi,\pi)$. The fits were based on $L\ge 8$ data, but the resulting curves also go through the $L=4$ and $6$ points. The dashed curves shows the 
behavior at $g=1.90948$ without the scaling corrections (i.e., with $\eta$ and $a$ kept at their values obtained in the fit including the corrections).}
\label{dimsx}
\end{figure}

In order to more clearly see the role of the subleading correction, $S(\pi,\pi)$ and $\chi(\pi,\pi)$ are graphed in Fig.~\ref{dimsx} with the dominant $L$ 
and $L^2$ factors divided out. The asymptotic $L \to \infty$ behavior is then in both cases $\sim L^{-\eta}$, where $\eta$ is expected to be small. The 
currently best estimate for the classical 3D Heisenberg universality class is $\eta=0.0375(5)$ \cite{series}. The fits to (\ref{sxfitforms}) give 
$\eta=0.029(2)$ for $S(\pi,\pi)$  and $\eta=0.020(3)$ for $\chi(\pi,\pi)$. In principle the forms (\ref{sxfitforms}) should of course work only exactly at 
$g_c$, but in practice, for a finite range of system sizes, they fit the data well in some window around the true critical point. As seen in Fig.~\ref{dimsx}, 
the interpolated values of the two quantities at $g_c$ plus and minus one error bar deviate visibly from those at the midpoint, but the fits are statistically
acceptable in all three cases. The statistical errors quoted above arise predominantly from the uncertainty in the critical coupling. The subleading 
exponents in (\ref{sxfitforms}) are $\omega=-0.2(2)$ for $S(\pi,\pi)$  and $\omega=0.6(2)$ for $\chi(\pi,\pi)$.

Fig.~\ref{dimsx} also shows the fitted functions with the subleading corrections disregarded (with the other parameters kept at their values obtained
in the fit with corrections). Clearly the corrections are quite significant, being completely responsible for the maximums in both curves at $L\approx 10$. 
One can of course obtain much better fits to the larger lattices without subleading corrections. As mentioned above, with the rather small error bars 
of the data used here, only systems with $L\ge 48$ can be included in such a fit. Even then, there must be some influence of the neglected corrections. The 
values of $\eta$ do in fact come out somewhat lower if no corrections are included. With the corrections, all the data ($L\ge 4$) can be included, but to 
be on the safe side only $L\ge 8$ data were included in the fits quoted here and graphed in the figure. These fits still do pass very closely through the 
$L=4$ and $6$ data points, which further reinforces the quality of the functional form used.

The $\eta$ values obtained here, and also in Ref.~\cite{wenzel1}, are a bit lower (by a few error bars) than the best available classical 3D Heisenberg 
value \cite{series} quoted above. Most likely, these small discrepancies are due to further scaling corrections, but it would still be good to push QMC 
calculations for various dimerized Heisenberg models to even higher precision (using larger lattices, a denser grid of lattices sizes, and also reducing the 
error bars of the computed quantities) in order to establish the agreement with the classical exponents for sure. This is particularly interesting and 
important in light of the fact that the staggered dimer model illustrated in Fig.\ref{dlattices}(c) (and also some other dimerization patterns) seems to 
show small but statistically significant deviations from the expected 3D Heisenberg exponents \cite{wenzel2,jiang09,wessel10}. These systems may still not be 
in a different universality class, as originally proposed \cite{wenzel2}, because the deviations could also originate from anomalously large scaling corrections 
\cite{jiang09,wessel10}. Regardless of the underlying reason for the deviations, their origin remains unclear and should be further examined (and such 
work is in progress \cite{wessel10}).

\paragraph{Quantum critical scaling at $T>0$} One of the most remarkable and important aspects of quantum criticality is that the properties (the universality 
class) of a $T=0$ critical point also can strongly influence a system at $T>0$, often up to rather high temperatures \cite{chubukov} and also if the system 
does not have exactly the couplings corresponding to a $T=0$ critical point (but is near such a point, in a sense which we will make more precise below). 
Critical fluctuations and scaling behavior can therefore be much more generic features of quantum systems than classical systems \cite{sondhirpm}, where 
critical behavior often can be observed only very close to the critical point. 

In a quantum system, critical behavior can be expected when the temperature is the dominant energy scale. In dimerized 2D Heisenberg models, the energy
scales on the N\'eel and nonmagnetic side of the transition are, respectively, the spin stiffness $\rho_s$ and the singlet-triplet excitation gap $\Delta$. 
One can therefore expect manifestations of quantum criticality for $T\gg \rho_s$ and $T\gg \Delta$, for $g<g_c$ and $g>g_c$, respectively. In continuum
field theories there is no upper-bound temperature for the quantum critical behavior, while for a lattice hamiltonian there has to be a cross-over to a 
different behavior above some temperature where the spins become essentially independent due to thermal fluctuations (which cannot be captured by continuum 
field theories, where independent discrete spins do not exist). The cross-over to the high-temperature limit could be defined, e.g., on the basis of the 
Curie (independent spin) susceptibility $\chi \to 1/(4T)$ obtaining for $T \gg J$. As we will see below, the quantum critical susceptibility is completely 
different. 

\begin{figure}
\includegraphics[width=8.75cm, clip]{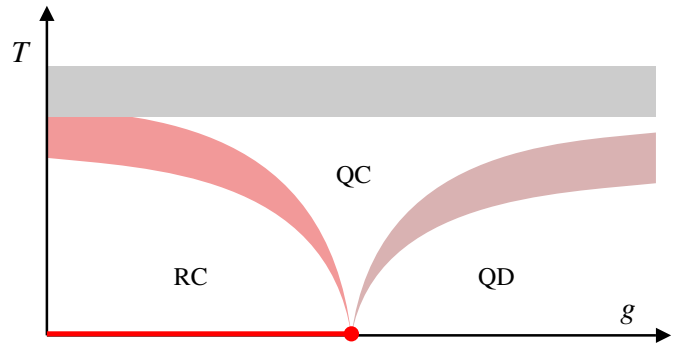}
\caption{Cross-overs (indicated by thick curves) in the coupling-temperature plane of a generic dimerized 2D Heisenberg model. The quantum-critical 
point (circle) controls the behavior in the $T\ge 0$ quantum critical (QC) regime, up to some temperature where lattice effects become important
(i.e., when $\xi \sim 1$). For couplings away from the critical point there is a low-temperature cross-over into either renormalized classical (RC) 
or quantum disordered behavior. For a 2D system there is N\'eel order only exactly at $T=0$. The thickness of the curves separating the different 
regimes reflect the fact that the cross-overs are not sharp, but can take place over significant temperature windows.}
\label{qcphases}
\end{figure}

Putting all this together, for a 2D dimerized Heisenberg model one expects a generic 
``cross-over diagram'' \cite{chn,chubukov}  of the type shown in Fig.~\ref{qcphases}. For temperatures below 
the high-$T$ cross-over, there are three different ``regimes'', in which the behavior is controlled be the corresponding largest energy scale in the problem; 
$\rho_s$, $\Delta$, or $T$. In the renormalized classical (RC) regime, the spin stiffness dominates, and the correlation length diverges as $T\to 0$ according 
to (\ref{corrlen2dform}), as we saw explicitly for the standard uniform Heisenberg model in Sec.~\ref{sec_2dheis}. In the quantum disordered (QD) regime, the 
correlation length saturates at some finite value as $T\to 0$, with $\xi\to{\rm constant}$ roughly below $T \approx \Delta$. In the quantum critical 
(QC) regime, the correlation length should diverge at $g_c$ when $T \to 0$ according to
\begin{equation}
\xi \sim 1/T^z.
\label{xidicgc}
\end{equation}
This behavior can be understood based on the path integral mapping of the quantum system in $d$ dimensions onto a classical ($d+z$)-dimensional system. 
There the coupling $g$ of the quantum system corresponds to $T$ of the ($d+z$)-dimensional system. We we can then think of temperature scaling in the quantum 
system as a kind of finite-size scaling in the imaginary time dimension, in which the thickness of the effective system is finite; $L_\tau \propto 1/T$ 
(i.e., we keep $T$ constant and just change the thickness of the effective system). By definition of the dynamic exponent, at $g_c$ we have
\begin{equation}
\xi \sim \xi_\tau^{1/z}.
\label{xixizinv}
\end{equation}
If the time dimension thickness $L_\tau=\infty$, both the correlation lengths $\xi$ and $\xi_\tau$ diverge at this point, but with finite $L_\tau$ (but 
infinite spatial size $L$) we should replace $\xi_\tau$ by $L_\tau \propto 1/T$ in scaling formulas (exactly as we do in classical finite-size scaling 
when $L < \xi$), whence (\ref{xixizinv}) gives (\ref{xidicgc}).

Quantum critical scaling forms have been derived for many quantities based on these ideas worked out quantitatively and in great detail \cite{chubukov} 
using renormalization group methods and large-$N$ expansions [with $N$ here being the number of components of SU($N$) symmetric spins, with $N=2$ for the 
physical $S=1/2$ spins] within the continuum field theory [the $(2+1)$-dimensional nonlinear sigma-model]. The main point is that one can observe 
scaling, with some corrections (the size of which depends on the quantity), also when $g\not=g_c$, in the QC regime illustrated in Fig.~\ref{dimsx}. Away from 
$g_c$, there is a cross-over into either RC or QD behavior. Here we only look at two important quantities; in addition to the correlation length we also
analyze the uniform susceptibility, which is perhaps the quantity for which QC behavior away from $g_c$ is the most robust. The predicted form is
linear in $T$ exactly at $g_c$, with a constant shift away from $g_c$;
\begin{equation}
\chi = \frac{a}{c^2}T+b,~~~~~~(b=0~ {\rm at}~ g=g_c),
\label{chitqc}
\end{equation}
where $c$ is the spin wave velocity. The constant $a$ is not known exactly, but its value computed based on the leading terms in an $1/N$ expansion is
believed to be close to the actual value \cite{chubukov}. The linear behavior should apply strictly only when $T\to 0$, since there are higher powers 
of $T$, with unknown prefactors, which are not included in (\ref{chitqc}).

\begin{figure}
\includegraphics[width=13cm, clip]{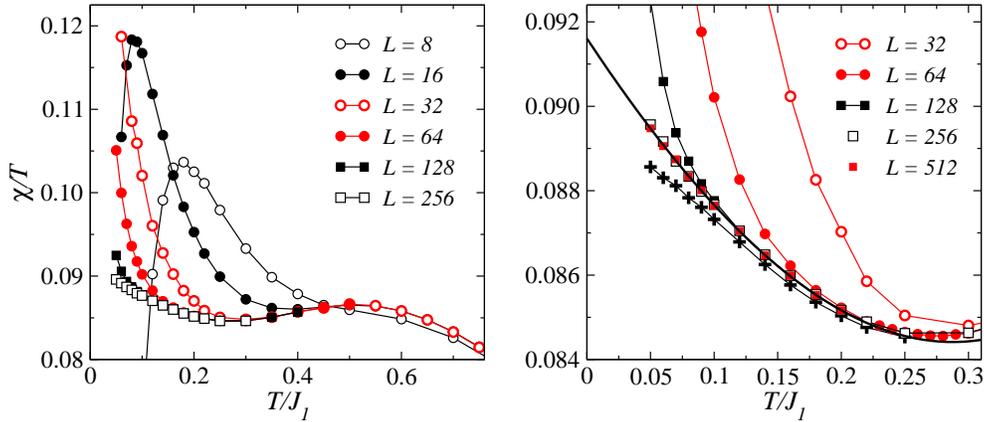}
\caption{Susceptibility divided by temperature for the dimerized system at the critical point ($g=1.9095$) computed using different lattice sizes. 
The peaks seen in the left panel for small systems is a finite-size feature which moves toward $T=0$ as $1/L$. The solid curve in the right panel is a 
cubic polynomial fit to the $L=512$ data ($T/J_1\le 0.20$), which gives the prefactor $a/c^2=0.0916(1)$ in Eq.~(\ref{chitqc}). For comparison, in the 
right panel, size-converged results for for $g=1.9090$ are also shown (the lowest curve, marked with ${\bf +}$). In this temperature range, quantum 
critical behavior is seen also at this coupling, which is slightly on the nonmagnetic side of the transition (and hence $\chi/T \to 0$ eventually 
as $T\to 0$)} 
\label{dimtsusc1}
\end{figure}

We begin by investigating the finite-size convergence of the susceptibility at $g_c$ (using $g_c=1.9095$, within the error bars of the critical
point determined above), in order to make sure that we can reliably obtain results reflecting the thermodynamic limit at low temperatures. The left panel
in Fig.~\ref{dimtsusc1} shows the temperature dependence of $\chi/T$ (which is easier to look at in graphs than $\chi$ itself) for several power-of-two 
system sizes up to $L=256$. As expected, the results converge quickly at high temperatures, with finite-size effects entering at approximately 
$T \sim 1/L$ (which can be expected on account of the dynamic exponent $z=1$). The right panel of Fig.~\ref{dimtsusc1} shows the low-$T$ results 
on a more detailed scale, including also results for $L=512$. Based on this comparison of results for different sizes, one can conclude that the 
thermodynamic limit can be studied with $L\le 512$ lattices down to $T/J_1=0.03$ (and probably even a bit below that).

The right panel of Fig.~\ref{dimtsusc1} also shows a fit of $\chi/T$ to a cubic polynomial at low temperatures. The corrections to the asymptotic linear 
form (\ref{chitqc}), is quite prominent, leading to an $\approx 8\%$ increase in $\chi/T$ [which can be considered as effective temperature dependent 
prefactor $a/c^2$ in (\ref{chitqc})] from the minimum around $T/J_1\approx 0.3$ to the eventual $T\to 0$ value. In principle the extrapolated $T=0$ value
$a/c^2=0.0916$ can be used to extract the spin wave velocity, but since the prefactor $a$ is not known exactly \cite{chubukov}, this estimate would likely
not be very precise. Another use of the result obtained here would be to use it in combination with an accurate estimate of $c$ obtained in some other
way, which would allow a test of the approximate calculation \cite{chubukov} of the prefactor. This is beyond the scope of the discussion here, however.

\begin{figure}
\includegraphics[width=13.5cm, clip]{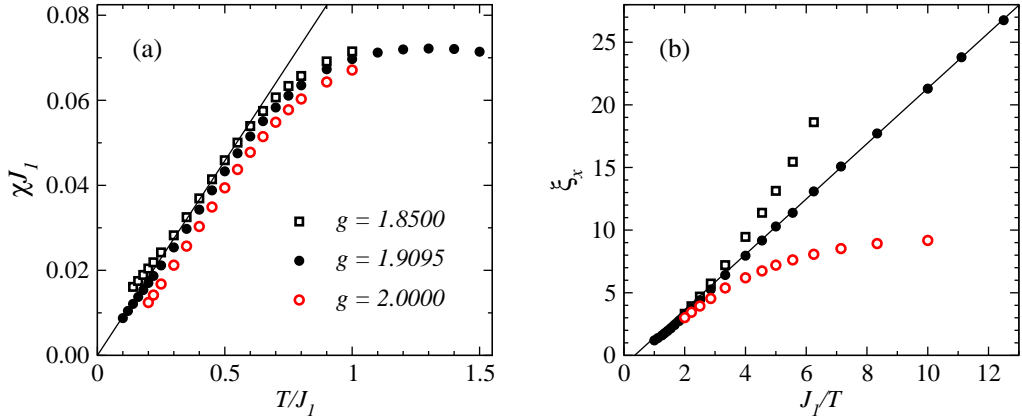}
\caption{Temperature dependence of the susceptibility (left) and the correlation length (right) in critical ($g=1.9095$) and near-critical systems. 
The solid lines show the asymptotic quantum-critical $T$-linear behavior of $\chi$ (from the $T=0$ value of the fit in Fig.~\ref{dimtsusc1}) and 
the $1/T$-linear behavior of $\chi_x$ [following the form (\ref{xixizinv}) with a small negative constant correction].}
\label{dimxk}
\end{figure}

Fig.~\ref{dimxk}(a) shows the temperature dependence of $\chi$ at $g=g_c$ and at two values some distance away on either side of the critical point. There
is a broad maximum at $T \approx J_1$, which corresponds to the cross-over into the eventual high-temperature Curie behavior. Below the maximum, these 
near-critical systems all exhibit an approximately $T$-linear susceptibility, in accord with the form (\ref{chitqc}) with $b<0$ and $b>0$ for $g>g_c$ 
and $g <g_c$, respectively. At still lower temperatures, there will be cross-overs into RC or QD behavior, which are not seen clearly here because the 
temperatures are still too high. In the QD regime $\chi \to 0$ exponentially for $T$ below the gap $\Delta$. The RC form is also linear, like the QC form, 
but the slope changes in the cross-over region. Such a cross-over can actually be seen in the standard 2D Heisenberg model, corresponding to $g=1$, which 
may appear to be too far away from the critical point. In the left panel of Fig.~\ref{hbsus2d} one can nevertheless see an approximately linear behavior of 
$\chi$ in the range $T/J \in [0.3 \sim 0.5]$, before the RC behavior sets in at lower temperatures. The slope of $\chi$ in the narrow window is 
in very good agreement with that obtained with the known spin wave velocity and the approximately calculated prefactor $a$ in (\ref{dimtsusc1}), which indicates 
that this behavior really is a manifestation of quantum critical behavior far from a quantum critical point \cite{chubukov,kim98} (although the very good 
agreement may be to some extent fortuitous, since, as we concluded above, there are significant corrections to the purely linear form exactly at $g_c$, 
at much lower temperatures than the QC window at $g=1$).  

Fig.~\ref{dimxk}(b) shows the correlation length in the $x$ direction (which is about $30\%$ larger than the $y$ correlation length at $g_c$) at the same
near-critical couplings as in Fig.~\ref{dimxk}(a). While the behavior is very linear, with a small constant correction to the asymptotic form (\ref{xixizinv}) 
with $z=1$, the results for the systems slightly off the critical point deviate significantly from linearity below $T/J_1\approx 0.3$. Being a divergent
quantity for $g\le g_c$, the correlation length has much larger corrections to the critical form than the non-divergent uniform susceptibility.

\subsubsection{N\'eel--VBS transitions in J-Q models}
\label{sec_jqresults}

With the J-Q models introduced in Sec.~\ref{jqmodelintro}, one can study quantum phase transitions at which not only the antiferromagnetic long-range 
order vanishes, but a different symmetry is broken in the nonmagnetic state as a VBS forms. Superficially, a VBS may seem rather similar to the 
nonmagnetic state of a dimerized Heisenberg model, because in both cases the system exhibits a pattern of strongly and weakly correlated nearest-neighbor 
pairs. These states are fundamentally very different, however, because in a ``manually'' dimerized Heisenberg model the hamiltonian itself breaks the 
translational symmetry, and the strongly correlated bonds also correspond to the strongly {\it coupled} spin pairs. In contrast, in J-Q models, the 
hamiltonian obeys all the symmetries of the lattice, and the translational symmetry is spontaneously broken in the VBS state. The quantum fluctuations 
of this VBS, especially close to the phase transition, are therefore very different (and much more interesting). 

On the square lattice, which we will consider here, and with the types of interactions we choose, the VBS can form in four equivalent patterns; 
hence the broken symmetry is $Z_4$. We will first discuss the standard J-Q model with four-spin couplings \cite{sandvik1} (and the term J-Q model 
will then refer just to that particular case). In this case the N\'eel--VBS transition appears to be continuous and shows many of the hallmarks 
of the proposed \cite{senthil1,senthil2} (and still controversial \cite{kuklov08}) deconfined quantum critical point. The VBS state in this 
case is most likely columnar, but the exact nature of the state is masked by an emergent U(1) symmetry in the VBS state (which in this model is 
always near-critical, with large fluctuations of the VBS order parameter). This is one of the salient features of the putative deconfined quantum 
critical point. Despite the many ways in which the behavior of the J-Q model agrees with the theory, there are also aspects of the N\'eel--VBS 
transition that remain unexaplained---although not necessarily in conflict with the theory, because detailed analytical calculations are very 
difficult \cite{senthil2,nogueira07} and many properties of the deconfined quantum critical point in SU(2) spin systems are simply not known
quantitatively based on the field theory proposed to describe the transition. We will here look at one example of strong corrections (possibly 
logarithmic) to the quantum critical scaling behavior, which had not been predicted by the theory of deconfined quantum critical points.

We will also discuss some results for a different kind of multi-spin coupling, which leads to a staggered VBS pattern [like the dimer pattern 
in Fig.~\ref{dlattices}(c)]. The phase transition in this case is clearly first-order \cite{arnab}. We will discuss the reasons for the qualitatively 
different nature of the transitions in the two models. 

\paragraph{SSE implementation for the J-Q model}

We first discuss some implementation issues for studies of the J-Q model with the SSE method.
The model was first studied with a projector QMC technique formulated in the valence bond basis (which we will briefly discuss in 
Sec.~\ref{sec_survey}). It is, however, also easy to implement the SSE method for it \cite{melko08a}, which was initially done using 
``directed loop'' updates (a generalized loop algorithm, which we will also summarize Sec.~\ref{sec_survey}). An SSE algorithm for the J-Q 
model can also be devised which is almost identical to the standard algorithm with operator-loops for the isotropic Heisenberg model. 
To see this, we can write the J-Q hamiltonian with four-spin interactions, Eq.~(\ref{jqham}), in terms of the diagonal and off-diagonal bond 
operators (\ref{heishb12_1})  and (\ref{heishb12_2}) used in the SSE algorithm for the Heisenberg model;
\begin{equation}
H = -J\sum_{[b]}(H_{1b}-H_{2b}) -Q\sum_{[bc]}(H_{1b}-H_{2b})(H_{1c}-H_{2c}).
\label{jqhamsse}
\end{equation}
As in the Heisenberg model $[b]$ is a bond with two interacting spins $[i(b),j(b)]$, while $[bc]$ denotes two parallel bonds $[i(b),j(b); k(c),l(c)]$,
corresponding to the singlet projector product $S_{ij}S_{kl}$ in (\ref{jqham}). We will consider these bonds arranged as in Fig.~\ref{jqops}, with the 
summation in (\ref{jqhamsse}) taken over all translations on the lattice and including horizontal as well as vertical bond orientations. The scheme 
is, however, independent on how the singlet projectors are arranged, and also the generalization to an arbitrary number of bonds in the Q term is 
trivial. The only constraint is that we have to avoid sign problems, which we do if both $J>0$ and $Q>0$ [with the minus signs in (\ref{jqhamsse}), 
which corresponds to energetically favoring singlets on the bonds included in both the J and Q terms]. The absence of sign problems was discussed
based on a sublattice rotation in \cite{melko08a}, and it can also be demonstrated using the simple operator counting arguments used for bipartite 
Heisenberg models in Sec.~\ref{sseheisenberg}.

\begin{figure}
\includegraphics[width=13cm, clip]{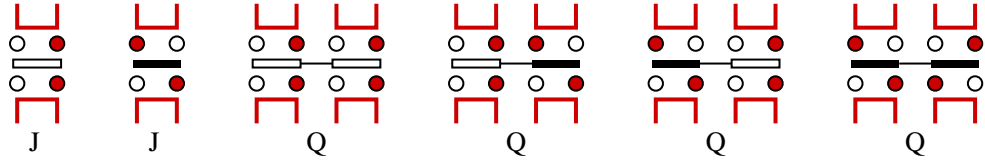}
\caption{Examples of vertices in the J-Q model. Only some of the allowed spin states of the vertex legs (open and solid circles) corresponding 
to the operators are shown. Open and solid bars indicate diagonal and off-diagonal bond operators, respectively. The J-vertices are identical 
to those in the Heisenberg model. The Q-terms in Eq.~(\ref{jqhamsse}) are products of two bond operators, which when expanded out include all 
combinations of diagonal and off diagonal factors. Allowed loops pass only through one of the operator factors in the case of Q-vertices, as 
illustrated here with loop segments at all the permissible leg pairs. Flipping loops can lead to any combination of allowed operators and spin 
states.}
\label{jqvertices}
\end{figure}

We now have J-vertices with four legs as well as Q-vertices with eight legs, as illustrated in Fig.~\ref{jqvertices}. The Q-vertices can be considered 
as two J-vertices joined together, in all possible combinations of diagonal and off-diagonal parts arising from the four operators in the Q term of 
(\ref{jqhamsse}). It is then clear that we can proceed in the same way as we did for the Heisenberg model, updating the operator string and a stored 
state using a combination of diagonal and loop updates. The key is here again that the matrix elements are the same for all J- and Q-vertices (the 
values being $J/2$ and $Q/4$), which means that a loop update in which the type of vertex (J or Q) is not changed can always be accepted. In the 
case of the Q-vertices, the loops satisfying this constraint enter and exit at the same operator factor, as illustrated with loop segments in 
Fig.~\ref{jqvertices}.

Both J and Q diagonal operators are inserted and removed in the diagonal update. The simplest way to insert diagonal operators is to choose completely 
randomly among all the possible single-bond $[b]$ and double-bond $[bc]$ instances in (\ref{jqhamsse}). There are $N$ each of horizontal and vertical 
bonds $[b]$ and also $N$ each of horizontal and vertical bond pairs $[bc]$, for a total of $4N$ cases to choose from. The spins in the current state 
have to be antiparallel on the bond or bonds acted on by the chosen operator, and if that is the case the acceptance probability is a simple 
modification of Eq.~(\ref{paccsseheis1}), with either $\beta_J=J/(2T)$ or $\beta_Q=Q/(4T)$ replacing $\beta$, depending on the type of operator inserted. 
The number of bonds $N_b$ is replaced by the total number of bonds and bond pairs, i.e., $4N$. The same modifications apply to the removal probability 
(\ref{paccsseheis1}) as well. If the ratio $Q/J$ is much different from $1$, which is the case in the VBS state and at the phase transition, it is 
better to take this ratio into account already when generating the bonds or bond pairs (but even the trivial random operator generation actually 
works very well). Note that $Q$ is much larger than $J$ in the parameter regime we are interested in, and it is therefore best to define the 
temperature in units of $Q$, i.e., setting $Q=1$ in the program. 

The loop update is identical to the one we developed for the Heisenberg model, with the simple extension that the linked vertex list (illustrated 
in Fig.~\ref{sseconfig2} for the Heisenberg model) now has eight elements for each vertex, instead of only four in the case of the Heisenberg model. 
Although the J-vertices have only four legs, the most practical way is to allocate eight storage elements for all vertices, using only the first 
four of them for the J-vertices (and filling the unused ones with a negative number, so that they are never visited in the loop update).

\paragraph{Continuous quantum phase transition  in the  J-Q model}

\begin{figure}
\includegraphics[width=13.5cm, clip]{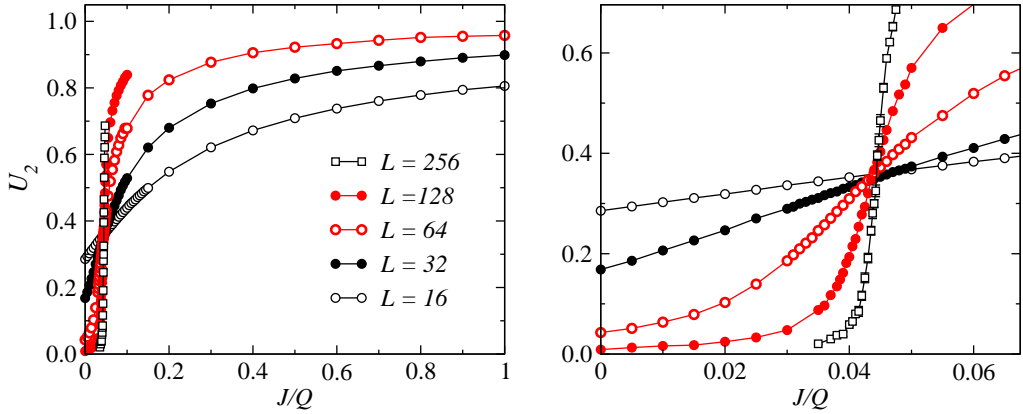}
\caption{Binder cumulant of the staggered magnetization of the J-Q model. The left panel shows results up to $J/Q=1$,
where $U_2$ approaches $1$, indicating a N\'eel state. For smaller $J/Q$ the cumulant vanishes with increasing size, as
shown in greater detail in the right panel, demonstrating a nonmagnetic state for $J/Q<0.045$ (below the crossing points,
which accumulate at the critical point). The cumulant remains positive for all $J/Q$, showing that there is no phase 
coexistence, i.e., based on these results the transition is continuous.}
\label{jqbinder}
\end{figure}

The initial study of the J-Q model \cite{sandvik1} established the existence of a VBS ground state for small $J/Q$ and a N\'eel--VBS transition
compatible with a $z=1$ quantum critical point. These results were soon thereafter confirmed in SSE studies of the $T>0$ quantum critical regime 
\cite{melko08a,melko08b}. A later world line QMC study claimed evidence for a first-order transition \cite{jiang2}, and this scenario was argued for 
also based on studies of other systems \cite{kuklov08}. Recent calculations on very large lattices (up to $L=256$) do not, however, find any indications 
of first-order behavior \cite{sandviklogs}. This is in violation of the ``Landau'' rule, according to which a direct transition between states breaking
unrelated symmetries should be generically discontinuous, but it agrees with the theory of deconfined quantum critical points \cite{senthil1}.
There are, however, significant unexpected scaling corrections in some quantities (which may have been mistaken for signs of a first-order transition
in \cite{jiang2}). Here we will first discuss some of the evidence for a single continuous N\'eel--VBS transition. 

The Binder cumulant is a useful quantity for analyzing both continuous and first-order phase transitions. As we discussed in Sec.~\ref{firstordertrans}, 
with an example in Fig.~\ref{j1j2binder}, phase coexistence at a first-order transition leads to a negative divergent peak in the cumulant at the
transition point. Fig.~\ref{jqbinder} shows results for the antiferromagnetic Binder cumulant $U_2$ of the J-Q model, calculated with the SSE method 
using $\beta=Q/T=L$ (motivated by the previous work showing that $z=1$). For $J/Q>0.05$, $U_2$ increases with $L$ and tends toward $1$, showing that 
the system has long-range N\'eel order. For smaller coupling ratios the cumulant decreases to $0$, with no indications of a negative peak developing. 
The crossing points are well behaved and indicate a critical point at $(J/Q)_c\approx 0.045$.

Next, we examine the correlation length, computed using the structure factor definition (\ref{xiadef}) for both the spin-spin and dimer-dimer correlations. 
While the wave-vector for the spin (N\'eel) order parameter is $(\pi,\pi)$, a columnar VBS corresponds to $(\pi,0)$ or $(0,\pi)$, which then is used
as the reference point $q=0$ in (\ref{xiadef}). To compute the dimer correlations, we use the definition (\ref{dxxdimorder}) but only include the diagonal 
correlations in the bond operator (\ref{bxxdef}) i.e.,
\begin{equation}
B_{xx}({\bf r})=S^z({\bf r})S^z({\bf r}+\hat {\bf x}). 
\label{bxxzzdef}
\end{equation}
Although the finite-size scaling properties of the correlation functions based on the rotational-symmetric and the $z$ component definitions may be different 
\cite{melko08a}, the correlation lengths extracted from these functions should still  both exhibit finite-size scaling $\sim L$ (up to scaling corrections) at 
the critical point. 

\begin{figure}
\includegraphics[width=13.5cm, clip]{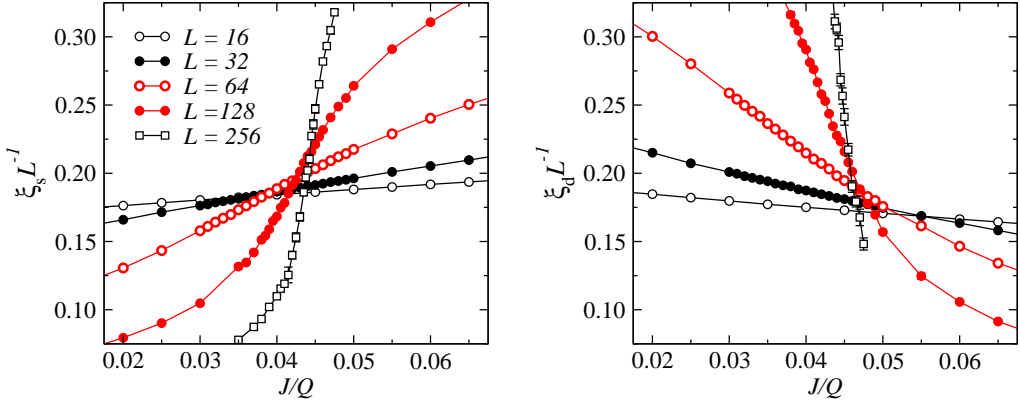}
\caption{Finite-size scaling of the spin (left panel) and dimer (right panel) correlation lengths of the J-Q model. Note that the crossing 
points for $\xi_s$ and $\xi_d$ move with increasing lattice size in different directions, toward each other.}
\label{jqcorrlen}
\end{figure}

We can use the scaled correlation lengths $\xi_s/L$ (spin) and $\xi_d/L$ (dimer) to test whether the two order parameters vanish at the same point, or whether 
there could possibly be two different transitions. Fig.~\ref{jqcorrlen} shows the raw SSE data, which again was generated using $\beta=L$. The curves for 
different system sizes cross each other, although there is some drift, both horizontally and vertically, even for the largest lattices. While the crossing 
$J/Q$ values for $\xi_s/L$ and $\xi_d/L$ are quite far apart for small lattices, they move toward each other with increasing system size, in a way 
consistent with a single critical point.

\begin{figure}
\includegraphics[width=9.6cm, clip]{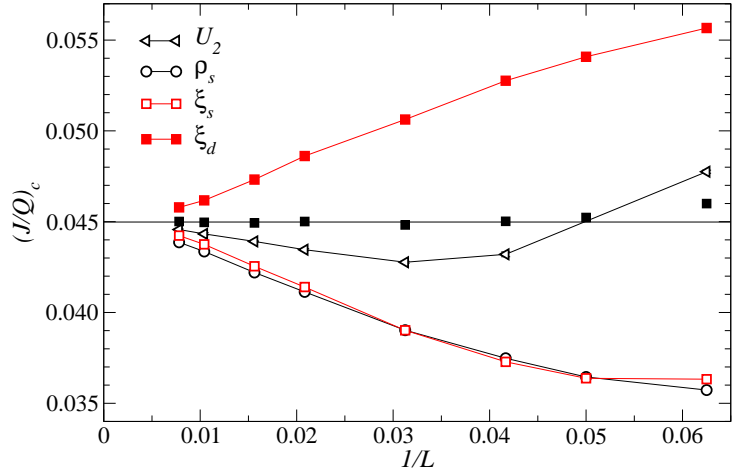}
\caption{Size dependent critical coupling estimates based on curve crossings for systems of size $L$ and $2L$, using several different quantities---the 
Binder cumulant $U_2$, the size-scaled spin stiffness $\rho_sL$, as well as the spin and dimer correlation lengths $\xi_s/L$ and $\xi_d/L$. The average
of the $\xi_s/L$ and $\xi_d/L$ estimates, shown with solid squares, exhibits almost no size dependence. The average of these points for the four 
largest $L$ gives the critical-point estimate $(J/Q)_c=0.04498(3)$, which is shown with the horizontal line.}
\label{jqcross}
\end{figure}

Fig.~\ref{jqcross} shows crossing points extracted from $L$ and $2L$ data for several different quantities (those discussed above as well as the 
spin stiffness). It is interesting to compare this graph with the corresponding one for the dimerized Heisenberg model, Fig.~\ref{dimcrossgc}. The 
convergence of the critical point is clearly much slower for the J-Q model. While the behavior for all quantities is almost linear in $1/L$ over some 
range of system sizes, for the largest systems the behavior starts to change, in a way which suggests an eventual convergence with some higher 
power of $L$ to a common critical point $(J/Q)_c\approx 0.045$. Recall that for the dimerized Heisenberg model the behavior is $\sim 1/L^\omega$ 
with $\omega$ in the range $2 \sim 2.5$, and this behavior applies already for rather small lattices ($L\approx 8$). Because of the slow convergence, 
it is difficult to estimate the critical point of the J-Q model precisely based on these individual crossing point estimates. There is, however,
a remarkable feature seen in Fig.~\ref{jqcross}: The $\xi_s/L$ and $\xi_d/L$ crossing points form curves that look very symmetric, approaching the 
apparent asymptotic value at the same rate but from different sides. Therefore, the average of these two estimates exhibits almost no size dependence 
at all (as also shown in the figure). Based on these average values one can therefore obtain a much better estimate of the critical point than what 
might initially have been expected. The result based on the four largest-$L$ points (which agree completely withn statistical errors) is 
$(Q/J)_c=0.04498(3)$. 

The slow convergence of the crossing points indicates large scaling corrections in the underlying physical quantities. Note that the Binder cumulant 
crossings have the weakest size dependence of the quantities analyzed in Fig.~\ref{jqcross}, which is also clear when comparing the raw data in 
Figs.~\ref{jqbinder} (cumulant) and \ref{jqcorrlen} (correlation lengths). Since the Binder cumulant involves a ratio of two similar quantities, 
anomalous scaling corrections in these may partially cancel each other. It is not at all clear why the finite-size corrections in the crossing
points for the two correlation lengths cancel each other almost completely. It may indicate some hidden symmetry between the near-critical VBS
and N\'eel states---perhaps some kind of duality inherent to deconfined quantum criticality [which is the case for the U(1) version of the theory 
\cite{senthil1}, but not explicitly in the SU(2) variant, which is the one of relevance for the J-Q model]. Note that it is not just the locations 
of the crossing points that are symmetric, but in fact the correlation length curves themselves look rather symmetric in Fig.~\ref{jqcorrlen}. By 
reflecting one of the graphs about the critical point, $J/Q-(J/Q)_c \to -[J/Q-(J/Q)_c]$, the spin and dimer correlation lengths for the largest 
system fall almost on top of each other, i.e., even the scaling functions for these correlation lengths appear to be the same.

\paragraph{Quantum critical susceptibility scaling at $T>0$}

\begin{figure}
\includegraphics[width=13.5cm, clip]{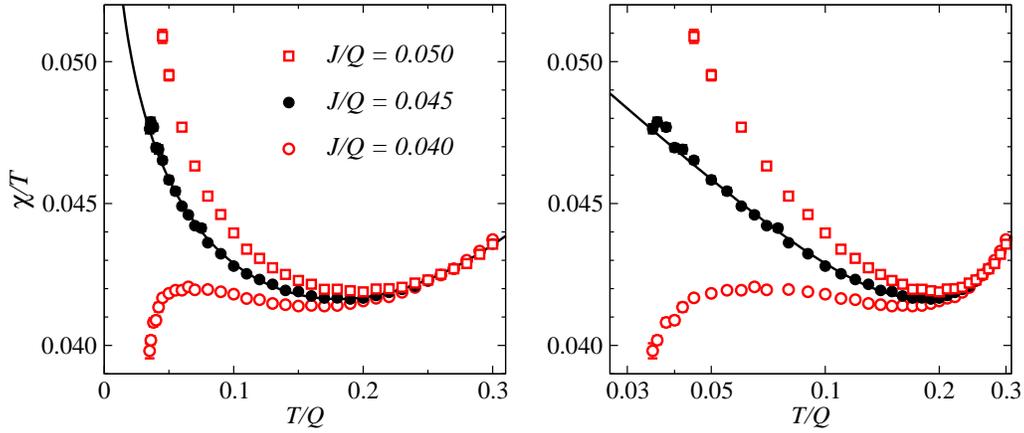}
\caption{Uniform susceptibility $\chi/T$ of the J-Q model graphed on linear (left) and logarithmic (right) temperature scales, for three coupling ratios 
in the neighborhood of the critical value, $(J/Q)_c = 0.04498(3)$. The solid curve is a fit to the form $\chi/T=a+b\ln(Q/T)+cT^2$ at $J/Q=0.045$
(which is within the error bar of the critical value). The SSE calculations were carried out using sufficiently large lattices (up to $L=512$ 
at the lowest temperatures) to obtain results representing the thermodynamic limit.}
\label{jqsusc}
\end{figure}

At a conventional quantum critical point, one would expect an asymptotic linear temperature dependence of the uniform magnetic susceptibility, as 
we saw in the dimerized Heisenberg model in Fig.~\ref{dimtsusc1}, where corrections in the form of higher powers of $T$ are also visible in
the ratio $\chi/T$. In the J-Q model there are much stronger corrections to the expected quantum critical form. As shown in Fig.~\ref{jqsusc},
there appears to be a multiplicative logarithmic correction, i.e., the asymptotic $T \to 0$ critical behavior may be of the form $\chi \sim T\ln(1/T)$.
The results are graphed both on linear and logarithmic temperature scales, in order to clearly convey how the behavior differs from that of 
the dimerized model in Fig.~\ref{dimtsusc1}. The behavior in the J-Q model does not appear to be compatible with a conventional higher power-law
correction, while a logarithm describes the behavior very nicely. 

Logarithmic corrections have not been suggested based on the present theory of deconfined quantum criticality. All analytical calculations within the 
non-compact CP$^1$ field theory proposed to describe the transition have been performed using large-$N$ expansions of the SU($N$) generalization of the 
system (the field theory generalized to CP$^{N-1}$, where $N=2$ is the physical number of spin components) \cite{senthil1,senthil1,nogueira07}. 
It is possible that different features appear for $N=2$. On the other hand, Monte Carlo calculations of lattice versions of the field theory 
\cite{motrunich04,takashima05,kuklov08} have not pointed to any anomalous scaling corrections (although one only found first-order transitions 
\cite{kuklov08}). Since these calculations are in disagreement with each other, the situation for the $N=2$ theory remains unsettled. Logarithmic 
corrections in the susceptibility and other quantities do appear in related fermionic gauge field theories \cite{kim97}. As we discussed in 
Sec.~\ref{sec_results1d} and \ref{sec_chainsse}, in field theory language logarithmic corrections can appear as a consequence of marginal 
operators, but other explanations are also possible. In principle the corrections may also not really be logarithmic---they could also be due 
to almost marginal (but in the end irrelevant) operators. Corresponding power-law corrections with a very small exponents may not be 
distinguishable from logs for the system sizes accessible. While this scenario is possible for some quantities, in the case of the uniform 
susceptibility that would also be highly unusual, since normally all corrections to quantum critical behavior in 2D antiferromagnets are 
integer powers of $T$ \cite{chubukov}.

\paragraph{Emergent U(1) symmetry in the VBS}

\begin{figure}
\includegraphics[width=13cm, clip]{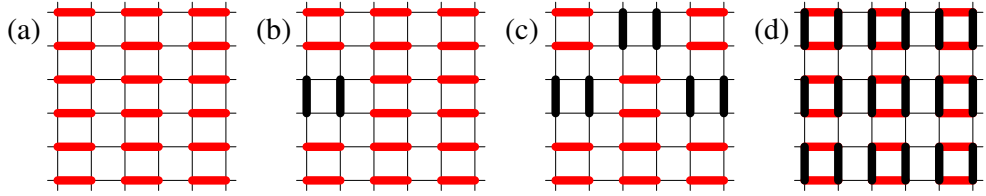}
\caption{Illustration of how local fluctuations of valence bond pairs can lead to a gradual change of a columnar state (a) into a plaquette 
state (d). In (b) and (c) a small number of bond pairs have been flipped. Local horizontal $\leftrightarrow$ vertical fluctuations of the 
maximum number of all bond pairs can take place with the plaquettes (bond superpositions) arranged as in (d) [one of four equivalent translated
patterns]. A fraction $p \in [0,1]$ of vertical bond pairs on these plaquettes corresponds to a VBS angle $\phi \in [0,\pi]$, defined using
the $x$ and $y$ dimer order parameters (\ref{dxdydef}). In these figures, angles $\phi \in [\pi,2\pi]$ correspond to shifting the vertical
bonds up by one step up.}
\label{vbsfluct}
\end{figure}

A prominent feature of the theory of deconfined quantum criticality is that, in addition to the normal correlation length $\xi$ (which can be taken
as the dimer correlation length $\xi_d$ in the N\'eel state and the spin correlation length $\xi_s$ in the VBS state), there is another length 
scale $\Lambda$ in the VBS phase. This is the distance scale on which excited $S=1/2$ spinons (which are defects, a kind of vortices, in the VBS 
\cite{levin04}) are confined into $S=1$ ``triplons'' (i.e., $\Lambda$ is the size of the spinon bound state). The same length scale should also 
be associated with an emergent U(1) symmetry, related to a particular kind of quantum fluctuations connecting two different kinds of VBS ordering
patterns---the columnar state and the plaquette state. These ``angular'' VBS fluctuations and the two kinds of VBS states are illustrated in
Fig.\ref{vbsfluct}. The angular fluctuations occur on length scales less than $\Lambda$ (i.e., on length scales less than $\Lambda$, the system is in
a superposition of the two different VBS states). The excited spinons can move freely away from each other up to distance $\Lambda$, but at
larger distances they become confined, due to an effective potential mediated by the VBS long-range order (and in this regard one can note
interesting connections to confinement/deconfinement in certain gauge theoreis in particle physics \cite{sachdev3}). As the critical point is 
approached, $\Lambda$ should diverge as a power of the correlation length, $\Lambda \sim \xi^a$, with the exponent $a>1$. 

The existence of an approximate U(1) symmetry in the VBS order parameter can be tested by examining the distribution $P(D_x,D_y)$ of order 
parameters $D_x$ and $D_y$, corresponding to columnar order with $x$- and $y$-oriented bonds and defined as
\begin{equation}
D_{x}=\sum_{i=1}^N {\bf S}({\bf r}_i)\cdot {\bf S}({\bf r}_i+\hat {\bf x})(-1)^{x_i} ,~~~~~~
D_{y}=\sum_{i=1}^N {\bf S}({\bf r}_i)\cdot {\bf S}({\bf r}_i+\hat {\bf y})(-1)^{y_i} .
\label{dxdydef}
\end{equation}
For a given Monte Carlo configuration, $D_x$ and $D_y$ can be evaluated at fixed imaginary time (propagated state in the SSE), using the improved 
estimator for the correlation function discussed in Sec.~\ref{improvedestimators}. The improved estimator is spin-rotational invariant, which can best be 
understood by considering the equivalence between the loop formulation and the valence bond basis \cite{nachtergale,evertz1,awshg} (which we briefly 
discuss in Sec.~\ref{sec_survey}). In fact, the improved loop estimator for an equal time correlation function is exactly equivalent to an expectation 
value of the operator in a particular valence bond state, provided that the temperature is sufficiently low for the simulation only to sample the 
singlet sector. 

While the calculation of $P(D_x,D_y)$ could be done with the SSE method, the results to be discussed below were instead generated with a ground state 
method in the valence bond basis (which in its most recent formulation \cite{awshg} actually is very similar to the SSE method, as discussed in 
Sec.~\ref{sec_survey}). The rotationally invariant estimator is not strictly needed here \cite{arnab}, and one can also just use the $z$ spin 
components in (\ref{dxdydef}), as we did in the correlation function Eq.~(\ref{bxxzzdef}). In the valence bond basis it is, however, natural to compute 
rotation-invariant quantities. Note that $P(D_x,D_y)$ is not an operator expectation value (i.e., it is not a normal physical observable), just a 
distribution of individual measurements in a simulation carried out in a particular basis. Its symmetry properties nevertheless reflect the actual
symmetries of the VBS order parameter of the system. 

\begin{figure}
\includegraphics[width=12.2cm, clip]{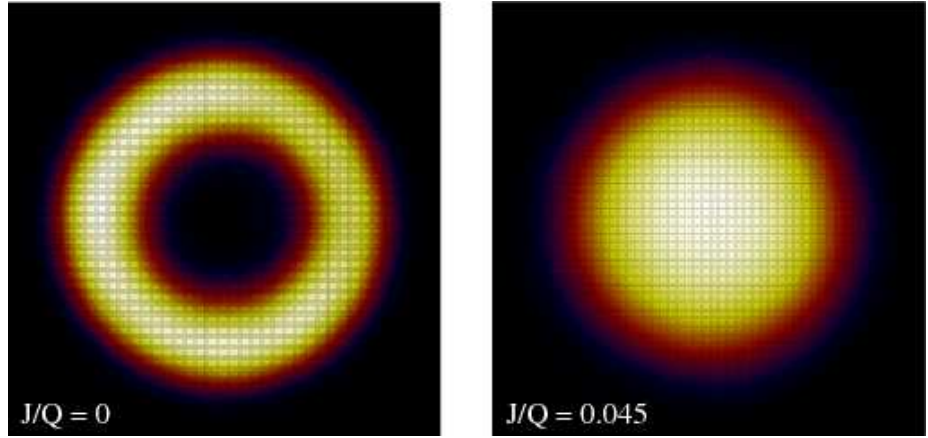}
\caption{VBS order parameter distributions $P(D_x,D_y)$ calculated for $L=128$ systems at $J/Q=0$ (left) and $0.045$ (right). The histograms are shown 
in regions $|D_x|,|D_x|\le D_{\rm cut}$, with $D_{\rm cut}$ chosen for the two coupling ratios to cover the region where the weight is significant. Brighter features 
correspond to higher probability density. The absolute scale is not important here, only the circular shapes of the distributions. The weigh is not 
distributed completely uniformly in these histograms [obeying neither perfect $Z_4$ nor U(1) symmetry], due to the inefficiency of the simulations to 
fluctuate the VBS angle for large systems.}
\label{vbhist}
\end{figure}

We have already looked at a classical analogy of $P(D_x,D_y)$; the distribution $P(M_x,M_y)$ of stripe order parameters in the frustrated Ising model, with 
results shown in Fig.~\ref{j1j2hist}. In that case the four-fold symmetry of the order parameter at and below the transition temperature is obvious, while going 
to very high temperatures (not shown in the figure) the distribution turns into a central peak with full rotational symmetry [i.e., U(1) symmetry] because of 
the independent Gaussian fluctuations of the short-range order in different parts of the system. In a similar way, in a VBS one should expect $P(D_x,D_y)$ 
to be U(1) symmetric in the N\'eel state (where there is only short-range VBS correlations), and become four-fold ($Z_4$) symmetric inside the VBS state. 
As seen in Fig.~\ref{vbhist} (which shows recent results for larger systems than in \cite{sandvik1,jiang2,lou1}), this is not quite the case, however. 
Inside the VBS phase, at $J/Q=0$, the distribution is ring shaped, with no weight at the center and no signs of a four-fold symmetry. This shows that the 
magnitude $D$, with $D^2=D_x^2+D_y^2,$ of the VBS order parameter has formed, $\langle D^2 \rangle >0$, but its angular symmetry has not yet been established, 
i.e., while the magnitude of the VBS order parameter has relatively small fluctuations, the fluctuations in the individual components $D_x$ and $D_y$ [which 
define an angle $\phi$ of the VBS in the $(D_x,D_y)$ plane] are very large. Close to the critical point, at $J/Q=0.045$ in the figure, the distribution 
is peaked at the center, as expected, but again there is no $Z_4$ symmetry.

In the continuum field theory, the operator responsible for the VBS formation is {\it dangerously irrelevant} \cite{senthil1,levin04,metlitski08}, which 
means that although it is a perturbation with $Z_4$ symmetry, which leads to a $Z_4$ symmetry-broken VBS state, this perturbation does not affect the 
critical point. The theory of deconfined quantum critical point has built-in U(1) symmetry, and this is unchanged in the presence of the VBS operator. 
The J-Q model has no explicit U(1) symmetry---what is built in here is instead the Z$_4$ discrete rotational symmetry of the square lattice, and naively 
one would expect the VBS forming on this lattice to exhibit $Z_4$ symmetry, exactly as the stripe order of the Ising model discussed in Sec.~\ref{firstordertrans}.
Instead, we see a U(1) symmetric order parameter distribution $P(D_x,D_y)$, which in light of the theory of deconfined quantum critical points should
be expected if the system is smaller than the deconfinement length scale $\Lambda$ (which therefore has to be large, $\Lambda > 128$ even at
$J/Q=0$ according to the results in Fig.~\ref{vbhist}). If we increase the system size, we should eventually, for $L>\Lambda$, obeserve a four-fold 
symmetric distribution (which has been seen in some related systems \cite{lou1,lou09,beach09}). A columnar state corresponds to peaks at $(0,\pm D)$ and 
$(\pm D,0)$. 

As discussed above and illustrated in Fig.~\ref{vbsfluct}, the emergent U(1) symmetry is a consequence of local fluctuations between columnar and plaquette 
VBS patterns \cite{levin04}, in such a way that all possible values of $D_x$ and $D_y$ compatible with a magnitude $D$ (weakly fluctuating) are sampled 
equally (or almost equally---perfect U(1) symmetry only applies at the critical point). There is an analogy in elementary quantum mechanics; 
a particle in a slightly assymmetric double well. If the barrier between the wells is very high, then the particle is localized in the deeper well, but if 
the barrier is not very high (relative to the kinetic energy), then the particle fluctates between both wells and the wave function also has weight inside 
the barrier. In the VBS, the wells correspond to the columnar and plaquette states, at angles $n\pi/2$ (columnar) and $n\pi/2+\pi/4$ (plaquette), with 
$n=0,1,2,3$, and fluctuations between them correspond to non-zero probability in the continuum of angles between these points. In the deconfinement 
theory, the columnar and plaquette VBS states are almost degenerate, becoming exactly degenerate at the critical point. Approaching the 
critical point, the barrier between them also is reduced. In the double-well analogy, the wells are becoming more degenerate and the barrier between 
them is reduced. There will then be significant fluctuations between columnar and plaquette states in finite systems (up to $L \sim \Lambda$), and 
this implies $P(D_x,D_y)$ with weight at all the angles and an  U(1) to $Z_4$ cross-over of the symmetry when $L > \Lambda$. This applies also 
to the order parameter computed on subsystems of finite length $L$ in an infinite (or very large) system (which has been studied in classical 
models with a $q$-fold order parameter which exhibits  U(1) to $Z_q$ symmetry cross-over \cite{oshikawa00,hove03}).

Since the $Z_4$ symmetry is not yet clearly manifested in the J-Q model up to the largest lattices studied so far, the issue of the nature of the VBS remains 
open. However, a J-Q model in which the Q term consists of products of three parallel singlet projectors, the VBS order is more robust and a $Z_4$ symmetric 
distribution corresponding to a columnar state is seen clearly \cite{lou1}. Simulations of SU($N$) generalizations of the standard J-Q model, with $N=3$ and 
$4$, also show columnar VBS ground states \cite{lou1}. Most likely, the standard SU(2) J-Q model also has a columnar state, and there is indeed some weak
hints of this in recent (ongoing) long simulations for $L=64$. 

For system sizes $L\gg\Lambda$, the VBS distribution should be four-fold symmetric, and by investigating how the change from U(1) to $Z_4$ symmetry takes place 
one can in principle determine the exponent $a$ relating the deconfinement length scale $\Lambda$ and the correlation length $\xi$. Classical examples of 
the cross-over phenomenon, which also inspired the theory \cite{senthil1}, were studied in Refs.~\cite{oshikawa00,carmona00,hove03,lou07b}. A finite-size 
scaling method for a quantity sensitive to $Z_4$ symmetry can be used to extract the exponent $a$ \cite{lou07b}. In the standard J-Q model, such an analysis 
cannot yet be performed, because the cross-over into a $Z_4$ symmetric order parameter is not seen clearly enough in the currently accessible lattices. 
The analysis has, however, been carried out in the J-Q model with columnar six-spin interactions, with the result that $a\approx 1.2$.

\paragraph{Other critical  exponents}

In addition to the dynamic exponent $z=1$, other critical exponents have been analyzed in several QMC studies \cite{sandvik1,melko08a,melko08b,lou1,sandviklogs}. 
The most noteworthy result is that $\eta_s$ (the exponent of the critical spin-spin correlation function) is anomalously large, $\eta_s \approx 0.3$, which 
is in qualitative agreement with the theory of deconfined quantum critical points. A ``large'' $\eta_s$ was argued for \cite{senthil1} based on the $N=\infty$ 
value $\eta_s=1$ in the $1/N$ expansion of the CP$^{N-1}$ field theory (i.e., different from the conventional mean-field value $\eta_s=0$). The $1/N$ corrections 
are difficult to compute, but  in general support an unusually large exponent \cite{nogueira07} (although it is not possible to extend the results reliably to 
$N=2$). The behavior of the dimer-dimer exponent $\eta_d$ obtained in QMC calculations for SU($N$) systems \cite{lou1} seems to follow the scaling with $N$ 
predicted in the theory \cite{metlitski08,murthy90} (where $\eta_d$ diverges with $N$). 

In light of the presence of significant scaling corrections, possibly logarithmic, in the susceptibility analyzed in Fig.~\ref{jqsusc}, and also in the spin 
stiffness \cite{sandviklogs}, an important question is whether the calculated  exponents also are affected. This has so far not been apparent, because 
standard data collapse procedures with reasonable values of the subleading exponents work very well \cite{lou1}. The best current estimates for $\eta_s$, 
based on both $T>0$ \cite{melko08a} and $T=0$ \cite{lou1} QMC calculations agree well with each other. On the other hand, the results discussed here 
and in \cite{sandviklogs} put the critcal point at slightly higher $J/Q$ than previously, and also the correlation length exponent $\nu$ analyzed with 
a logarithmic correction included \cite{sandviklogs} is significantly smaller than other estimates \cite{melko08a,lou1}. It would therefore be good to repeat 
the studies of all the exponent, using several different quantities and larger lattices (which is work in progress).

There is also evidence for logarithmic corrections in the J-Q model from studies of its response to impurities. Introducing a vacancy into the
system and studying the spatial distribution of the resulting magnetization imbalance, one can normally, at conventional quantum critical points 
in dimerized models, observe scaling with the system size in this distribution \cite{hoglund07,banerjee10}. In the J-Q model, such scaling analysis
again requires the introduction of a multiplicative logarithmic correction \cite{banerjee10}.

\paragraph{Applications of other methods to the J-Q model}

One interesting and puzzling aspects of the J-Q model is that it seems very difficult to capture its properties correctly using 
analytical many-body techniques, even ones that can rather accurately locate the N\'eel--VBS transition in the 2D frustrated J$_{\rm 1}$-J$_{\rm 2}$ 
Heisenberg model. A mean-field treatment starting from a columnar dimer state gives a critical $J/Q$ very far from the QMC 
result \cite{kotov09}. This approach can be improved to better take into account some of the local fluctuations on plaquettes, which
improves the value of the critical point but seems to results in a strongly first-order transition \cite{kotov09}. Cluster mean-field 
calculations converge very poorely with the cluster size \cite{isaev09}. These results point to unusually strong non-local quantum fluctuations 
\cite{kotov09}, which cannot be easily captured with local approaches starting from small clusters or conventional fluctuations 
around a fixed dimer pattern. The reason for these difficulties to capture the VBS state may be the emergent U(1) symmetry, which makes 
it difficult to obtain both the correct long-distance behavior (likely columnar order) as well as the strong fluctuations between 
columnar and plaquette order on shorter length scales.

\paragraph{First-order transition in a staggered J-Q model}

One way to test the link between emergent U(1) symmetry and a continuous N\'eel--VBS transition is to construct a model in which the local
fluctuations responsible for rotating the VBS angle are suppressed. Intuition for how to accomplish this comes from the Rokhsar-Kivelson (RK)
quantum dimer model \cite{rk88,fradkin04}, which can be regarded as an effective model for an extreme nonmagnetic system dominated by short valence bonds 
(for which the internal singlet structure is also neglected---the bond configurations are regarded as orthogonal states). The RK hamiltonian on 
the square lattice can be written as $H_{\rm RK}=vV-kK$, where $V$ is the diagonal (potential-energy) operator, which counts the number of flippable 
plaquettes [parallel bond pairs, exactly as in Fig.~\ref{vbsfluct}(b,c)], and $K$ is an off-diagonal (kinetic) term which flips such a pair. This 
model has a critical point at $k=v$ which separates a plaquette VBS state [similar to the one in Fig.~\ref{vbsfluct}(d)] for $v<k$ and a 
staggered VBS state [with the bond pattern exactly as in Fig.~\ref{dlattices}(c)]. While the plaquette state is destroyed continuously by quantum 
fluctuations as $v \to k^-$, the staggered state (of which there are four symmetry-related equivalent ones) has no fluctuations, because it has 
no flippable plaquettes. The transition upon $v\to k^+$ is therefore first-order. 

\begin{figure}
\includegraphics[width=8cm, clip]{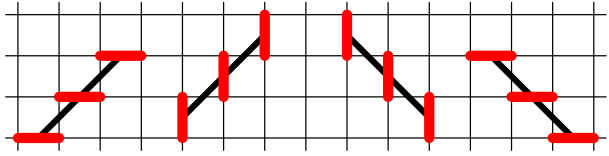}
\caption{In the J-Q$_{\rm 3}$ model studied here, three singlet projectors are arranged in a staggered pattern. All distinct orientations
(as shown) and translations of the projector products are included.}
\label{jq3oper}
\end{figure}

This simple picture of the RK model suggests that an actual staggered VBS in a spin model also should have strongly suppressed local fluctuations, 
and therefore should not be associated with an emergent U(1) symmetry. Due to the suppression of local fluctuations (and therefore also of
large-scale fluctuations), the transition between it and the N\'eel state should be first-order. The picture is not complete, however, because 
clearly there must be some fluctuations in the staggered VBS state, considering that a reasonable spin hamiltonian will be quite far from a dimer 
model and the valence bond configurations making up the gound state should also include some fraction of longer bonds (in the presence of which some
local fluctuations are always possible). It is therefore worth testing this picture in simulations of a model whose ground state can be tuned 
from a N\'eel antiferromagnet to a staggered VBS. Such a model can easily be constructed within the J-Q framework, by arranging the singlet projectors 
of the Q term in a staggered fashion, instead of within a $2\times 2$ plaquette (or in columns of three or more projectors, as in \cite{lou1}). It turns 
out, however, that a Q term consisting of two staggered projectors is not sufficient for destroying the N\'eel state. With three projectors, arranged 
as in Fig.~\ref{jq3oper} (and the coupling of which we call $Q_3$), a staggered VBS is stabilized, however. We here briefly summarize the 
evidence for a first-order N\'eel---VBS found in a recent SSE study of the staggered J-Q$_{\rm 3}$ model \cite{arnab} .

\begin{figure}
\includegraphics[width=13.5cm, clip]{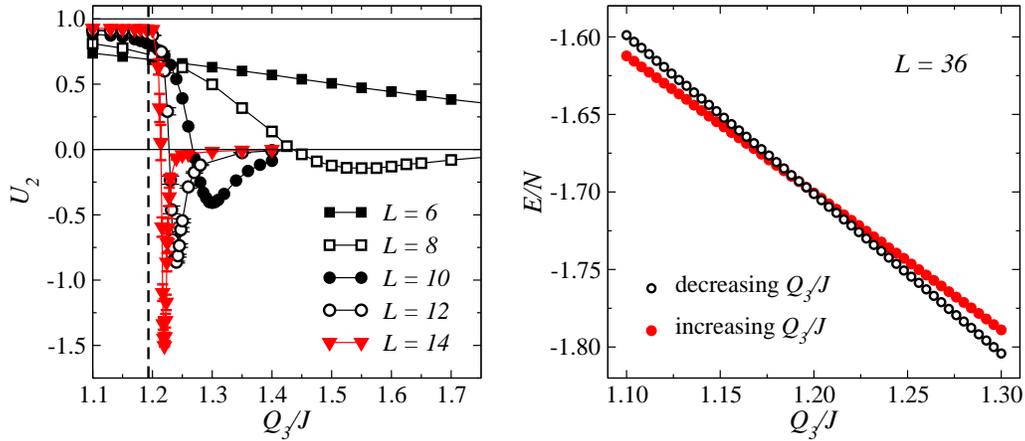}
\caption{Evidence for a first-order transition in the saggered J-Q$_{\rm 3}$ model \cite{arnab}. The Binder cumulant of the sublattice magnetization 
(a) exhibits a negative peak, which grows with the system size. (b) Hysteresis effects are observed: simulations in which the coupling ratio is
slowly increased and decreased give different properties in large systems, with the energies crossing each other at a point which can be
taken as a size-dependent transition point. Extrapolating these points for different system sizes gives $(Q_3/J)_c=1.1933(1)$, which is
marked with the vertical dashed line in the left panel. All these results were obtained with SSE simulations at inverse temperature
$Q_3/T=L$.}
\label{jq3data}
\end{figure}

Some results are shown in Fig.~\ref{jq3data}. The Binder cumulant of the N\'eel order parameter exhibits a negative peak, which grows
with increasing system size. This is a strong indication of a first-order transition, as discussed in Sec.~\ref{firstordertrans} (and
the results can be compared with those of the classical frustrated Ising model in Fig.~\ref{j1j2binder}). The data become very noisy for $L>12$, 
because of the inefficiency of the simulations in transitioning between the two coexisting states. 
For $L>24$ the system can get completely trapped in a metastable state. On the VBS side close to the transition, if the simulation 
happens to enter a N\'eel like configuration during the equilibration, it will stay in a metastable N\'eel state for a very long time 
(and vice versa on the other side of the transition). One can use this trapping to investigate the transition in other ways. By doing
a series of runs at closely spaced values of $Q_3/J$, starting either far inside the VBS or N\'eel phase, and starting each new
run from the last configuration of the previous run, one can follow the ground state into its corresponding metastable state at the
``wrong'' side of the transition. The energies computed in two such runs, with $Q_3/J$ either increasing or decreasing, are shown
in Fig.~\ref{jq3data}. The two branches cross each other, which clearly shows the first-order nature of the transition. The level
crossing is most likely an avoided one, between two states with the same quantum numbers, but the spacing between the levels should
be extremely small here. This spacing should be inversely related to the tunneling time between the two states at the coexistence 
point (but note that the relationship between simulation time and the real time dynamics of the system is not known). 

The crossing point of the two energy branches can be taken as a finite-size definition of the critical coupling. It shifts with the 
system size approximately as $1/L^3$, which is consistent with a first-order transition with dynamic exponent $z=1$ \cite{continentino}. 
The extrapolated transition point is $Q_3/J=1.1933(1)$. At the transition, the VBS order parameter is about $75\%$ of its maximum
value (in a perfect staggered valence bond state), which is rather large and motivates the classification of the transition as strongly
first-order.

\section{Survey of related computational methods}
\label{sec_survey}

In these lecture notes we have discussed exact diagonalization methods and the SSE QMC method in some detail and also looked 
at some illustrative calculations with results for $S=1/2$ models. There are many issues that were left untouched, regarding 
methods as well as physics, by restricting the discussion to $S=1/2$ systems, and within that class also to spin-isotropic 
systems. The exact diagonalization approach can be easily used for any spin model, but the rapidly growing size of the Hilbert 
space with $S$ imposes even more severe restrictions on the lattices sizes for $S>1/2$ systems. The world line and SSE QMC methods 
can be generalized to any spin model, as long as there is no frustration leading to sign problems. In this section we briefly 
summarize some QMC schemes for more general spin models, as well as some further recent developments for isotropic $S=1/2$ systems. 

There are several other important computational methods that go beyond the scope of these lecture notes, such as series 
expansions techniques (high-temperature expansions and expansions around some solvable limit of the ground state, 
e.g., the Ising limit of an anisotropic Heisenberg model) \cite{domb,oitmaabook} and the DMRG method \cite{white1,schollwock2}. 
While series expansions can in principle be applied to any model, there are issues with extrapolating the series in a controllable 
manner, and one cannot, in most cases of interest, expect to reach the same level of accuracy as in QMC simulations of sign problem 
free models (e.g., in studies of quantum phase transitions). Series expansion methods are nevertheless one of the most powerful 
classes of methods currently available for 2D frustrated quantum spin systems \cite{singh}, where QMC methods are limited by the sign 
problem (but note recent progress in controlling the sign problem at high temperatures \cite{nyfeler08}). More powerful series 
expansion schemes are also still being actively developed \cite{rigol07}, and one can expect progress to continue on this front. 
The DMRG method is very powerful for 1D systems and can also be used for 2D systems of moderate size \cite{chernyshev}. Extensions 
of the DMRG method and the related matrix-product states \cite{aklt,ostlund95,verstraete1} to higher-dimensional 
``tensor-network'' states \cite{verstraete1,murg1,vidal07} are currently being explored very intensely. This is an exciting line 
of research, that may eventually lead to viable schemes for unbiased studies of frustrated spins (and even fermionic systems). 
It appears unlikely, however, that these methods will, for the foreseeable future, be able to probe quantum spin systems at the 
level of precision possible with existing QMC methods for unfrustrated systems. 

In light of the above advantages and disadvantages of different methods, the author advocates a two-pronged approach in research on 
quantum spin models and related systems: (i) Explore new and improved schemes (including also QMC schemes) that may be useful to study 
models beyond the reach of currently available QMC techniques. (ii) Study interesting physics with state-of-the-art QMC methods for sign 
problem free models. It is far from true that everything that can be done has already been done with sign problem free models (although 
one can some times hear such claims)! With the QMC methods available now, such as the SSE algorithm described in Sec.~\ref{sec_sse}, 
and with the power of modern computers (which still is increasing at an amazing pace, e.g., with a steadily increasing numbers of cores 
per CPU), it is possible to access very interesting physics that was beyond reach only a few years ago. The discussion of the dimerized
and J-Q classes of models in Sec.~\ref{sec_jqresults}, in particular, has hopefully convinced the reader that there is much to explore 
in such systems, and beyond (e.g., random systems exhibiting unusual states and quantum phase transitions 
\cite{dimdiluted,yu06,roscilde07,wangperc}). 

\paragraph{Directed loop QMC algorithms}

\begin{figure}
\includegraphics[width=11.25cm, clip]{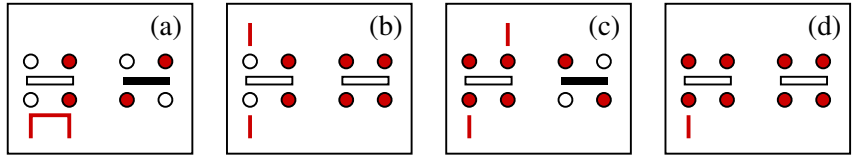}
\caption{Examples of the four possible paths through vertices in the anisotropic $S=1/2$ Heisenberg model (which may include also an external 
magnetic field) and the new vertices generated when moving along the paths in the directed loop algorithm. The vertical lines at the 
vertices to the left in each box indicate the entrance and exit legs (with either direction of movement being possible). Given an entrance 
leg, the exit must be chosen such that a new allowed vertex is generated when the leg spins are flipped (the vertices to the right, 
without the path indicated). In the ``bounce'' process (d), the exit is at the same leg as the entrance and the vertex is not modified. 
There are six allowed vertices in the generic anisotropic Heisenberg model---those in which the spin ($z$ component) is conserved. 
In the ``deterministic'' loop algorithm discussed in Sec.~\ref{sec_sse}, there are only four vertices and only paths of type (a) are allowed}
\label{vertpaths}
\end{figure}

The directed loop algorithm is a generalization of the SSE operator-loop scheme, introduced in \cite{syljuasen02} (building on a previous
less efficient formulation \cite{sandvik99a}) and applied there to $S=1/2$ Heisenberg systems with interaction anisotropies (Ising and XY) 
and external magnetic field. The main difference between loops and directed loops (in both SSE and world line formulations) is that the path 
through the vertices is not unique in the directed loop scheme. The path (the exit leg, given an entrance leg to a vertex) is chosen according 
to certain probabilities, constructed to maintain detailed balance in a space with two defects---the open loop ends (in a way similar in spirit 
to the ``worm'' algorithm \cite{prokofev96}). For $S=1/2$ systems, there are four types of vertex paths, corresponding to the location of 
the exit relative to the entrance, as illustrated and discussed in more detail in Fig.~\ref{vertpaths}. The path probabilities, which should 
be solutions of the corresponding {\it directed loop equations} (for detailed balance) are in general not unique. For some models, eliminating 
the ``bounce'' process in Fig.~\ref{vertpaths}(d) reduces the directed loop scheme to one of the standard loop updates previously developed for 
$S=1/2$ (as in the isotropic system discussed in Sec.~\ref{sec_sse}) and higher-$S$ models \cite{evertz93,evertz1,kawashima94}. In other cases the 
directed loops (and worm) algorithms allow for efficient simulations where the standard loop algorithms are not applicable 
\cite{syljuasen02,syljuasen03,kawashima04,melko05}. Here it can also be mentioned that it is some times useful in SSE directed loop algorithms
to work with operators defined on cells larger than those containing the elementary operators of the Hamiltonian (e.g., the two-spin bond 
operators of the Heisenberg model)  \cite{louis04}. This allows for more options in choosing how operators are reconfigured when vertices
are traversed by the loops, which some times can make simulations much more efficient.

\paragraph{QMC algorithms in the valence-bond basis}

In Sec.~\ref{vbsrvb} we briefly discussed the valence bond basis, in which a basis state is a product of two-spin singlets, as in Eq.~(\ref{psivbprod}), 
and in which any total singlet singlet state can be expanded. The basis is overcomplete and non-orthogonal, which implies that such an expansion is not 
unique. One way to work with the valence bond basis is to construct and optimize variational states, the simplest type of which is an {\it amplitude 
product states} \cite{liang1}. Such a state is a superposition 
\begin{equation}
|\Psi\rangle = \sum_\alpha \Psi_\alpha |V_\alpha\rangle,
\label{psivbexpand}
\end{equation}
where $|V_\alpha\rangle$ is a valence bond product state of the form (\ref{psivbprod}) and the expansion coefficients are products of amplitudes 
$h({\bf r}_{\alpha,i})$ corresponding to the ``shapes'' of the bonds (the bond lengths in the $x$ and $y$ direction in the case of a 2D system), 
with $i=1,\ldots,N/2$ referring to the $N/2$ valence bonds in the configuration (bond tiling) labeled by $\alpha$;
\begin{equation}
\Psi_\alpha =\prod_{i=1}^{N/2} h({\bf r}_{\alpha,i}).
\end{equation}
Amplitude product states of this kind can closely reproduce the ground states of many bipartite Heisenberg systems, for which each singlet 
$(a,b)=(|\up_a\dn_b\rangle-|\dn_a\up_b\rangle)/\sqrt{2}$ should be defined with site $a$ on sublattice $A$ and site $b$ on sublattice $B$. With 
all positive expansion coefficients $\Psi_\alpha$ in (\ref{psivbexpand}), this convention for the singlet sign corresponds to {\it Marshall's sign 
rule} for the ground state wave function of a bipartite system \cite{marshall55}, which in the standard $\up,\dn$ spin basis can be written as
\begin{equation}
{\rm sign}[\Psi_\alpha]=\Psi_\alpha/|\Psi_\alpha|=(-1)^{n_{A\dn}},
\end{equation}
where $n_{A\dn}$ is the number of $\dn$ spins on sublattice $A$. 

The properties of these bipartite amplitude product states can be studied by Monte Carlo sampling of the valence bonds \cite{liang1,lou07,awshg}. For the 
2D Heisenberg model, the state with all the amplitudes $h(x,y)$ variationally optimized is a N\'eel state with properties in very good agreement with 
the true ground state \cite{lou07}, e.g., the energy is within $0.1\%$ and the sublattice magnetization within $1\%$ of the values obtained in QMC 
calculations. For frustrated systems, appropriate sign rules are not known (and may be too complicated to write down in practice).

\begin{figure}
\includegraphics[width=7cm, clip]{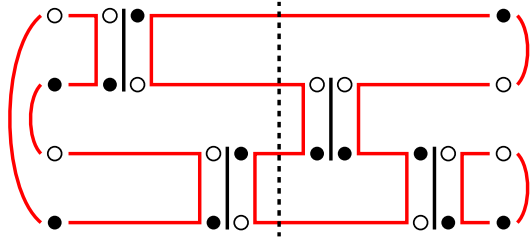}
\caption{Example (from Ref.~\cite{awshg}) of a loop configuration in the valence-bond projector QMC method formulated in a combined 
space of spins (open and solid circles) and valence bonds (the arcs capping the loops on the left and right boundaries). The bars with
four spins are diagonal and off-diagonal vertices with the same meaning and function as in the SSE operator-loop method (illustrated 
in Fig.~\ref{sseloop}). Loops form according to the connectivity of the vertices, and also through the valence bonds at the boundaries,
and can be flipped without changing the configuration weight. The sum over all possible loop configurations corresponds exactly to the 
formulation of the projector scheme in the pure valence bond basis (i.e., without using the spins at all) \cite{vbmethod1}. Using the 
spins allows for a more efficient sampling of the configurations. Expectation values are evaluated using loop estimators at the mid-point 
indicated with a vertical dashed line, and this is analogous to the improved estimators (discussed in Sec.~\ref{improvedestimators}) 
in the finite-temperature versions of the loop algorithm.}
\label{vbloops}
\end{figure}

The idea to use amplitude product states as a starting point for {\it projector QMC} simulations is quite old \cite{liang90b,santoro99}, but was only recently 
developed into a generic and efficient tool \cite{vbmethod1,vbmethod2,awshg}. The general idea of a projector scheme is the same as we discussed in the context 
of Krylov space methods in Sec.~\ref{lanczos}: By acting with a high power $H^\Lambda$ of the hamiltonian on an arbitrary state $|\Psi\rangle$, only the 
component of that state with the largest energy eigenvalue---normally the ground state (and if not this can be arranged by subtracting a constant from 
$H$)---survives in the limit $\Lambda \to \infty$, as demonstrated by Eq.~(\ref{hmexpansion}). Expectation values of the form 
\begin{equation}
\langle A\rangle = \frac{\langle \Psi|H^\Lambda A H^\Lambda |\Psi\rangle}{\langle \Psi|H^{2\Lambda} |\Psi\rangle},
\end{equation}
where $|\Psi\rangle$ is a valence bond state or superposition (e.g., an amplitude product state), can be sampled using Monte Carlo simulations. In
the original formulations of this approach \cite{liang90,vbmethod1}, $H$ is written as a sum of singlet projector operators (as in the Heisenberg 
antiferromagnet), or products of singlet operators (as in J-Q models), and strings of such operators are sampled (along with the valence bond
configurations of $|\Psi\rangle$). The singlet projectors $S_{ij}$, defined in Eq.~(\ref{singproj}), lead to simple reconfigurations of bond pairs 
when acting on valence bond states. When acting on a valence bond, this operator is diagonal with eigenvalue $1$,
\begin{equation}
S_{ab}(a,b)=(a,b),
\label{cijdia}
\end{equation} 
while acting on a pair of different valence bonds leads to a simple reconfiguration of those bonds, with matrix element $1/2$;
\begin{equation}
S_{bc}(a,b)(c,d)=\half (c,b)(a,d),
\label{cijop}
\end{equation}
This can be shown easily by going back to the basis of $\up$ and $\dn$ spins. Note the order of the indices within the singlets, which reflects 
consistently the convention corresponding to Marshall's sign rule discussed above; $a,c$ are on sublattice $A$ and $b,d$ on sublattice $B$.
These rules allow for a well-defined path integral (or SSE) like propagation of the states, and these propagation paths can be sampled using
a Monte Carlo scheme.

In a more recent formulation of the valence bond projector method \cite{awshg}, the $\up$ and $\dn$ spin basis is reintroduced, by expressing all singlets 
in terms of their sums over antiparallel spin pairs. This leads to an algorithm very similar to SSE and world line loop methods. Essentially, the periodic
time boundaries used when simulating a system at a fixed temperature are cut open and replaced by two separate boundaries at which the valence bond states 
live. This is illustrated and discussed in more detail with a simple configuration for a four-site Heisenberg chain in Fig.~\ref{vbloops}.

The valence bond basis (or its translation to loop methods \cite{nachtergale,awshg})  has some unique aspects which makes it possible to access observables that are 
normally out of reach or difficult to calculate. In an extended valence bond basis for the triplet sector, one can study some properties of excited states 
\cite{vbmethod1,vbmethod2,wangperc}. One can also generalize the valence bond basis to include one \cite{banerjee10b} or several \cite{wangperc} 
unpaired spins, which is useful for studies of, e.g., the magnetization distribution in systems with unequal sublattice occupation of the spins. 
Simulations in the valence bond basis have also recently found applications in studies of entanglement entropy 
\cite{hastings1,alet07,chhajlany07,tran09,kallin09}. One can also extend valence bond projector methods to SU(N) spins \cite{lou1} (including
even non-integer $N$ generalizations \cite{beach09}) and other related symmetry groups \cite{tran09}.

\begin{theacknowledgments}

These lecture notes build largely on material to be published as a book, {\it Quantum Spin Systems---A Computational Perspective}, by Cambridge 
University Press. The opportunity to lecture at the XIV Training Course in the Physics of Strongly Correlated Systems (Vietri sul Mare, Salerno,
Italy, in October 2009) prompted me to prepare a condensed version of this as of yet unfinished work (which will discuss a larger set of models and 
computational methods). These lecture notes cover the topics I discussed at Vietri sul Mare, but I have added more background and details. Some 
of the applications have been complemented with recent results. As in my lectures, the target audience for these notes is graduate students and 
postdocs who are entering the field of quantum magnetism, or some related area in where quantum spin models are of interest, and need a source 
for elementary concepts and calculations. I believe that this material could also form the core of a one-semester advanced topics course in 
quantum magnetism (focused on concepts and numerical calculations).

I would like to thank the organizers of the Salerno Training Course, Professors Ferdinando Mancini and Adolfo Avella, for the invitation to lecture and 
publish these notes. I also thank the students at Vietri sul Mare for their enthusiasm. I am grateful to Ying Tang for her careful reading of
the manuscript. I thank Cambridge University Press and the American Institute of Physics for agreeing on the copyright issues related to dual 
publication of most of the material presented here. Excerpts are printed with the permission of Cambridge University Press. Some of this work 
was supported  by the NSF under Grant No.~DMR-0803510.

\end{theacknowledgments}

\end{document}